\pdfoutput=1

\newcommand{\thesistitlenormal}{Mitigating The Memory Bottleneck with Machine Learning-Driven and Data-Aware Microarchitectural Techniques}
\newcommand{\thesisTitleFrontmatter}{MITIGATING THE MEMORY BOTTLENECK\\[\titleFrontBaselineskip-1ex]WITH MACHINE LEARNING-DRIVEN AND DATA-AWARE\\[\titleFrontBaselineskip-1.245ex]MICROARCHITECTURAL TECHNIQUES}

\newcommand{\thesisTitlePlain}{\thesistitlenormal}
\newcommand{\thesisDissNumber}{31461}
\newcommand{\thesisAuthor}{Rahul Bera}
\newcommand{\thesisUni}{\protect{ETH Z\"urich}}

\newcommand{\thesisYear}{2025}
\newcommand{\thesisVersion}[0]{\theversion.0}

\documentclass[12pt,twoside,a4paper,openright]{ethzthesis}

\newif\ifcameraready
\camerareadytrue

\newif\ifhidechange
\hidechangefalse

\newcommand{\thesisversionnum}[0]{7}

\usepackage{dblfloatfix}
\usepackage{multirow}
\usepackage{booktabs}
\usepackage{array}
\usepackage[dvipsnames]{xcolor}
\usepackage{algorithm}
\usepackage[noend]{algpseudocode}
\usepackage[rightComments=true,commentColor=gray]{algpseudocodex}
\usepackage{eucal}
\usepackage{datetime}
\usepackage{enumitem} %
\usepackage{tikz}
\usepackage{pifont}
\usepackage{graphicx}
\usepackage{fancyhdr}
\usepackage{emptypage} %
\usepackage{titlesec}
\usepackage{datetime2}
\usepackage{fontawesome5} %
\usepackage{hyperref}
\usepackage{cleveref}

\hypersetup{
    bookmarks=true,
    breaklinks=true,
    colorlinks=true,
    linkcolor=blue,
    citecolor=blue,
    urlcolor=urlblue,
    pdfpagelabels=false,
}

\usepackage{wrapfig}
\usepackage{float}
\usepackage[newfloat]{minted}

\usepackage{flushend}
\usepackage{xspace}

\usepackage{tabularx}
\usepackage{caption}
\usepackage{rotating}
\usepackage{nicefrac}
\usepackage{amsmath} %

\usepackage{xhfill}

\DeclareMathOperator*{\argmax}{argmax}

\newcolumntype{L}[1]{>{\raggedright\let\newline\\\arraybackslash\hspace{0pt}}m{#1}}
\newcolumntype{C}[1]{>{\centering\let\newline\\\arraybackslash\hspace{0pt}}m{#1}}
\newcolumntype{R}[1]{>{\raggedleft\let\newline\\\arraybackslash\hspace{0pt}}m{#1}}

\makeatletter
\def\thickhline{%
  \noalign{\ifnum0=`}\fi\hrule \@height \thickarrayrulewidth \futurelet
   \reserved@a\@xthickhline}
\def\@xthickhline{\ifx\reserved@a\thickhline
               \vskip\doublerulesep
               \vskip-\thickarrayrulewidth
             \fi
      \ifnum0=`{\fi}}
\makeatother
\newlength{\thickarrayrulewidth}
\newcommand*\circled[1]{\tikz[baseline=(char.base)]{
            \node[shape=circle,fill,inner sep=1pt] (char) {\textcolor{white}{#1}};}}

\definecolor{freakishgreen}{HTML}{0A982B}
\definecolor{urlblue}{HTML}{319dd6}
\hypersetup{
    colorlinks=true,
    linkcolor=black,
    filecolor=magenta,
    citecolor=violet,
    urlcolor=urlblue,
    pdfpagemode=FullScreen,
}

\crefname{section}{§\hspace{-2pt}}{§§}
\Crefname{section}{§}{§§}

\newcommand{\xmark}{\ding{54}}

\newcommand\Tstrut{\rule{0pt}{2ex}}             %
\newcommand\Bstrut{\rule[-1ex]{0pt}{0pt}}       %
\newcommand{\Tabval}[1]{{\Tstrut #1 \Bstrut}}   %

\newcommand{\paraheading}[1]{\vspace{1em}\noindent \textbf{#1}}

\definecolor{chapteraccent}{HTML}{0B5AA2}
\newcommand{\chapterlabelword}{\chaptername}
\makeatletter
\g@addto@macro\appendix{\gdef\chapterlabelword{\appendixname}}
\makeatother
\newcommand{\chapterBanner}[1]{%
  \tikz[baseline=(num.base)]{
    \coordinate (base) at (0,0);
    \fill[chapteraccent!90!black] ([xshift=-1.5cm,yshift=1.6cm]base) rectangle ++(0.9cm,-2.6cm);
    \node[anchor=base west,font=\sffamily\bfseries\LARGE\scshape] (label) at (base) {\chapterlabelword};
    \node[anchor=base west,font=\rmfamily\bfseries\fontsize{32}{32}\selectfont] (num) at ([xshift=0.1em]label.base east) {#1};
  }%
}
\titleformat{\chapter}[display]
  {\normalfont}
  {\chapterBanner{\thechapter}}
  {0pt}
  {\vspace{5ex}\raggedright\rmfamily\bfseries\fontsize{26}{30}\selectfont}
  [\vspace{0.5ex}]
\titleformat{name=\chapter,numberless}[display]
  {\normalfont}
  {}
  {0pt}
  {\vspace{5ex}\raggedright\rmfamily\bfseries\fontsize{26}{30}\selectfont}
  [\vspace{0.5ex}]
\titlespacing*{\chapter}{0pt}{-40pt}{32pt}

\newcommand{\chapterRB}[2]{%
  \chapter[#2]{#2}%
  \chaptermark{#1}%
}

\newcommand{\sectionRB}[3]{%
  {%
    \let\oldsectionmark\sectionmark
    \def\sectionmark##1{\markright{\MakeUppercase{\thesection:\ #1}}}%
    \section{#2}\label{#3}
    \let\sectionmark\oldsectionmark
  }%
}

\newcommand\kon[1]{}

\newcommand{\rbc}[1]{{#1}}
\newcommand{\rbd}[1]{{#1}}
\newcommand{\rbe}[1]{{#1}}
\newcommand{\rbf}[1]{{#1}}
\newcommand{\rbg}[1]{{#1}}
\newcommand{\rbh}[1]{{#1}}
\newcommand{\rbi}[1]{{#1}}
\newcommand{\rbcd}[1]{{#1}}
\newcommand{\rbcb}[1]{{#1}}
\newcommand{\rbcc}[1]{{#1}}

\newcommand{\thesisversiontext}[0]{\textcolor{blue}{Thesis v\thesisversionnum~---~\today, \xxivtime \ UTC}}

\newcommand{\rbone}[1]{{#1}}
\newcommand{\rbtwo}[1]{{#1}}
\newcommand{\rbthr}[1]{{#1}}
\newcommand{\rbfor}[1]{{#1}}
\newcommand{\rbfiv}[1]{\textcolor{BurntOrange}{#1}}

\renewcommand{\chaptermark}[1]{%
  \markboth{\MakeUppercase{\chapterlabelword\ \thechapter:\ #1}}{\MakeUppercase{#1}}%
}

\ifcameraready
    \renewcommand{\rbone}[1]{{#1}}
    \renewcommand{\rbtwo}[1]{{#1}}
    \renewcommand{\rbthr}[1]{{#1}}
    \renewcommand{\rbfor}[1]{{#1}}
    \renewcommand{\rbfiv}[1]{{#1}}

    \fancypagestyle{headings}{%
        \fancyhf{}%
        \fancyhead[LE]{\small\sffamily\thepage}%
        \fancyhead[RO]{\small\sffamily\thepage}%
        \fancyhead[LO]{\small\sffamily\itshape\rightmark}%
        \fancyhead[RE]{\small\sffamily\itshape\leftmark}%
        \fancyfoot{}%
    }
    \fancypagestyle{plain}{%
        \fancyhf{}%
        \fancyfoot[C]{\thepage}%
    }
    
\else
    \ifhidechange
        \renewcommand{\rbone}[1]{{#1}}
        \renewcommand{\rbtwo}[1]{{#1}}
        \renewcommand{\rbthr}[1]{{#1}}
        \renewcommand{\rbfor}[1]{{#1}}
        \renewcommand{\rbfiv}[1]{{#1}}
    \fi

    \fancypagestyle{plain}
    {
    	\fancyhead[C]{\thesisversiontext}
    	\fancyfoot[C]{\thepage}
    }
    \fancypagestyle{headings}
    {
    	\fancyhead[C]{\thesisversiontext}
    	\fancyfoot[C]{\thepage}
    }
\fi

\newcommand{\pred}[0]{POPET\xspace}

\definecolor{colorV1}{rgb}{0.0, 0.0, 1.0}    %
\definecolor{colorV2}{rgb}{1.0, 0.0, 1.0}    %
\definecolor{colorV3}{rgb}{0.0, 0.5, 0.0}    %
\definecolor{colorV4}{rgb}{0.6, 0.4, 0.2}    %

\newcounter{CurrentDraftVersion}
\ExplSyntaxOn

\cs_new:Npn \revcol:n #1
  {
    \int_case:nnF {#1}
      {
        {1}{blue}
        {2}{magenta}
        {3}{orange}
        {4}{cyan}
        {5}{violet}
        {6}{teal}
        {7}{purple}
        {8}{brown}
        {9}{olive}
        {10}{red}
      }
      {blue}
  }

\NewDocumentCommand{\edit}{ O{\value{CurrentDraftVersion}} m }
  {
    \int_compare:nNnTF { \value{CurrentDraftVersion} } = {1000}
      { #2 }
      {
        \int_compare:nNnTF { \value{CurrentDraftVersion} } = {100}
          { \textcolor{blue}{#2} }
          {
            \int_compare:nNnTF { #1 } = { \int_eval:n { \value{CurrentDraftVersion} - 1 } }
              { \textcolor{\revcol:n {#1}}{#2} }
              { #2 }
          }
      }
  }

\ExplSyntaxOff

\newcommand*\bluecircled[1]{\tikz[baseline=(char.base)]{
            \node[shape=circle,fill,color=blue,inner sep=1pt] (char) {\textcolor{white}{#1}};}}
\newcommand*\redcircled[1]{\tikz[baseline=(char.base)]{
            \node[shape=circle,fill,color=red,inner sep=1pt] (char) {\textcolor{white}{#1}};}}

\def\thickhline{\noalign{\hrule height.8pt}}
\newcolumntype{?}{!{\vrule width 0.8pt}}

\crefformat{section}{\S#2#1#3} %
\crefformat{subsection}{\S#2#1#3}
\crefformat{subsubsection}{\S#2#1#3}

\newcommand\pnm{\cite{farmahini2015nda,babarinsa2015jafar,devaux2019true,ghiasi2022genstore,gomez2021benchmarkingcut,gomezluna2021benchmarking,gomez2022benchmarking,syncron,singh2020nero,skhynixpim,ke2021near,giannoula2022sparsep,shin2018mcdram,cho2020mcdram,denzler2021casper,asghari2016chameleon,IRAM_Micro_1997,C_RAM_1999,CASES_MVX,Xi_2015,sun2021abc,matam2019graphssd,gokhale1995processing,hall1999mapping,MEMSYS_MVX,lockerman2020livia,ahn2016scalable,Nai2017GraphPIMEI,boroumand2018google,lazypim, top-pim, gao2016hrl, kim2018grim,kim2017grim,drumond2017mondrian, RVU, NIM, PEI, gao2017tetris,Kim2016,gu2016leveraging, boroumand2019conda, hsieh2016transparent, cali2020genasm, NDC_ISPASS_2014,pattnaik2016scheduling,akin2015data,hsieh2016accelerating,lee2015bssync,boroumand2021mitigating,boroumand2021google,boroumand2022polynesia,boroumand2021polynesia,amiraliphd,besta2021sisa,fernandez2020natsa,singh2019napel,kwon202125,lee2021hardware,niu2022184qps,Sparse_MM_LiM,azarkhish2016logic,azarkhish2018neurostream,guo20143d,de2018design,akin2014hamlet,huang2020heterogeneous,dai2018graphh,liu2018processing,tsai:micro:2018:ams,gu2020ipim,DRAMA_CAL_2014,Asghari-Moghaddam_2016,huang2019active,kersey2017lightweight,li2019pims,kim2017grim,boroumand2017lazypim,zhuo2019graphq,zhang2018graphp,lim2017triple,smc_sim,HIVE,jang2019charon,IBM_ActiveCube,hadidi2017cairo,santos2018processing,ghose.ibmjrd19,eden,gomez2022benchmarking,gomezluna2021benchmarking,gomez2021benchmarkingcut,oliveira2021damov,gomez2023evaluating,diab2023framework,ghiasi2024megis,ghiasi2023metastore,chen2025reis,item2023transpimlib,kautz1969cellular,mutlu2019processing,oliveira2025proteus,oliveira2025dappadataparallelprogrammingframework,stone1970logic,execube,kogge1995combined,elliott1992computational,oliveira2025thesis,rosenfeld2014performance,fracdram,nika2023alp,hashemi2016continuous,wang2020figaro}\xspace}

\newcommand\pum{\cite{Chi2016, Shafiee2016, Seshadri2017AmbitMemoryAccelerator, seshadri2019indram, li2017drisa, seshadri2018rowclone, seshadri2016processing, deng2018dracc, xin2020elp2im, song2018graphr, song2017pipelayer,gao2019computedram, eckert2018neural, aga2017compute,dualitycache,besta2021sisa,Seshadri2016BuddyRAMPerformance,seshadri.bookchapter17,seshadri2018rowclone,Seshadri2015BulkBitwiseDRAM,li2016pinatubo,ferreira2021pluto,ferreira2022pluto,imani2019floatpim,he2020sparse,flashcosmos,truong2022adapting,truong2021racer,olgun2021quactrng,kim2019d,kim2018dram,bostanci2022dr,olgun2022pidram,ali2019memory,angizi2019graphide,li2018scope,subramaniyan2017parallel,zha2020hyper,fujiki2018memory,orosa2021codic,sharad2013ultra,rezaei2020nom,mutlu2024memory,hajinazarsimdram,oliveira2024mimdram}\xspace}

\makeatletter
\newcommand\requiredelimiter[2][########]{%
  \ifdefined#2%
    \def\@temp{\def#2#1}%
    \expandafter\@temp\expandafter{#2}%
  \else
    \@latex@error{\noexpand#2undefined}\@ehc
  \fi
}
\@onlypreamble\requiredelimiter
\makeatother

\begin{document}
\frenchspacing
\raggedbottom
\selectlanguage{english}
\pagenumbering{roman}
\pagestyle{plain}

\bstctlcite{IEEEexample:BSTcontrol}
\setbiblabelwidth{1000} %

\bstctlcite{IEEEexample:BSTcontrol}

\providecommand{\titleFrontFontSize}{16.5pt} %
\providecommand{\titleFrontBaselineskip}{16pt}

\begin{titlepage}
    \large
    \begin{center}
        
        \ifcameraready
        \else
            \thesisversiontext \\
        \fi
        
        \begingroup
        \MakeUppercase{Diss. ETH No. -}
        \thesisDissNumber{}
        \endgroup
    
        \hfill

        \vfill

        \begingroup
            {\fontsize{\titleFrontFontSize}{\titleFrontBaselineskip}\selectfont\textbf{\thesisTitleFrontmatter}}
        \endgroup

        \vfill

        \begingroup
            A thesis submitted to attain the degree of\\
            \vspace{0.5em}
            \MakeUppercase{Doctor of Sciences}\\
            \vspace{0.5em}
            (Dr. sc. \thesisUni) \\
            
        \endgroup

        \vfill

        \begingroup
            presented by\\
            \vspace{1.5em}
            RAHUL BERA\\
            \vspace{0.5em}
            born on 12.09.1992
        \endgroup

        \vfill

        \begingroup
            \vspace{2em}
            accepted on the recommendation of\\
            \vspace{1em}
            Prof.\ Dr.\ Onur Mutlu, examiner\\
            \vspace{0.5em}
            Dr. Aamer Jaleel, co-examiner \\
            \vspace{0.5em}
            Prof. Dr. Boris Grot, co-examiner \\
            \vspace{0.5em}
            Chris Wilkerson, co-examiner \\
            \vspace{0.5em}
            Prof. Dr. Daniel Jim\'enez, co-examiner \\
            \vspace{0.5em}
            Dr. Pradip Bose, co-examiner \\
            
        \endgroup

        \vfill

        \thesisYear%

        \vfill
    \end{center}
\end{titlepage}

\thispagestyle{empty}

\hfill

\vfill

\noindent\thesisAuthor: \textit{\thesisTitlePlain,}
\textcopyright\ \thesisYear

\thispagestyle{empty}

\vspace*{\stretch{1}}
\begingroup
\centering
\textit{In the memory of my beloved baba, \\[2pt]
Mr. Rebati Raman Bera \raisebox{0.5ex}{\scalebox{0.7}{\faDove}} \\ 
(1960 - 2020)}
\par
\endgroup
\vspace*{\stretch{1}}

\setstretch{1.1}

\chapter*{Acknowledgments}
\addcontentsline{toc}{chapter}{Acknowledgments}

This dissertation is the outcome of more than half a decade's worth of effort, which would not be possible without the help, advice, motivation, and encouragement from numerous people in my life. While it may not be possible to acknowledge all of them in this short section, I am truly thankful to everyone who shaped my PhD journey, knowingly or unknowingly.

First and foremost, I am thankful to my advisor, Prof. Onur Mutlu, for believing in me and providing the opportunity, resources, and continuous enthusiasm during the entire course of the PhD.
Without his invaluable guidance and deep expertise, this dissertation would not have been possible.
The scientific rigor he instilled in me during this journey has not only shaped me as a researcher, but also as a problem solver in real life.

I would extend my sincere gratitude to all other mentors in my life who had contributed profoundly to making me what I am today. 
My interest in computer architecture started from the lectures of Prof. Atal Chaudhuri at Jadavpur University. 
Then Prof. Mainak Chaudhuri at IIT Kanpur introduced me to the cutting-edge research in computer architecture during my master's studies.
During this period, I also had the opportunity to work on challenging problems in industry under the mentorship of Dr. Kanishka Lahiri at AMD.
Lastly, I will be forever grateful to my Intel mentors---Anant V. Nori, Dr. Shankar Balachandran, and Sreenivas Subramoney---for their invaluable guidance and mentorship before, during, and after the PhD.
In particular, I am especially thankful to Anant who not only helped me grow professionally as my mentor, but also in my personal life as an elder brother.

I also thank my doctoral exam committee members---Dr. Aamer Jaleel, Prof. Boris Grot, Chris Wilkerson, Prof. Daniel Jimenez, and Dr. Pradip Bose---for taking time out of their busy schedules to provide thoughtful feedback and constructive criticism that immensely helped strengthen this dissertation. I thank Prof. Kaveh Razavi for chairing my doctoral exam.

My sincere gratitude goes to all the funding partners: ETH Future Computing Lab (EFCL), Futurewei, Google, Huawei, Intel, Microsoft, Semiconductor Research Corporation (SRC), and VMware. Their support was instrumental in making this dissertation possible.

The role of the SAFARI Research Group has been quintessential during my PhD journey. 
I thank all current and past SAFARI members for providing such a unique, stimulating, yet friendly environment for doing research. 
I am especially indebted to my PhD buddies---Konstantinos Kanellopoulos and Nika Mansouri Ghiasi---for always being there to share the many joys, sorrows, and frustrations along this journey.
Their warm friendship, solidarity, and support helped me to function as a human during this long marathon, which otherwise would have surely driven me to despair.
I feel deeply fortunate that the three of us embarked on this journey nearly at the same time and now stand at its conclusion nearly together. 
When I look back, I could not be more proud of what we have achieved together and the person we have grown into.
I also extend my sincere gratitude to my mentees, especially Konstantinos Sgouras, Liana Koleva, and Zhenrong Lang, for teaching me how to become a good mentor.
I am immensely proud of the work we did together and I cannot wait to see their future endeavors.
I also sincerely thank Andreas Kosmas Kakolyris, Ataberk Olgun, Christina Giannoula, Gagandeep Singh, Geraldo Francisco de Oliveira Junior, A. Giray Yağlıkçı, Harsh Songara, Harshita Gupta, Ismail E. Yuksel, Konstantina Koliogeorgi, Mayank Kabra, Mohammad Sadrosadati, Nisa Bostanci, and Rakesh Nadig,  for their invaluable friendship and presence that made SAFARI a family away from home.
I thank Tracy Ewen and Tulasi Blake for their administrative support along the PhD journey.

Outside SAFARI, I thank all the friends I have met and made during my PhD who eventually became the family outside work: Ahalya, Akash, Akriti, Ankita, Arun, Ashabari, Blake, Dimos, Maria, Martina, Neethu, Saurav, Shimony, Siddharth, and Vaisakh. 
It is because of their friendship that Zürich felt like home.

Finally, I owe everything to my family. 
I express my profound gratitude to my mother, Sukla Bera, whose lifelong sacrifices created the foundation upon which I built my path. She consistently placed my growth and aspirations above her own comfort and interests. Without her enduring love, unwavering support, and boundless affection, I would not have become the person I am today.
I express my deepest gratitude to my wife, Moumita Dey, for her unwavering love, steadfast support, and remarkable patience. The doctoral journey did not always proceed along a smooth or predictable path. Her constant presence and encouragement provided the mental resilience required to confront uncertainty with confidence and perseverance. This dissertation stands as much a testament to her strength and sacrifice as it does to my own efforts.

And lastly, I would like to thank my late father, Rebati Raman Bera, who is the reason I embarked on this journey in the first place. I still fondly recall the day 
I was preparing to depart for Zürich, uncertain whether it was right to move far away from my parents at a stage in life when they might need me by their side.
My father assured me that I should stop worrying about them and chase my dream.
I lost him within a year of my PhD and since then, everyday I contemplate the decision I took.
Wherever he may be, I hope he would be proud to see me here today.
This dissertation is dedicated to my father for his unwavering support, encouragement, and love.

\vspace{3em}
\noindent Rahul Bera\\
\noindent March 8, 2026

\clearpage
\chapter*{Abstract}
\addcontentsline{toc}{chapter}{Abstract}

Modern applications operate on massive amounts of data, easily overwhelming the storage and retrieval capabilities of modern memory systems. 
As a result, memory is (and will likely continue to be) the key performance and \rbtwo{energy} efficiency bottleneck of computing systems. 
To alleviate this bottleneck, architects have proposed and employed numerous microarchitectural techniques in general-purpose processors that aim to hide or tolerate the long latency of memory accesses. 
These techniques have consistently pushed the boundaries of performance and energy efficiency of state-of-the-art general-purpose processors. 
Yet, as the growth in data footprints continues to far outpace the scaling of process technology nodes, we are in constant need of better microarchitectural techniques that can extract ever-more performance and \rbtwo{energy} efficiency scaling.

This dissertation shows that even though modern microarchitectures observe a large amount of \rbtwo{application data (e.g., an application's control-flow information such as program counter value and branch outcome, or data-flow information such as memory address and data value) and system-generated data (e.g., memory bandwidth usage, cache pollution)} during their course of operation, the decisions they make online are \rbtwo{often} agnostic to the data they are observing. 
Using four case studies of state-of-the-art microarchitectural techniques employed at various parts of modern processors, we quantitatively show that this data-agnosticism of the microarchitectural techniques cost them a severe opportunity loss of performance and \rbtwo{energy} efficiency improvements.

To alleviate this, this dissertation advocates to fundamentally shift microarchitecture design from being data-agnostic to data-informed.
We posit that by enabling microarchitectural techniques (1) to adapt their policies by continuously learning from the data (i.e., making them \emph{data-driven}), and (2) to tailor their decisions exploiting characteristics/semantics of the application data (i.e., making them \emph{data-aware}), we can significantly improve performance and \rbtwo{energy} efficiency of state-of-the-art processors that were otherwise untapped by the conventional data-agnostic microarchitectural techniques.

To substantiate this hypothesis, we exploit various forms of lightweight and practical machine learning (ML) techniques, as well as previously-underexplored application data characteristics, and drive the architectural decision making in four key components of a state-of-the-art processor. 
First, we exploit online reinforcement learning (RL) to enable a hardware data prefetcher to autonomously learn patterns in an application's memory access stream and prefetch adaptively by taking various system-level feedback into account.
Second, we exploit perceptron learning to accurately identify memory requests that would likely go off-chip even after deploying state-of-the-art prefetching mechanisms and reduce their memory latency. 
Third, we study the interference between data prefetching and off-chip prediction and propose a lightweight RL-based mechanism to synergize the two speculative techniques by autonomously learning from their behavior online.
Fourth, we exploit the consistent repeatability in the memory address and the data value loaded by certain memory instructions present in modern applications to propose an efficient instruction execution mechanism that safely eliminates execution of certain memory instructions.
In all four cases, we show that our proposed ML-driven and data-aware microarchitectural techniques significantly improve performance and/or energy efficiency compared to the best prior conventional techniques.
\rbthr{
We further show that our proposed ML-driven techniques are capable of generalizing their performance benefits beyond the workloads considered at design-time by evaluating them on a large corpus of previously-unseen workload traces from emerging AI, graph mining, and real-world datacenter workloads collected as part of the 4th Data Prefetching Championship (DPC4).
}

Overall, this dissertation builds a comprehensive understanding of how the data-agnostic nature of state-of-the-art microarchitectural techniques limits their effectiveness in mitigating the memory bottleneck and proposes novel data-driven and data-aware microarchitectural techniques. We hope and believe that the insights and techniques presented in this dissertation will encourage the next-generation of data-driven and data-aware microarchitectures to mitigate the ever-growing memory bottleneck problem. Such architectures will not only improve the performance and \rbtwo{energy} efficiency of computing systems, but will also greatly reduce the system architect's burden in designing sophisticated policies.

\chapter*{Zusammenfassung}
\addcontentsline{toc}{chapter}{Zusammenfassung}

Moderne Anwendungen verarbeiten enorme Datenmengen, die die Speicher- und Abrufkapazitäten heutiger Speichersysteme leicht überfordern. Der Speicher stellt daher gegenwärtig – und voraussichtlich auch künftig – den zentralen Leistungs- und Energieeffizienzengpass moderner Rechnersysteme dar. Um diesen Engpass zu entschärfen, wurden zahlreiche mikroarchitektonische Techniken für allgemeine Prozessoren entwickelt, die darauf abzielen, die hohe Latenz von Speicherzugriffen zu verbergen oder zu tolerieren. Diese Techniken haben die Leistungsfähigkeit und Energieeffizienz moderner Prozessoren erheblich verbessert. Da jedoch das Wachstum der Datenmengen die Fortschritte der Halbleiterskalierung weiterhin deutlich übertrifft, besteht weiterhin Bedarf an neuen mikroarchitektonischen Ansätzen, die zusätzliche Leistungs- und Energieeffizienzgewinne ermöglichen.

Diese Dissertation zeigt, dass moderne Mikroarchitekturen während ihres Betriebs zwar große Mengen an Anwendungsdaten – etwa Kontrollflussinformationen wie Programmzählerwerte und Verzweigungsergebnisse sowie Datenflussinformationen wie Speicheradressen und Datenwerte – und systemgenerierten Daten, beispielsweise zur Speicherbandbreitennutzung oder Cache-Verschmutzung, beobachten. Dennoch treffen viele mikroarchitektonische Mechanismen ihre Entscheidungen weitgehend unabhängig von diesen Daten. Anhand von vier Fallstudien zu modernen mikroarchitektonischen Techniken in verschiedenen Komponenten eines Prozessors zeigen wir quantitativ, dass diese Datenunabhängigkeit zu erheblichen ungenutzten Potenzialen bei Leistungs- und Energieeffizienzverbesserungen führt.

Diese Dissertation plädiert daher für einen grundlegenden Paradigmenwechsel im Mikroarchitekturdesign: weg von datenunabhängigen hin zu dateninformierten Ansätzen. Leistungsfähigkeit und Energieeffizienz moderner Prozessoren lassen sich deutlich steigern, wenn mikroarchitektonische Techniken (1) ihre Strategien kontinuierlich durch Lernen aus beobachteten Daten anpassen (datengetrieben) und (2) ihre Entscheidungen unter Ausnutzung der Eigenschaften beziehungsweise Semantik der Anwendungsdaten treffen (datenbewusst). Dadurch können Leistungs- und Energieeffizienzpotenziale erschlossen werden, die von konventionellen datenunabhängigen Techniken bislang ungenutzt bleiben.

Zur Überprüfung dieser Hypothese nutzt diese Dissertation leichtgewichtige und praxisnahe Verfahren des maschinellen Lernens (ML) sowie bislang wenig untersuchte Eigenschaften von Anwendungsdaten, um architektonische Entscheidungen in vier zentralen Komponenten moderner Prozessoren zu steuern. Erstens verwenden wir Online-Reinforcement-Learning (RL), um einem Hardware-Prefetcher zu ermöglichen, Muster im Speicherzugriffsverhalten einer Anwendung autonom zu erlernen und sein Prefetching adaptiv unter Berücksichtigung verschiedener Systemrückmeldungen anzupassen. Zweitens nutzen wir Perzeptron-Lernen, um Speicheranforderungen zu identifizieren, die selbst bei Einsatz moderner Prefetching-Mechanismen mit hoher Wahrscheinlichkeit den Hauptspeicher erreichen würden, und reduzieren dadurch ihre effektive Speicherlatenz. Drittens untersuchen wir die Interferenz zwischen Datenprefetching und Off-Chip-Vorhersage und schlagen einen leichtgewichtigen RL-basierten Mechanismus vor, der beide spekulativen Techniken durch Online-Lernen ihres Verhaltens koordiniert. Viertens nutzen wir die konsistente Wiederholbarkeit sowohl der Speicheradresse als auch des geladenen Datenwerts bestimmter Speicherinstruktionen moderner Anwendungen und entwickeln einen effizienten Ausführungsmechanismus, der die Ausführung solcher Instruktionen sicher vermeiden kann. In allen vier Fällen zeigen wir, dass die vorgeschlagenen ML-basierten und datenbewussten mikroarchitektonischen Techniken Leistung und/oder Energieeffizienz gegenüber den besten bisherigen konventionellen Ansätzen deutlich verbessern.

Darüber hinaus zeigen wir, dass die vorgeschlagenen ML-basierten Techniken ihre Leistungsgewinne auch über die während der Entwurfsphase betrachteten Workloads hinaus verallgemeinern können. Hierzu evaluieren wir sie anhand eines umfangreichen Korpus zuvor ungesehener Workload-Traces aus aufkommenden KI-, Graph-Mining- und realen Rechenzentrums-Workloads, die im Rahmen der 4. Data Prefetching Championship (DPC4) gesammelt wurden.

Insgesamt vermittelt diese Dissertation ein umfassendes Verständnis dafür, wie die datenunabhängige Natur moderner mikroarchitektonischer Techniken ihre Wirksamkeit bei der Abschwächung des Speicherengpasses begrenzt, und stellt neuartige datengetriebene sowie datenbewusste mikroarchitektonische Ansätze vor. Die in dieser Arbeit gewonnenen Erkenntnisse können die Entwicklung einer nächsten Generation datengetriebener und datenbewusster Mikroarchitekturen fördern, die dem stetig wachsenden Speicherengpass wirksam begegnen und gleichzeitig die Leistungsfähigkeit sowie Energieeffizienz moderner Rechnersysteme verbessern.

\pagestyle{headings}
\cleardoublepage
\tableofcontents
\newpage
\listoffigures
\newpage
\listoftables

\cleardoublepage
\pagenumbering{arabic}%

\chapter{Introduction}
\label{chap:intro}

Modern applications generate and process massive amounts of data. 
Applications spanning key domains such as high-performance computing, large-scale graph analytics, machine learning, and genome analysis, exhibit massive data footprints that easily overwhelms the storage and retrieval capabilities of state-of-the-art memory systems~\cite{warehouse,cloudsuite,wang2014bigdatabench,boroumand2018google,boroumand2021google,alser2020accelerating,cali2020genasm}.
Multiple prior works have shown that such applications, irrespective of whether they execute on cloud servers or mobile systems, spend a significant fraction of their total execution time simply waiting for the data to be retrieved from the memory system~\cite{mutlu2003runahead,mutlu2003runahead2,cloudsuite,warehouse,jain2023optimizing,asmdb,Nai2017GraphPIMEI,oliveira2021damov,schall2022lukewarm,ayers2018memory,schall2023warming,hsia2020cross,Subramaniyan2021GenomicsBench,Luo2024HarvestingMC,Addisie2018Heterogeneous,Gao2018Data,Liang2023Ditto,Gao2018BigDataBenchAS,Samudrala2024Performance,Awan2015Performance,Sirin2020Micro,dcperf,Jia2017Understanding,Makrani2018comprehensive,gem5_profile,sen2023microarchitectural,Bavarsad2021HosNa,Awan2016Micro,Mahar2023WorkloadBD,Chilukuri2023Analyzing,Asheim2022Impact,Awan2017Identifying,Corda2020Near,Sirin2017methodology}.
As such, memory constitutes (and will likely continue to be) the key performance and energy efficiency bottleneck of modern computing systems~\cite{mutlu2013memory,mutlu2025memory,mutlu2021intelligent,mutlu2022modern}.

To mitigate this memory bottleneck, \rbone{computer} architects have proposed and employed numerous microarchitectural techniques that aim to \emph{hide} or \emph{tolerate} long memory access latency~\cite{teresa_cache_bypassing,teresa_cache_bypassing2,tyson1995modified,tyson1997managing,adaptive_cache_hierarchy,jg,ship,hawkeye,jimenez2010dead,jimenez2017multiperspective,sdbp,rivers1996reducing,rivers1998utilizing,hu2002timekeeping,kharbutli2008counter,liu2008cache,piquet2007exploiting,chi1989improving,dybdahl2006enhancing,wong2000modified,markov,stems,somogyi_stems,wenisch2010making,domino,isb,misb,triage,wenisch2005temporal,chilimbi2002dynamic,chou2007low,ferdman2007last,hu2003tcp,bekerman1999correlated,cooksey2002stateless,karlsson2000prefetching,stride,streamer,baer2,jouppi_prefetch,ampm,fdp,footprint,sms,spp,vldp,sandbox,bop,dol,dspatch,bingo,mlop,ppf,ipcp,dundas,mutlu2003runahead,mutlu2003runahead2,mutlu2006efficient,mutlu2005address,mutlu2005techniques,precise_runahead,vector_runahead,decoupled_vector_runahead,hashemi2016continuous,bfetch,fdp,ebrahimi2009coordinated,ebrahimi_paware,ebrahimi2009techniques,lee2008prefetch,sandbox,panda2016expert,seshadri2015mitigating,wu2011pacman,memory_bank,seq_stream,hur,tjaden1970detection,austin1995zero,austin_load_addr,rfp,elar,lipasti1996vp,sazeides_vp,sazeides_vp2,mendelson1997speculative,dfcm,last_n,selective_vp,perais2012revisiting,perais2014eole,perais2014practical,perais2015bebop,fvp,perais2021ssr,eves,rami_ap,rami_composite,kalaitzidis2019value,sakhuja2019combining,mrn,mrn2,mrn_classifying,moshovos1997streamlining,moshovos1999speculative,store_set,data_preload,mowry1995thesis,charney,charneyphd,fields2006using,SiNUCA_Thesis_2014,lee2016reducingthesis,chang2017understandingphd,o2021energy,mineshphd,hasanphd,chou2004microarchitecture,chou2005store,connors2000eliminating,suleman2009accelerating,suleman2010data,chakraborty2006computation,tyson1994effects,zilles2000understanding}.\footnote{Along with hiding and tolerating memory latency, literature is also rich with techniques that aim to fundamentally \emph{reduce} the memory latency. These techniques, while largely orthogonal to this dissertation, are summarized in~\cref{sec:rw_lat_red}.} 
Some key examples of latency hiding techniques include efficiently caching of data in faster on-chip caches~\cite{wilkes1965slave,liptay1968structural,cache_og}, and prefetching data to on-chip caches before the application demands it~\cite{jouppi_prefetch,Anderson1967TheIS,smith1978sequential}. 
Some key examples of latency tolerance techniques include out-of-order (OoO) execution~\cite{tomasulo1967efficient,Anderson1967TheIS,patt1985hps,patt1985critical}, multi-threading~\cite{Papadopoulos1991MultithreadingAR,Pearce1978AnalysisOA,tullsen1996exploiting,Smith1986APS,smith1982architecture,jordan1983performance,yamamoto1995increasing,Hirata1992AnEP}, and speculative execution via data-dependence prediction~\cite{lipasti1996vp,sazeides_vp,sazeides_vp2,mendelson1997speculative,mrn,mrn2,mrn_classifying,moshovos1997streamlining,moshovos1999speculative,store_set}.
These microarchitectural techniques have collectively pushed the boundaries of performance and energy efficiency of state-of-the-art general-purpose processors.
Yet, as the growth in application data footprint continues to far outpace the scaling of logic and memory technology nodes, 
we are in a constant need for better microarchitectural designs 
to extract ever-more performance and energy efficiency scaling.

\section{Key Problem}
\label{sec:intro.key_problem}

Even though modern \rbone{microarchitectures observe} a massive amount of \emph{data}\footnote{\rbtwo{The data observed by a microarchitecture can be broadly classified into two categories: \textit{application data} and \textit{system-generated data}. 
Application data consists of information related to an application's own execution, such as its control-flow information (e.g., program counter value, instruction type, branch outcome) and data-flow information (e.g., memory addresses touched, data value loaded by a memory instruction).
System-generated data, on the other hand, consists of information related to the behavior of the system when executing given application(s), such as memory bandwidth usage, cache pollution, and energy consumption.}}
during \rbone{its} course of operation, the decisions \rbone{the microarchitecture makes} online are, by and large, agnostic to the data \rbone{it observes}. 
\rbone{The} decisions are often dictated by simple static rules that we, as human architects, typically select at design time, \rbone{which offers limited flexibility for dynamic adaptation at runtime}.
We observe that this data-agnosticism severely limits the effectiveness of many state-of-the-art microarchitectural techniques and leaves large potential for both performance and energy efficiency improvement on the table.

We identify two concrete types of data-agnosticism typically found in microarchitectural techniques and demonstrate how these impact the overall effectiveness of such techniques.

\subsubsection{Inadequate Adaptation to \rbtwo{the Observed Data}}

Many microarchitectural techniques \rbtwo{inadequately adapt their policies} based on the data they observe during their online operation. 
These techniques \rbone{make} decisions based on rigid, and often myopic, human-crafted heuristics selected at design time, completely disregarding the massive amounts of data available during runtime.
While effective for the set of applications considered during the design, these techniques often fail to adapt dynamically \rbone{to} diverse application behaviors and/or complex system configurations.\footnote{\rbtwo{Although this limitation manifests in a wide range of microarchitectural techniques, we note that prior works (most notably in branch prediction literature~\cite{jimenez_ml2,jimenez_ml3,jimenez_ml4,jimenez_ml5,tarsa,branchnet,perceptron,BranchNetV2_2020,Zouzias2021BranchPA,jimenez2016multiperspective,tarjan2005merging,jimenez2003fast,jimenez2002neural} and, to a lesser extent, in other domains~\cite{rlmc,morse,teran2016perceptron,jimenez2016multiperspective,singh2022sibyl,rakesh2025harmonia}) have proposed policies that adaptively learn online, thereby relying less on human-designed heuristics. 
In this dissertation, we either integrate such techniques into our evaluation baseline,
or directly compare our proposed mechanisms against them to demonstrate the advancement of the state-of-the-art.}}

To demonstrate this observation, we show an example using hardware data prefetching techniques.
A hardware data prefetcher predicts the addresses of memory requests and fetches the corresponding data to the on-chip caches before
the application demands it~\cite{jouppi_prefetch,stride,baer2,streamer}.
Modern prefetchers typically exploit human-crafted program context information (we call this a program \emph {feature}; e.g., the address of the instruction generating the memory request, the physical/virtual address of the memory request) to identify patterns in the addresses of memory requests.
This program feature is selected at design time and remains fixed throughout the course of the prefetcher's operation, regardless of the positive or negative impact of the prefetcher's decisions on the overall system~\cite{stride,streamer,baer2,stride_vector,jouppi_prefetch,ampm,fdp,footprint,sms,sms_mod,spp,vldp,sandbox,bop,dol,dspatch,mlop,ppf,ipcp,mutlu2005address}. 
\rbone{As a result, while a given prefetcher may successfully capture patterns in certain applications, it often fails to dynamically adapt to others, resulting in inconsistent gains (or even \textit{degradation}) in performance}.

\begin{figure}[!h]
\centering
\includegraphics[width=5.5in]{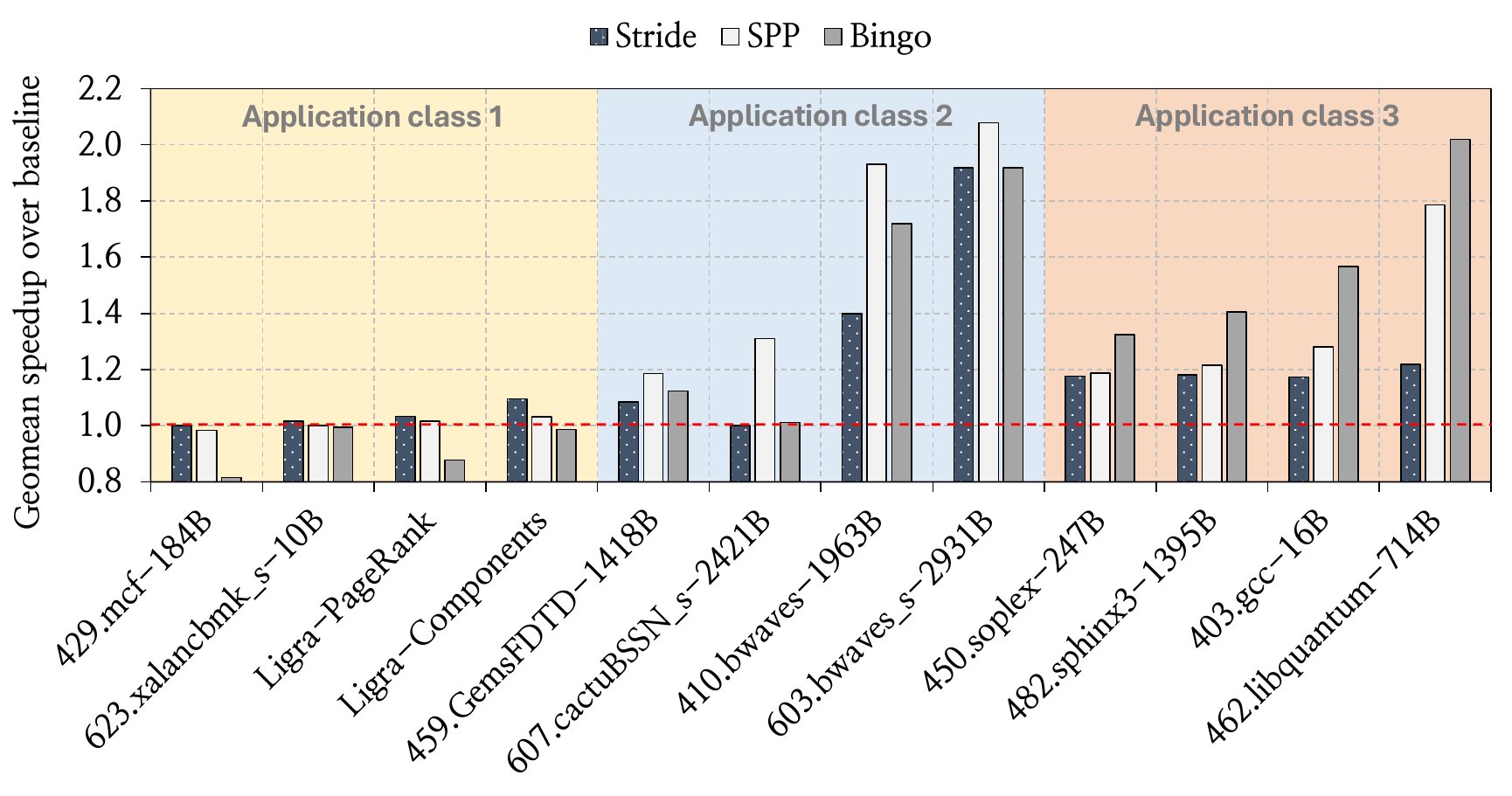}
\caption{Performance comparison of three contemporary prefetchers - stride~\cite{stride}, SPP~\cite{spp}, and Bingo~\cite{bingo} - across diverse applications. 
No prefetcher provides the highest performance across all applications.
}
\label{fig:thesis_intro1}
\end{figure}

\Cref{fig:thesis_intro1} shows the performance of three contemporary prefetchers, stride~\cite{stride}, SPP~\cite{spp}, and Bingo~\cite{bingo}, across a diverse set of applications.\footnote{\Cref{sec:pythia_methodology} discusses the evaluation methodology in detail.}
Each prefetcher exploits a different program feature to identify patterns in memory addresses.
As we can see, \emph{no single prefetcher} provides the highest performance across \emph{all} applications. 
This is primarily due to the static nature of the program feature exploited by each prefetcher.
For example, SPP provides the highest performance in application class 2, where the program feature it exploits (i.e., the sequence of last-four deltas in cacheline addresses) successfully predicts the future memory addresses. Yet, it does not perform well in other application classes (or even degrades performance in application class 1), where the same program context information cannot identify patterns in memory addresses.
\rbone{This demonstrates that a prefetcher's reliance on a single, static program feature for finding address patterns inherently limits its adaptability to diverse workloads, leaving significant potential for performance improvement behind.}

\subsubsection{\rbtwo{Insufficient Exploitation} of Data Characteristics}

\rbone{Beyond their limited ability to learn, many microarchitectural techniques \rbtwo{insufficiently} exploit the inherent characteristics and semantics of the application data they process. 
By design, these techniques often treat data uniformly, even though different data can exhibit widely varying properties (e.g., criticality, value locality, compressibility, approximability) that may influence the overall performance and energy efficiency of the system}.\footnote{\rbone{While this limitation applies to many microarchitectural policies and techniques (e.g., the policy to allocate resources (like reorder buffer entry, reservation station, load/store queue entry) in an out-of-order processor, cache management policy, branch prediction), there exist proposals that incorporate certain forms of data-awareness, like, awareness to data criticality (e.g., criticality-driven instruction fetch~\cite{deshmukh2021criticality}, instruction execution~\cite{fields2001focusing,fields2006using,Srinivasan2001LocalityVC}, and prefetching~\cite{clip,litz2022crisp}), and data repeatability (e.g., load value prediction~\cite{lipasti1996vp,sazeides_vp,sazeides_vp2,mendelson1997speculative}).
In this dissertation, we build on top of such policies as baseline and demonstrate that by exploiting previously underexplored characteristics of data, one can unlock further performance and \rbtwo{energy} efficiency improvements}.}
Neither the programmer conveys these data characteristic information to the underlying microarchitectural techniques, nor the techniques themselves try to infer and exploit such characteristics in the hardware.
This unawareness \rbone{of} data characteristics leaves significant potential for performance and \rbtwo{energy} efficiency improvement on the table.

\begin{figure}[!h]
\centering
\includegraphics[width=5in]{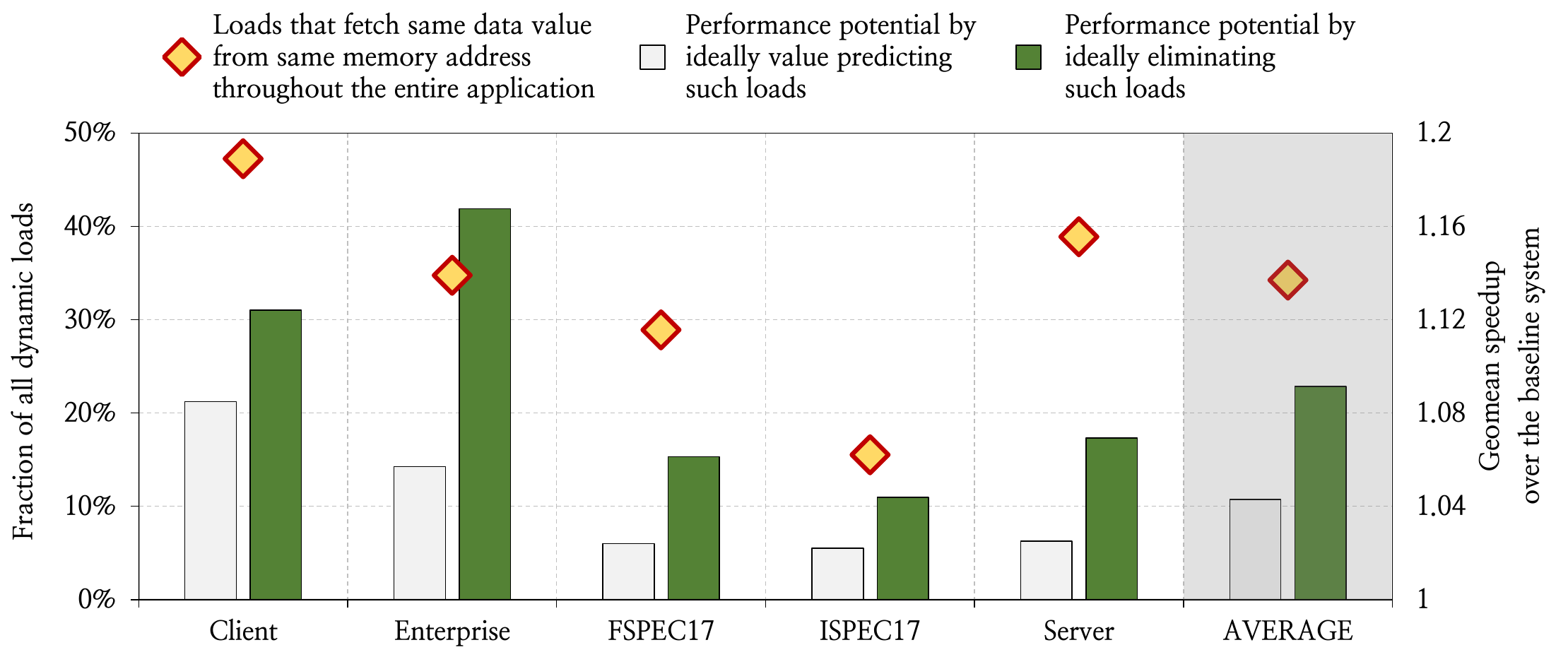}
\caption{Fraction of dynamic loads that fetch the same data value from the same load address throughout the entire execution of an application (on the primary y-axis), and the performance improvement potential by ideally value-predicting and eliminating the execution of such loads (on the secondary y-axis).
}
\label{fig:thesis_intro2}
\end{figure}

\rbone{For instance, we find that nearly one-third of all dynamic load instructions, on average across a diverse range of real-world applications, repeatedly fetch \textit{the same value from the same memory address} (as plotted in the primary y-axis of \Cref{fig:thesis_intro2}).\footnote{\Cref{sec:cst_constable_methodology} details the applications and the simulation methodology used in this study.}}
\rbone{Prior techniques such as load value prediction (LVP)~\cite{lipasti1996vp,sazeides_vp,sazeides_vp2,mendelson1997speculative} can exploit only the repeatable-value aspect of such loads to mitigate data dependence by speculating their values.
However, as LVP ignores the repeatable-address aspect, it still requires executing such loads to verify the prediction, thereby consuming scarce pipeline resources.
As a result, LVP mitigates data dependence but not resource dependence.
By fully exploiting both value and address repeatability, these loads can be safely eliminated, thereby removing not only their data dependence but also their resource dependence.
We show that ideally eliminating such load instructions provides a $9.1\%$ performance improvement potential (as shown in the secondary y-axis in \Cref{fig:thesis_intro2}), which is more than double of the performance potential provided by ideally value predicting all such loads.
This highlights how overlooking data characteristics can leave a significant potential of performance and \rbtwo{energy} efficiency improvement on the table.}

\noindent \rbone{In summary}, we observe that the data-agnostic design of many current microarchitectural techniques significantly limits their effectiveness in addressing the memory bottleneck. 
Their failure to learn from runtime data and their lack of awareness of data characteristics together prevent them from fully leveraging the abundant information inherently available during execution. 
As the data footprint of modern (and likely future) applications continues to grow, these limitations will only get more pronounced.

\section{Dissertation Objective and Vision}

The primary objective of this dissertation is to design fundamentally-better microarchitectural techniques that can inherently exploit the ever-growing data footprint to mitigate the memory bottleneck.
We advocate for a fundamental shift from a data-agnostic approach toward \rbtwo{a more} \textbf{\emph{data-driven}} approach that can continuously learn from runtime data to dynamically adapt their policies, and a \textbf{\emph{data-aware}} approach that can tailor their actions by exploiting the characteristics and semantics of data.

\section{Thesis Statement}

To this end, this dissertation posits, and later quantitatively substantiates, the following \rbone{thesis statement}:
\vspace{1pt}

\emph{By designing microarchitectural techniques that (1) adapt their policies through continuous learning from application and system-generated data, and (2) tailor their decisions by exploiting the characteristics and semantics of the data, we can significantly improve the performance and \rbtwo{energy} efficiency of state-of-the-art processors that was otherwise untapped by the conventional data-agnostic microarchitectural techniques.}

\section{Our Approach}

To substantiate our \rbone{thesis statement}, we conduct the research in \rbone{two} broad stages.
\rbone{First, we analyze various state-of-the-art microarchitectural techniques to build a deeper understanding of how the data-agnostic nature of their design limits their effectiveness in mitigating the memory bottleneck.}
Then, we exploit our understanding of the limitations in current microarchitectures to design novel data-driven and data-aware techniques.
For data-driven microarchitecture design, we exploit, design, and evaluate various lightweight \emph{machine learning} (ML) techniques to guide architectural decision making at various points in the memory hierarchy.\footnote{\rbtwo{Our approach of using ML for guiding or managing microarchitectural techniques at runtime fundamentally differs from the \rbfor{orthogonal} research direction that applies ML to explore microarchitecture, processor, and system design space (e.g., \rbfor{processor design and performance modeling~\cite{ipek2008efficient,Nasr2025Concorde}}, network-on-chip design~\cite{fettes2018dynamic,zheng2019energy,lin2020deep,yin2020experiences}, chip \rbfor{floorplanning} optimization~\cite{mirhoseini2021,mirhoseini2018hierarchical,mirhoseini2017device,cheng_rebuttal,markov_rebuttal,goldie2024chipsailedcritiqueunfounded}). Nonetheless, we provide a comprehensive literature review of applied ML in processor/system design in~\cref{subsec:ml_driven_design}.}} For data-aware microarchitecture design, we explore and evaluate previously-underexplored characteristics of application data, that are seemingly abundant in a wide variety of workloads, to propose new microarchitectural techniques for improving performance and energy efficiency.

In the remainder of this section, we briefly introduce the four techniques developed in this dissertation.

\subsection{\rbone{Machine-Learning-Driven} Microarchitecture}

\subsubsection{Efficient Hardware Data Prefetching via Online Reinforcement Learning}

Prefetching is a key speculation technique for designing high-performance processors that hides the long memory access latency by predicting the addresses of memory requests and fetching their data to on-chip caches before the processor demands it.
As we discuss in~\cref{sec:intro.key_problem}, a prefetcher typically exploits human-crafted program context information (also called a \emph{feature}; e.g., the address of the load instruction) for identifying patterns in the addresses of memory requests.
Even though prior research has proposed numerous prefetching mechanisms, we observe that most of them exhibit three key limitations. 
First, they typically rely on a single, static program feature fixed at design time, which significantly limits their applicability across diverse workloads~\cite{stride,streamer,baer2,stride_vector,jouppi_prefetch,ampm,fdp,footprint,sms,sms_mod,spp,vldp,sandbox,bop,dol,dspatch,mlop,ppf,ipcp}. 
Second, they either lack awareness to the inherent system-level feedback (e.g., memory bandwidth usage, cache pollution)~\cite{stride,streamer,baer2,stride_vector,jouppi_prefetch,ampm,footprint,sms,sms_mod,spp,vldp,sandbox,bop,dol,bingo,mlop,ppf,ipcp}, 
or incorporate system awareness as \rbc{an afterthought \rbc{(i.e., a separate control component)}} to the underlying system-unaware prefetch algorithm~\cite{fdp,ebrahimi2009coordinated,ebrahimi2009techniques,ebrahimi_paware,dspatch,mutlu2005,wflin,zhuang,wflin2,charney,memory_bank,lee2008prefetch}. 
Third, they lack the ability to be customized on-the-fly to make them adapt to diverse workloads and system configurations.

To alleviate these limitations, we aim to design a single prefetching framework that 
(1) can holistically learn to prefetch using both \emph{multiple different types of program features} and \emph{system-level feedback} \rbc{information}, 
and (2) can be \emph{easily customized} in silicon via simple configuration registers to exploit \rbc{different types of program features and/or} to change the objective of the prefetcher \rbc{(e.g., increasing/decreasing coverage, accuracy, or timeliness)} without any changes to the underlying hardware.

To this end, we formulate prefetching from the grounds-up as a reinforcement learning (RL) problem where
the prefetcher acts as the RL agent, the processor and memory system act as the environment, and the agent (i.e., the prefetcher) \emph{autonomously learns} to prefetch by interacting with its environment.
More specifically, for every demand request, the prefetcher extracts a set of program features and uses it as a \emph{state} information to take a prefetch \emph{action} based on its prior experience. 
For every prefetch action, the prefetcher receives a numerical \emph{reward}
\rbc{which} evaluates the efficacy of the prefetch action (e.g., accuracy and timeliness of the prefetch)
under various system-level feedback \rbc{information} (e.g., memory bandwidth usage).
The prefetcher uses this reward to reinforce the \rbc{correlation} between various program features and the prefetch action. 
By formulating prefetching in this way, we enable the prefetcher to \emph{autonomously learn} from its own experience and \emph{continuously adapt} to changing workloads and system configurations for generating effective prefetch requests, without relying on rigid, and often myopic, human-crafted heuristics.
Our extensive evaluation using both hardware simulation and synthesis 
shows that our RL-based prefetcher (called \textbf{\emph{Pythia}}~\cite{pythia,pythia_github}) consistently outperforms multiple state-of-the-art prefetchers in single-core, multi-core, and bandwidth-constrained core configurations across a diverse set of workloads, while incurring a modest area and power overhead.
\rbthr{We further show that Pythia is capable of generalizing its performance benefits beyond the workloads considered at the design time by evaluating it across a large corpus of previously-unseen workload traces collected as part of the 4th Data Prefetching Championship (DPC4)~\cite{dpc4}.}
We open-source our artifact-evaluated implementation of Pythia, and all necessary evaluation infrastructure, to facilitate future research: \url{https://github.com/CMU-SAFARI/Pythia}. 

\rbfor{An earlier version of Pythia is published at MICRO 2021~\cite{pythia} and an extended version on arXiv~\cite{pythia_extended}.}
\rbone{Pythia has influenced multiple subsequent works both as a state-of-the-art baseline~\cite{mab,duong2024new,pmp,eris2022puppeteer,lin2024pars,micro_mama,dpc4_bertigo,dpc4_edp,dpc4_emender,dpc4_gberti,dpc4_sberti,dpc4_sppam,dpc4_umama,dpc4_vip}, as well as a reference framework for modeling architectural decision making using machine learning~\cite{singh2022sibyl,rakesh2025harmonia,chrome,chen2025gaze,gogineni2024swiftrl,long2023deep,zhang2022resemble,zhou2023efficient,pandey2023neurocool,yi2025artmem,alkassab2024deepref,huang2023rlop,Xue2025AugurST,Zeng2025ChartenOnlineRL,Xing2025ProactiveDP}}.

\subsubsection{Accelerating Loads via Perceptron Learning-Based Off-Chip Load Prediction}

While Pythia advances the state-of-the-art in prefetching, it only successfully predicts and prefetches nearly half of the memory load requests that would have otherwise gone to the off-chip main memory.
For the remaining off-chip loads, we observe that, a large fraction of their latency is spent in accessing the on-chip cache hierarchy to solely determine that they need to go off-chip.
As the size of on-chip caches continue to increase to cater to the ever-growing data footprint of workloads, the latency overhead stemming from on-chip cache access is going to exacerbate further in next-generation processors.

To address this, we aim to accelerate off-chip load requests by removing the on-chip cache access latency from their critical path.
\rbc{To this end}, we propose a new technique (called \textbf{\emph{Hermes}}~\cite{hermes,hermes_github}), \rbc{whose} key idea is to: (1) accurately predict which load requests might go off-chip, and (2) \rbc{speculatively fetch} the data required by the predicted off-chip loads directly from the main memory, \rbc{while also concurrently accessing the cache hierarchy for such loads}.
\rbc{To enable Hermes, we develop} a new lightweight, perceptron learning-based off-chip load prediction \rbc{technique} that learns to identify off-chip load requests using multiple program features (e.g., sequence of program counters, byte offset of a load request). 
For every load request generated by the \rbc{processor}, the predictor observes a set of program features to predict whether \rbc{or not} the load would go off-chip. 
If the load is predicted to go off-chip, Hermes issues a speculative load request directly to the main memory controller once the load's physical address is generated. 
If the prediction is correct, the load eventually misses the cache hierarchy and waits for the ongoing speculative load request to finish, 
\rbc{and} thus \rbc{Hermes} completely hides the on-chip cache \rbc{hierarchy} access latency from the critical path \rbc{of the \rbe{correctly-predicted off-chip} load}.
\rbc{Our} extensive evaluation using a wide range of workloads \rbc{shows} that Hermes provides consistent performance improvement on top of a \rbc{state-of-the-art} baseline system \rbc{across} a wide range of configurations with varying core count, main memory bandwidth, high-performance data prefetchers, and on-chip cache hierarchy access latencies, while incurring \rbc{only modest} storage overhead.
\rbthr{We further show that Hermes is capable of generalizing its performance benefits beyond the workloads considered at the design time by evaluating it across the large, previously-unseen DPC4 trace corpus.}
We open-source our artifact-evaluated implementation of Hermes, and all necessary evaluation infrastructure, to facilitate future research: \url{https://github.com/CMU-SAFARI/Hermes}.

\rbfor{An earlier version of Hermes is published at MICRO 2022~\cite{hermes} and an extended version on arXiv~\cite{hermes_extended}.}
\rbone{Hermes has already influenced multiple subsequent works as a state-of-the-art baseline~\cite{jamet2024tlp,clip,krause2022hbpb,fang2025dhcm,wu2025concurrency,sato2024learning}}. One follow-up work~\cite{prefetchx} has revealed the existence of a similar speculation mechanism already employed in Intel 3rd generation Xeon processors~\cite{intel_xeon_3rdgen}.

\subsubsection{Synergizing Data Prefetching and Off-Chip Load Prediction via Lightweight Reinforcement Learning}

Prefetching and off-chip prediction, while differing in their speculation mechanism, share the same goal of hiding memory access latency. 
A prefetcher hides latency by predicting addresses of future memory requests and proactively fetching their data closer to the core, whereas an off-chip predictor (OCP) hides latency by predicting loads that would likely go off-chip and removing the on-chip cache access latency from their critical path.
However, we demonstrate that: 
(1) a prefetcher and an OCP often provide complementary performance benefits, 
yet (2) naively combining these two mechanisms often fails to realize their full performance potential, 
and (3) existing prefetcher control policies (both heuristic- and learning-based) leave significant room for performance improvement behind.

To address this, our goal is to design a holistic framework that can autonomously learn to coordinate an off-chip predictor with multiple prefetchers employed at various cache levels, delivering \emph{consistent} performance benefits across a wide range of workloads and system configurations.
To this end, we propose a new technique called Athena, which models the coordination between prefetchers and off-chip predictor (OCP) as a reinforcement learning (RL) problem.
Athena acts as the RL agent that observes multiple system-level features (e.g., prefetcher/OCP accuracy, bandwidth usage) over an epoch of program execution, and uses them as \emph{state} information to select a coordination \emph{action} (i.e., enabling the prefetcher and/or OCP, and adjusting prefetcher aggressiveness). 
At the end of every epoch, Athena receives a numerical \emph{reward} that measures the change in multiple system-level metrics (e.g., number of cycles taken to execute an epoch).
Athena uses this reward to autonomously and continuously learn a policy to coordinate prefetchers with OCP.
Athena makes a key observation that using performance improvement as the sole RL reward, as used in prior work, is unreliable, as it confounds the effects of the agent's actions with inherent variations in workload behavior.
To address this limitation, Athena introduces a composite reward framework that separates 
(1) system-level metrics directly influenced by Athena's actions (e.g., last-level cache misses) from 
(2) metrics primarily driven by workload phase changes (e.g., number of mispredicted branches).
This allows Athena to autonomously learn a coordination policy by isolating the true impact of its actions from inherent variations in workload behavior.
Our extensive evaluation using a diverse set of 100 memory-intensive workloads shows that Athena \emph{consistently} outperforms multiple prior state-of-the-art coordination policies across a wide range of system configurations with various combinations of underlying prefetchers at various cache levels, OCPs, and main memory bandwidths, while incurring only modest storage overhead and design complexity.
\rbthr{We further demonstrate Athena's capability of generalizing its performance benefits beyond the workloads considered at the design time by evaluating it across the large, previously-unseen DPC4 trace corpus.}
We open-source our artifact-evaluated implementation of Athena, and all necessary evaluation infrastructure, to facilitate future research: \url{https://github.com/CMU-SAFARI/Athena}.
\rbfor{An earlier version of Athena is published at HPCA 2026~\cite{athena} and an extended version on arXiv~\cite{athena_extended}.}

\subsection{Data-Aware Microarchitecture}

\subsubsection{Improving Performance and Power Efficiency by Eliminating Load Instruction Execution}

While prefetching and off-chip prediction are very effective in \emph{hiding} memory access latency, modern processors also employ numerous mechanisms inside the processor core that \emph{tolerate} latency by preventing stalls caused by long-latency instructions (e.g., a memory load instruction). 
Two such widely studied techniques are load value prediction (LVP) and memory renaming (MRN), both of which break the data dependency on a load instruction by speculating the value fetched by the load. 
However, even when their predictions are correct, these techniques still execute the predicted load instructions, like any other load instruction, to verify the result, thereby consuming scarce pipeline resources that could otherwise be used by other load instructions. 
This limitation leaves a significant performance and power efficiency improvement opportunity untapped.

Our analysis reveals that a substantial fraction of dynamic load instructions in real workloads \emph{repeatedly fetch the same value from the same memory address throughout the execution}. 
This recurring \emph{stability} in load behavior represents a valuable characteristic that is often underexplored by conventional techniques.
We aim to exploit this stability to not only break load data dependence but completely eliminate the entire execution of the load instruction, thereby mitigating \emph{both} data and resource dependence.

To this end, we propose \textbf{\emph{Constable}}~\cite{constable}, a data-aware microarchitectural technique that dynamically detects \emph{likely-stable} loads (i.e., loads that have historically fetched the same data from the same address) and safely eliminates their execution.
For every likely-stable load, Constable (1) tracks modifications to its source architectural registers and memory location via lightweight hardware structures, and (2) eliminates the execution of subsequent instances of the load instruction until there is a write to its source register or a store or snoop request to its load address.
In doing so, Constable treats likely-stable loads differently from other loads, tailoring the execution policy based on the inherent characteristics of the load instructions.
Our extensive evaluation shows that Constable simultaneously improves performance and reduces core dynamic power by over a strong superscalar baseline processor design, with larger benefits in processors with simultaneous multithreading support.
We open-source a binary instrumentation tool that we use to identify load instructions that repeatedly fetch the same value from the same memory address in any off-the-self x86-64 binary at \url{https://github.com/CMU-SAFARI/Load-Inspector}. 

\rbfor{An earlier version of Constable is published at ISCA 2024~\cite{constable} and an extended version on arXiv~\cite{constable_extended}.}
Constable has influenced both academic research and industrial product development.
A later work has independently verified Constable's key observation of loads with repeated address-value stability in off-the-shelf programs~\cite{gsl}. To the best of our knowledge, the key idea of Constable is in the process of getting transferred to a real-world commercial processor design, and has been the subject matter of a patent application filed by Intel Corporation~\cite{constable_patent}.

\section{Contributions}

This dissertation makes the following contributions:

\begin{itemize}
    \item We build a comprehensive understanding of how the data-agnostic nature of conventional microarchitecture design restricts the effectiveness of state-of-the-art mechanisms deployed in high-performance general-purpose processors. Chapters~\ref{chap:pythia} to~\ref{chap:constable} dissect four case studies of state-of-the-art microarchitectural techniques employed at various parts of modern processors and quantify the opportunity loss of performance and \rbtwo{energy} efficiency improvements due to data-agnostic designs.
    \item We advocate and quantitatively substantiate a fundamental shift from data-agnostic techniques towards \textbf{\emph{machine-learning (ML)-driven}} techniques that can continuously learn from runtime information to dynamically adapt their policies, and \textbf{\emph{data-aware}} techniques that can tailor their actions by exploiting the characteristics and semantics of data.
    \item \rbfor{We design and release a large-scale, community-wide workload traces and evaluation infrastructure through the 4th Data Prefetching Championship (DPC4), comprising $610$ traces from emerging artificial intelligence (AI), graph mining, and real-world datacenter workloads, $483$ of which are previously unseen by the architecture research community (see~\Cref{chap:dpc4}). 
    We further develop a rigorous stress-testing methodology using these traces to systematically validate the performance generalization capability of ML-driven microarchitectural techniques without workload-specific tuning.}
    \item We propose Pythia (see~\Cref{chap:pythia}), the first practical reinforcement-learning-based hardware data prefetching framework. We show that Pythia dynamically learns correlations between program features and system-level feedback to generate accurate, timely, and bandwidth-aware prefetches, achieving robust performance across diverse workloads and configurations.
    \item We propose Hermes (see~\Cref{chap:hermes}), that employs the first perceptron-learning-based off-chip load predictor. Hermes formulates the off-chip prediction as a binary classification problem, learning to accurately identify loads that would go off-chip early and initiating speculative requests directly to the main memory, thereby effectively hiding the long main memory access latency.
    \item We propose Athena (see~\Cref{chap:athena}), the first reinforcement-learning-based framework designed for prefetcher–OCP coordination. Athena autonomously synergizes prefetching and off-chip prediction by observing system-level metrics and learning to control both mechanisms in unison, consistently outperforming both mechanisms individually across a diverse set of workloads and system configuration.
    \item We propose Constable (see~\Cref{chap:constable}), the first data-aware technique that exploits consistent repeatability in both the address and data value of a load instruction to safely eliminate its entire execution in today's multi-core processors, providing simultaneous improvement in performance and power efficiency.
    \item \rbfor{We open-source the implementations, workload infrastructures, and/or accompanying toolchains for all proposals to ensure reproducibility, enable rigorous comparison, and facilitate future research. By releasing complete implementations and evaluation artifacts, this dissertation aims to promote transparency and accelerate community-driven progress in ML-driven and data-aware microarchitecture design.}
\end{itemize}

\section{Dissertation Outline}

This dissertation is organized into 8 chapters. 
\Cref{chap:related_work} provides a comprehensive view of various microarchitectural techniques to improve processor performance and \rbtwo{energy} efficiency by mitigating the memory bottleneck. 
\Cref{chap:ml_primer} provides a primer on two types of machine learning methods relevant to this dissertation: reinforcement learning and perceptron learning, and their prior applications in processor design.
\rbfor{\Cref{chap:dpc4} introduces a \rbfiv{new set of previously-unseen workload traces and a} rigorous and systematic methodology \rbfiv{using those traces} to empirically validate the performance generalization capability of the ML-driven techniques proposed in this dissertation.}
\Cref{chap:pythia} introduces and evaluates Pythia, a reinforcement-learning-based hardware data prefetcher.
\Cref{chap:hermes} introduces and evaluates Hermes, a perceptron-learning-based off-chip load predictor.
\Cref{chap:athena} introduces and evaluates Athena, a reinforcement-learning-based prefetcher-OCP coordinator.
\Cref{chap:constable} introduces and evaluates Constable, a purely-microarchitectural technique to safely eliminate load instruction execution.
\Cref{chap:conc} summarizes this dissertation and discusses the possibilities of designing numerous other microarchitectural techniques using machine-learning-driven and data-aware principles.

\chapter{Related Work}
\label{chap:related_work}

Long-latency memory accesses remain a fundamental performance and \rbtwo{energy} efficiency bottleneck in modern general-purpose processors. 
To address this challenge, researchers have proposed a wide range of microarchitectural techniques, which can be broadly categorized into four classes: (1) \emph{latency-hiding} techniques, (2) \emph{latency-tolerance} techniques, (3) \emph{latency-reduction} techniques, and (4) techniques that improve processor \rbtwo{energy and/or power} efficiency. 
This section surveys these classes of techniques and positions them in relation to the approaches developed in this dissertation.

\section{Techniques to Hide Memory Latency}
\label{sec:rw_lat_hide}

\subsection{Caching}
\label{subsec:caching}

Caching~\cite{cache_og,wilkes1965slave,liptay1968structural} hides memory access latency by exploiting temporal and spatial locality in memory accesses. 
Most modern computing systems employ caches, which are often organized in a hierarchical way with varying capacities and latencies.
Researchers have proposed numerous cache management policies~\cite{teresa_cache_bypassing,teresa_cache_bypassing2,tyson1995modified,tyson1997managing,adaptive_cache_hierarchy,jg,ship,hawkeye,jimenez2010dead,jimenez2017multiperspective,sdbp,rivers1996reducing,rivers1998utilizing,hu2002timekeeping,kharbutli2008counter,liu2008cache,piquet2007exploiting,chi1989improving,dybdahl2006enhancing,wong2000modified,jimenez_cache1,jimenez_cache2,jimenez_cache3,jimenez_cache4,jimenez_cache5,jimenez_cache6,jimenez_cache7,jimenez_cache8,jimenez_cache9,jimenez_cache10,jimenez_cache11,jimenez_cache12,jimenez_cache13,jimenez_cache14,boris_cache1,boris_cache2,boris_cache3,boris_cache4,aamer_cache1,aamer_cache2,aamer_cache3,aamer_cache4,aamer_cache5,aamer_cache6,seshadri2015mitigating,seshadri2012evicted,alameldeen2004adaptive,alameldeen2004frequent,chen2009c,dusser2009zero,villa2000dynamic,yang2000frequent,zhang2000frequent2,pekhimenko2012base,alper2025enterprise,pekhimenko2016thesis,pekhimenko2016case,pekhimenko2015exploiting,tyson1995modified,vijaykumar2015case,pa_haweye,hardavellas2009reactive,seznec1993twoway,seznec1993skewed,seznec1993set,bodin1997skewed,drach1995associativity,seznec1997newcase,michaud2003statistical,seznec2003tlb,seznec1994sectored,seznec1997sectored,seznec1996pointer,dusser2009zero,dusser2011decoupled,sardashti2014skewed,drach1993semi,seznec1995dasc,seznec1995misspenalty,verma2025leakyrand,chaudhuri2021zeroinclusion,chaudhuri2019bandwidth,gaur2013gpu,natarajan2013sharing,chaudhuri2012hierarchy,gaur2011exclusive,chaudhuri2009pseudolifo,chaudhuri2009pagenuca,basu2007scavenger,chandarlapati2007lemap,sweta2025drishti,panda2018synergistic,raghavendra2016pbc,raghavendra2015mbzip,panda2012csharp,Lee2010DRAMAwareLC,Lee2010DRAMAwareLC2} to improve cache performance.
Prior works have also proposed various mechanisms that reduce memory access latency in deep cache hierarchies by predicting the cache level at which a given memory request is likely to hit~\cite{d2d,d2m,lp,peir_bloom_filter,mnm,loh_miss_map,alloy_cache}.
We evaluate each proposed technique presented in this dissertation on a processor configuration that incorporates a deep cache hierarchy and state-of-the-art cache management policies~\cite{ship,jaleel2010high}. 
Moreover, 
Hermes introduces off-chip prediction that works orthogonally to cache management policies and bypasses intermediate cache levels during cache fills~\cite{rivers1996reducing,rivers1998utilizing,hawkeye,jimenez2010dead,jimenez2017multiperspective,sdbp,kharbutli2008counter,liu2008cache,piquet2007exploiting,dybdahl2006enhancing,wong2000modified,chi1989improving,tyson1995modified,tyson1997managing,nt_load_intel_manual,movnti}.

\subsection{Prefetching}
\label{subsec:prefetching}

Prefetching~\cite{jouppi_prefetch,Anderson1967TheIS,smith1978sequential,baer2} is a well-studied speculation technique that predicts the addresses of long-latency memory requests and fetches the corresponding data to on-chip caches before the program demands it, thus hiding long memory access latency.
Prior works on prefetching can be broadly categorized into three groups: software, hardware, and pre-computation-based prefetching.

\subsubsection{Software Prefetching}

In software prefetching, the programmer or compiler inserts explicit prefetch instructions into the program to fetch data before it is required~\cite{Klaiber1991AnAF,software_pref,Cher2004SoftwarePF,Jamilan2022APTGETPT,Ainsworth2018SoftwarePF,Wu2013APS,Li2023HoPPHC,Ainsworth2017SoftwarePF,Lipasti1995SPAIDSP,mowry1995thesis,Snchez1999SoftwareDP,Zhang2025HierarchicalPA,Ainsworth2016GraphPU,asmdb,luk1996compiler,mowry1997predicting,luk2000optimizing}. 
This approach is effective for workloads with regular access patterns that can be statically predicted, but it is generally ineffective for workloads with irregular or hard-to-predict access behavior.

\subsubsection{Hardware Prefetching}

In hardware prefetching, dedicated microarchitectural structures detect patterns in the stream of observed memory addresses and issue prefetch requests transparently to the programmer and compiler. 
Based on the underlying pattern-matching algorithm, hardware prefetchers can be broadly classified into two categories.

\textbf{\emph{Temporal prefetchers}}~\cite{markov,stems,somogyi_stems,wenisch2010making,domino,isb,misb,triage,wenisch2005temporal,chilimbi2002dynamic,chou2007low,ferdman2007last,hu2003tcp,bekerman1999correlated,karlsson2000prefetching,varkey2017rctp} 
\rbc{memorize} long sequences of cacheline addresses \rbc{demanded by the processor}.
\rbc{When} a previously-seen address is encountered \rbc{again}, a temporal prefetcher \rbc{issues prefetch requests} \rbc{to} addresses that previously followed the currently-seen cacheline address. 
However, temporal prefetchers \rbc{usually} have \rbc{high storage requirements (often multi-megabytes of \rbc{metadata} storage, which necessitates storing metadata in memory~\cite{stems,somogyi_stems,isb}).}

\textbf{\textit{Spatial prefetchers}}~\cite{stride,streamer,baer2,jouppi_prefetch,ampm,fdp,footprint,sms,spp,vldp,sandbox,bop,dol,dspatch,bingo,mlop,ppf,ipcp,jimenez_cache_pref1,jimenez_pref1,jimenez_pref2,jimenez_pref3,jimenez_pref4,Zhang2025HierarchicalPA,kpc,cooksey2002stateless,ebrahimi2009techniques,charney1995,charneyphd,iacobovici2004effective,panda2012prefetchers,panda2014xstream}, on the other hand, predict addresses of future memory requests by learning program access patterns over \rbc{different} spatial \rbc{memory} \rbc{regions}. Spatial prefetchers provide high-accuracy prefetches, \rbc{usually} with \rbc{lower} storage overhead \rbc{than temporal prefetchers}. 
Our proposed prefetcher, Pythia, belongs to this category. 
We extensively evaluate Pythia with other spatial prefetchers~\cite{mlop,bingo,spp,ppf,dspatch} in this dissertation and \rbc{demonstrate that Pythia delivers state-of-the-art performance benefits}.

\subsubsection{Pre-Computation-Based Prefetching}

Pre-computation-based prefetching executes a program’s own code (either the complete code, or a \emph{slice} of it) ahead of the actual execution for the sole purpose of generating future memory requests.
Since prefetch requests are generated by pre-executing the code, these techniques can achieve high prefetch accuracy even when the program’s memory access stream lacks easily-identifiable patterns.
Based on the mechanism for constructing the code for pre-execution, these techniques can be classified into two broad categories.

\textbf{\emph{Thread-based pre-computation}}~\cite{ssmt,precompute,Luk2001ToleratingML,sohi_slice,helper_thread,bfetch,runahead_threads,helper_thread2,zhang2007accelerating,dubois1998assisted,collins2001speculative,solihin2002using,Luk2001ToleratingML,Roth2001SpeculativeDM,Sundaramoorthy2000SlipstreamPI,hashemi2016continuous,Moshovos2001SliceprocessorsAI,Chappell2002DifficultpathBP,zhou2005dual,srinivasan2004continual,Garg2008APE,Parihar2014AcceleratingDL,Purser2000ASO,RothMS99} mechanisms explicitly construct a simplified version of the program's own code (denoted as a \emph{helper thread}), often derived from slices of the original instruction stream, to generate memory prefetches ahead of demand.
Software-directed approaches~\cite{Luk2001ToleratingML,collins2001speculative,ssmt,sohi_slice,Luk2001ToleratingML,Chappell2002DifficultpathBP,Garg2008APE,Parihar2014AcceleratingDL,roth1998dependence} rely on the compiler or programmer to explicitly extract and spawn helper threads. In contrast, hardware-generated approaches~\cite{precompute,bfetch,Moshovos2001SliceprocessorsAI,Sundaramoorthy2000SlipstreamPI,hashemi2016continuous,zhou2005dual,srinivasan2004continual,Purser2000ASO,roth1999effective,roth1998dependence} transparently derive and execute pre-computation slices in hardware, thus reducing programmer burden while still exposing substantial memory-level parallelism.

\textbf{\emph{Runahead execution}}~\cite{dundas,mutlu2003runahead,mutlu2003runahead2}, while fundamentally following pre-computation-based prefetching principle, does not require extracting and spawning explicit helper threads. 
Instead, it checkpoints the architectural state of the processor in an event of a full-window stall (i.e., when the reorder buffer neither can retire instructions due to a long-latency operation blocking the retirement, nor can it accept new instructions because it is full) and enters into runahead mode. 
In this mode, the processor speculatively continues to fetch, decode, and execute program's own instructions without updating the architectural state, thereby exposing additional memory-level parallelism.
While runahead execution provides performance benefits with relatively modest hardware overhead, its effectiveness depends on the accuracy of the speculative execution and the ability to generate useful prefetches. As such, subsequent works have refined runahead policies to maximize coverage while minimizing unnecessary speculation~\cite{mutlu2006efficient,mutlu2005address,mutlu2006address,precise_runahead,vector_runahead,decoupled_vector_runahead,hashemi2016continuous,bfetch,roelandts2024svr,naithani2022rar}.

\subsubsection{Prefetcher Coordination}

Along with the pattern-matching algorithm, prefetcher coordination policies also play a crucial role in designing a high-performance prefetcher.
Several prior works incorporate prefetching metrics such as coverage, accuracy, and bandwidth consumption to selectively throttle or discard prefetch requests, aiming to reduce memory bandwidth usage and prefetch-induced cache pollution~\cite{fdp,ebrahimi2009coordinated,ebrahimi_paware,ebrahimi2009techniques,lee2008prefetch,sandbox,panda2016expert,seshadri2015mitigating,wu2011pacman,memory_bank,seq_stream,hur,eris2022puppeteer}.
In this dissertation, we show that prefetcher coordination policies are often applied as an afterthought to an otherwise system-unaware prefetch algorithm. Moreover, many coordination policies rely on static, and often myopic, human-designed heuristics and thresholds.
These two caveats together significantly limit the effectiveness a prefetcher.
Pythia addresses these challenges by incorporating system-level feedback inherently to its design and autonomously learning from the workload behavior, without relying on static heuristics.

\section{Techniques to Tolerate Memory Latency}
\label{sec:rw_lat_tolerate}

Although caching and prefetching hide memory access latency, processors can still stall while waiting for memory (in general, any long-latency) operations to complete. 
To address this, prior works have introduced a variety of latency-tolerance techniques that identify and execute useful instructions during such stalls. This section summarizes such techniques.

\subsection{Out-of-Order Execution}
\label{subsec:ooo}

Out-of-order (OoO) execution~\cite{Anderson1967TheIS,tomasulo1967efficient,patt1985hps,patt1985critical} is a fundamental latency-tolerance technique employed in modern processors. 
By dynamically reordering instructions at runtime, OoO execution enables independent instructions to execute by bypassing stalled ones, thereby exploiting instruction-level parallelism that would otherwise remain hidden. 
This mechanism significantly mitigates the impact of long-latency operations, such as memory accesses, on overall performance. 
However, the scalability of OoO execution is constrained by the complexity and energy cost of maintaining large instruction windows and supporting structures.
All proposed mechanisms in this dissertation are evaluated on top of an aggressive OoO baseline processor.

\subsection{Multithreading}
\label{subsec:mt}

Multithreading~\cite{Papadopoulos1991MultithreadingAR,Pearce1978AnalysisOA,tullsen1996exploiting,thornton1964parallel,Kang2004SpeculationCF,Wallace1998ThreadedMP,Hirata1992AnEP,Tullsen1995SimultaneousMM,Smith1986APS,smt,Snavely2000SymbioticJF,Tullsen2001HandlingLL,Olukotun1996TheCF,jordan1983performance,smith1982architecture,yamamoto1995increasing,Hirata1992AnEP} is a well-established technique to tolerate long memory access latencies by overlapping the execution of multiple threads, thereby improving overall processor utilization and performance. 
The literature distinguishes three major forms of multithreading: fine-grained, coarse-grained, and simultaneous. 
\textit{\textbf{Fine-grained multithreading}}~\cite{Smith1986APS,thornton1964parallel,sun_niagra,nemirovsky1991} switches between threads at every cycle to issue instructions from a ready thread, thus hiding latency at the cost of reduced single-thread throughput. 
\textit{\textbf{Coarse-grained multithreading}}~\cite{oehler1991ibm} switches only on long-latency events (e.g., a cache miss), which reduces the overhead of frequent context switching but provides less opportunity to overlap latency. 
\textit{\textbf{Simultaneous multithreading}} (SMT)~\cite{tullsen1996exploiting,Tullsen1995SimultaneousMM,yamamoto1995increasing,Hirata1992AnEP,yamamoto1994performance,serrano1994amodel} issues instructions from multiple threads in the same cycle, enabling the processor to better utilize its wide issue width, but increasing contention for shared pipeline resources. 
While these forms of multithreading can improve performance by exploiting otherwise idle cycles, they also exacerbate competition for scarce hardware resources. Constable mitigates such resource pressure and demonstrates higher performance gains when deployed in SMT configurations.

\subsection{Speculative Execution via Data Dependence Prediction}
\label{subsec:dep_pred}

Load data dependence significantly limits instruction-level parallelism (ILP)~\cite{tjaden1970detection,patt1985hps,jouppi1989available,smith1989limits,austin1995zero,Srinivasan1999LoadProcessors}. 
A major class of latency-tolerance techniques extracts ILP by exploiting speculative execution to overcome the data dependence induced by load instructions.
Load Value Prediction (LVP)~\cite{lipasti1996vp,lipasti1996value,sazeides_vp,sazeides_vp2,mendelson1997speculative,dfcm,last_n,selective_vp,perais2012revisiting,perais2014eole,perais2014practical,perais2015bebop,fvp,perais2021ssr,eves,rami_ap,rami_composite,kalaitzidis2019value,sakhuja2019combining,sazeides1999limits,zhou2003detecting,zhou2003enhancing} breaks load data dependence by predicting the value of a load instruction and speculatively executing load-data-dependent instructions with the predicted value. The predicted value is later verified by executing the load instruction. A correct prediction increases ILP, but an incorrect prediction leads to re-execution of load-dependent instructions, incurring both performance and power overheads.
Memory Renaming (MRN)~\cite{mrn,mrn2,mrn_classifying,moshovos1997streamlining,moshovos1999speculative} learns \rbd{the} dependency \rbd{relationship} between a store-load instruction pair, and speculatively executes the load-dependent instructions by forwarding the data directly from the associated store instruction. 
\rbd{The forwarded data is later verified by executing the load. A correct data forwarding increases ILP, whereas an incorrect forwarding incurs both performance and power overhead due to re-execution of load-dependent instructions.}
Memory Dependence Prediction (MDP)~\cite{store_set,yoaz1999speculation,Sha2006NoSQSC,Subramaniam2006StoreVF,Perais2018CostES,Kim2024EffectiveCM,MoshovosS00,jilp/MoshovosS00} extends speculation further by predicting whether a load is dependent on a preceding store with an unresolved address; if predicted independent, the load executes early, exposing additional instruction-level parallelism. These mechanisms collectively reduce stalls by overlapping useful computation with pending memory operations, thus significantly improving tolerance to long-latency loads.

However, all such speculative techniques still require the predicted load instruction to execute for correctness verification, which consumes critical hardware resources (e.g., reservation station, load port, and cache access bandwidth). As a result, while they effectively mitigate data dependence, they do not eliminate resource dependence, leaving untapped performance potential.
This shortcoming directly motivates Constable, which aims to mitigate both load data and resource dependence to deliver higher \rbtwo{power} efficiency.

\section{Techniques to Reduce Memory Latency}
\label{sec:rw_lat_red}

While prior techniques primarily aim to hide or tolerate memory latency, they do not fundamentally \emph{reduce} the latency experienced by a memory instruction. 
A complementary line of research targets direct latency reduction by shortening the latency of either of the two component operations of a memory instruction, i.e., memory address generation and memory access.
We classify such works into two broad categories: (1) techniques that reduce memory address generation latency, and (2) techniques that reduce memory access latency by exploiting device-level characteristics and optimizations.

\subsection{Reducing Memory Address Generation Latency}

Prior works \rbd{propose} both speculative and non-speculative techniques to accelerate the address generation of a load memory instruction. 
\cite{austin_load_addr} uses a fast, speculative carry-free addition to speed up load address computation.
\cite{austin1995zero} caches the values of recently-used registers to speculatively compute load address.
Register file prefetching (RFP)~\cite{rfp} predicts the load address of an instruction to prefetch its data to the register file.
Early load address resolution (ELAR)~\cite{elar}, on the other hand, tracks stack register \rbd{values} using a small computation unit in the decode stage to \emph{safely and non-speculatively} compute the load address of most stack loads immediately after \rbd{the} decode stage. 
While Constable bears resemblance to these works, it significantly differs from them in a major way: These prior works necessitate \rbd{the execution of} the load instruction whose load address has been computed early. 
Works that \emph{speculatively} compute load address~\cite{austin_load_addr,austin1995zero,rfp} need to execute the load to verify the speculation. ELAR, which employs a safe technique to \emph{non-speculatively} compute the address of stack loads, still needs to fetch the load data from the memory hierarchy. 
Constable safely eliminates \emph{both the address computation and the data \rbc{fetch} operations} of a load execution altogether.
We evaluate Constable against ELAR and  RFP demonstrating their performance \rbd{benefits}.

\subsection{Reducing Memory Access Latency by Exploiting Device-Level Characteristics}
\label{subsec:dram_arch}

In this section, we review a substantial body of prior work aimed at reducing memory access latency by exploiting device-level characteristics of memory technologies.
Given that dynamic random-access memory (DRAM) serves as the de-facto main memory technology in modern computing systems, we first discuss techniques that reduce DRAM access latency by leveraging its microarchitectural organization and device-level properties. 
The techniques we propose in this dissertation are largely orthogonal to the latency reduction techniques mentioned in this section.

\paraheading{Exploiting DRAM Microarchitecture.}
A significant corpus of research has reduced DRAM access latency by exploiting its internal microarchitectural structure~\cite{arp,lisa,Choi2015MultipleCloneRow,Gulur2012MultipleSubRow,Hidaka1990CacheDRAMArchitecture,salp,lee2013tiered,Lu2015DRAMLatencyDynamic,O2014RowBufferDecoupling,rowclone,Seshadri2015BulkBitwiseDRAM,Seshadri2015GatherScatterDRAM,Seshadri2016BuddyRAMPerformance,Seshadri2017AmbitMemoryAccelerator,Son2013MemoryAccessLatency,Zhang2014HalfDRAMBandwidth,mineshphd,hasanphd}. 
Subarray-Level Parallelism (SALP)~\cite{salp} overlaps operations within different subarrays of a bank, alleviating bank-level serialization and thereby speeding up access latencies with minimal hardware additions.
Tiered-Latency DRAM (TL-DRAM)~\cite{lee2013tiered} divides long bitlines into “near” and “far” segments using isolation transistors, allowing near-segment accesses at much lower latency without substantially increasing cost per bit.
Low-Cost Inter-Linked Subarrays (LISA)~\cite{lisa} introduces inexpensive isolation transistors between adjacent subarrays to enable ultra-fast inter-subarray bulk data movement.
Copy-Row DRAM (CROW)~\cite{crow} duplicates selected rows using an in-DRAM bulk copy mechanism~\cite{rowclone} and accesses them by simultaneously activating the regular and copied row. This dual-row activation drives the sense amplifiers faster, thereby reducing access latency.
CLR-DRAM~\cite{clr_dram} enables dynamic per-row reconfiguration between high-capacity and low-latency modes, significantly reducing DRAM timing parameters with modest area overhead.

\paraheading{Exploiting DRAM Device-Level Characteristics.}
DRAM manufacturers specify conservative timing margins to guarantee correct operation under worst-case conditions such as high temperature, low voltage, and process variation. 
These margins are often overly pessimistic for common-case operation, and multiple works show that they can be safely reduced while still ensuring correctness. 
Prior research has exploited diverse DRAM device-level characteristics (e.g., operating temperature~\cite{lee2015adaptive,date1}, DRAM cell's charge retention behavior~\cite{chargecache,yaohua2018micro}, and latency variation across cells and subarrays~\cite{chang2016understanding, solar_dram, diva}) to achieve this goal.
Other studies also leverage voltage–latency trade-offs~\cite{voltron}, application error tolerance~\cite{eden}, and  refresh/access parallelization~\cite{arp,hira} to reduce DRAM latency.

\section{Techniques to Improve Processor Efficiency}
\label{sec:rw_eff}

\subsection{Memoization}
\label{subsec:memoization}

Memoization~\cite{michie1968memo} caches computed results from \rbd{repeated} code executions, \rbd{enabling a} program or a microarchitecture to skip redundant computations when encountering identical input sets.
Memoization has been applied in both software~\cite{michie1968memo,richardson_function_memo,conners1999compiler,suresh2015intercepting,suresh2017compile,connors2000hardware} and in hardware at various program granularities, including instruction-level~\cite{harbison1982architectural,richardson1993exploiting,citron1998accelerating,sodani1997dynamic,sodani1998empirical,molina1999dynamic}, basic-block-level~\cite{huang1999exploiting}, trace-level~\cite{gonzalez1999trace}, and function-level~\cite{citron2000hardware}.
Early works on instruction-level memoization aim to accelerate long-latency operations (e.g., floating point multiplication and division) by storing their operands and results in value caches~\cite{richardson1993exploiting,oberman1996reducing}.
Sodani and Sohi propose a PC-indexed \emph{reuse buffer} to store the results of (multiple) dynamic instances of every static instruction~\cite{sodani1997dynamic}.
Molina et al. improve upon the reuse buffer to capture reuse of results across dynamic instances of different static instructions~\cite{molina1999dynamic}.

Constable, \rbd{in principle}, resembles instruction-level memoization with three key differences that make Constable more performant, lightweight, and usable in today's high-performance multi-core processors.
First, prior works aim to memoize (multiple) results of \emph{every} static instruction, irrespective of whether or not the results would be useful for instruction elimination~\cite{sodani_vp_reuse,sodani1997dynamic}. This requires a large memoization buffer, often as large as L1 data cache~\cite{citron2002revisiting,zhang2017leveraging,zhang2017leveraging2}, to capture elimination opportunities across long inter-occurrence distances (see \Cref{sec:cst_headroom_stable_load_charac}). Gonzalez et al. have shown that, while such a large memoization buffer may provide a significant performance benefit, the benefits reduce significantly when the latency to access the buffer is considered~\cite{gonzalez1998performance}.
Constable, on the other hand, (a) only targets loads, and (b) employs a confidence-based mechanism to filter out likely-stable load instructions from all loads.
This significantly reduces Constable's storage overhead and design complexity (e.g., port requirements of SLD as discussed in ~\Cref{sec:cst_design_design_decisions_sld}) while providing high elimination coverage.
\rbd{Second, prior works may delay retrieving the memoized instruction output until its source register values are available~\cite{harbison1982architectural,richardson1993exploiting,citron1998accelerating,molina1999dynamic}.
As a result, these works may not alleviate resource dependence on hardware structures like the reservation station.}
\rbd{Constable, however, explicitly monitors changes in source architectural registers of likely-stable load instructions and eliminates them early in the pipeline, alleviating resource dependence from both load reservation station and load execution unit.}
Third, prior works may not be applied in today's high-performance multi-core processors as they do not address challenges related to (a) keeping memoization buffer coherent across multiple cores, and (b) maintaining program correctness in presence of out-of-order load issuing. Constable addresses both these challenges (\Cref{sec:cst_design_in_flight_stores} and \Cref{sec:cst_design_multi_core_coherence}) and we extensively verify its correctness via functional simulation (\Cref{sec:cst_methodology_func_veri}).

\subsection{Dead Instruction Elimination}
\label{subsec:elimination}

Another line of research aims to improve processor \rbtwo{energy} efficiency by eliminating instructions that do not contribute to the final program outcome. 
Such \textit{dead instructions}~\cite{dead_instr} may arise due to compiler limitations, aggressive speculation, or dynamic program behavior, and they unnecessarily consume execution resources and energy. 
Prior works propose both compile-time analyses to statically remove dead code~\cite{Knoop1994PartialDC,Xi1999DeadCE,Damiani2000AutomaticUE} and runtime mechanisms to dynamically detect and eliminate instructions~\cite{dead_instr,micro/ButtsS02} whose results are never used. 
These approaches reduce pipeline pressure, free scarce microarchitectural resources, and improve overall energy efficiency without compromising correctness. 
In contrast, Constable targets redundant but \emph{live} load instructions, whose results are used in the program but repeatedly identical, and eliminates their execution by exploiting address-value stability.

\subsection{Dynamic Instruction Optimization}
\label{subsec:dyn_opt}

Prior works on trace-cache-based optimizations~\cite{trace_cache,trace_processor,rotenberg1997trace,rotenberg1999control,smith2002trace} and rePLay framework~\cite{replay,replay2} enable a wide range of runtime code optimizations (e.g., move elimination, zero-idiom elimination) that can eliminate instructions in microarchitecture. 
Continuous optimization (CO)~\cite{cont_opt} builds \rbd{on} these works and enables removing redundant instructions (including loads) using the register renaming logic.
Constable differs from these works in two key ways.
First, these schemes learn optimizations offline on a per-trace (or frame) basis. Constable learns optimization online and applies directly to the program's dynamic instruction stream.
Second, \rbd{unlike Constable}, CO \rbd{does not} eliminate a load instruction in a multi-core system \rbd{in order to maintain coherence}.

\section{Summary}
\label{sec:rw_summary}

In summary, the literature encompasses a rich body of techniques that aim to mitigate the memory bottleneck through diverse approaches, including caching and prefetching to hide latency, out-of-order and speculative execution to tolerate latency, microarchitectural and device-level optimizations to reduce latency, and instruction elimination mechanisms to improve \rbtwo{energy} efficiency. 
These works have substantially advanced the state-of-the-art in addressing the long-standing memory bottleneck problem and its impact on processor performance and energy efficiency. 
Building on this foundation, this dissertation proposes new ML-driven and data-aware microarchitectural techniques that further extend these directions and push the state-of-the-art in alleviating the ever-growing memory bottleneck.

\chapter{A Primer on Reinforcement and Perceptron Learning}
\label{chap:ml_primer}

Reinforcement learning~\cite{rl_bible} and perceptron learning~\cite{rosenblatt1958perceptron} are two powerful machine learning (ML) methods that have found increasing adoption in various aspects of processor design.
In this chapter, we first provide a brief background on these two learning methods, and then survey prior works that exploit them for processor design—ranging from managing microarchitectural mechanisms (e.g., caching, prefetching) to higher-level design tasks (e.g., floorplanning).

\section{Reinforcement Learning}
\label{sec:background_rl}

\begin{figure}[!ht]
\centering
\includegraphics[width=4.5in]{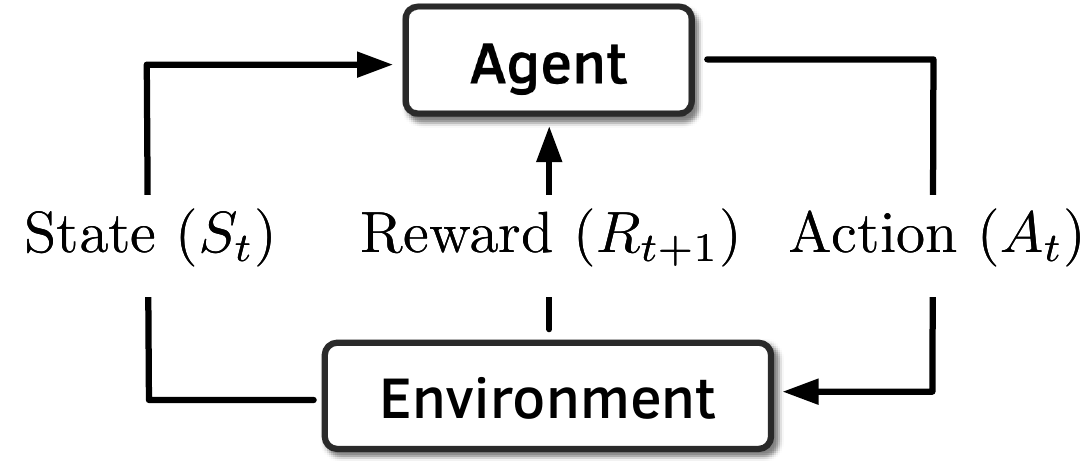}
\caption{Interaction between an agent and the environment in a reinforcement learning system.}
\label{fig:rl_basics}
\end{figure}

Reinforcement learning (RL)~\cite{rl_bible,rlmc}, in its simplest form, is the algorithmic approach to learn how to take \rbc{an} \emph{action} in a given \emph{situation} to maximize a numerical \emph{reward} signal. 
A typical RL system \rbc{comprises} of two main components: \emph{the agent} and \emph{the environment}, as shown in \Cref{fig:rl_basics}. The agent is the entity \rbc{that takes actions}.
\rbc{The agent resides in the environment and interacts with it in discrete timesteps.}
At each timestep $t$, the agent observes the current \textbf{\emph{state}} of the environment \textbf{$S_t$} and takes \textbf{\emph{action}} \textbf{$A_t$}. Upon receiving the action, the environment transitions to a new state $\mathbf{S_{t+1}}$, and emits an immediate \textbf{\emph{reward}} $R_{t+1}$, which is delivered immediately or later to the agent. The reward scheme encapsulates \rbc{the} agent's objective and drives the agent \rbc{towards taking} optimal actions.

\rbc{The \textbf{\emph{policy}} of the agent dictates it to take a certain action in a given state. \emph{The agent's goal is to find the optimal policy that maximizes the cumulative reward collected from the environment over time.}}
The expected cumulative reward by taking an action $A$ in a given state $S$ is defined as the \textbf{\emph{Q-value}} of the state-action pair (denoted as $Q(S,A)$).
\rbc{At} every timestep $t$, the agent iteratively optimizes its policy in two steps: (1) the agent updates the Q-value of a state-action pair using \rbc{the reward} collected in the current timestep, and (2) the agent \rbc{optimizes its current policy} using the newly updated Q-value.

\subsubsection{Updating \rbc{Q-values}} 
Prior works have proposed numerous algorithms to update the Q-values with varying computational complexities. 
SARSA~\cite{sarsa} is one such algorithm that strikes a good trade-off between the learning accuracy and the computational complexity, which makes it suitable for \emph{online} architectural decision making.
In SARSA, if at a given timestep $t$, the agent observes a state $S_t$, takes an action $A_t$, while the environment transitions to a new state $S_{t+1}$ and emits a reward $R_{t+1}$ and the agent takes action $A_{t+1}$ in the new state, the Q-value of the old state-action pair $Q(S_t, A_t)$ is iteratively optimized using \rbc{the} SARSA~\cite{sarsa, rl_bible} \rbc{algorithm,} as shown in Eqn.~\eqref{eq:sarsa}\rbc{:}

\begin{equation}\label{eq:sarsa}
\begin{aligned}
Q\left(S_t, A_t\right) & \gets Q\left(S_t, A_t\right)\\
&+ \alpha\left[R_{t+1}+\gamma Q\left(S_{t+1}, A_{t+1}\right) - Q\left(S_t, A_t\right)\right]
\end{aligned}
\end{equation} 

$\alpha$ is the \emph{learning rate} parameter \rbc{that} controls the convergence rate of Q-values.
$\gamma$ is the \emph{discount factor}, \rbc{which is used} to assign more weight to the immediate reward received by the agent at any given timestep than to the delayed future rewards. A $\gamma$ value closer to 1 gives a ``far-sighted" planning capability to the agent, i.e., the agent can trade off a low immediate reward to gain higher rewards in the future. This is particularly useful in creating an autonomous agent that can anticipate the \rbc{long-term} \rbc{effects} of taking an action to optimize its policy \rbc{that gets closer to optimal over time}.

\subsubsection{Policy Optimization} 
\label{subsubsec:rl_policy}

To find a policy that maximizes the cumulative reward collected over time, a purely-greedy agent always exploits the action $A$ in a given state $S$ that provides the highest Q-value $Q(S,A)$. However, greedy exploitation can leave the state-action space under-explored.
Thus, in order to strike a balance between exploration and exploitation, an $\epsilon$-greedy agent \emph{stochastically} takes a random action with a low probability of $\epsilon$ (called \emph{exploration rate}); otherwise, it selects the action that provides the highest Q-value~\cite{rl_bible}. 

In short, the Q-value serves as the cornerstone of reinforcement learning.
By iteratively learning Q-values of state-action pairs, an RL-agent \rbc{continuously optimizes its policy} to take actions \rbc{that get closer to optimal over time}.

\section{Perceptron Learning}
\label{sec:ml_primer_perceptron_learning}

\begin{figure}[!h]
\centering
\includegraphics[width=5in]{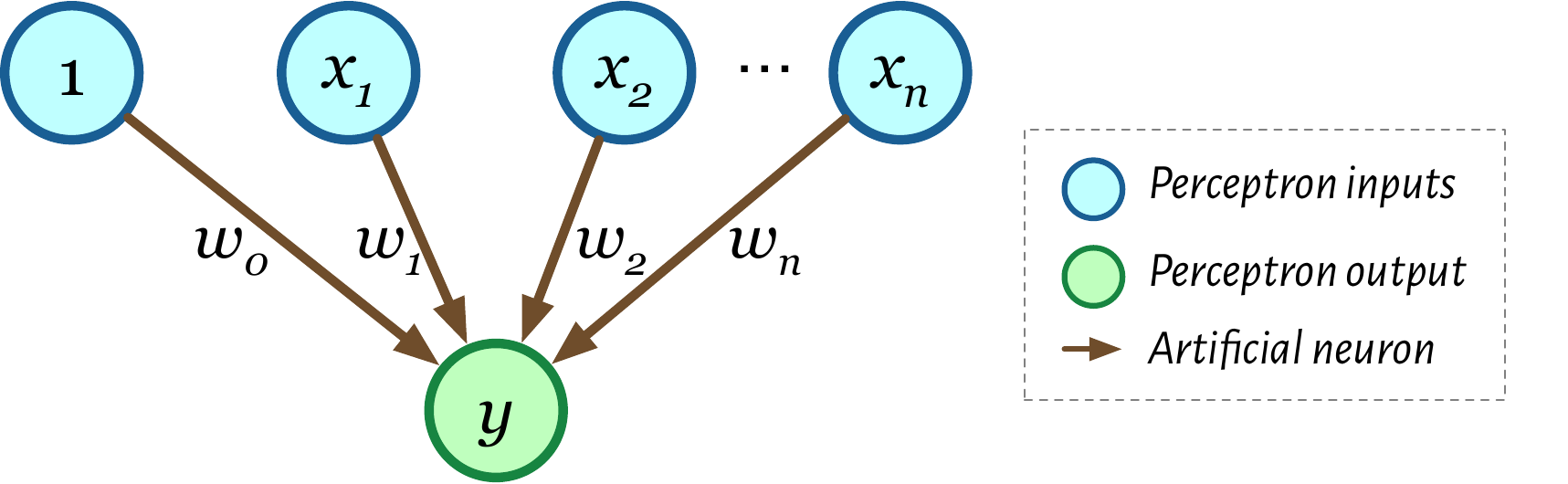}
\caption{Overview of a single-layer perceptron model. Each blue circle denotes an input and the green circle denotes the output of the perceptron.}
\label{fig:perc_basic}
\end{figure}

Perceptron learning, \rbe{whose roots \rbfor{can be traced} back to~\cite{mcculloch1943logical} and which was} demonstrated by Rosenblatt~\cite{rosenblatt1958perceptron}, is a simplified learning model that mimics biological neurons.
\Cref{fig:perc_basic} shows a \emph{single-layer perceptron} network where each \emph{input} is connected to the \emph{output} \rbe{via an} \emph{artificial neuron}. Each artificial neuron is represented by a numeric value, called \emph{weight}. 

\rbe{The perceptron network as a whole iteratively learns a binary classification function $f(x)$ (shown in Eq.~\ref{eq:perceptron}), a function that maps the input $X$ (a vector of $n$ values) to a single binary output.}

\begin{equation} \label{eq:perceptron}
f(x)=\left\{\begin{array}{ll}
1 & \text { if } w_{0}+\sum_{i=1}^{n} w_{i} x_{i}>0 \\
0 & \text { otherwise }
\end{array}\right.
\end{equation}

\rbe{The perceptron learning algorithm starts by initializing the weights of all neurons and iteratively trains the weights using each input vector from the training dataset in two steps.
First, for an input vector $X$, the perceptron network computes a binary output using Eq.~\ref{eq:perceptron} and the current weight values of its neurons. Second, if the computed output differs from the desired output for that input vector provided by the dataset, the weight of each neuron is updated~\cite{rosenblatt1958perceptron}. This iterative process is repeated until the error between the computed and desired output falls below a user-specified threshold.}

\section{Application of Machine Learning in Processor Design}

The works that apply ML-based methods in processor design can be broadly classified into two categories: (1) works that use ML to manage microarchitectural decision making, and (2) works that use ML for high-level design-space exploration.
We briefly summarize such works in this section.

\subsection{ML-Managed Microarchitectures} 
\label{subsec:ml_managed_uarch}

Researchers have proposed ML-based algorithms for various microarchitectural decision-making tasks. Prominent examples include memory scheduling~\cite{rlmc,morse}, cache \rbc{management}~\cite{glider, imitation,rl_cache,teran2016perceptron,balasubramanian2021accelerating}, branch direction prediction~\cite{tarsa,branchnet,perceptron,BranchNetV2_2020,Zouzias2021BranchPA,jimenez2016multiperspective,tarjan2005merging,jimenez2003fast,jimenez2002neural,gao2005adaptive}, branch target prediction~\cite{garza2019bit}, branch confidence estimation~\cite{akkary2004perceptron}, \rbc{address translation~\cite{margaritovTranslationNIPS18}}, cache reuse prediction~\cite{jimenez2017multiperspective,teran2016perceptron}, hardware prefetching~\cite{peled2018neural,peled_rl,hashemi2018learning,voyager,shi2019learning,voyager,zeng2017long,duong2024new}, and prefetch usefulness prediction~\cite{ppf,jamet2024tlp}.

\subsection{Application of ML in High-Level Processor/System Design}
\label{subsec:ml_driven_design}

Researchers have also explored ML techniques to explore the large microarchitectural design space, e.g., 
system modeling~\cite{dong2013circuit,deng2017memory,lin2019design,dai2016block2vec},
GPU performance modeling~\cite{baldini2014predicting,ardalani2019static,ardalani2015cross,jia2012stargazer,jooya2019multiobjective,o2017halwpe,wu2015gpgpu},
CPU performance modeling~\cite{joseph2006construction,joseph2006predictive,lee2006accurate,eyerman2011mechanistic,zheng2015learning,zheng2016accurate,Nasr2025Concorde},
data-center-scale system modeling~\cite{roy2011efficient,calheiros2014workload,khan2012workload,gong2010press,islam2012empirical,cortez2017resource,mahdisoltani2017proactive,xu2018improving},
NoC design~\cite{fettes2018dynamic,zheng2019energy,lin2020deep,yin2020experiences,ebrahimi2012haraq,ditomaso2016dynamic,clark2018lead,ditomaso2017machine,van2018extending,yin2018toward}, chip \rbfor{floorplanning}~\cite{mirhoseini2021,mirhoseini2018hierarchical,mirhoseini2017device,cheng_rebuttal,markov_rebuttal,goldie2024chipsailedcritiqueunfounded}, 
hardware resource assignment and task allocation~\cite{kao2020confuciux,dubach2010predictive,ganapathi2009case,gomez2001neuro,jain2016machine,lu2015reinforcement,wu2012inferred,esteves2018adaptive,yazdanbakhsh2021apollo},
power management~\cite{aboughazaleh2007integrated,won2014up,pd2015q,cochran2011pack,ma2012greengpu,imes2018energy,bai2017voltage,ravi2017charstar,pan2014scalable,juan2012power,hoffmann2015jouleguard},
scheduling~\cite{fedorova2007operating,vengerov2009reinforcement,whiteson2004adaptive},
and various aspects of chip design~\cite{rotman2020electric,rosa2020using,settaluri2020autockt,liu2020closing,wang2020gcn,zhu2019geniusroute,zacharopoulos2018machine,mehrabi2020prospector,wu2021ironman,neto2019lsoracle,haaswijk2018deep,hosny2020drills,pasandi2019approximate,wu2016flip,lu2019gan,cheng2018replace,lin2019dreamplace,ren2021nvcell,zhuo2017accelerating}.

\section{Summary and Further Reading}
In summary, this chapter outlines the foundations of reinforcement and perceptron learning and surveys a broad set of works that employ machine learning for processor design, ranging from microarchitectural mechanisms to higher-level design tasks. 
Together, these discussions underscore the increasing importance of data-driven methods in modern architecture, complementing the dissertation’s broader vision of moving beyond data-agnostic design. 
We refer interested readers other works that give a \rbfor{more} comprehensive view of machine-learning-aided techniques in processor and system design~\cite{ml_survey1, ml_survey2,mutlu2021intelligent}.
The next chapters build on the background presented here to progressively propose, design, and evaluate machine-learning-based microarchitectural techniques.

\chapterRB{Validating Generalization of ML-Driven Techniques}{Validating Generalization of ML-Driven Microarchitectural Techniques to Emerging Artificial Intelligence (AI) and Industrial Workloads}
\label{chap:dpc4}

\section{The Generalization Challenge}

The design of conventional microarchitectural techniques has predominantly relied on rigid, rule-based heuristics. 
These heuristics, often derived from a computer architect’s intuition or exhaustive search over a narrow set of workloads, aim to identify a \textit{general-case} behavior that yields performance gains across common benchmarks. 
While effective for well-characterized workloads, such static designs lack the intrinsic flexibility required to adapt to rapidly-evolving modern workloads, as we show in~\Cref{chap:intro}.

In this dissertation, we propose a fundamental shift in this design paradigm by replacing human-crafted heuristics with machine learning (ML)-driven microarchitectural policies. By exploiting ML methods, such as reinforcement learning (RL) in Pythia~\cite{pythia} and Athena~\cite{athena} and perceptron learning in Hermes~\cite{hermes}, we enable the microarchitecture to autonomously and continuously learn from application and system-generated data.
However, such ML-driven policies necessitate a rigorous verification of their ability to \textit{generalize} their performance benefits across a broad range of workloads beyond those used at the design time.

We observe that the evaluation of many prior microarchitectural techniques often suffers from design-time bias, i.e., these techniques are often iteratively tuned and evaluated to maximize the performance on the same set of well-known standard benchmarks (e.g., SPEC CPU workloads)~\cite{jimenez2010dead,jimenez2017multiperspective,sdbp,hawkeye,d2d,d2m,lp,loh_miss_map,alloy_cache,isb,misb,triage,baer2,jouppi_prefetch,ampm,fdp,footprint,bop,sandbox,dol,ipcp,mlop,hashemi2016continuous,fdp,ebrahimi2009coordinated,ebrahimi_paware,ebrahimi2009techniques,lee2008prefetch,panda2016expert,seshadri2015mitigating,wu2011pacman,memory_bank,seq_stream,hur,eris2022puppeteer,mab,micro_mama,mutlu2008parallelism,kim2010atlas,ebrahimi_fst,ebrahimi2011parallel,subramanian2016bliss,rlmc,mutlu2007stall,Moscibroda2007FairQueuing,ausavarungnirun2012staged,Ghose2013ImprovingMS,seshadri2015mitigating,seshadri2012evicted,alameldeen2004adaptive,alameldeen2004frequent,chen2009c,dusser2009zero,villa2000dynamic,yang2000frequent,zhang2000frequent2,pekhimenko2012base}.
We argue that this conventional evaluation methodology is inadequate for validating an ML-driven microarchitectural technique for two reasons.
First, while standard benchmark suites provide a strong baseline, prior works have shown that they often do not represent complex workload \rbfiv{behaviors} of contemporary production workloads, such as emerging artificial intelligence (AI) applications~\cite{llama2,whisper,vit,sd1,biogpt,clip_ml} and workloads running in large-scale datacenter \rbfor{fleets}~\cite{warehouse,asmdb,song2022thermometer,khan2022whisper,khan2021ripple,khan2021twig,ranganathan2021warehouse,schall2024last,schall2024last2,cloudsuite,amiraliphd,wenisch2008temporal,ferdman2012quantifying,ferdman2014case,lotfi2012scale,palit2016demystifying,falsafi2024server}.
Second, using the same workloads for both iterative tuning and final evaluation risks overfitting the ML-driven policy.
Thus, to truly demonstrate the benefits of the ML-driven microarchitectural techniques presented in this dissertation, we must demonstrate that the proposed techniques do not merely \emph{memorize} the characteristics of well-known standard benchmarks, but instead \emph{generalize} and \emph{adapt} to \rbfor{previously-unseen} emerging workloads.

\paraheading{The goal} of this chapter is to establish a rigorous and systematic methodology to validate the generalization of the ML-driven microarchitectural techniques introduced in this dissertation.
To this end, we leverage the unique infrastructure and workload traces we developed for the \textit{4th Data Prefetching Championship} (DPC4)~\cite{dpc4} to serve as an independent, industrial-grade \emph{stress test} for our mechanisms.

\section{The 4th Data Prefetching Championship (DPC4)}
\label{sec:dpc4_infra}

We organized the 4th Data Prefetching Championship (DPC4)~\cite{dpc4}, in conjunction with the 32nd IEEE International Symposium on High-Performance Computer Architecture (HPCA) on February 1, 2026, in Sydney, Australia. 
The goal of the championship was to evaluate innovative data prefetching ideas under a common evaluation framework using a broad range of diverse emerging workloads. 

While prior data prefetching championships primarily relied on conventional benchmark suites, most notably SPEC CPU workloads, as the basis for evaluation~\cite{dpc2,dpc3}, we substantially expanded the workload corpus used in DPC4. 
We collected and open-sourced $610$ single-core workload traces spanning multiple emerging application domains. 
Of these, $483$ traces have \emph{never been publicly released or evaluated} by the computer architecture research community prior to DPC4. This includes (1) $81$ traces from emerging state-of-the-art AI applications (e.g., Llama2~\cite{llama2}, ViT~\cite{vit}, CLIP~\cite{clip_ml}, Stable Diffusion~\cite{sd1}, Whisper~\cite{whisper}, and BioGPT~\cite{biogpt}), (2) $43$ traces from graph mining workloads~\cite{gms}, and (3) $359$ traces captured from real-world workloads running on Google's datacenter fleet~\cite{gtrace_v2}.
These workloads exhibit diverse and complex memory access characteristics that differ significantly from those observed in traditional SPEC CPU benchmarks.
By incorporating these traces into a unified evaluation framework, DPC4 provided a substantially more representative stress-testing environment for modern data prefetching mechanisms.
All DPC4 competitors \rbfiv{were} evaluated using the entire trace corpus across three processor configurations to tangibly improve the state-of-the-art in data prefetching~\cite{dpc4_bertigo,dpc4_edp,dpc4_emender,dpc4_gberti,dpc4_sberti,dpc4_sppam,dpc4_umama,dpc4_vip}.
All traces can be downloaded freely from the DPC4 repository~\cite{dpc4_github}.

\sectionRB{Exploiting DPC4 Infrastructure for Validation}{Exploiting DPC4 Infrastructure to Validate Generalization of ML-Driven Microarchitectural Techniques}{sec:dpc4_meth}

\noindent We leverage the DPC4 infrastructure to establish a rigorous evaluation methodology to validate the generalization capability of the three ML-driven microarchitectural techniques proposed in this dissertation (i.e., Pythia~\cite{pythia} in~\Cref{chap:pythia}, Hermes~\cite{hermes} in~\Cref{chap:hermes}, and Athena~\cite{athena} in~\Cref{chap:athena}).

More specifically, we use the corpus of $483$ previously-unseen single-core DPC4 workload traces that span across three application domains of AI, graph mining, and Google datacenter, to evaluate all three ML-driven techniques. 
To further increase our evaluation rigor, we also construct $966$ random four-core trace mixes from the $483$ single-core traces using both intra-domain (i.e., picking four traces at random from the same application domain) and inter-domain (i.e., picking four traces at random from any application domain) trace mixing strategies.
\Cref{table:dpc4_traces} summarizes the traces considered in this evaluation methodology.

\begin{table}[htbp]
  \centering
  \small
    \begin{tabular}{c||L{20em}||r}
    \thickhline
          & \textbf{Workload type} & \multicolumn{1}{l}{\textbf{\# traces}} \\
    \thickhline
    \multirow{3}[6]{*}{Single-core} & \Tabval{AI/ML applications (\rbfor{AIML})} & \Tabval{81} \\
        \cline{2-3}          & \Tabval{Graph Mining Suite (GMS)} & \Tabval{43} \\
        \cline{2-3}          & \Tabval{Google datacenter workloads (Google)} & \Tabval{359} \\
        \cline{2-3}          & \Tabval{\textbf{\textit{Total}}} & \Tabval{\textbf{483}} \\
    \hline
    \hline
    \multirow{4}[8]{*}{Four-core} & \Tabval{AI application mix (AIML-mix)} & \Tabval{81} \\
        \cline{2-3}          & \Tabval{Graph Mining application mix (GMS-mix)} & \Tabval{43} \\
        \cline{2-3}          & \Tabval{Google workload mixes (Google-mix)} & \Tabval{359} \\
        \cline{2-3}          & \Tabval{Random application mixes (Random-mix)} & \Tabval{483} \\
        \cline{2-3}          & \Tabval{\textbf{\textit{Total}}} & \Tabval{\textbf{966}} \\
    \thickhline
    \end{tabular}%
  \caption{DPC4 workload traces used for evaluation.}
  \label{table:dpc4_traces}%
\end{table}%

\paraheading{Rationale \rbfor{Behind} the \rbfor{Methodology}. }
As all three ML-driven techniques proposed in this dissertation - Pythia, Hermes, and Athena - have been designed, implemented, and finalized long \emph{before} the collection of DPC4 workload traces, this trace corpus never contributed to the design-space exploration (e.g., feature selection, hyperparameter optimization, reward design, policy structure selection, or architectural parameter tuning) of Pythia, Hermes, or Athena.
Moreover, we do not perform any additional hyperparameter search, feature engineering, or architectural adjustment for these traces. 
Instead, \rbfor{we evaluate each mechanism with its baseline configuration}. 
Thus, this methodology enables us to empirically answer a central research question: do ML-driven microarchitectural techniques merely \emph{memorize} the characteristics of familiar benchmark suites, or do they \emph{learn} policies that \emph{generalize} across fundamentally different and previously-unseen workloads? By demonstrating sustained performance improvements under this ``blind'' evaluation methodology (as we show in~\Cref{subsec:pythia_dpc4},~\Cref{subsec:hermes_dpc4}, and~\Cref{subsec:athena_dpc4}), we provide strong empirical evidence that the proposed mechanisms generalize beyond their design-time workloads and adapt effectively to emerging AI and industrial applications.

\section{Summary}

This chapter addresses the methodological challenge of validating whether ML-driven microarchitectural techniques generalize beyond their design-time workloads. We argue that conventional evaluation practices, which rely heavily on standard benchmark suites such as SPEC CPU, risk design-time bias and potential overfitting. To address this limitation, we introduce the 4th Data Prefetching Championship (DPC4) infrastructure, which expands the evaluation space to emerging AI applications, graph mining workloads, and real-world workloads running on Google's datacenter fleet. The DPC4 trace corpus contains $610$ traces, including $483$ traces previously unseen by the research community and exhibiting memory behaviors distinct from traditional benchmarks.

We leverage this corpus to construct a blind stress-testing methodology for Pythia, Hermes, and Athena. We evaluate \rbfor{all three} mechanisms on the $483$ unseen single-core traces and on $966$ randomly generated four-core mixes derived from them. 
Importantly, we finalized \rbfor{the three} mechanisms prior to the collection of these traces and performed no additional hyperparameter tuning, feature engineering, or architectural adjustments. The results presented in subsequent chapters demonstrate that the proposed ML-driven techniques sustain their performance benefits under \rbfiv{these} rigorous and previously-unseen workloads, thereby providing strong empirical evidence of their generalization capability.

\chapterRB{Hardware Prefetching using Reinforcement Learning}{Hardware Prefetching using Online Reinforcement Learning}
\label{chap:pythia}

\noindent Prefetching is a key memory latency-hiding technique employed in most high-performance processor. A prefetcher predicts the addresses of long-latency memory requests and fetches the corresponding data from main memory to on-chip caches before the \rbc{program executing on the processor} demands it. 
In this chapter, we study the behavior of a wide range of prefetchers proposed in the literature, build a comprehensive understanding of how the data-agnostic nature of their design restricts their effectiveness, and propose a novel data-driven hardware prefetcher exploiting reinforcement learning that advances the state-of-the-art in prefetching.

\sectionRB{Pythia: Motivation and Goal}{Motivation and Goal}{sec:pythia_motivation}

\noindent \rbc{As a program repeatedly accesses over its data structures, it \rbfiv{creates} patterns in its memory request addresses}.
A prefetcher tries to identify such memory access patterns from past memory requests to predict \rbc{the addresses of} future memory requests. To quickly identify a memory access pattern, a prefetcher typically uses some program context information to examine only \rbc{a subset} of memory requests. We call this program context a \emph{feature}. 
\rbc{The prefetcher associates a memory access pattern with a feature and generates prefetches following the same pattern \rbc{when} the feature reoccurs \rbc{during program execution}}.

\rbc{Past research has} proposed numerous prefetchers that consistently pushed the limits of prefetch coverage (i.e., the fraction of memory requests predicted by the prefetcher) and accuracy (i.e., the fraction of \rbc{prefetch} requests that are \rbc{actually} demanded by the program) by exploiting various program features, e.g., program counter (\texttt{PC}), cacheline address (\texttt{Address}), page offset of a cacheline (\texttt{Offset}), or \rbc{a simple} combination of such features using simple operations like concatenation (\texttt{+}) ~\cite{stride,streamer,baer2,stride_vector,jouppi_prefetch,ampm,fdp,footprint,sms,sms_mod,spp,vldp,sandbox,bop,dol,dspatch,bingo,mlop,ppf,ipcp}. 
For example, a PC-based stride prefetcher~\cite{stride,stride_vector,jouppi_prefetch} uses PC as the feature to learn the constant stride between two consecutive memory accesses \rbc{caused} by the same PC. VLDP~\cite{vldp} and SPP~\cite{spp} use a sequence of cacheline address deltas as the feature to predict the next cacheline \rbc{address} delta. Kumar and Wilkerson~\cite{footprint} use \texttt{PC+Address} of the first access \rbc{in} a \rbc{memory} region as the feature to predict the \rbc{spatial} memory access footprint \rbc{in} the entire memory region. SMS~\cite{sms} empirically finds \texttt{PC+Offset} of the first access \rbc{in a memory region} to be a better feature to predict the \rbc{memory access} footprint. Bingo~\cite{bingo} combines the features from~\cite{footprint} and SMS and uses \rbc{\texttt{PC+Address} and \texttt{PC+Offset}} \rbc{as its features}.

\rbc{Accurate and timely prefetch requests reduce} the long memory access latency \rbc{experienced by} the \rbc{processor}, \rbc{thereby} improving overall \rbc{system} performance. \rbc{However}, \rbc{speculative prefetch requests} \rbc{can} cause \rbc{undesirable} effects on the system (e.g., increased memory bandwidth consumption, cache pollution, \rbc{memory access interference,} etc.), which can reduce or negate the performance improvement gained by hiding memory access latency~\cite{fdp,ebrahimi2009coordinated}. \rbc{Thus, a good prefetcher aims to maximize its benefits while minimizing its undesirable effects on the system.}

\subsection{Key Observations} 
\label{subsec:pythia_ovservation}

\rbc{Even though there is a large number of prefetchers proposed in the literature, we observe 
three key shortcomings in almost every prior prefetcher design that significantly limits \rbc{its} performance benefits over a wide range of workloads and system configurations:
(1) \rbc{the use of \rbc{mainly} a single program feature for prefetch prediction}, (2) lack of inherent system awareness, and (3) \rbc{lack of ability to customize \rbc{the} prefetcher design to seamlessly adapt to a wide range of workload \rbc{and system configurations}}}.

\subsubsection{Prefetching using a Single Program Feature} 

\begin{figure}[!ht]
\centering
\includegraphics[width=5.75in]{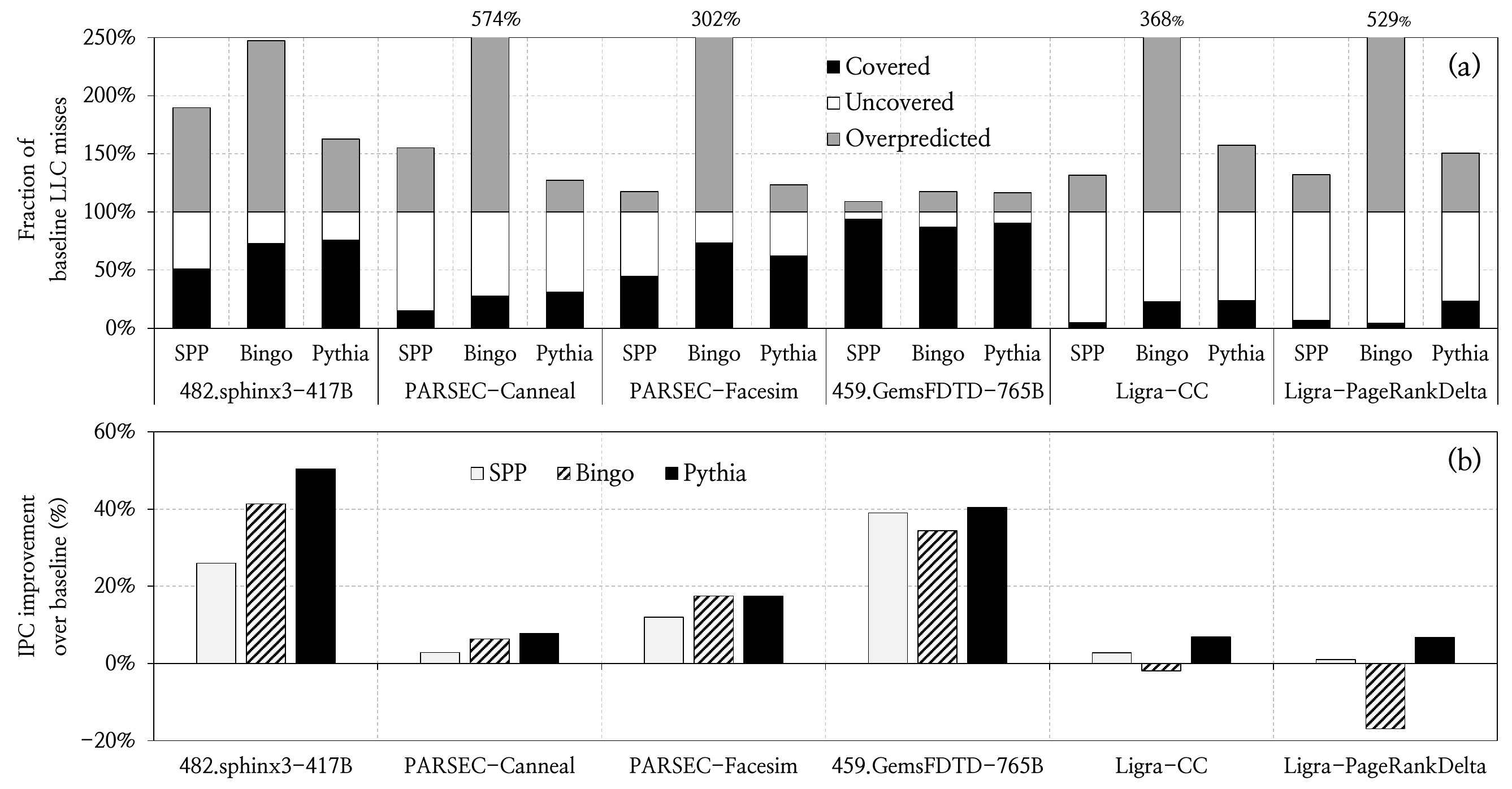}
\caption{Comparison of \rbc{(a) coverage, overprediction, and (b) performance of two recently-proposed prefetchers, SPP~\cite{spp} and Bingo~\cite{bingo}, and our new proposal, Pythia.}}
\label{fig:py_intro_perf_cov_acc}
\end{figure}

\rbc{Almost every prior prefetcher relies on \emph{only one} program feature to correlate with the program memory access pattern and generate prefetch requests~\cite{stride,streamer,baer2,stride_vector,jouppi_prefetch,ampm,fdp,footprint,sms,sms_mod,spp,vldp,sandbox,bop,dol,dspatch,mlop,ppf,ipcp}. As a result, a prefetcher typically \rbc{provides} good \rbc{(or poor)} performance benefits in \rbc{mainly} those workloads where the correlation between the feature used by the prefetcher and program's memory access pattern is dominantly present \rbc{(or absent)}.}
\rbc{To demonstrate this, we show the coverage and overpredictions (i.e., \rbc{prefetched} memory requests that do \emph{not} get demanded by the processor) of two recently proposed prefetchers, SPP~\cite{spp} and Bingo~\cite{bingo}, and our new proposal Pythia (see~\Cref{sec:py_design}) for six \rbc{example} workloads (\Cref{sec:pythia_methodology} discusses our experimental methodology) in \Cref{fig:py_intro_perf_cov_acc}(a). \Cref{fig:py_intro_perf_cov_acc}(b) shows the performance of SPP, Bingo and Pythia on the same workloads.}
As we see in \Cref{fig:py_intro_perf_cov_acc}(a), Bingo provides higher prefetch coverage than SPP in \texttt{sphinx3}, \texttt{PARSEC-Canneal}, and \texttt{PARSEC-Facesim}, where the correlation exists between the first access \rbc{in} a \rbc{memory} region and the other accesses \rbc{in} the same region. \rbc{As a result, Bingo performs better than SPP in these workloads (\Cref{fig:py_intro_perf_cov_acc}(b)).} 
\rbc{In contrast}, for workloads like \texttt{GemsFDTD} that have regular access patterns within a physical page, SPP's \emph{sequence of deltas} feature \rbc{provides} better coverage and performance than Bingo.

\subsubsection{Lack of Inherent System Awareness}
\rbc{All prior prefetchers either completely neglect their undesirable effects on the system (e.g., memory bandwidth usage, cache pollution, memory access interference, system energy consumption, etc.)~\cite{stride,streamer,baer2,stride_vector,jouppi_prefetch,ampm,footprint,sms,sms_mod,spp,vldp,sandbox,bop,dol,bingo,mlop,ppf,ipcp} or incorporate system awareness as \rbc{an afterthought \rbc{(i.e., a separate control component)}} to the underlying system-unaware prefetch algorithm~\cite{fdp,ebrahimi2009coordinated,ebrahimi2009techniques,ebrahimi_paware,dspatch,mutlu2005,wflin,zhuang,wflin2,charney,memory_bank,lee2008prefetch}. Due to the lack of inherent system awareness,}
a prefetcher often loses its performance gain in resource-constrained scenarios.
\rbc{For example, as shown in} \Cref{fig:py_intro_perf_cov_acc}(a), Bingo achieves similar \rbc{prefetch} coverage in \texttt{Ligra-CC} as compared to \texttt{PARSEC-Canneal}, while generating significantly lower overpredictions in \texttt{Ligra-CC} than \texttt{PARSEC-Canneal}.
However, 
Bingo loses performance in \texttt{Ligra-CC} by $1.9$\% compared to a no-prefetching baseline, whereas it improves performance by $6.4$\% in \texttt{PARSEC-Canneal} (\Cref{fig:py_intro_perf_cov_acc}(b)).
\rbc{
This contrasting outcome is due to Bingo's lack of awareness \rbc{of} the memory bandwidth usage.
\rbc{Without prefetching, }\texttt{Ligra-CC} consumes higher memory bandwidth than \texttt{PARSEC-Canneal}. As a result, 
\rbc{each overprediction}
made by Bingo in \texttt{Ligra-CC} wastes more precious \rbc{memory} bandwidth and is more detrimental to performance than 
\rbc{that in}
\texttt{PARSEC-Canneal}.}

\subsubsection{Lack of Online Prefetcher Design Customization}
\rbc{The high design complexity of architecting a multi-feature, system-aware prefetcher has traditionally compelled architects to statically select only one program feature at design time. With every new prefetcher, architects design new rigid hardware structures to exploit the selected program feature.}
To exploit a new program feature for higher performance benefits, one must design a new prefetcher from scratch \rbc{and} extensively evaluate and verify \rbc{it both \rbc{in} pre-silicon and post-silicon realization}.
Due to the rigid design-time decisions, the hardware structures proposed by prior prefetchers cannot be customized \rbc{online} in silicon either to exploit any other program feature or to change the prefetcher’s objective (\rbc{e.g.}, to increase/decrease coverage, accuracy, or timeliness) \rbc{so that it can seamlessly adapt to} varying workloads and system configurations.

\subsection{Our Goal}

Our goal is to design a single prefetching framework that (1) can holistically learn to prefetch using both \emph{multiple different types of program features} and \emph{system-level feedback} \rbc{information that is inherent to the design}, and (2) can be \emph{easily customized} in silicon via simple configuration registers to exploit 
\rbc{different types of program features and/or}
to change the objective of the prefetcher \rbc{(e.g., increasing/decreasing coverage, accuracy, or timeliness)} without any changes to the underlying hardware.

\section{Formulating Prefetching using Reinforcement Learning}

To this end, we formulate prefetching from the grounds-up as a reinforcement learning (RL) problem. More specifically, we propose a new hardware data prefetcher, named \emph{\textbf{Pythia}},\footnote{Pythia, according to Greek mythology, is the oracle of Delphi who is known for
accurate prophecies~\cite{pythia_wiki}.} based on RL, where the prefetcher itself acts as the RL agent, the processor and memory system act as the environment, and the agent (i.e., the prefetcher) \emph{autonomously learns} to prefetch by interacting with its environment. 

\subsection{Why \rbc{is} RL a Good Fit for Modeling Prefetching?} \label{sec:py_background_rl_prefetching}
RL \rbc{has} been recently successfully demonstrated to solve complex problems like mastering human-like control \rbc{on} Atari~\cite{deepmind_atari} and Go~\cite{alpha_go, alpha_zero}.
\rbc{We argue that RL is an inherent fit to model a hardware prefetcher for three key reasons.}

\paraheading{Adaptive Learning in \rbc{a Complex State Space}.}
As we show in \cref{sec:pythia_motivation}, the \rbc{benefits} of a prefetcher not only \rbfiv{depend} on its coverage and accuracy but also on its \rbc{undesirable effects} on the system, like \rbc{memory bandwidth usage}.
In other words, \emph{it is not sufficient for a prefetcher only to make highly accurate predictions}. Instead, a prefetcher should be \emph{performance-driven}. 
A prefetcher should have the capability to adaptively trade-off coverage for higher accuracy (and vice-versa) depending on its impact on the overall system to provide a robust performance improvement with varying \rbc{workloads and system configurations}. This adaptive and performance-driven nature of prefetching \rbc{in a complex state space} makes RL a good fit for modeling a prefetcher \rbc{as an autonomous agent who learns to prefetch by interacting with the system.}
    
\paraheading{Online Learning.}
\rbc{An RL agent \emph{does not} require an expensive offline training phase. Instead, it can \emph{continuously} learn \emph{online} by iteratively optimizing its policy using the rewards received from the environment. A hardware prefetcher, similar to an RL agent, also needs to continuously learn from the changing workload behavior and system \rbc{conditions} to provide consistent performance benefits. The online learning requirement of prefetching makes RL an inherent fit to model a hardware prefetcher.}

\paraheading{Ease of Implementation.}
Prior works have evaluated many sophisticated \rbc{machine} learning models like simple neural \rbc{networks}~\cite{peled2018neural}, \rbc{LSTMs}~\cite{hashemi2018learning,voyager}, and Graph Neural \rbc{Networks} (GNNs)~\cite{shi2019learning} as \rbc{models for} hardware \rbc{prefetching}. \rbc{Even though} these techniques show encouraging results in accurately predicting memory accesses, they fall short \rbc{especially} in two major aspects. First, these models' \rbc{sizes} often exceed even the \rbc{largest} \rbc{caches in traditional processors}~\cite{peled2018neural,hashemi2018learning,voyager,shi2019learning}, making them impractical \rbc{(or at best very difficult) to implement}. 
Second, due to the vast amount of computation \rbc{they require for inference}, these models’ inference latency is much higher than an acceptable latency of a prefetcher at any cache level.
On the other hand, we can efficiently implement an RL-based model, as we \rbc{demonstrate} in this work \rbc{(see~\Cref{sec:py_design})}, that can \emph{quickly} make predictions and can be \rbc{relatively} easily adopted in a real processor.

\sectionRB{Pythia: Overview}{Pythia: Overview}{sec:py_key_idea_formulation}

\begin{figure}[!h]
\centering
\includegraphics[width=5in]{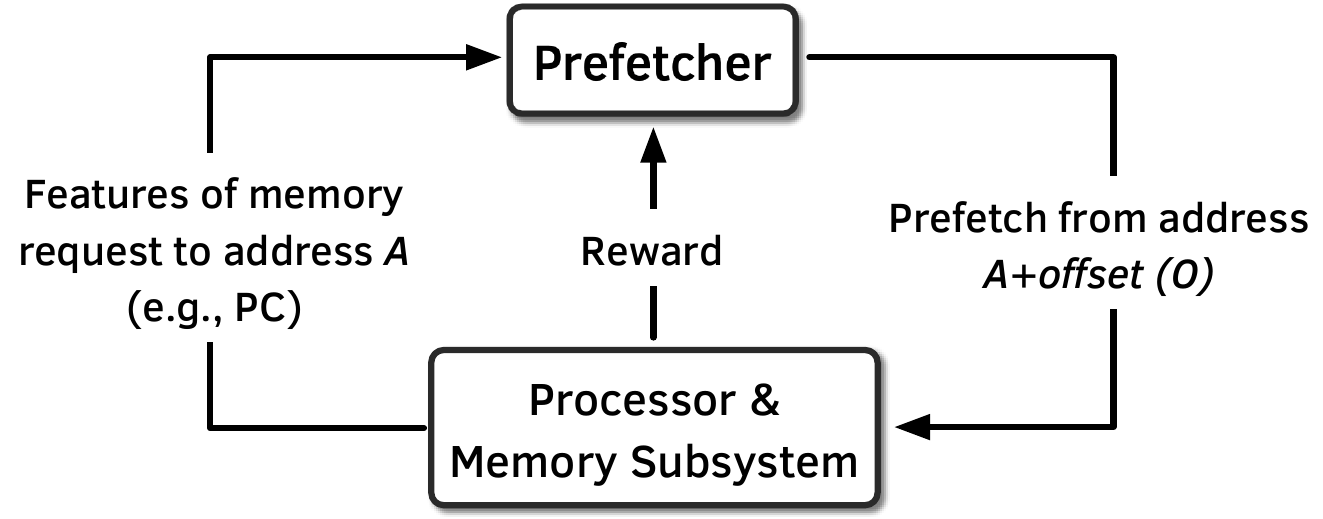}
\caption{Formulating \rbc{the} prefetcher as an RL-agent.}
\label{fig:py_rl_as_prefetcher}
\end{figure}

\noindent Pythia formulates prefetching as a reinforcement learning problem, as shown in~\Cref{fig:py_rl_as_prefetcher}. Specifically, Pythia as the RL-agent that learns to make accurate, timely, and system-aware prefetch decisions by interacting with the environment, i.e., the processor and the memory subsystem.
Each timestep corresponds to a new demand request seen by Pythia. With every new demand request, Pythia observes the state of the processor and the memory subsystem and takes a prefetch action.
\rbc{For every prefetch action (including \rbc{\emph{not to prefetch}}), Pythia receives a numerical reward \rbc{that} evaluates the accuracy and timeliness of the prefetch action \rbc{taking into account} various system-level feedback information. 
\emph{Pythia's goal is to find the optimal prefetching policy that would maximize the number of accurate and timely prefetch requests, \rbc{taking} system-level feedback information \rbc{into account}}. While Pythia's framework is general enough to incorporate any type of system-level feedback \rbc{into its decision making}, in this \rbfor{work} we demonstrate Pythia using \rbc{\emph{memory bandwidth usage}} as the system-level feedback information.}

We now formally define the three pillars of our RL-based prefetcher: the state \rbc{space}, \rbc{the} actions, and the \rbc{reward} scheme.

\subsection{State} 
We define the state as a $k$-dimensional vector of program features. 
\begin{equation}\label{eq:py_state}
    S \equiv \{ \phi_{S}^{1}, \phi_{S}^{2}, \ldots, \phi_{S}^{k} \}
\end{equation}
Each program feature is composed of \rbc{at most} two components: (1) program control-flow component, and (2) program data-flow component. \rbc{The} control-flow component is further made \rbc{up} of simple information like \rbc{load-PC (i.e., the PC of a load instruction)} or \rbc{branch-PC (i.e., the PC of a branch instruction that immediately precedes a load instruction)}, and a history that denotes whether this information is extracted only from the current demand request or a series of past demand requests.  
Similarly, the data-flow component is made \rbc{up} of simple information like cacheline address, \rbc{physical} page \rbc{number}, page offset, cacheline delta, and its corresponding history. 
\Cref{table:py_features} shows some example program features. 
\rbc{Although} Pythia can \rbc{theoretically} learn to prefetch \rbc{using} \emph{any} number of such program features, we fix the state-vector dimension (i.e., $k$) at design time \rbc{given a limited storage budget in hardware}. However, the exact selection of $k$ program features out of all possible program features is configurable \rbc{online} using simple configuration registers. 
In~\Cref{sec:py_tuning_feature_selection}, we \rbc{provide} an \emph{automated feature selection method} to find a vector of program features \rbc{to be used at design time}.

\begin{table}[htbp]
  \centering
  \small
    \begin{tabular}{m{8.8em}C{3em}C{4em}C{6.7em}C{4em}}
    \toprule
    \multicolumn{1}{l}{\multirow{2}[4]{*}{\textbf{Feature}}} & \multicolumn{2}{c}{\textbf{Control-flow}} & \multicolumn{2}{c}{\textbf{Data-flow}} \\
    \cmidrule{2-5} & \textbf{Info.} & \textbf{History} & \textbf{Info.} & \textbf{History} \\
    \midrule
    Last 3-PCs & PC & last 3 & \xmark  & \xmark \\
    Last 4-deltas & \xmark  & \xmark  & Cacheline delta & last 4 \\
    PC+Delta & PC & current & Cacheline delta & current \\
    Last 4-PCs+Page \rbc{no.} & PC & last 4 & Page \rbc{no.} & current \\ 
    \bottomrule
    \end{tabular}%
  \caption{Example program features.}
  \label{table:py_features}%
\end{table}%

\subsection{Action} 
We define the action of the RL-agent as selecting a \rbc{\emph{prefetch offset}} (i.e., \rbc{a} delta, \rbc{"O" in~\Cref{fig:py_rl_as_prefetcher}}, between the predicted and the \rbc{demanded} cacheline address) from a set of candidate prefetch offsets.
As every post-L1\rbc{-cache} prefetcher generates prefetch requests within a physical page~\cite{ampm,fdp,footprint,sms,sms_mod,spp,vldp,sandbox,bop,dol,dspatch,bingo,mlop,ppf,ipcp}, the list of prefetch offsets only contains values in the range of $[-63,63]$ for a system with a traditionally-sized $4$KB page and $64$B cacheline. 
Using prefetch offsets as \rbc{actions} (instead of full cacheline addresses)
drastically reduces the action space size. We further reduce the action space size by fine tuning, as described in\Cref{sec:py_tuning_action_selection}.
\rbc{A prefetch offset of} zero means no \rbc{prefetch is generated}.

\subsection{Reward}
The reward structure defines the prefetcher's objective.
We define five different reward levels as follows.
\begin{itemize}
    \item \textbf{\emph{Accurate and timely}} ($\mathcal{R}_{AT}$). This reward is assigned to an action whose corresponding prefetch address gets demanded \emph{after} the prefetch fill.
    \item \textbf{\emph{Accurate but \rbc{late}}} ($\mathcal{R}_{AL}$). This reward is assigned to an action whose corresponding prefetch address gets demanded \emph{before} the prefetch fill.
    \item \textbf{\emph{Loss of coverage}} ($\mathcal{R}_{CL}$). This reward is assigned to an action whose corresponding prefetch address \rbc{is to a different physical page than the demand access that led to the prefetch.}
    \item \textbf{\emph{Inaccurate}} ($\mathcal{R}_{IN}$). This reward is assigned to an action whose corresponding prefetch address does \emph{not} get demanded in a temporal window. The reward is classified into two sub-levels: inaccurate given low bandwidth \rbc{usage} ($\mathcal{R}_{IN}^{L}$) and inaccurate given high bandwidth \rbc{usage} ($\mathcal{R}_{IN}^{H}$).
    \item \textbf{\emph{No-prefetch}} ($\mathcal{R}_{NP}$). This reward is assigned when Pythia decides not to prefetch. This reward level is also classified into two sub-levels: no-prefetch given low bandwidth \rbc{usage} ($\mathcal{R}_{NP}^{L}$) and no-prefetch given high bandwidth \rbc{usage} ($\mathcal{R}_{NP}^{H}$).
\end{itemize}

\rbc{By increasing (decreasing) a reward level value, we reinforce (deter) Pythia to collect such rewards from the environment in the future.}
$\mathcal{R}_{AT}$ and $\mathcal{R}_{AL}$ \rbc{are} used to guide Pythia to generate more accurate and timely prefetch requests.
$\mathcal{R}_{CL}$ is used to guide Pythia to generate prefetches within the physical page of the triggering demand request.
$\mathcal{R}_{IN}$ and $\mathcal{R}_{NP}$ are used to define Pythia's prefetching strategy with respect to memory bandwidth usage feedback.
In~\Cref{sec:py_tuning_reward_hyp_selection}, we \rbc{provide} an \emph{automated method} to configure the reward values.
The reward values can be easily customized further for target workload suites to extract higher performance gains (see~\Cref{sec:py_eval_pythia_custom}).

\sectionRB{Pythia: Detailed Design}{Pythia: Detailed Design}{sec:py_design}

\begin{figure}[!ht]
\centering
\includegraphics[width=5in]{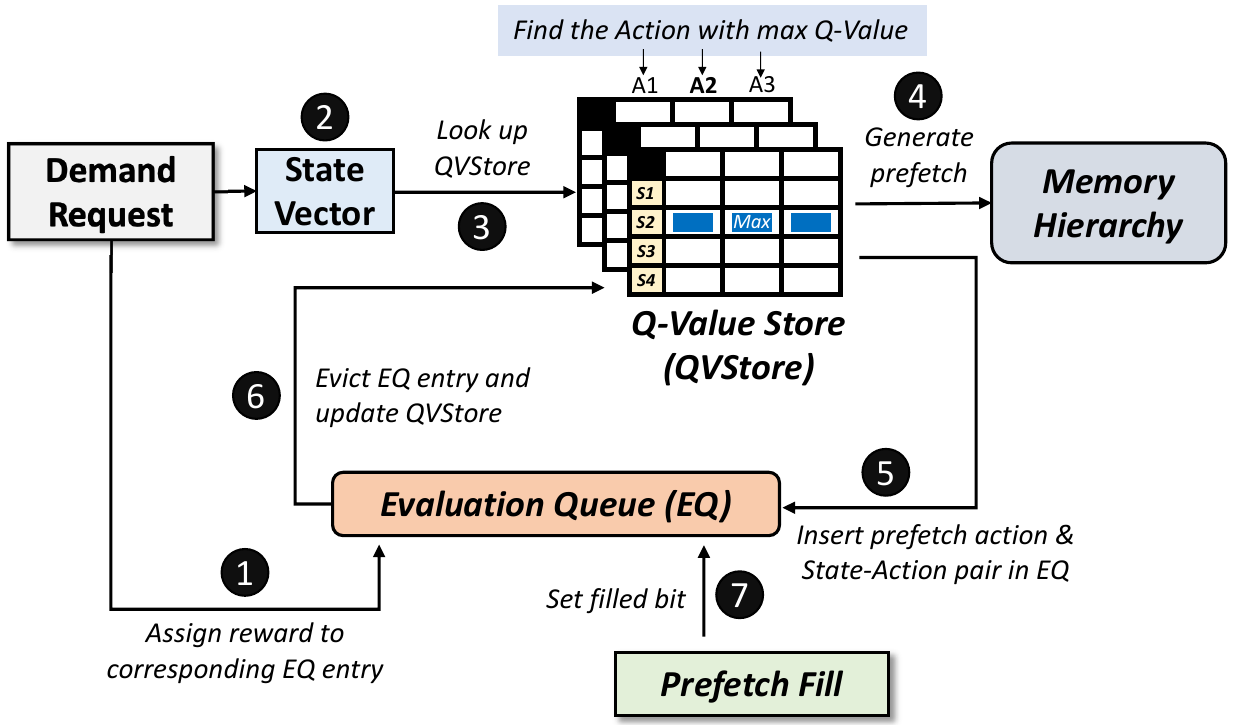}
\caption{Overview of Pythia.}
\label{fig:py_overview}
\end{figure}

\noindent \Cref{fig:py_overview} shows a \rbc{high-level} \rbc{overview} of Pythia. 
Pythia is mainly comprised of two hardware structures: \emph{Q-Value Store} (QVStore) and \emph{Evaluation Queue} (EQ).
The purpose of QVStore is to \rbc{record} Q-values for all state-action pairs that are \rbc{observed} by Pythia.
The purpose of EQ is to maintain a first-in-first-out list of Pythia's recently-taken actions.\footnote{Pythia keeps track of recently-taken actions because it cannot always \emph{immediately} assign a reward to an action, as the \rbc{usefulness} of the generated prefetch request (i.e., \rbc{if and when} the prefetched address is demanded by the processor) is not immediately known while the action is being taken.
During EQ residency, \rbc{if the address of a demand request matches with the prefetch address stored in an EQ entry}, the corresponding action is considered to \rbc{have generated a useful prefetch request}. 
}
\rbc{Every EQ entry holds three \rbc{pieces of} information: \rbc{(1)} the \rbc{taken} action, \rbc{(2)} the prefetch address generated for the corresponding action, and \rbc{(3)} a \emph{filled} bit. A set filled bit indicates that the prefetch request has been filled into the cache.}

For every new demand request, Pythia first checks the EQ with the demanded memory address (\circled{1}). \rbc{If the address is present in the EQ (i.e., Pythia has issued a prefetch request for this address in the past)},
\rbc{it signifies that the prefetch action corresponding to the EQ entry \rbc{has generated a useful prefetch request}. As such, Pythia assigns a reward (either $\mathcal{R}_{AT}$ or $\mathcal{R}_{AL}$) to the EQ entry, based on whether \rbc{or not} the EQ entry's filled bit is set.}
\rbc{Next, Pythia extracts the state-vector from the attributes of the demand request (e.g., PC, address, cacheline delta, etc.) (\circled{2}) and looks up QVStore to find the action with the maximum Q-value for the given state-vector (\circled{3}).}
\rbc{Pythia selects the action with the maximum Q-value to generate prefetch request and issues the request to the memory hierarchy (\circled{4}).}
\rbc{At the same time, Pythia inserts the selected prefetch action, its corresponding prefetched memory address, and the state-vector into EQ (\circled{5}).}
Note that, a \emph{no-prefetch} action or an action that prefetches an address beyond the current physical page
is also inserted into EQ. The reward for such \rbc{an} action is instantaneously assigned to the EQ entry.
\rbc{When an EQ entry gets evicted,} the state-action pair and the reward stored in the evicted EQ entry are used to update the Q-value in the QVStore (\circled{6}).
For every prefetch fill in cache, Pythia looks up EQ with the prefetch address \rbc{and sets the \emph{filled} bit in the matching EQ entry indicating that the prefetch request has been filled \rbc{into the} cache (\circled{7}).}
\rbc{Pythia uses this filled bit in \circled{1} to classify actions that generated timely or late prefetches.}\footnote{In this \rbfor{work}, we define prefetch timeliness as a binary value due to its measurement simplicity. One can easily make the definition non-binary by storing three timestamps per EQ entry: (1) when the prefetch is issued ($t_{issue}$), (2) when the prefetch is filled ($t_{fill}$), and (3) when a demand is generated for the same prefetched address ($t_{demand}$).}

\begin{algorithm*}[!ht]
  \footnotesize
  \begin{algorithmic}[1]
  \Procedure{Initialize}{}
    \State initialize QVStore: $Q(S,A)\gets \frac{1}{1-\gamma}$ 
    \State clear EQ
  \EndProcedure
  \State
  \Procedure{Train\_and\_Predict}{Addr} \Comment{Called for every demand request}

    \State $entry \gets search\_EQ(Addr)$ \Comment{Search EQ with the demand address}
    \If{entry is valid}
        \If {$entry.filled == true$}
            \State $entry.reward \gets \mathcal{R}_{AT}$ \Comment{Prefetch already filled means accurate and timely prefetch}
        \Else
            \State $entry.reward \gets \mathcal{R}_{AL}$ \Comment{Otherwise, the prefetch is accurate but late}
        \EndIf
    \EndIf
    \State $S\gets get\_state()$ \Comment{Extract the state-vector from current demand request}
    \If{$rand()\le \epsilon$}
        \State $action\gets get\_random\_action()$ \Comment{Explore state-action space with a small probability}
    \Else
        \State $action\gets \argmax_a Q(S,a)$ \Comment{Otherwise, select the action with the highest Q-value}
    \EndIf
    \State $prefetch(Addr+Offset[action])$ \Comment{Generate prefetch address from the selected offset}
    \State $entry \gets create\_EQ\_entry(S,action,Addr+Offset[action])$ \Comment{Create a new EQ entry}
    \If {no prefetch action}
        \State $entry.reward \gets \mathcal{R}_{NP}^{H}$ or $\mathcal{R}_{NP}^{L}$ \Comment{Immediately assign reward (${R}_{NP}^{H}$ or ${R}_{NP}^{L}$) for no-prefetch}
    \ElsIf{out-of-page prefetch}
        \State $entry.reward \gets \mathcal{R}_{CL}$ \Comment{Immediately assign reward ${R}_{CL}$ for out-of-page prefetch}
    \EndIf
    \State $dq\_entry\gets insert\_EQ(entry)$ \Comment{Insert the new EQ entry and get the evicted EQ entry.}
    \If{$has\_reward(dq\_entry) == false$}
        \State $dq\_entry.reward \gets \mathcal{R}_{IN}^{H}$ or $\mathcal{R}_{IN}^{L}$ \Comment{Unused prefetch; assign $\mathcal{R}_{IN}^{H}$ or $\mathcal{R}_{IN}^{L}$ based on bandwidth}
    \EndIf
    \State $R \gets  dq\_entry.reward$ \Comment{Get the reward stored in the evicted entry}
    \State $S_1\gets dq\_entry.state;\ \ A_1\gets dq\_entry.action$ \Comment{Get state and action from evicted EQ entry}
    \State $S_2\gets EQ.head.state;\ \ A_2\gets EQ.head.action$ \Comment{Get state and action from EQ head}
    \State $Q(S_1, A_1)\gets Q(S_1, A_1)+\alpha[R+\gamma Q(S_2, A_2)-Q(S_1, A_1)]$ \Comment{Perform the SARSA update}
  \EndProcedure
  \State
  \Procedure{Prefetch\_Fill}{Addr}
    \State $search\_and\_mark\_EQ(Addr,\ \texttt{FILLED})$ \Comment{Mark prefetch fill bit}
  \EndProcedure
  \end{algorithmic}
\caption{Pythia's \rbc{reinforcement learning} based prefetching algorithm}\label{algo:Pythia}
\end{algorithm*}

\subsection{RL-Based Prefetching Algorithm}
\Cref{algo:Pythia} shows Pythia's RL-based prefetching algorithm. \rbc{Initially,} all entries in QVStore are
reset to the highest possible Q-value ($\frac{1}{1-\gamma}$) and the EQ is cleared (lines 2-3). 
For every demand \rbc{request} to \rbc{a} cacheline address $Addr$, Pythia searches for $Addr$ in EQ (line $6$). If a matching entry is found, Pythia \rbc{assigns a reward (either $\mathcal{R}_{AT}$ or $\mathcal{R}_{AL}$) based on the \emph{filled} bit in the EQ entry (lines 8-11)}. Pythia then extracts the state-vector to \emph{stochastically} select a prefetching action that provides the highest Q-value (lines 13-16). \rbc{Pythia uses the selected action to generate the prefetch request (line 17) and creates a new EQ entry with the current state-vector, the selected action, and its corresponding prefetched address (line 18).}
\rbc{In case of a no-prefetch action, or an action that prefetches beyond the current physical page, Pythia immediately assigns the reward to the newly-created EQ entry (lines 19-22).}
\rbc{The EQ entry is then inserted, which evicts an entry from EQ. If the evicted EQ entry does not already have a reward assigned (\rbc{indicating that} the corresponding prefetch address is \emph{not} demanded \rbc{by} the processor \rbc{so far}), Pythia assigns the reward $\mathcal{R}_{IN}^{H}$ or $\mathcal{R}_{IN}^{L}$ based on the current memory bandwidth usage (lines 25).}
\rbc{Finally, the Q-value of the evicted state-action pair is updated \rbc{via} the SARSA algorithm (see~\Cref{sec:background_rl}), using the reward stored in the evicted EQ entry and the Q-value of the state-action pair in the head of the EQ-entry (lines 26-29).}

\subsection{Detailed Design of Pythia}
\rbc{We describe the organization of QVStore \rbc{(\Cref{sec:py_ke_design})}, how Pythia searches QVStore to get the action with the maximum Q-value for a given state-vector (\circled{3}) \rbc{(\Cref{sec:py_design_config_pipeline})}, how Pythia assigns rewards to each taken action and how it updates Q-values (\circled{6}) \rbc{(\Cref{sec:py_ke_update})}.
}

\subsubsection{\textbf{Organization of QVStore}} \label{sec:py_ke_design}
\rbc{The purpose of QVStore is to record Q-values for all state-action pairs that Pythia observes.}
Unlike prior real-world applications of RL~\cite{deepmind_atari,alpha_go,alpha_zero}, which use deep neural networks to \emph{approximately} \rbc{store} Q-values of every state-action pair, we propose a \rbc{new}, table-based, hierarchical QVStore organization that is \rbc{custom-designed to} our RL-agent.
 
\begin{figure}[!ht]
\centering
\includegraphics[width=5.75in]{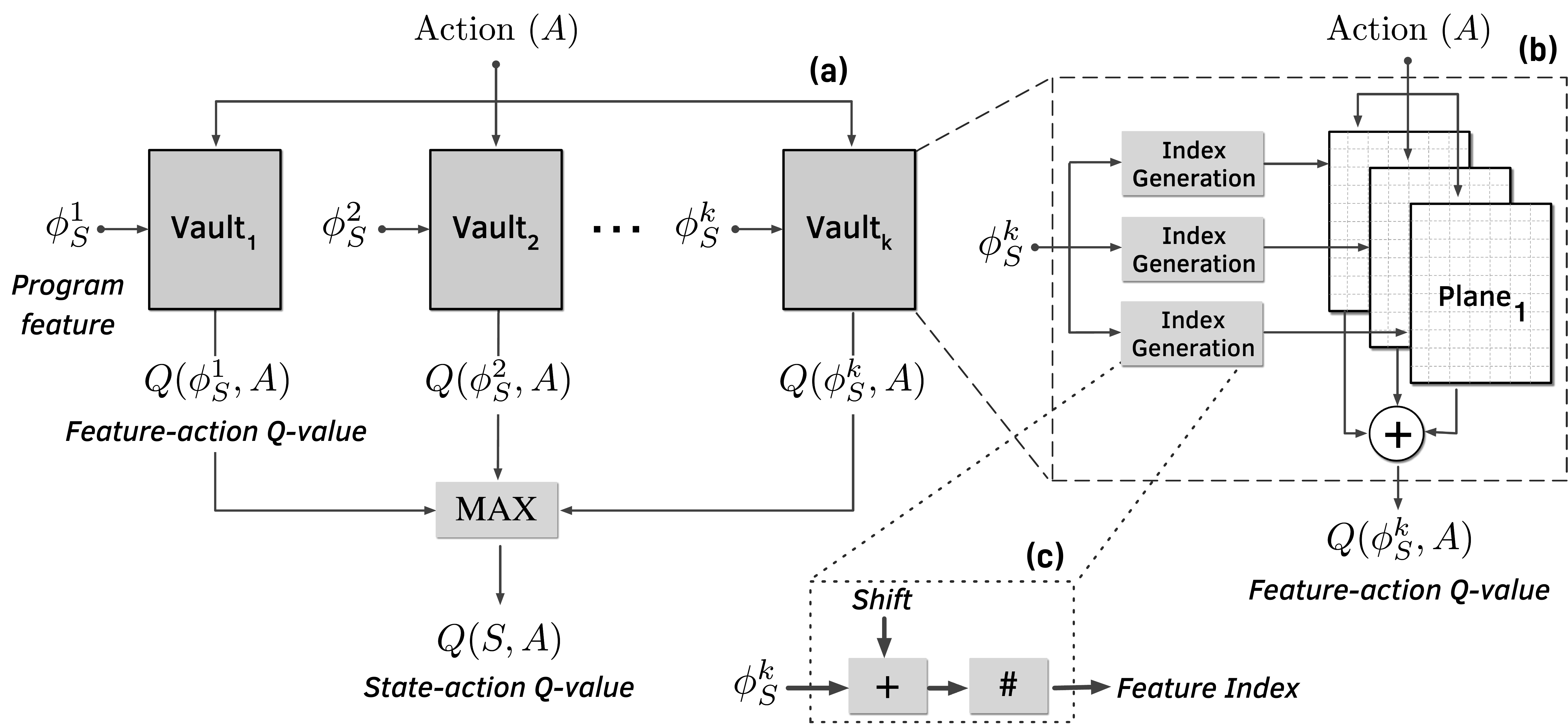}
\caption{(a) The QVStore is comprised of multiple vaults. (b) Each vault is comprised of multiple planes. (c) Index generation from feature value.}
\label{fig:py_ke_design}
\end{figure}

\rbc{\Cref{fig:py_ke_design}(a) shows the high-level organization of QVStore and how the Q-value is retrieved from QVStore for a given state $S$ (which is a k-dimensional vector of program features, $\{\phi_{S}^{1}, \phi_{S}^{2}, \ldots, \phi_{S}^{k}\}$) and an action $A$. As the state space grows rapidly with the state-vector dimension ($k$) and the bits used to represent each feature, we employ a hierarchical organization for QVStore. 
We organize QVStore in $k$ partitions, each \rbc{of which} we call a \emph{vault}. Each vault corresponds to one constituent feature of the state-vector and records the Q-values for the feature-action pair, $Q(\phi_{S}^{i},A)$.
}
During the Q-value retrieval for a given state-action pair $Q(S,A)$, Pythia queries each vault in parallel to retrieve the Q-values of constituent feature-action pairs $Q(\phi_{S}^{i},A)$. The final Q-value of the state-action pair $Q(S,A)$ is computed as the \emph{maximum} of all constituent feature-action Q-values, \rbc{as}~\Cref{eq:py_q_value} shows).
\rbc{The maximum operation ensures that the state-action Q-value is driven by the constituent feature of the state-vector that has the highest feature-action Q-value.}
\rbc{The vault organization enables QVStore to efficiently scale up \rbc{to} higher state-vector \rbc{dimensions: one} can increase the state-vector dimension by simply adding a new vault to the QVStore.}
\begin{equation} \label{eq:py_q_value}
    Q(S, A) = \max_{i \in (1, k)} Q(\phi_{S}^{i}, A)
\end{equation}

\rbc{\rbc{\Cref{fig:py_ke_design}(a) shows the organization of QVStore as a collection of multiple vaults.} 
The purpose of a vault is to record Q-values of all feature-action pairs that Pythia observes for a specific feature type.} A vault can be conceptually visualized as a monolithic two-dimensional table (as shown in~\Cref{fig:py_ke_design}(a)), indexed by the feature and action values, that stores Q-value for every feature-action pair. 
\rbc{However, the key challenge in \rbc{implementing} vault as a monolithic table}
is that \rbc{the size of the table increases exponentially with a} linear increase in the number of bits used to represent the feature.
This not only makes the monolithic table \rbc{organization} impractical for implementation but also increases the \rbc{design} complexity to satisfy \rbc{its} latency and power requirements. 

One way to address this challenge is \rbc{to quantize} the feature space into \rbc{a} small number of tiles.
\rbc{Even though} feature space quantization can achieve a drastic reduction in the monolithic table size, 
\rbc{it requires a compromise between the resolution of a feature value and the generalization of feature values.}
We draw inspiration from \emph{tile coding}~\cite{rl_bible, cmac, rlmc} to strike a balance between resolution and generalization.
\rbc{Tile coding uses} \emph{multiple overlapping} hash functions to quantize a feature value into smaller \emph{tiles}. The quantization achieves generalization of similar feature values, whereas multiple hash functions increase resolution to represent a feature value.

We leverage the idea of tile coding to organize a vault as a collection of $N$ small two-dimensional tables, each \rbc{of which} we call a \emph{plane}. 
Each plane entry stores a \emph{partial} Q-value of a feature-action pair.\footnote{\rbc{Our application of tile coding is similar to that used in \rbc{the} self-optimizing memory controller (RLMC)~\cite{rlmc}. The key difference is that RLMC uses a hybrid combination of feature and action values to index \emph{single-dimensional} planes, whereas Pythia uses feature and action values \rbc{\emph{separately}} to index \emph{two-dimensional} planes.}} As~\Cref{fig:py_ke_design}(c) shows, \rbc{to retrieve a feature-action Q-value $Q(\phi_{S}^{i},A)$,} the given feature is first shifted by a shifting constant \rbc{(which is randomly selected at design time)}, followed by a hashing to get the feature index for the given plane. This feature index, along with the action index, is used to \rbc{retrieve} the partial Q-value from the plane. The final feature-action Q-value is computed \rbc{as the \emph{sum of all}} the partial Q-values from \rbc{all} planes, \rbc{as shown in~\Cref{fig:py_ke_design}(b)}.
\rbc{The use of tile coding provides two key advantages to Pythia. First, the tile coding of a feature enables the sharing of partial Q-values between similar feature values, which shortens prefetcher training time. Second, multiple planes reduces the chance of sharing partial Q-values between widely different feature values.}

\subsubsection{\textbf{Pipelined Organization of QVStore Search}}\label{sec:py_design_config_pipeline}
\rbc{To generate a} prefetch request, Pythia has to \rbc{(1)} look up the QVStore with the state-vector extracted from the current demand request, and (2) search for the action that has the maximum state-action Q-value (\circled{3} in~\Cref{fig:py_overview}). As a result, the search operation lies on Pythia's critical path and directly impacts Pythia's prediction latency. To improve the prediction latency, we pipeline the search operation.

\begin{figure}[!ht]
\centering
\includegraphics[width=5in]{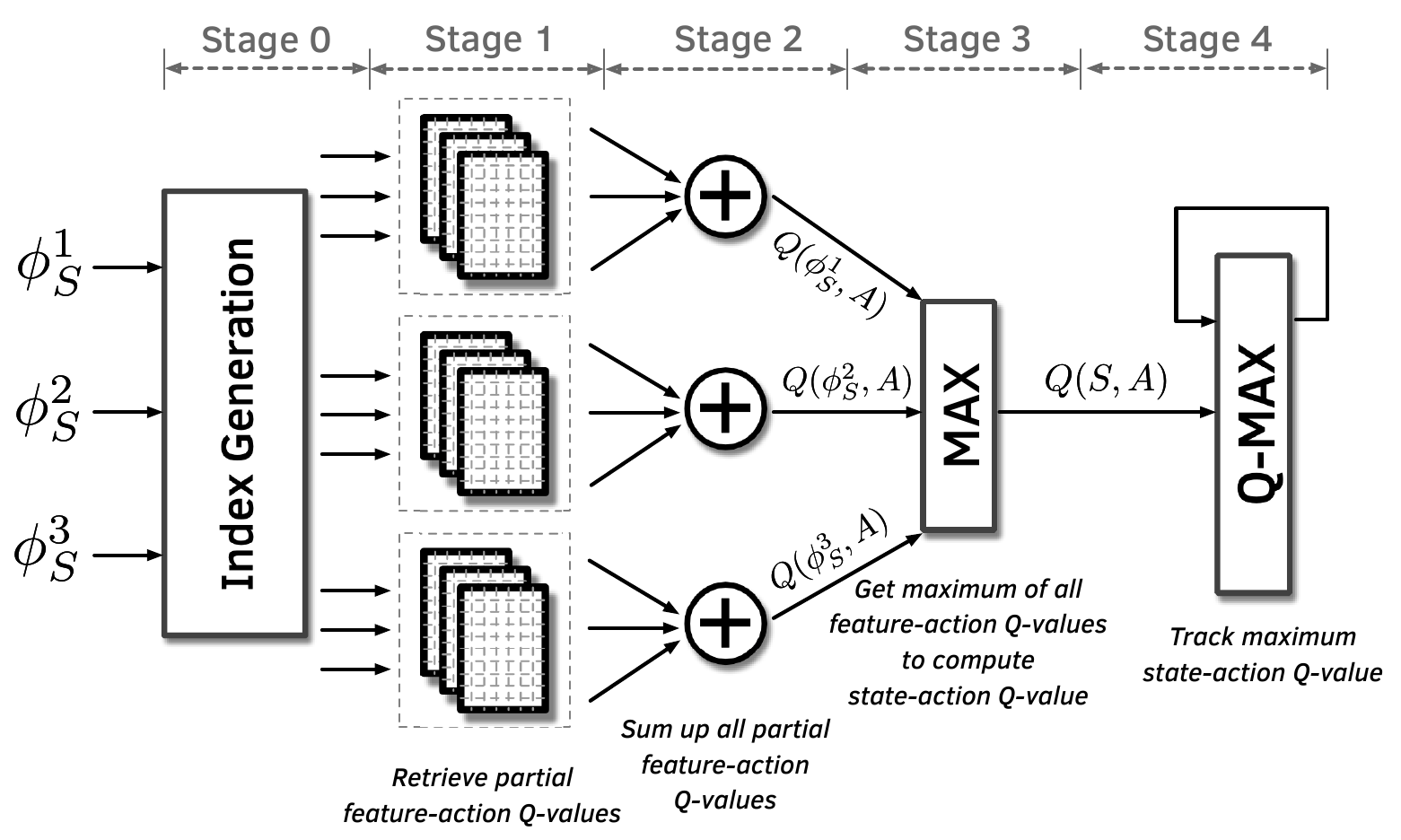}
\caption{\rbc{Pipelined} organization of QVStore search operation. The illustration depicts three program features, each having three planes. 
}
\label{fig:py_pythia_pipeline}
\end{figure}

\rbc{The Q-value search operation is implemented in the following way. For a given state-vector, Pythia iteratively retrieves the Q-value of each action. Pythia also maintains a variable, $Q_{max}$, that tracks the maximum Q-value found so far. $Q_{max}$ gets compared \rbc{to} every retrieved Q-value. The search operation concludes when Q-values for \emph{all} \rbc{possible} actions \rbc{have} been retrieved.}
We pipeline the search operation into five stages \rbc{as~\Cref{fig:py_pythia_pipeline} shows}.
Pythia first computes the index for each plane and each constituent feature \rbc{of the given state-vector} \rbc{(Stage 0)}. \rbc{\rbc{In \rbc{Stage} 1}, Pythia uses the feature indices} and an action index to retrieve \rbc{the} partial Q-values from each plane. In \rbc{Stage 2}, Pythia \rbc{sums up} the partial Q-values to get the feature-action Q-value for each constituent feature. \rbc{In Stage 3}, Pythia computes \rbc{the maximum of all feature-action Q-values to get the state-action Q-value}. In \rbc{Stage} 4, \rbc{the maximum state-action Q-value found so far} is compared against the retrieved state-action Q-value, and the maximum Q-value is updated. Stage 2 (i.e., the partial Q-value summation) \rbc{is} the longest stage \rbc{of the pipeline and thus it} dictates the pipeline's throughput. 
\rbc{We accurately measure the area and power overhead of the pipelined implementation of the search operation by modeling Pythia using Chisel~\cite{chisel} hardware design language and synthesize the model using Synopsys design compiler~\cite{synopsys_dc} and 14-nm library from GlobalFoundries~\cite{global_foundries} (see~\Cref{sec:py_eval_overhead}).}

\subsubsection{\textbf{Assigning Rewards and Updating Q-values}} \label{sec:py_ke_update}
\rbc{
To track usefulness of the prefetched requests, Pythia maintains a first-in-first-out list of recently taken actions, along with their corresponding prefetch addresses in EQ.
\emph{Every} prefetch action 
is inserted into EQ. A reward gets assigned to every EQ entry before or \rbc{when} it gets evicted from EQ.
During eviction, the reward and the state-action pair \rbc{associated with} the evicted EQ entry are used to update \rbc{the corresponding} Q-value in QVStore (\circled{6} in~\Cref{fig:py_overview}). 
}

\rbc{We describe how Pythia appropriately assigns rewards to each EQ entry.
We divide the reward assignment into three classes based on \emph{when} the reward gets assigned to an entry: 
(1) immediate reward assignment during EQ insertion, (2) reward assignment during EQ residency, and (3) reward assignment during EQ eviction.
}
\rbc{If Pythia selects \rbc{the} action \rbc{\emph{not to prefetch}} or one that generates a prefetch request beyond the current physical page, Pythia immediately assigns \rbc{a} reward to the EQ entry. \rbc{For} out-of-page prefetch action, Pythia assigns $\mathcal{R}_{CL}$. \rbc{For the action \emph{not to prefetch}}, Pythia assigns $\mathcal{R}_{NP}^{H}$ or $\mathcal{R}_{NP}^{L}$, \rbc{based on whether} the current system memory bandwidth usage is high or low.
If \rbc{the} address of a demand request matches with the prefetch address stored in an EQ entry during its residency,
Pythia assigns $\mathcal{R}_{AT}$ or $\mathcal{R}_{AL}$ based on the \emph{filled} bit of the EQ entry. If the filled bit is set, it indicates that the demand request is generated \emph{after} the prefetch fill. Hence the prefetch is accurate and timely, and Pythia assigns the reward $\mathcal{R}_{AT}$. Otherwise, Pythia assigns the reward $\mathcal{R}_{AL}$.
If a reward does not get assigned to an EQ entry \rbc{until it is going to be evicted}, it signifies that the corresponding prefetch address is not yet demanded by the processor. Thus, Pythia assigns a reward $\mathcal{R}_{IN}^{H}$ or $\mathcal{R}_{IN}^{L}$ to the entry during eviction \rbc{based on whether} the current system memory bandwidth usage is high or low.  
}

\subsection{Automated Design-Space Exploration}\label{sec:py_tuning}

\rbc{We} propose an automated, performance-driven approach to systematically explore Pythia's vast design space and 
derive a \rbcb{basic} configuration\footnote{Using a compute-grid with ten $28$-core machines, the automated exploration across $150$ workload traces (mentioned in detail in~\Cref{sec:pythia_methodology}) takes $44$ hours to complete.} with appropriate program features, action set, reward and hyperparameters.
\rbc{\Cref{tab:pythia_config} shows the \rbcb{basic} configuration.}

\begin{table}[h]
    \centering
    \small 
    \begin{tabular}{m{9.5em}m{18em}}
         \thickhline
         \Tabval{\textbf{Features}} & \Tabval{\texttt{PC+Delta}, \texttt{Sequence of last-4 deltas}}\\
         \hline
         \Tabval{\textbf{Prefetch Action List}} & \Tabval{\{-6,-3,-1,0,1,3,4,5,10,11,12,16,22,23,30,32\}} \\
         \hline
         \Tabval{\textbf{Reward Level Values}} & \Tabval{$\mathcal{R}_{AT}$=$20$, $\mathcal{R}_{AL}$=$12$, $\mathcal{R}_{CL}$=$-12$, $\mathcal{R}_{IN}^{H}$=$-14$, $\mathcal{R}_{IN}^{L}$=$-8$, $\mathcal{R}_{NP}^{H}$=$-2$, $\mathcal{R}_{NP}^{L}$=$-4$}\\
         \hline
         \Tabval{\textbf{Hyperparameters}} & \Tabval{$\alpha=0.0065$, $\gamma=0.556$, $\epsilon=0.002$} \\
         \thickhline
    \end{tabular}
    \caption{\rbc{Basic} Pythia \rbc{configuration} derived from \rbc{our} automated design-space exploration.}
    \label{tab:pythia_config}
\end{table}

\subsubsection{\textbf{Feature Selection}}\label{sec:py_tuning_feature_selection}
\rbc{We derive a list of possible program features for feature-space exploration in \rbc{four} steps. First, we derive a list of $4$ control-flow components, and $8$ data-flow components, which are mentioned in~\Cref{table:py_list_of_features}. \rbc{Second,} we combine each control-flow component with each data-flow component \rbc{with the} concatenation operation, to \rbc{obtain} a total of $32$ possible program features.}
\rbc{Third,} we use \rbc{the} linear regression technique~\cite{lr1,lr2,lr3} to create any-one, any-two, and any-three feature-combinations from the set of $32$ initial features, each providing a different state-vector. \rbc{Fourth,} we run Pythia with every state-vectors across all single-core workloads (\Cref{sec:pythia_methodology}) and select the \rbc{winning} state-vector that provides the highest performance gain over no-prefetching baseline. \rbc{As~\Cref{tab:pythia_config} shows}, the two constituent features of the winning state-vector are \texttt{PC+Delta} and \texttt{Sequence of last-4 deltas}.

\begin{table}[htbp]
  \centering
  \small
    \begin{tabular}{m{16em}m{16em}}
    \thickhline
    \Tabval{\textbf{\small Control-flow Components}} & \Tabval{\textbf{\small Data-flow Components}} \\
    \hline
    \Tstrut \begin{minipage}{16em}
      \small
      \begin{enumerate}[leftmargin=1.5em]
        \setlength{\itemsep}{1pt}
        \setlength{\parskip}{0pt}
        \setlength{\parsep}{0pt}
        \item PC of load request
        \item PC-path (XOR-ed last-3 PCs)
        \item PC XOR-ed branch-PC
        \item None
      \end{enumerate}
    \end{minipage} \Bstrut & 
    \Tstrut
    \begin{minipage}{16em}
      \small
      \vskip 4pt
      \begin{enumerate}[leftmargin=1.5em]
        \setlength{\itemsep}{1pt}
        \setlength{\parskip}{0pt}
        \setlength{\parsep}{0pt}
        \item Load cacheline address
        \item Page number
        \item Page offset
        \item Load address delta
        \item Sequence of last-4 offsets
        \item Sequence of last-4 deltas
        \item Offset XOR-ed with delta
        \item None
      \end{enumerate}
    \end{minipage} \Bstrut\\
    \thickhline
    \end{tabular}%
    \caption{List of program control-flow and data-flow components used to derive the list of features for exploration.}
    \label{table:py_list_of_features}%
\end{table}%

\textbf{Rationale behind the \rbc{winning} state-vector.} The \rbc{winning} state-vector is intuitive as \rbc{its} constituent features \texttt{PC+Delta} and \texttt{Sequence of last-4 deltas} closely match with the program features exploited by \rbc{two prior state-of-the-art prefetchers}, Bingo~\cite{bingo} and SPP~\cite{spp}, respectively.
However, concurrently running SPP and Bingo \rbc{as a hybrid prefetcher does not provide} the same \rbc{performance} benefit as Pythia, as we show in~\Cref{sec:py_eval_perf_1T}. This is because combining SPP with Bingo not only \rbc{improves} their prefetch coverage, but also combines \rbc{their} prefetch overpredictions, \rbc{leading to performance degradation, especially in resource-constrained systems}. \rbc{In contrast}, Pythia's RL-based learning strategy \rbc{that inherently uses} the same two features successfully \rbc{increases} prefetch coverage, while maintaining \rbc{high prefetch accuracy}. As a result, \emph{Pythia not only outperforms SPP and Bingo individually, but also outperforms \rbc{the} combination of \rbc{the two} prefetchers}.

\subsubsection{\textbf{Action Selection}}\label{sec:py_tuning_action_selection}
In a system with \rbc{conventionally}-sized $4$KB pages and $64$B cachelines, Pythia's list of actions \rbc{(i.e., the list of possible prefetch offsets)} contains \emph{all} prefetch offsets in the range of $[-63,63]$. \rbc{However, such} a long action list poses two drawbacks. First, a long action list requires \rbc{more online exploration} to find the \rbc{best} prefetch offset \rbc{given a state-vector, \rbc{thereby} reducing Pythia's performance benefits}.
Second, a longer \rbc{action} list increases Pythia's storage \rbc{requirements}. 
\rbc{To avoid these problems, we prune the action list. We drop each action individually from the full action list $[-63,63]$ and measure the performance improvement relative to the performance improvement with the full action list, across all single-core workload traces. We prune any action that \emph{does not} have any significant impact on the performance.}
\rbc{\Cref{tab:pythia_config} shows} the final pruned action list.

\subsubsection{\textbf{Reward and Hyperparameter Tuning}}\label{sec:py_tuning_reward_hyp_selection}
\begin{sloppypar}
We separately tune \rbc{seven} reward \rbc{level values (i.e., $\mathcal{R}_{AT}$, $\mathcal{R}_{AL}$, $\mathcal{R}_{CL}$, $\mathcal{R}_{IN}^{H}$, $\mathcal{R}_{IN}^{L}$, $\mathcal{R}_{NP}^{H}$, and $\mathcal{R}_{NP}^{L}$) and three hyperparameters (i.e., learning rate $\alpha$, discount factor $\gamma$, and exploration rate $\epsilon$)} 
in three steps. 
\rbc{First,} we create a test trace suite by \emph{randomly} selecting $10$ \rbc{workload} traces from \rbc{all of our} $150$ \rbc{workload} traces (\Cref{sec:pythia_methodology}). 
\rbc{Second}, \rbc{we create a list of tuning configurations} using \rbc{the} uniform grid search technique~\cite{grid_search1,grid_search2}. \rbc{To do so,} we first define a value range for each parameter to be tuned and divide the \rbc{value} range into uniform grids. For example, each of the three hyperparameters ($\alpha$, $\gamma$, and $\epsilon$) can take \rbc{a} value in the range of $[0,1]$. We divide each hyperparameter \rbc{range} into ten \rbc{exponentially}-sized grids (i.e., $1e^0$, $1e^{-1}$, $1e^{-2}$, etc.) to \rbc{obtain} $10\times10\times10=1000$ possible tuning configurations. 
\rbc{For each tuning configuration,} we run Pythia on the test trace suite and select the top-$25$ highest-performing configurations for the third step.
\rbc{Third}, we run Pythia on all single-core \rbc{workload} traces using each of the $25$ selected configurations. 
We select the \rbc{winning} configuration that provides the highest \rbc{average} performance gain.
\rbc{\Cref{tab:pythia_config} provides}
\rbc{reward level and hyperparameter values of \rbc{the \rbcb{basic} Pythia}.}
\end{sloppypar}

\subsection{Storage Overhead}
\rbc{\Cref{table:pythia_storage} shows the storage overhead of Pythia in its \rbcb{basic} configuration. Pythia requires only $25.5$KB of metadata storage. QVStore consumes $24$KB to store all Q-values. The EQ consumes only $1.5$KB.}

\begin{table}[h]
  \centering
  \small
    \begin{tabular}{|m{5em}||m{28em}|R{4em}|}
    \hline
    \Tabval{\textbf{Structure}} & \Tabval{\textbf{Description}} & \Tabval{\textbf{Size}} \\
    \hline
    \hline
    \Tabval{\textbf{QVStore}} &
    \Tstrut
    \begin{minipage}{28em}
      \small
      \vskip 4pt
      \begin{itemize}[leftmargin=1.5em]
        \setlength{\itemsep}{1pt}
        \setlength{\parskip}{0pt}
        \setlength{\parsep}{0pt}
        \item \# vaults = $2$
        \item \# planes/vault = $3$
        \item \# entries/plane = feature dimension ($128$) $\times$ action dimension ($16$)
        \item Entry size = Q-value width ($16$b)
      \end{itemize}
    \end{minipage} \Bstrut &
    \Tabval{\textbf{24 KB}} 
    \\
    \hline
    \Tabval{\textbf{EQ}} &
    \Tstrut
    \begin{minipage}{28em}
    \vskip 4pt
      \small
      \begin{itemize}[leftmargin=1.5em]
        \setlength{\itemsep}{1pt}
        \setlength{\parskip}{0pt}
        \setlength{\parsep}{0pt}
        \item \# entries = 256
        \item Entry size = state ($21$b) + action index ($5$b) + reward ($5$b) + filled-bit ($1$b) + address ($16$b)
      \end{itemize}
    \end{minipage} \Bstrut &
    \Tabval{\textbf{1.5 KB}} 
    \\    
    \hline
    \hline
    \Tabval{\textbf{Total}} &      & \Tabval{\textbf{25.5 KB}} \\
    \hline
    \end{tabular}%
  \caption{Storage overhead of Pythia.}
  \label{table:pythia_storage}%
\end{table}

\sectionRB{Pythia: Evaluation Methodology}{Methodology}{sec:pythia_methodology}

\noindent We use \rbc{the trace-driven} ChampSim simulator~\mbox{\cite{champsim}} to evaluate Pythia \rbc{and compare it to five} prior prefetching proposals.
We simulate an Intel Skylake~\mbox{\cite{skylake}}-like multi-core processor that supports \rbc{up to} 12 cores. \rbc{\Cref{table:py_sim_params} provides the \rbc{key system} parameters.}
For single-core simulations ($1C$), we warm up the core using $100$~M instructions \rbc{from each workload} and simulate the next $500$~M instructions. For multi-core multi-programmed simulations ($nC$), we use $50$~M and $150$~M instructions \rbc{from each workload} respectively to warmup and simulate. If a core finishes early, the workload is replayed \rbc{until} every core finishes simulating $150$~M instructions. 
We also \rbc{implement} Pythia using \rbc{the} Chisel~\cite{chisel} \rbc{hardware design language (HDL)} and functionally verify the resultant \rbc{register transfer logic (RTL) design} to accurately measure Pythia's \rbc{chip} area and power overhead.
The source-code of Pythia is freely available at~\cite{pythia_github}.

\begin{table}[!ht]
    \centering
    \small
    \begin{tabular}{L{4.5em}||L{31em}}
         \thickhline
         \Tabval{\textbf{Core}} & \Tabval{{1-12 cores, 4-wide OoO, 256-entry ROB, \rbc{72/56-entry LQ/SQ}, Perceptron-based branch predictor~\cite{perceptron}, 20-cycle misprediction penalty}}\\
         \hline
         \Tabval{\textbf{L1/L2 Caches}} & \Tabval{{Private, 32KB/256KB, 64B line, 8 way, LRU, 16/32 MSHRs, 4-cycle/14-cycle round-trip latency}} \\
         \hline
         \Tabval{\textbf{LLC}} & \Tabval{{2MB/core, 64B line, 16 way, SHiP~\cite{ship}, 64 MSHRs per LLC Bank, 34-cycle round-trip latency}}\\
         \hline
         \Tabval{\textbf{Main Memory}} & \Tabval{{\textbf{1C:} Single channel, 1 rank/channel; \textbf{4C:} Dual channel, 2 ranks/channel; \textbf{8C:} Quad channel, 2 ranks/channel; 8 banks/rank, 2400 MTPS, 64b data-bus per channel, 2KB row buffer/bank, tRCD = tRP = 15ns, tCAS = 12.5ns}}\\
         
         \thickhline
    \end{tabular}
    \caption{Simulated system parameters.}
    \label{table:py_sim_params}
\end{table}

\subsection{Workloads}
\begin{sloppypar}
\rbc{We evaluate Pythia using a diverse set of memory-intensive workloads} spanning \texttt{SPEC CPU2006}~\cite{spec2006}, \texttt{SPEC CPU2017}~\cite{spec2017}, \texttt{PARSEC 2.1}~\cite{parsec}, \texttt{Ligra}~\cite{ligra}, and \texttt{Cloudsuite}~\cite{cloudsuite} \rbc{benchmark suites}.
\rbc{For \texttt{SPEC CPU2006} and \texttt{SPEC CPU20017} workloads, we reuse the instruction traces provided by \rbc{the} 2nd and \rbc{the} 3rd data prefetching championships (DPC)~\cite{dpc2, dpc3}. For \texttt{PARSEC} and \texttt{Ligra} workloads, we collect the instruction traces using \rbc{the} Intel Pin dynamic binary instrumentation tool~\cite{intel_pin}.
We do not consider workload traces that \rbc{have} lower than 3 last-level cache misses per kilo instructions (MPKI) in the baseline system with no prefetching. In all, we present results for $150$ workload traces spanning $50$ workloads. \Cref{table:py_workloads} shows a categorized view of all the workloads \rbc{evaluated} in this \rbfor{work}.}
For multi-core multi-programmed simulations, we create \rbc{both homogeneous and heterogeneous} trace mixes from our single-core trace list. For an $n$-core homogeneous trace mix, we run $n$ copies of \rbc{a} trace from our single-core trace list, one in each core. For \rbc{a} heterogeneous trace mix, we \emph{randomly} select $n$ traces from our single-core trace list and run one trace in every core. All the single-core traces and multi-programmed trace mixes used in our evaluation are freely available online~\cite{pythia_github}.
\end{sloppypar}

\begin{table}[htbp]
  \centering
  \small
    \begin{tabular}{L{7em}L{5em}L{5em}m{15em}}
    \toprule
    \textbf{Suite} & \multicolumn{1}{l}{\textbf{\# Workloads}} & \multicolumn{1}{l}{\textbf{\# Traces}} & \textbf{Example Workloads} \\
    \midrule
    \Tabval{SPEC06} & \Tabval{16}    & \Tabval{28}    & \Tabval{gcc, mcf, cactusADM, lbm, ...} \\
    \Tabval{SPEC17} & \Tabval{12}    & \Tabval{18}    & \Tabval{gcc, mcf, pop2, fotonik3d, ...} \\
    \Tabval{PARSEC} & \Tabval{5}     & \Tabval{11}    & \Tabval{canneal, facesim, raytrace, ...} \\
    \Tabval{Ligra} & \Tabval{13}    & \Tabval{40}    & \Tabval{BFS, PageRank, Bellman-ford, ...} \\
    \Tabval{Cloudsuite} & \Tabval{4}     & \Tabval{53}    & \Tabval{cassandra, cloud9, nutch, ...} \\
    \bottomrule
    \end{tabular}%
  \caption{Workloads used for evaluation.}
  \label{table:py_workloads}%
\end{table}%

\subsection{Prefetchers}
\begin{sloppypar}
We \rbc{compare} Pythia \rbc{to five state-of-the-art} prior prefetchers: SPP~\cite{spp}, SPP+PPF~\cite{ppf}, SPP+DSPatch~\cite{dspatch}, Bingo~\cite{bingo}, and MLOP~\cite{mlop}.
We model each competing prefetcher using the source-code provided by their respective authors and fine-tune them in our environment to extract the \rbc{highest} performance gain \rbc{across} all single-core traces.
\rbc{\mbox{\Cref{table:py_pref_params}} shows the parameters of all evaluated prefetchers.}
Each prefetcher is trained on \rbc{L1-cache} misses and fills prefetched lines into L2 and LLC. 
\end{sloppypar}
We also \rbc{compare} Pythia against multi-level prefetchers found in commercial processors (e.g., stride prefetcher at L1-cache and streamer at L2~\cite{intel_prefetcher}) and IPCP~\cite{ipcp} in~\Cref{sec:py_eval_multi_level}. \rbc{For fair comparison, we add a simple PC-based stride prefetcher~\cite{stride,stride_vector,jouppi_prefetch} \rbc{at the} L1 \rbc{level}, along with Pythia at \rbc{the} L2 \rbc{level} for such \rbc{multi-level comparisons}.}

\begin{table}[h]
    \centering
    \small
    \begin{tabular}{L{6em}||L{24em}||R{5em}}
    \thickhline
    \Tabval{\textbf{SPP}~\cite{spp}} & \Tabval{256-entry ST, 512-entry 4-way PT, 8-entry GHR} & \Tabval{\textbf{6.2~KB}} \\
    \hline
    \Tabval{\textbf{Bingo}~\cite{bingo}} & \Tabval{2KB region, 64/128/4K-entry FT/AT/PHT} & \Tabval{\textbf{46~KB}} \\
    \hline
    \Tabval{\textbf{MLOP}~\cite{mlop}} & \Tabval{128-entry AMT, 500-update, 16-degree} & \Tabval{\textbf{8~KB}} \\
    \hline
    \Tabval{\textbf{DSPatch}~\cite{dspatch}} & \Tabval{Same configuration as in~\cite{dspatch}} & \Tabval{\textbf{3.6~KB}} \\
    \hline
    \Tabval{\textbf{PPF}~\cite{ppf}} & \Tabval{Same configuration as in~\cite{ppf}} & \Tabval{\textbf{39.3~KB}} \\
    \thickhline
    \Tabval{\textbf{Pythia}} & \Tabval{2 features, 2 vaults, 3 planes, 16 actions} & \Tabval{\textbf{25.5~KB}}\\
    \thickhline
    \end{tabular}
    \caption{Configuration of evaluated prefetchers.}
    \label{table:py_pref_params}
    \vspace{-1em}
\end{table}

\sectionRB{Pythia: Evaluation}{Evaluation}{sec:py_evaluation}

\subsection{Prefetch Coverage and Overprediction Analysis}\label{sec:py_eval_cov_acc}
\begin{figure}[!ht]
\centering
\includegraphics[width=5.75in]{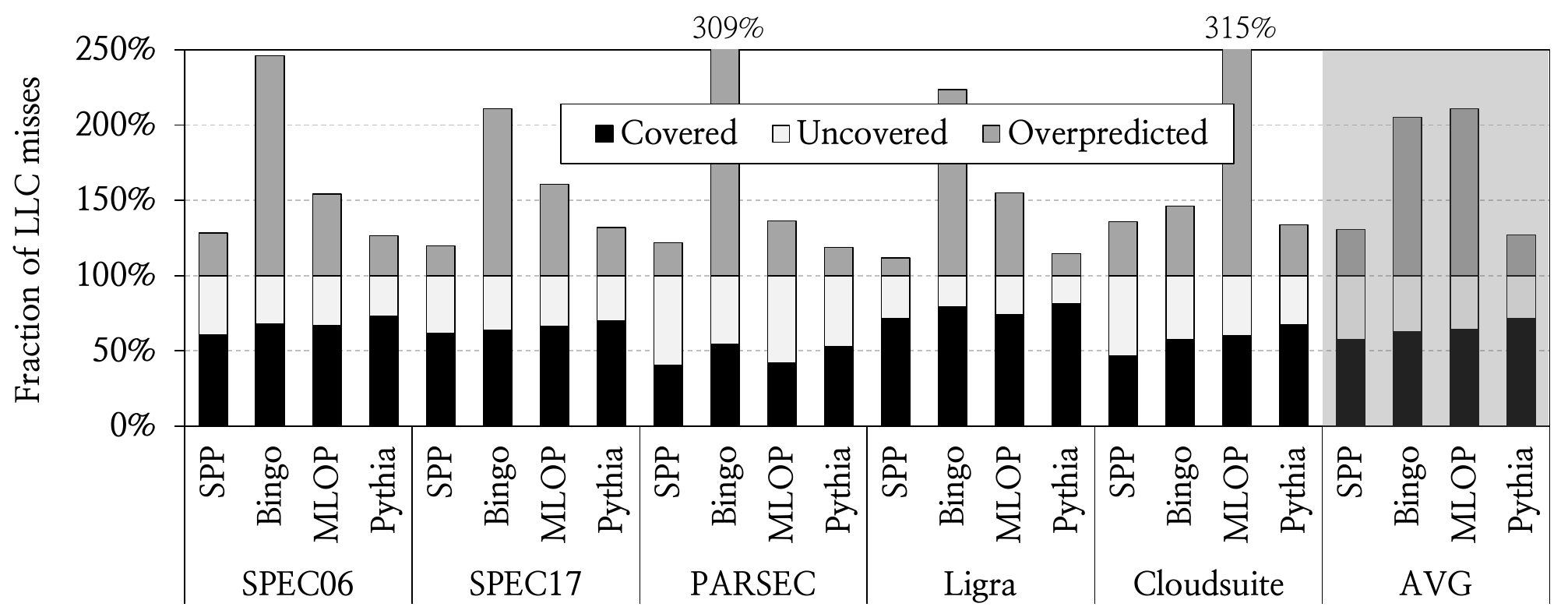}
\caption{Coverage and overprediction with respect to the baseline LLC misses in \rbc{the} single-core \rbc{system}. 
}
\label{fig:py_cov_acc}
\end{figure}

\Cref{fig:py_cov_acc} shows the coverage and overprediction of each prefetcher in \rbc{the} single-core \rbc{system}, \rbc{as} measured at the LLC-main memory boundary. 
The key takeaway is that Pythia improves prefetch coverage, while simultaneously reducing overprediction \rbc{compared to} state-of-the-art prefetchers. On average, Pythia provides $6.9$\%, $8.8$\%, and $14$\% \rbc{higher} coverage than MLOP, Bingo, and SPP respectively, while generating $83.8$\%, $78.2$\%, and $3.6$\% \rbc{fewer} overpredictions.   

\subsection{Performance Analysis}\label{sec:py_eval_perf}

\subsubsection{\textbf{Performance in Single-Core System}}\label{sec:py_eval_perf_1T}
\begin{figure}[!ht]
\centering
\includegraphics[width=5.75in]{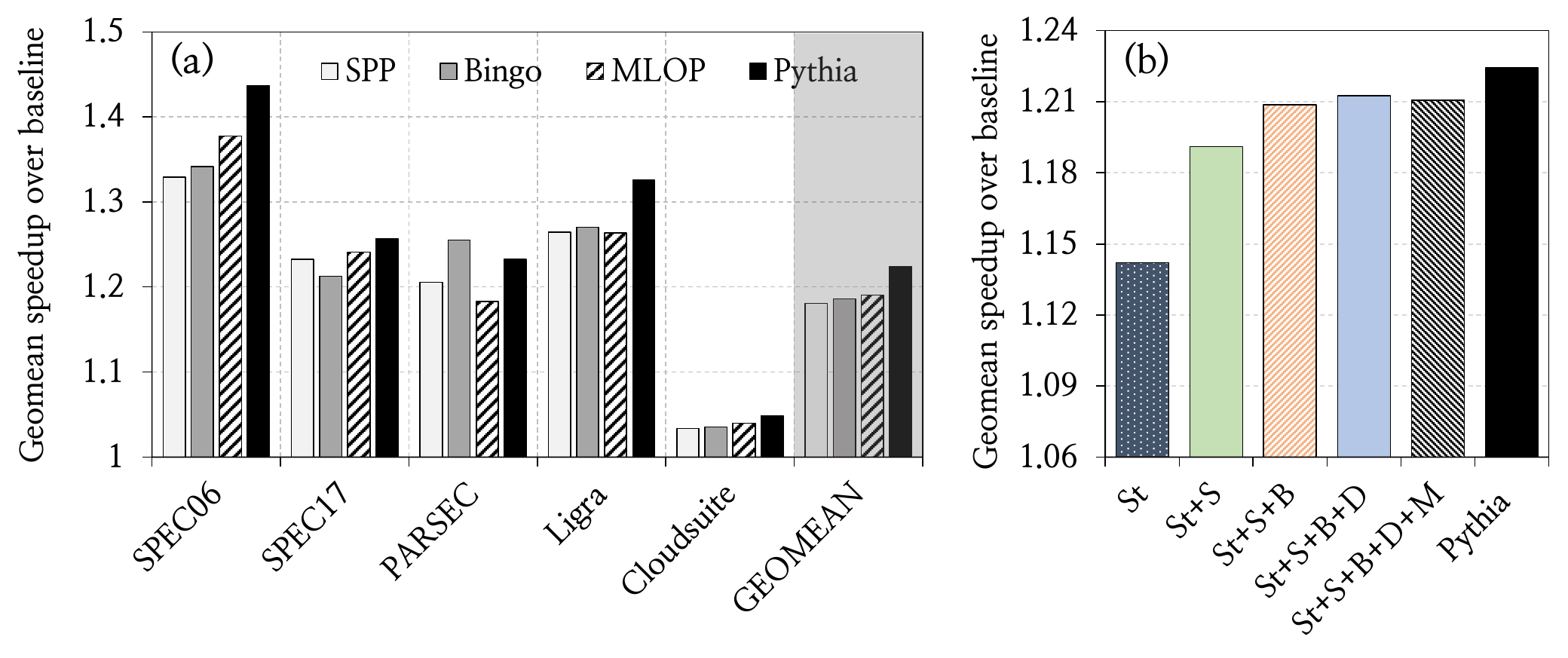}
\caption{Performance improvement in single-core workloads. St=Stride, S=SPP, B=Bingo, D=DSPatch, and M=MLOP.}
\label{fig:py_perf_1T}
\end{figure}

\Cref{fig:py_perf_1T}(a) \rbc{shows} the performance improvement of each individual prefetcher in each workload category in the single-core \rbc{system}. We make two major observations.
First, Pythia improves performance by $22.4$\% on average over a no-prefetching baseline. Pythia outperforms MLOP, Bingo, and SPP by $3.4$\%, $3.8$\%, and $4.3$\% on average, respectively. 
Second, \rbc{only} Bingo outperforms Pythia \rbc{only} in \rbc{the} \texttt{PARSEC} suite, by $2.3$\%. However, Bingo's performance comes at \rbc{the} cost of \rbc{a} high overprediction \rbc{rate}, which hurts performance in multi-core \rbc{systems (see~\Cref{sec:py_eval_perf_4T})}.

To demonstrate \rbc{the novelty of Pythia's RL-based prefetching approach using multiple program features}, \Cref{fig:py_perf_1T}(b) compares Pythia's performance improvement with the performance improvement of \rbc{various} combinations of prior prefetchers.
Pythia not only outperforms \rbc{all prefetchers (stride, SPP, Bingo, DSPatch, and MLOP)} individually, but also outperforms \rbc{their combination} by $1.4$\% on average, with less than half of the combined storage size \rbc{of the five prefetchers}. 
\rbc{We conclude that Pythia's RL-based prefetching approach using multiple program features under one single framework provides \rbc{higher} performance \rbc{benefit} than combining multiple prefetchers, each exploiting only one program feature.}

\begin{figure}[!ht]
\centering
\includegraphics[width=5.5in]{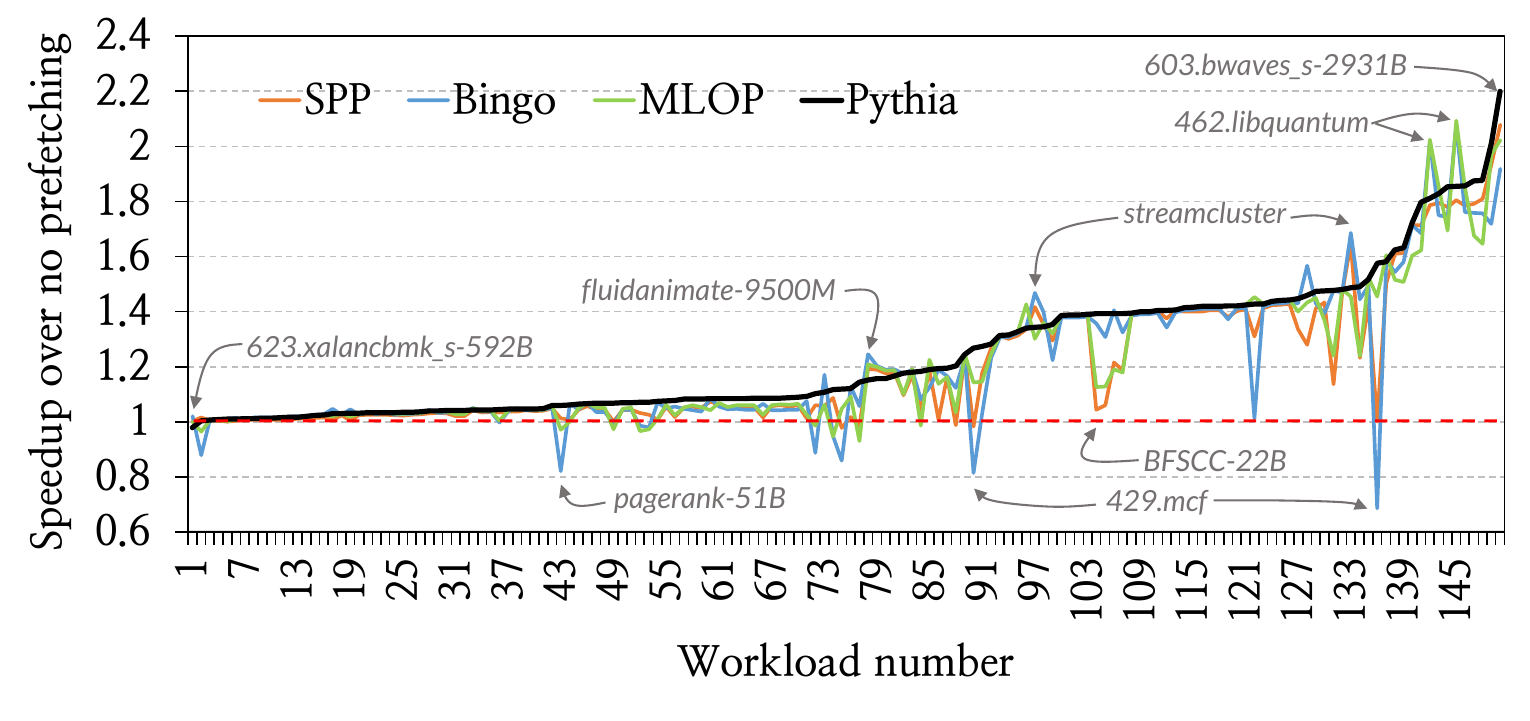}
\caption{Performance line graph of 150 single-core traces.}
\label{fig:py_perf_1T_scurve}
\end{figure}

\Cref{fig:py_perf_1T_scurve} shows the performance line graph of all prefetchers \rbcc{for the $150$} single-core workload traces. The workload traces are sorted in ascending order of performance improvement \rbcc{of} Pythia over the baseline without prefetching. We make three key observations. First, Pythia outperforms the no-prefetching baseline in every single-core trace, except \texttt{623.xalancbmk-592B} (where it underperforms the baseline by $2.1$\%). \texttt{603.bwaves-2931B} enjoys the highest performance improvement of $2.2\times$ over the baseline. \rbcc{Performance of the} top $80$\% of traces improve by at least $4.2$\% \rbcc{over the baseline}. Second, Pythia underperforms Bingo in workloads like \texttt{libquantum} 
due to \rbcc{the heavy streaming} nature of memory accesses. As \texttt{libquantum} streams through all physical pages, Bingo simply prefetches all cachelines of a page \rbcc{at once} just by seeing the first access \rbcc{to} the page. As a result Bingo achieves higher timeliness and higher performance than Pythia. Third, Pythia significantly outperforms every competing \rbcc{prefetcher} in workloads with irregular access patterns (e.g., \texttt{mcf}, \texttt{pagerank}).
We conclude that Pythia provides consistent performance \rbcc{gains} over the no-prefetching baseline and multiple prior state-of-the-art prefetchers over a wide range of workloads. 

\subsubsection{\textbf{Performance in Four-Core System}}\label{sec:py_eval_perf_4T}
\begin{figure}[!ht]
\centering
\includegraphics[width=5.75in]{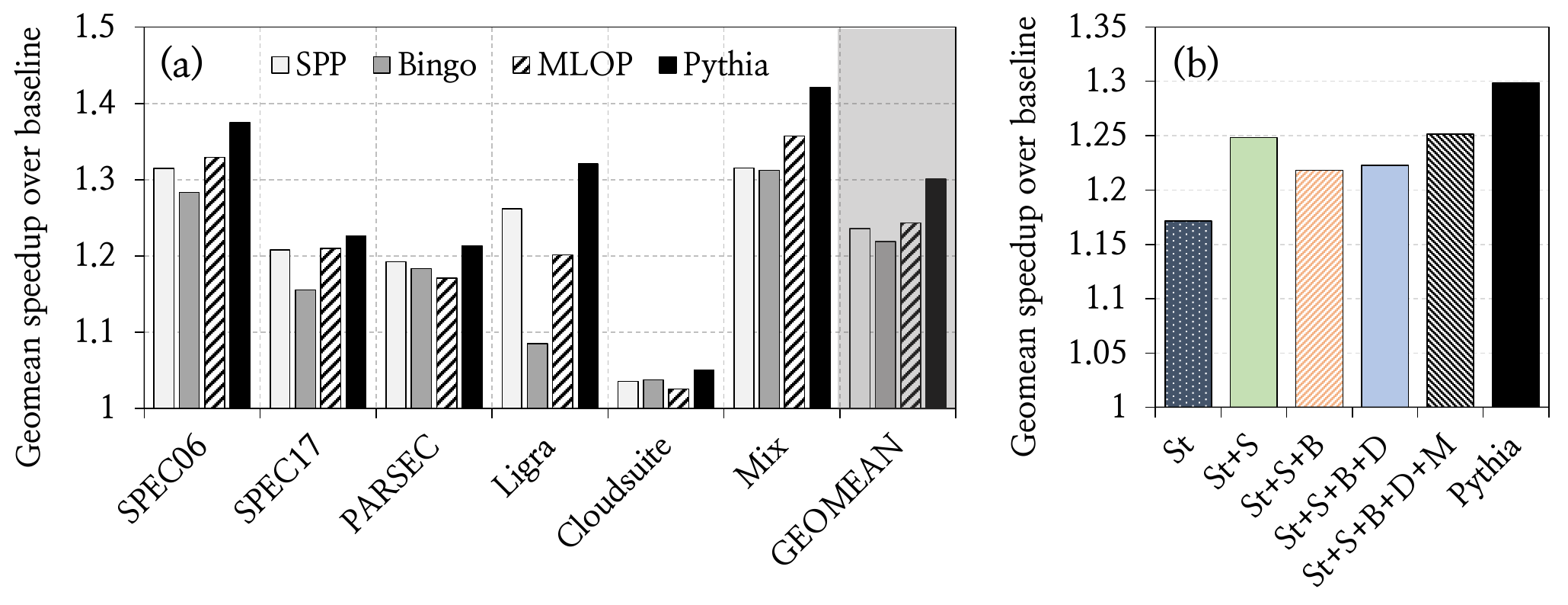}
\caption{Performance in \rbc{the} four-core \rbc{system}. 
}
\label{fig:py_perf_4T}
\end{figure}

\Cref{fig:py_perf_4T}(a) \rbc{shows} the performance improvement of each individual prefetcher in each workload category in the four-core \rbc{system}. We make \rbc{two} major observations. 
First, Pythia provides significant performance improvement over all prefetchers in \emph{every} workload category in \rbc{the} four-core \rbc{system}. On average, Pythia outperforms MLOP, Bingo, and SPP by $5.8$\%, $8.2$\%, and $6.5$\% respectively.
Second, unlike \rbc{in the} single-core \rbc{system}, Pythia outperforms Bingo in \texttt{PARSEC} by $3.0$\% in \rbc{the} four-core \rbc{system}. \rbc{This is due to Pythia's ability to dynamically increase prefetch accuracy during high DRAM bandwidth usage.}

\rbc{\Cref{fig:py_perf_4T}(b) shows that} Pythia outperforms the combination of stride, SPP, Bingo, DSPatch, and MLOP prefetchers by $4.9$\% on average.
Unlike \rbc{in the} single-core system, combining \rbc{more} prefetchers on top of stride+SPP in four-core system lowers the overall performance gain. 
This is due to the additive increase in the overpredictions \rbc{made} by each individual prefetcher, \rbc{which leads to performance degradation in the bandwidth-constrained four-core system.} Pythia's RL-based framework holistically learns to prefetch using multiple program features \rbc{and} \rbc{generates} \rbc{fewer} overpredictions, outperforming \rbc{all} combinations of all individual prefetchers.

\begin{figure}[!ht]
\centering
\includegraphics[width=5.5in]{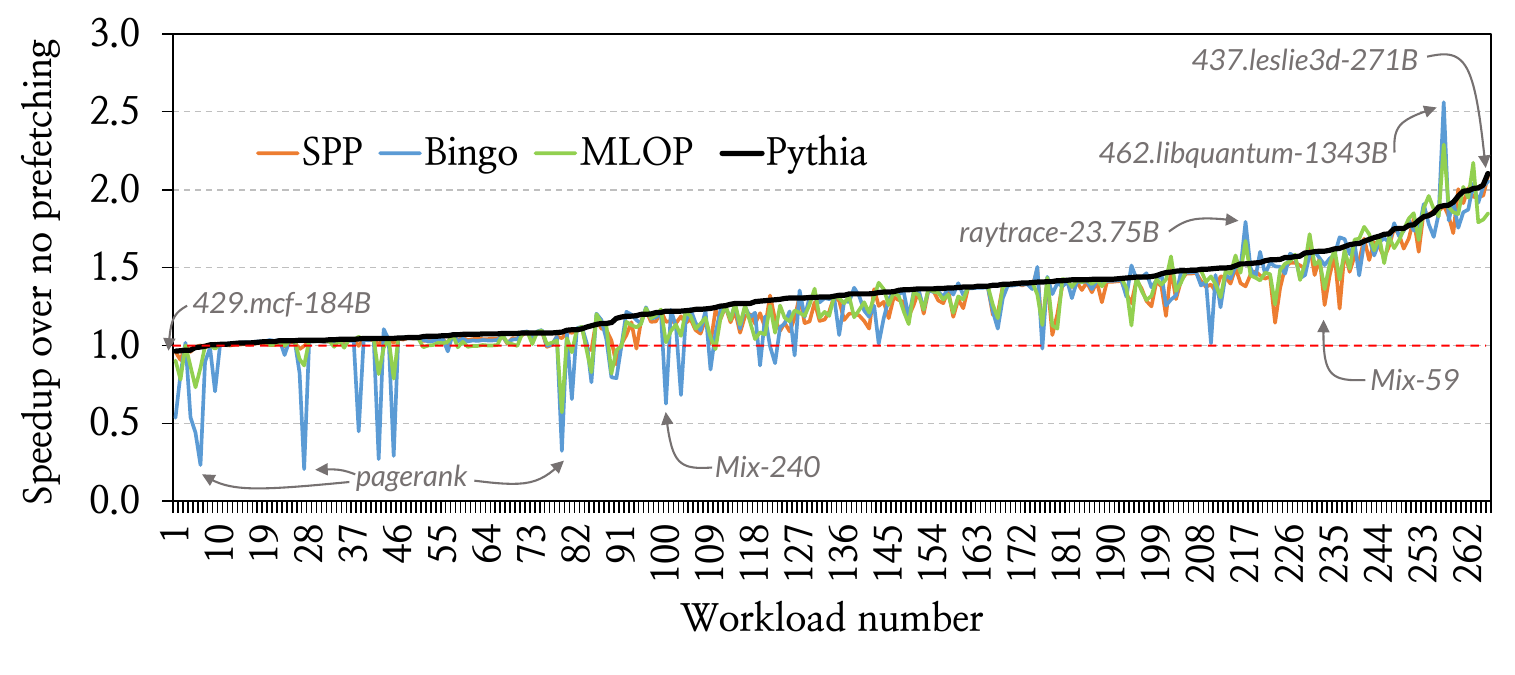}
\caption{Performance line graph of 272 four-core trace mixes.}
\label{fig:py_perf_4T_scurve}
\end{figure}

\Cref{fig:py_perf_4T_scurve} shows the performance line graph of all prefetchers \rbcc{for 272} four-core workload trace mixes (including both homogeneous and heterogeneous mixes). The workload mixes are sorted in ascending order of performance improvement \rbcc{of} Pythia over the baseline without prefetching. We make two key observations. First, Pythia outperforms the baseline without prefetching in \rbcc{all but one} four-core trace mix. Pythia provides the highest performance gain in \texttt{437.leslie3d-271B} ($2.1\times$) and lowest performance gain in \texttt{429.mcf-184B} (-$3.5$\%) over the no-prefetching baseline. Second, Pythia also outperforms (or matches performance) all competing prefetchers in \rbcc{majority of} trace mixes. Pythia underperforms Bingo in \rbcc{the} \texttt{462.libquantum} homogeneous trace mix due to \rbcc{the very} regular streaming access pattern. On the other hand, Pythia significantly outperforms \rbcc{Bingo} in \texttt{Ligra} workloads (e.g., \texttt{pagerank}) due to its adaptive prefetching strategy to trade-off coverage for accuracy in high memory bandwidth usage. We conclude that Pythia provides a consistent performance gain over multiple prior state-of-the-art prefetchers over a wide range of workloads even in bandwidth-constrained \rbcc{multi-core} systems.

\subsubsection{Performance \rbc{on Unseen} Traces} \label{sec:py_eval_unknown_traces}
To demonstrate Pythia's ability to \rbc{provide performance gains} across \rbc{workload traces that are not used \rbc{at all} to tune Pythia}, we evaluate Pythia using an additional $500$ traces from the second value prediction championship~\mbox{\cite{cvp2}} \rbc{on} both single-core and four-core \rbc{systems}. These traces \rbc{are classified into} floating-point, integer, crypto, and server \rbc{categories} and each of them has at least 3 LLC MPKI in \rbc{the} baseline \rbc{without prefetching}. 
No prefetcher, including Pythia, has been tuned on these traces. 
\rbc{In \rbc{the} single-core system, Pythia outperforms MLOP, Bingo, and SPP on average by $8.3$\%, $3.5$\%, and $4.9$\%, respectively, across these traces. In \rbc{the} four-core system, Pythia outperforms MLOP, Bingo, and SPP on average by $9.7$\%, $5.4$\%, and $6.7$\%, respectively.}
\rbc{We conclude that, Pythia, tuned on a set of workload traces, provides \rbc{equally high} (or even better) performance benefits \rbc{on} unseen traces \rbc{for} which it has not \rbc{been} tuned.}

\begin{figure}[!ht]
\centering
\includegraphics[width=5.75in]{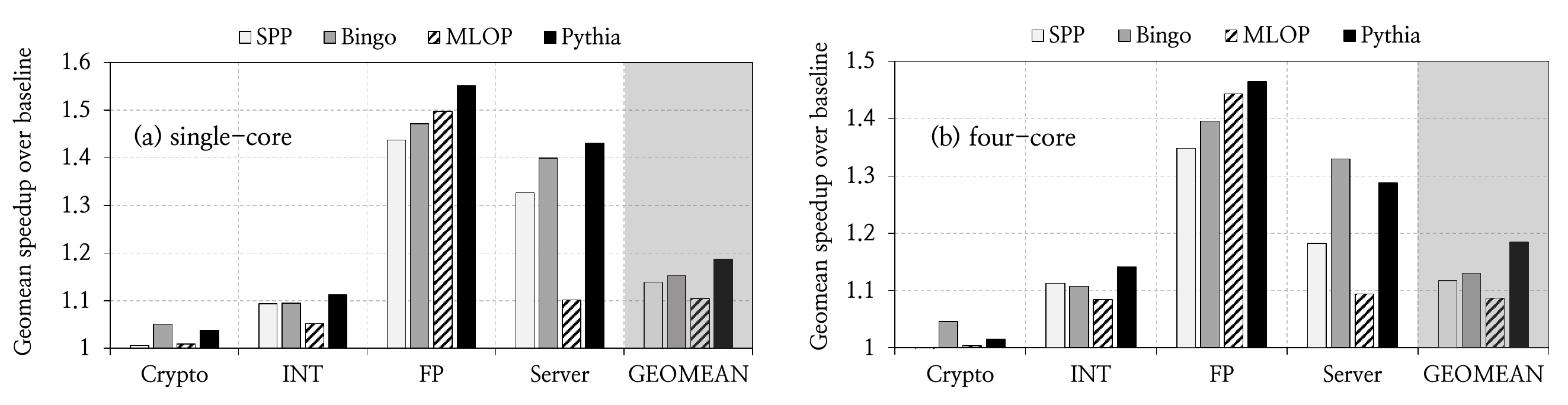}
\caption{Performance on unseen traces.}
\label{fig:py_perf_cvp1}
\end{figure}

\subsubsection{\textbf{Benefit of Awareness to \rbc{Memory} Bandwidth \rbc{Usage}}} \label{sec:py_eval_bw_awareness}
\begin{figure}[!ht]
\centering
\includegraphics[width=5.75in]{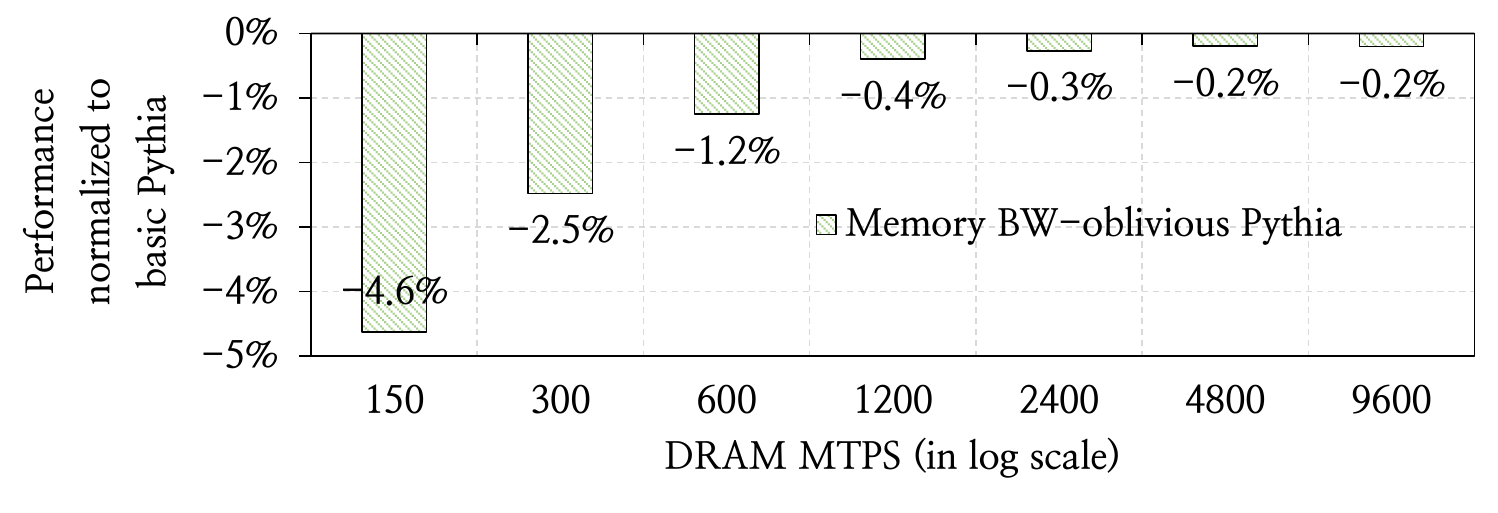}
\caption{Performance of \rbc{memory} bandwidth-oblivious Pythia \rbc{versus} the \rbcb{basic} Pythia.}
\label{fig:py_bw_awareness}
\end{figure}

To demonstrate the benefit of Pythia's awareness \rbc{of} system memory bandwidth usage, we compare the performance of the full-blown Pythia with a new version of Pythia that is oblivious to system memory bandwidth usage. We create this bandwidth-oblivious version of Pythia by \rbc{setting the high and low bandwidth usage variants of the rewards $\mathcal{R}_{IN}$ and $\mathcal{R}_{NP}$ to the same value (\rbc{i.e.,} essentially removing the bandwidth usage distinction from the reward values). More specifically, we set $\mathcal{R}_{IN}^{H}=\mathcal{R}_{IN}^{L}=-8$ and $\mathcal{R}_{NP}^{H}=\mathcal{R}_{NP}^{L}=-4$.}
\Cref{fig:py_bw_awareness} shows the performance benefit of the \rbc{memory} bandwidth-oblivious Pythia normalized to the \rbcb{basic} Pythia \rbc{as we vary the DRAM bandwidth}. \rbc{The key takeaway is that} 
\rbc{the bandwidth-oblivious Pythia loses performance by up to $4.6$\% on average across all \rbc{single-core} traces when the available memory bandwidth is low ($150$-MTPS to $600$-MTPS configuration).}
However, \rbc{when the available memory bandwidth is high ($1200$-MTPS to $9600$-MTPS)}, the \rbc{memory} bandwidth-oblivious Pythia \rbc{provides} similar performance improvement \rbc{to} the \rbcb{basic} Pythia.
We conclude that, memory bandwidth awareness gives Pythia the ability to provide robust performance \rbc{benefits} across a wide range of system \rbc{configurations}.

\subsection{Performance Sensitivity Analysis}\label{sec:py_eval_scalability}
\subsubsection{\textbf{Varying Number of Cores}}

\begin{figure*}[!ht]
\centering
\includegraphics[width=\columnwidth]{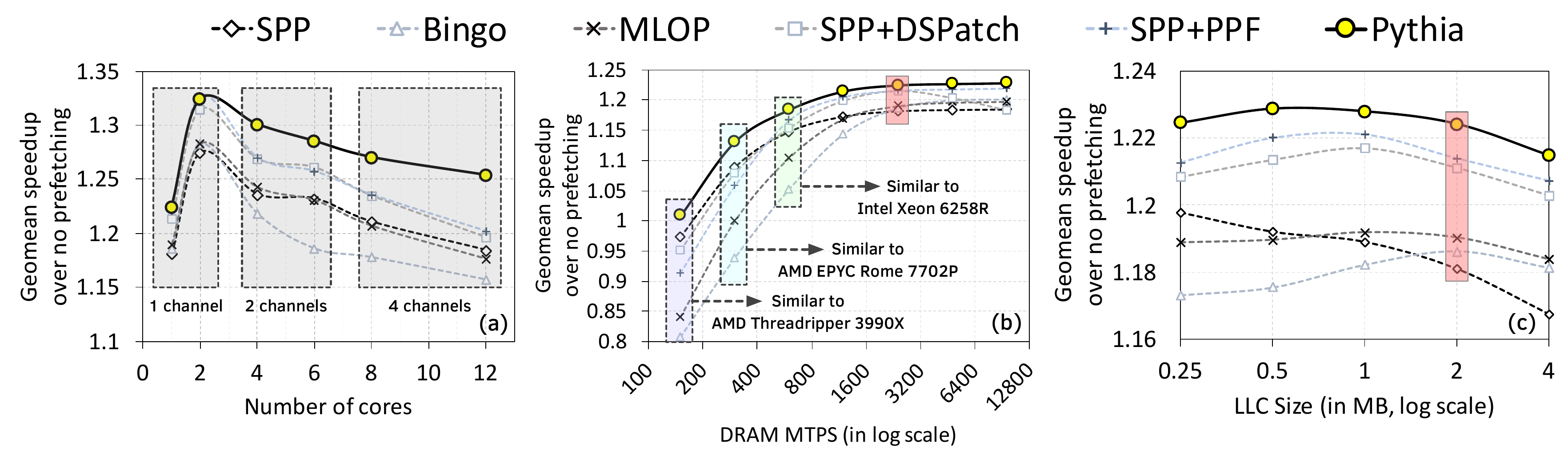}
\caption{Geomean performance improvement of prefetchers in \rbc{systems} with varying (a) number of cores, (b) DRAM \rbc{million transfers per second (MTPS)}, and (c) LLC size. Each DRAM bandwidth configuration roughly matches MTPS/core of various commercial processors~\cite{intel_xeon_gold,zen_epyc,zen_threadripper}. The baseline bandwidth/LLC configuration is marked in red.}
\label{fig:py_scalability_master}
\end{figure*}

\Cref{fig:py_scalability_master}(a) shows \rbc{the} performance improvement of all prefetchers averaged across all traces in single-core to 12-core \rbc{systems}. 
\rbc{To realistically model} modern commercial multi-core processors, we simulate $1$-$2$ core, $4$-$6$ core, and $8$-$12$ core \rbc{systems} with one, two, and four DDR4-$2400$ DRAM~\cite{ddr4} channels, respectively.
We make two key observations from~\Cref{fig:py_scalability_master}(a). First, Pythia consistently outperforms MLOP, Bingo, and SPP in \emph{all} \rbc{system} configurations. Second, Pythia's performance improvement over prior prefetchers increases \rbc{as} core count \rbc{increases}. 
In \rbc{the} single-core \rbc{system}, Pythia outperforms MLOP, Bingo, SPP, and an aggressive SPP with perceptron filtering (PPF~\mbox{\cite{ppf}}) by $3.4$\%, $3.8$\%, $4.3$\%, and $1.02$\% respectively.
In four (and twelve) core \rbc{systems}, Pythia outperforms MLOP, Bingo, SPP, and SPP+PPF by $5.8$\% ($7.7$\%), $8.2$\% ($9.6$\%), $6.5$\% ($6.9$\%), and $3.1$\% ($5.2$\%), respectively.

\subsubsection{\textbf{Varying DRAM Bandwidth}}
To evaluate Pythia in band-width-constrained, highly-\rbc{multi-}threaded commercial server-class processors, where each core \rbc{can have} only a fraction of a channel's bandwidth, we simulate the single-core single-channel configuration by scaling the DRAM bandwidth (\Cref{fig:py_scalability_master}(b)). 
Each bandwidth configuration roughly \rbc{corresponds} to the available \rbc{per-core} DRAM bandwidth \rbc{in} various commercial processors \rbc{(e.g., Intel Xeon Gold~\cite{intel_xeon_gold}, AMD EPYC Rome~\cite{zen_epyc}, and AMD Threadripper~\cite{zen_threadripper})}.
The key takeaway is that Pythia \emph{consistently} outperforms all competing prefetchers in \emph{every} DRAM bandwidth configuration from $\frac{1}{16}\times$ to $4\times$ bandwidth of the baseline system.
Due to \rbc{their large overprediction \rbc{rates}}, \rbc{the} performance \rbc{gains} of MLOP and Bingo \rbc{reduce} sharply \rbc{as} DRAM bandwidth \rbc{decreases}. \rbc{By} actively trading off prefetch coverage for higher accuracy based on \rbc{memory} bandwidth usage, Pythia outperforms MLOP, Bingo, SPP, and SPP+PPF by $16.9$\%, $20.2$\%, $3.7$\%, and $9.5$\% respectively in the most bandwidth-constrained configuration with 150 million transfers per second (MTPS).
In \rbc{the} $9600$-MTPS configuration, every prefetcher enjoys ample DRAM bandwidth. Pythia \rbc{still} outperforms MLOP, Bingo, SPP, and SPP+PPF by $3$\%, $2.7$\%, $4.4$\%, and $0.8$\%, respectively.

\subsubsection{\textbf{Varying LLC Size}}
\begin{sloppypar}
\Cref{fig:py_scalability_master}(c) shows performance of all prefetchers averaged across all traces in \rbc{the} single-core \rbc{system} \rbc{while} varying \rbc{the} LLC size from $\frac{1}{8}\times$ to $2\times$ of the baseline 2MB LLC.
\rbc{The key takeaway is that Pythia consistently outperforms all prefetchers in \emph{every} LLC size configuration.}
For $256$KB (and $4$MB) LLC, Pythia outperforms MLOP, Bingo, SPP, and SPP+PPF by $3.6$\% ($3.1$\%), $5.1$\% ($3.4$\%), $2.7$\% ($4.8$\%), and $1.2$\% ($0.8$\%), respectively.
\end{sloppypar}

\subsubsection{\textbf{Varying Hyperparameters}}
\begin{figure}[!ht]
\centering
\includegraphics[width=5.5in]{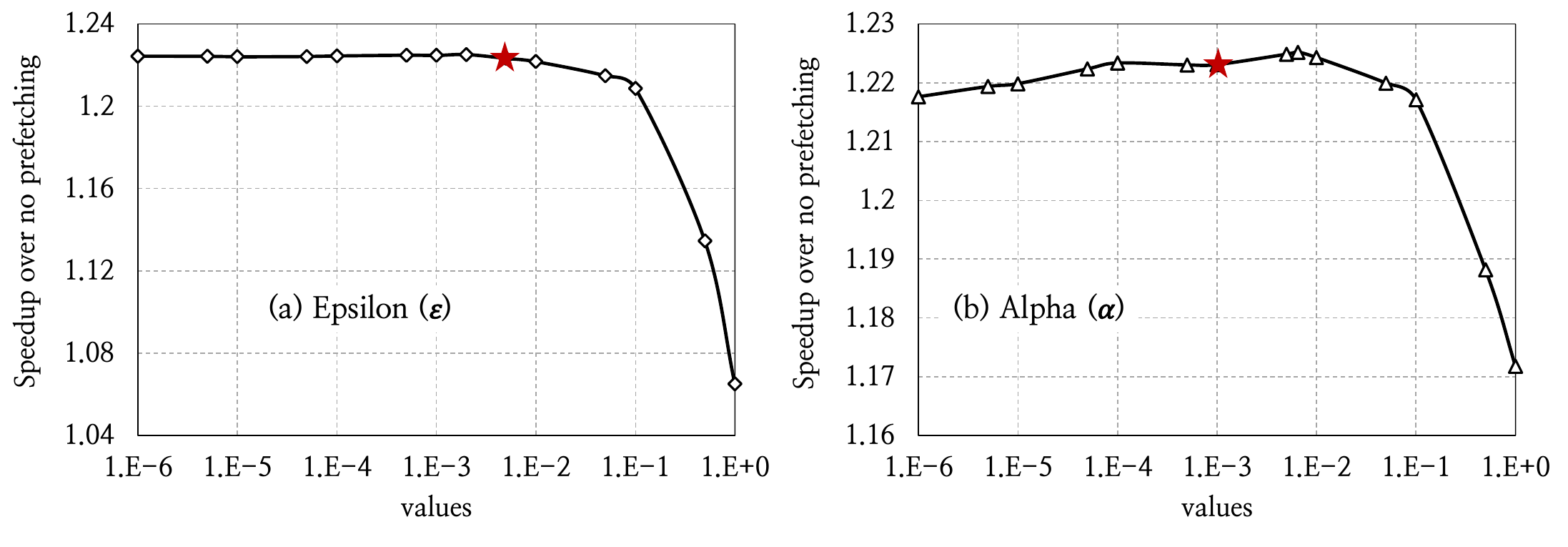}
\caption{Performance sensitivity of Pythia towards (a) the exploration rate ($\epsilon$), and (b) the learning rate ($\alpha$) hyperparameter values. The values in basic Pythia configuration are marked in red.}
\label{fig:py_hyp_sensitivity}
\end{figure}

\Cref{fig:py_hyp_sensitivity}(a) shows Pythia's performance sensitivity \rbcc{to the} exploration rate ($\epsilon$) averaged across all single-core traces. The key takeaway from~\Cref{fig:py_hyp_sensitivity}(a) is that Pythia's performance improvement drops sharply if the underlying RL-agent heavily \emph{explores} the state-action space \rbcc{as opposed to exploiting the learned policy}. Changing the $\epsilon$-value from $0.002$ to $1.0$ \rbcc{reduces Pythia's} performance improvement by $16.0$\%.
\Cref{fig:py_hyp_sensitivity}(b) shows Pythia's performance sensitivity \rbcc{to} learning rate \rbcc{parameter} ($\alpha$), averaged across all single-core traces. The key takeaway from \Cref{fig:py_hyp_sensitivity}(b) is \rbcc{that Pythia's performance improvement reduces for both increasing or decreasing the learning rate parameter. Increasing the learning rate reduces the hysteresis in Q-values (i.e., Q-values change significantly with the immediate reward received by Pythia), which reduces Pythia's performance improvement. Similarly, decreasing the learning rate also reduces Pythia's performance as it increases the hysteresis in Q-values.}
Pythia achieves optimal performance improvement for $\alpha=0.0065$. 

\subsubsection{Varying Program Features}
\begin{figure}[!ht]
\centering
\includegraphics[width=5.5in]{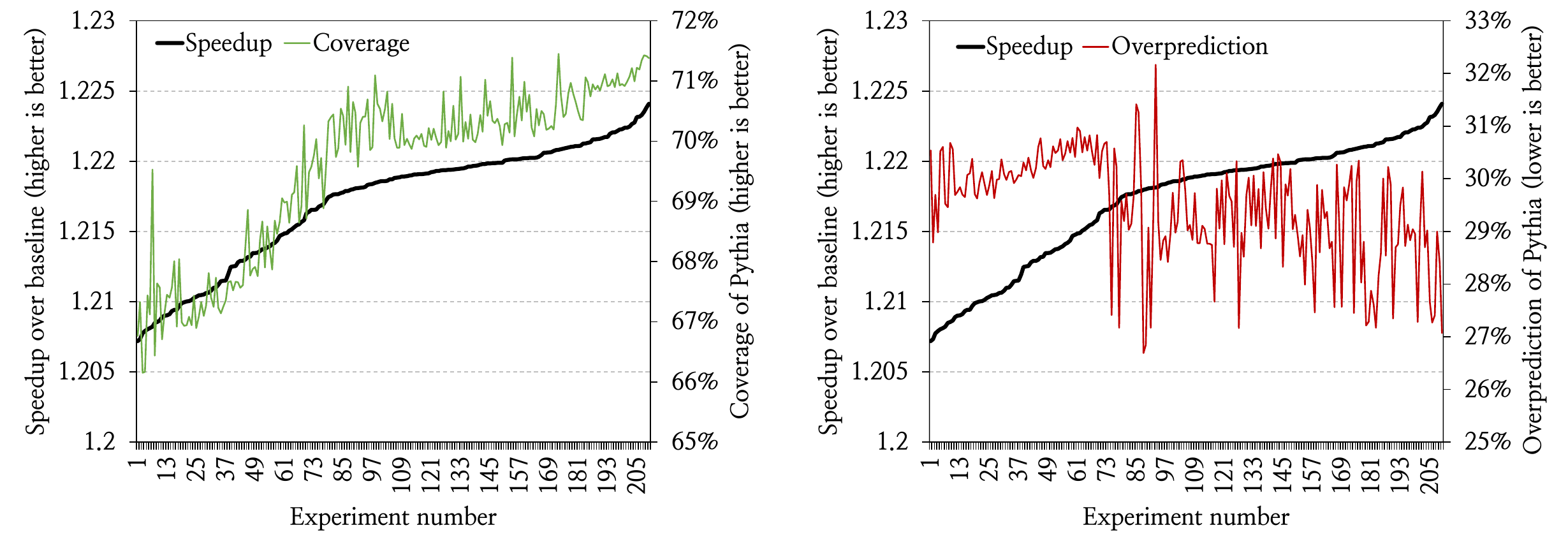}
\caption{Performance, coverage, and overprediction of Pythia with different feature combinations. \rbcc{The x-axis shows experiments with different feature combinations.}}
\label{fig:py_feature_sel}
\end{figure}

\Cref{fig:py_feature_sel} shows the performance, coverage, and overprediction of Pythia averaged across all single-core traces with different feature combinations during automated feature selection (\Cref{sec:py_tuning_feature_selection}). For brevity, we show results for all experiments with any-one and any-two combinations of $20$ features taken from the full list of $32$ features. Both graphs are sorted in ascending order of performance improvement \rbcc{of Pythia} over the baseline without prefetching. We make three key observations. First, Pythia's performance gain over the no-prefetching baseline improves from $20.7$\% to $22.4$\% \rbcc{by varying} the feature combination. We select the feature combination that provides the highest performance gain as the basic Pythia configuration (\Cref{tab:pythia_config}). Second, Pythia's coverage and overprediction also change significantly with \rbcc{varying} feature \rbcc{combination}. Pythia's coverage improves from $66.2$\% to $71.5$\%, whereas overprediction improves from $32.2$\% to $26.7$\% by changing feature \rbcc{combination}. Third, Pythia's performance gain positively correlates \rbcc{with} Pythia's coverage in single-core configuration. \rbcc{We conclude that automatic design-space exploration can significantly optimize Pythia's performance, coverage, and overpredictions.}

\subsubsection{Varying Number of Warmup Instructions}
\begin{figure}[!ht]
\centering
\includegraphics[width=5.75in]{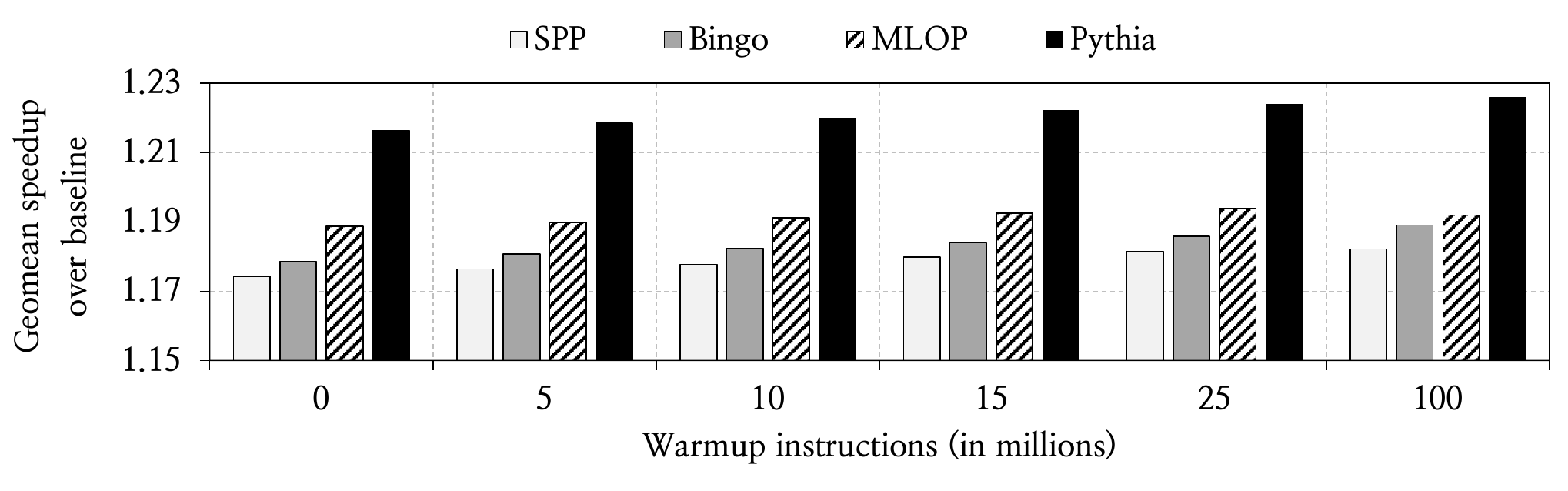}
\caption{Performance sensitivity of all prefetchers to number of warmup instructions.}
\label{fig:py_warmup_sen}
\end{figure}

\rbcd{\Cref{fig:py_warmup_sen} shows performance sensitivity of all prefetchers \rbcc{to the} number of warmup instructions averaged across all single-core traces. Our baseline simulation configuration uses $100$ million warmup instructions. The key takeaway from \Cref{fig:py_warmup_sen} is that Pythia consistently outperforms prior prefetchers in a wide range of simulation configurations using different number of warmup instructions. In the baseline simulation configuration using $100$M warmup instructions, Pythia outperforms MLOP, Bingo, and SPP by $3.4$\%, $3.8$\%, and $4.4$\% respectively. In a simulation configuration with no warmup instruction, Pythia continues to outperform MLOP, Bingo, and SPP by $2.8$\%, $3.7$\%, and $4.2$\% respectively. We conclude that, Pythia can quickly learn to prefetch from a program's memory access pattern and provides higher performance than other heuristics-based prefetching techniques over a wide range of simulation configurations using different number of warmup instructions.}

\subsection{Performance Comparison Against Hybrid Prefetching Schemes}\label{sec:py_eval_multi_level}

\subsubsection{Comparison to Multi-Level Prefetching}

\begin{figure}[!ht]
\centering
\includegraphics[width=5.5in]{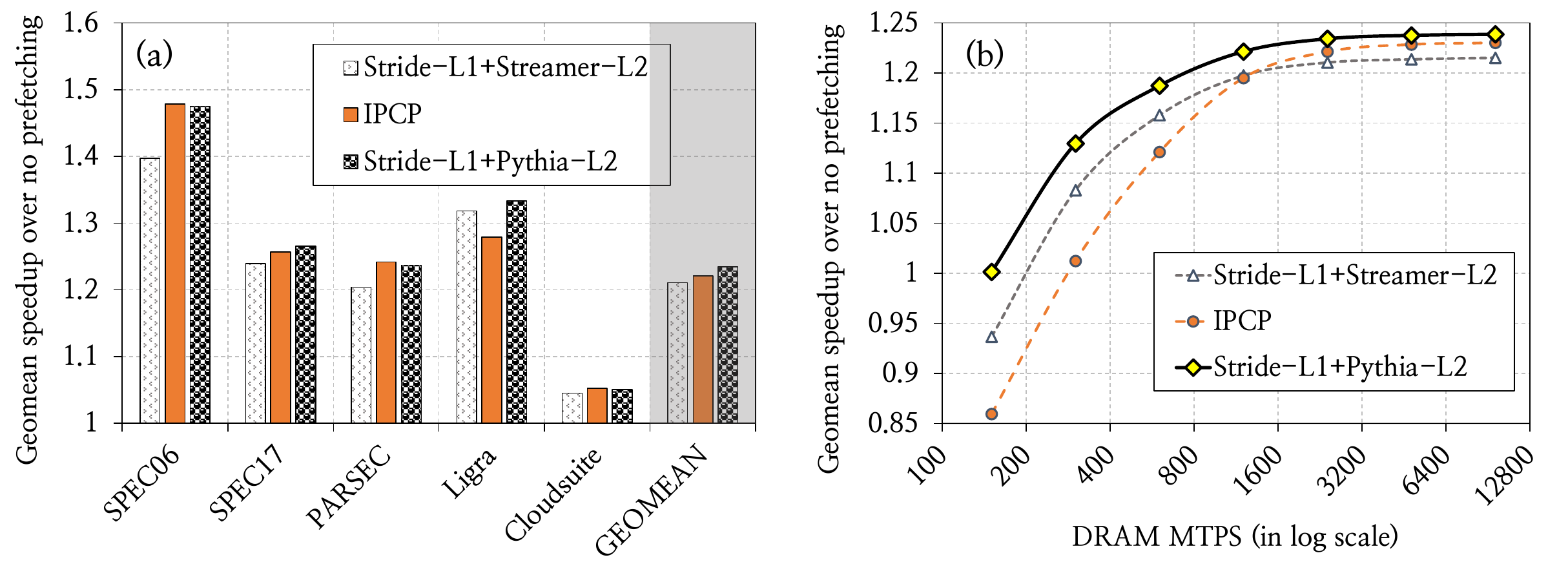}
\caption{Performance of multi-level prefetching schemes (a) in the baseline 2400-MTPS configuration, and (b) with varying main memory bandwidth.}
\label{fig:py_multi_level_pref}
\end{figure}

\Cref{fig:py_multi_level_pref}(a) shows the performance comparison of Pythia in single-core system with the baseline 2400-MTPS main memory bandwidth against two state-of-the-art \emph{multi-level} prefetching schemes: (1) stride prefetcher~\cite{stride,stride_vector,jouppi_prefetch} at L1 and streamer~\cite{streamer} at L2 cache found in commercial Intel processors~\cite{intel_prefetcher}, and (2) IPCP, the winner of \rbc{the} third data prefetching championship~\cite{dpc3}. 
For \rbc{fair} comparison, we add a stride prefetcher in \rbc{the} L1 cache along with Pythia in \rbc{the} L2 cache for this experiment and measure performance over the no prefetching baseline.
The key takeaway is that Stride+Pythia outperforms Stride+Streamer and IPCP by $2.4\%$ and $1.3\%$, respectively. 
To demonstrate Pythia's adaptiveness to memory bandwidth usage, we also show the average performance of Stride+Streamer, IPCP, and Stride+Pythia with varying main memory bandwidth in \Cref{fig:py_multi_level_pref}(b).
\rbc{The key takeaway is that \rbc{Stride+Pythia} consistently outperforms Stride+Streamer and IPCP in \emph{every} DRAM bandwidth configuration. Stride+Pythia outperforms \rbc{Stride+Streamer} and IPCP by $6.5$\% and $14.2$\% in \rbc{the} $150$-MTPS configuration and \rbc{by} $2.3$\% and $1.0$\% in \rbc{the} $9600$-MTPS configuration, respectively.}

\subsubsection{Comparison to the Context Prefetcher}
\begin{figure}[!ht]
\centering
\includegraphics[width=5.75in]{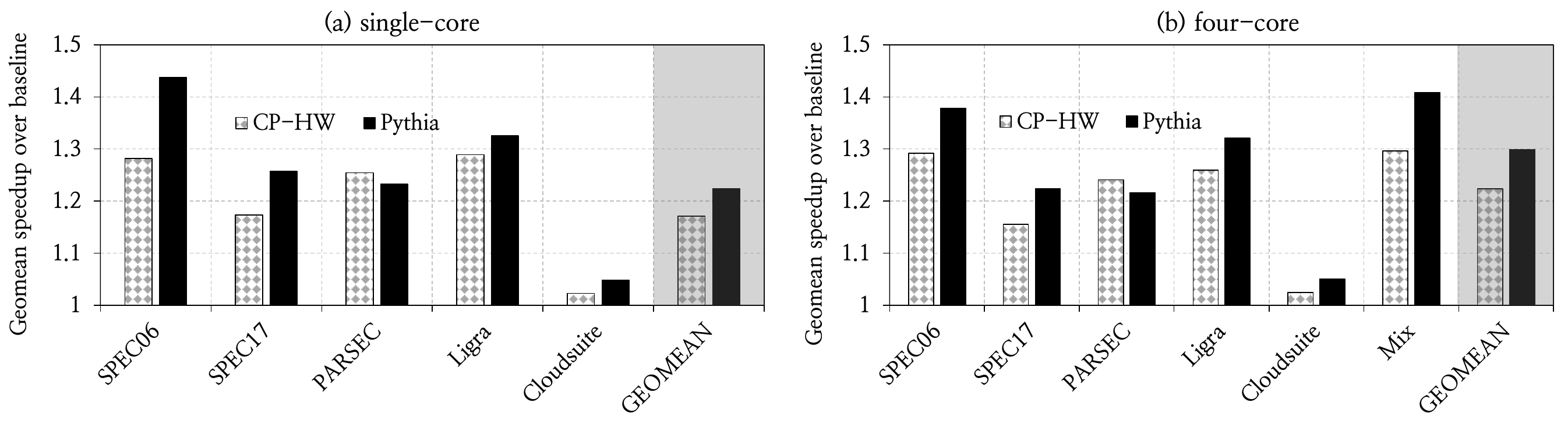}
\caption{Performance of Pythia vs. the context prefetcher~\cite{peled_rl} using hardware contexts.}
\label{fig:py_perf_context_pref}
\end{figure}

Context Prefetcher (CP~\cite{peled_rl}) is the first work that explored the application of RL for prefetching.
However, unlike Pythia, CP relies on both hardware and software contexts. A tailor-made compiler needs to encode the software contexts using \rbcc{special} NOP instructions, which are decoded by the core front-end to pass the context to the CP.
For a fair comparison, we implement the context prefetcher using only hardware contexts (CP-HW) and show the performance comparison \rbcc{of Pythia and CP-HW} in~\Cref{fig:py_perf_context_pref}. \rbcc{The key takeaway is that} Pythia outperforms the CP-HW prefetcher by $5.3$\% and $7.6$\% in single-core and four-core configurations, respectively.  \rbcc{Pythia's performance improvement over CP-HW mainly comes from two key aspects: (1) Pythia's ability to take memory bandwidth usage into consideration while taking prefetch actions, and (2) the far-sighted predictions made by Pythia as opposed to myopic predictions by CP-HW.}

\subsubsection{Comparison to \rbcc{the IBM POWER7} Adaptive Prefetcher}
\begin{figure}[!ht]
\centering
\includegraphics[width=5.75in]{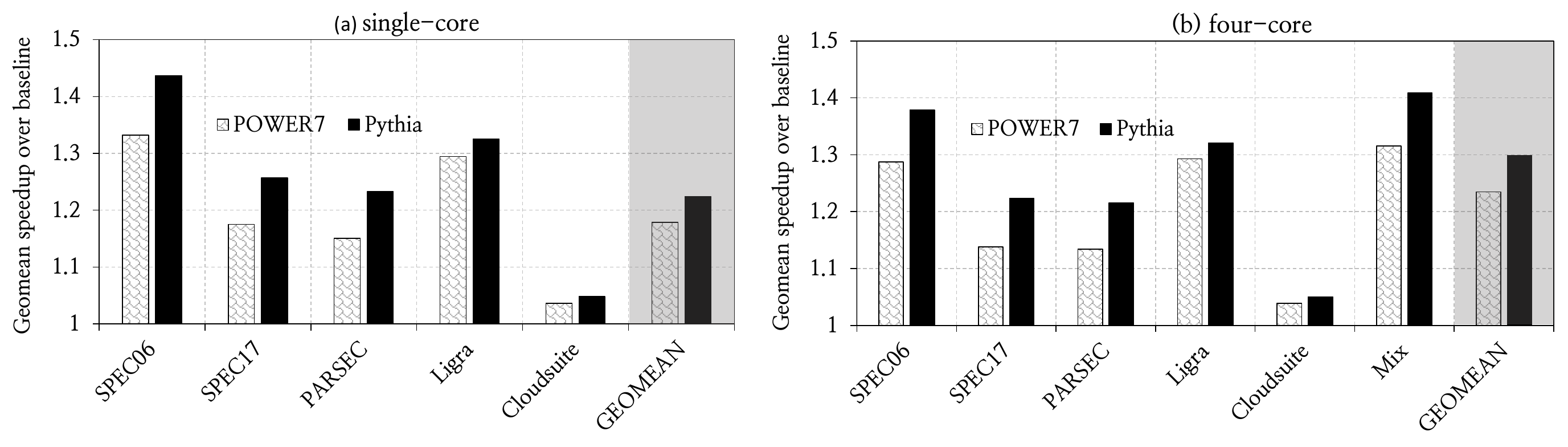}
\caption{Performance comparison against IBM POWER7 prefetcher~\cite{ibm_power7}.}
\label{fig:py_perf_power7}
\end{figure}

\rbcc{\Cref{fig:py_perf_power7} compares Pythia against the IBM POWER7 adaptive prefetcher~\mbox{\cite{ibm_power7}}. The POWER7 prefetcher dynamically tunes its prefetch aggressiveness (e.g., selecting prefetch depth, enabling stride-based prefetching) by monitoring program performance.}
We make two observations from \Cref{fig:py_perf_power7}. First, Pythia outperforms \rbcc{the} POWER7 prefetcher by $4.5$\% in single-core system. This is mostly due to Pythia's ability \rbcc{to} \rbcc{capture} different types of address patterns than just streaming/stride patterns. Second, Pythia outperforms POWER7 prefetcher by $6.5$\% in four-core and $6.1$\% in eight-core systems (not \rbcc{plotted}), respectively. The increase in performance improvement from single to four (or eight) core configuration suggests \rbcc{that} Pythia \rbcc{is more adaptive} than \rbcc{the} POWER7 prefetcher.

\subsection{Understanding Pythia using a Case Study}\label{sec:py_eval_case_study}
\begin{figure}[!ht]
\centering
\includegraphics[width=5.5in]{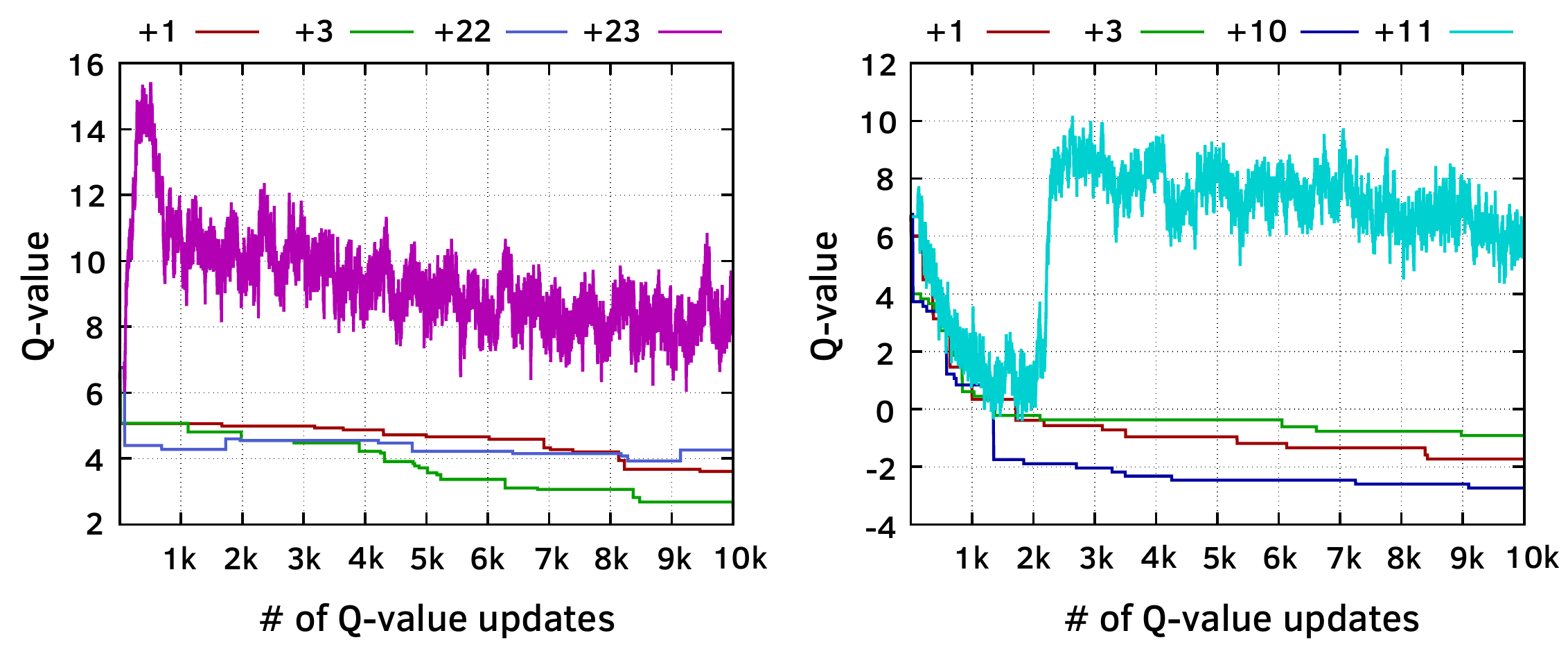}
\caption{Q-value curves of \texttt{PC+Delta} feature \rbc{values} (a) \texttt{0x436a81+0} and (b) \texttt{0x4377c5+0} in \texttt{459.GemsFDTD-1320B}.}
\label{fig:py_deepdive_gems}
\end{figure}

\begin{sloppypar}
\rbc{We} delve deeper into \rbc{an example} workload trace, \texttt{459.GemsFDTD-1320B}, from \texttt{SPEC CPU2006} suite \rbc{to provide more insight into Pythia's prefetching strategy and benefits}. 
In this trace, the top two most selected \rbc{prefetch} offsets by Pythia are $+23$ and $+11$, which cumulatively account for nearly $72$\% of all offset selections. 
For each of these offsets, we examine the program feature \rbc{value} that \rbc{selects} that offset \rbc{the most}. For simplicity, we only focus on the \texttt{PC+Delta} feature here.
The \texttt{PC+Delta} feature \rbc{values} \texttt{0x436a81+0} and \texttt{0x4377c5+0} select the \rbc{offsets} $+23$ and $+11$ the most, respectively. \rbc{\Cref{fig:py_deepdive_gems}(a) and (b)} show the Q-value curve of different actions for these feature. The x-axis shows the number of Q-value updates to the corresponding feature. Each color-coded line represents the Q-value of the respective action.
\end{sloppypar}

\begin{sloppypar}
As \Cref{fig:py_deepdive_gems}(a) shows, \rbc{the} Q-value of action $+23$ for feature \rbc{value} \texttt{0x436a81+0} consistently stays higher than \rbc{all} other actions (only three other \rbc{representative} actions are shown in~\Cref{fig:py_deepdive_gems}(a)). This means Pythia actively \rbc{favors} to prefetch \rbc{using} $+23$ offset whenever the PC \texttt{0x436a81} generates the first load to a physical page \rbc{(hence the delta $0$)}. By dumping the program trace, we indeed find that whenever PC \texttt{0x436a81} generates the first load to a physical page, there is only one more \rbc{address demanded} in that page \rbc{that} is $23$ cachelines ahead from the first loaded cacheline. \rbc{In} this case, the positive reward for generating \rbc{a} correct prefetch with offset $+23$ drives the Q-value of $+23$ \rbc{much} higher than \rbc{those of} other offsets and Pythia \rbc{successfully uses} the offset $+23$ for \rbc{prefetch request generation given} the feature \rbc{value} \texttt{0x436a81+0}.
We see similar \rbc{a} trend for the feature \rbc{value} \texttt{0x4377c5+0} with offset $+11$ (\Cref{fig:py_deepdive_gems}(b)).
\end{sloppypar}

\subsection{Performance Benefits via Customization}\label{sec:py_eval_pythia_custom}
\rbc{In this section, we show two examples of Pythia's \rbc{online} customization ability to extract \rbc{even} higher performance gain than the baseline Pythia configuration in target workload suites. First, we customize Pythia's reward level values for \rbc{the} \texttt{Ligra} \rbc{graph processing} workloads. Second, we customize the program features used by Pythia for \rbc{the} \texttt{SPEC CPU2006} workloads.}

\subsubsection{\textbf{Customizing Reward Levels}}\label{sec:py_eval_pythia_custom_reward}
\begin{sloppypar}
For workloads from \rbc{the} Ligra \rbc{graph processing} suite, we observe a general trend that a prefetcher with higher prefetch accuracy typically provides higher performance benefits. This is because any incorrect prefetch request wastes precious main memory bandwidth, which is already heavily used by the \rbc{demand requests of the} workload.
Thus, to improve Pythia's performance benefit in \rbc{the} Ligra suite, we create a new \emph{strict configuration} of Pythia that favors \rbc{\emph{not to prefetch}} \rbc{over} \rbc{generating inaccurate prefetches}. We create this strict configuration by simply \rbc{reducing} the reward level values for inaccurate prefetch (i.e., $\mathcal{R}_{IN}^{H}=-22$ and $\mathcal{R}_{IN}^{L}=-20$) and \rbc{increasing the reward level values for} no prefetch (i.e., $\mathcal{R}_{NP}^{H}=\mathcal{R}_{NP}^{L}=0$). 
\end{sloppypar}

\begin{figure}[!ht]
\centering
\includegraphics[width=5.75in]{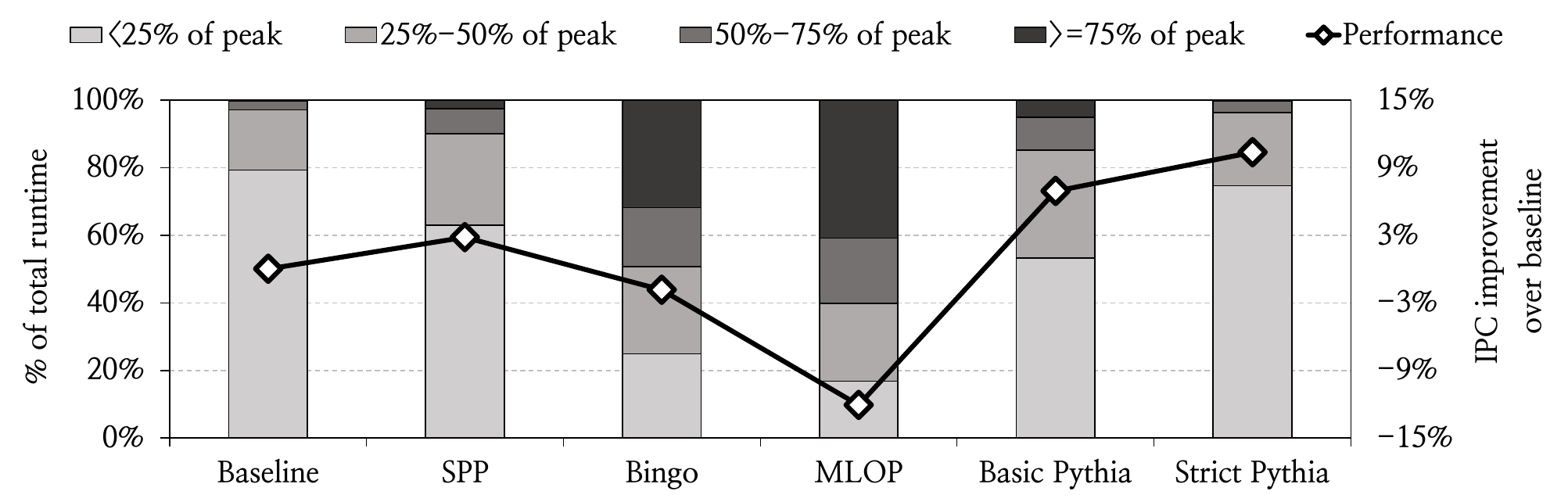}
\caption{Performance and main memory bandwidth usage of prefetchers in \texttt{Ligra-CC}.}
\label{fig:py_deepdive_ligra_cc}
\end{figure}

\begin{figure}[!ht]
\centering
\includegraphics[width=5.75in]{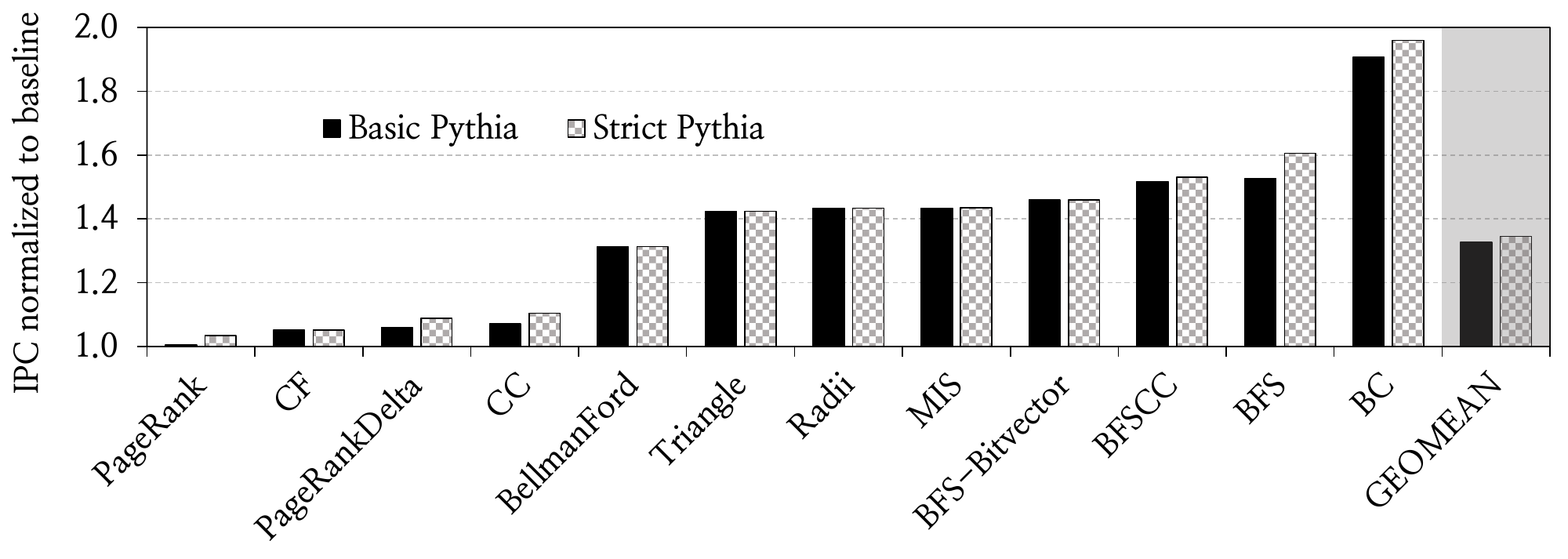}
\caption{Performance of the \rbc{basic} and strict \rbc{Pythia} configurations \rbcb{in} the \texttt{Ligra} workloads.}
\label{fig:py_ligra}
\end{figure}

\rbc{\Cref{fig:py_deepdive_ligra_cc} shows the percentage of the total runtime the workload spends in different bandwidth usage buckets in primary y-axis and the overall performance improvement in the secondary y-axis for each competing prefetcher \rbc{in one example workload} from \rbc{the} \texttt{Ligra} suite, \texttt{Ligra-CC}.}
\rbc{We make two key observations.}
\rbc{First, with MLOP and Bingo prefetchers enabled, \texttt{Ligra-CC} spends \rbc{a much} higher percentage of runtime consuming more than half of the peak DRAM bandwidth than in the no prefetching baseline. As a result, MLOP and Bingo underperforms the no prefetching baseline by $11.8\%$ and $1.8\%$, respectively. In contrast, \rbcb{basic} Pythia \rbc{leads to only} a modest memory bandwidth usage overhead, and outperforms the no prefetching baseline by $6.9$\%.}
\rbc{Second,} in \rbc{the} strict configuration, Pythia \rbc{has even less} \rbc{memory} bandwidth \rbc{usage} overhead, and provides $3.5$\% \rbc{higher} performance \rbc{than} the \rbcb{basic} Pythia configuration ($10.4$\% over \rbc{the no prefetching} baseline), without any hardware changes.
\Cref{fig:py_ligra} shows the performance \rbc{benefits} of the \rbcb{basic} and strict Pythia configurations for all \rbc{workloads from} \texttt{Ligra}. The key takeaway is that by simply changing the reward level values via configuration registers on the silicon, \rbc{strict} Pythia \rbc{provides} up to $7.8$\% ($2.0$\% on average) \rbc{higher} performance \rbc{than} \rbcb{basic} Pythia.
\rbc{We conclude that the objectives of Pythia can be easily customized via simple configuration registers for target workload suites to extract \rbc{even} higher performance benefits, without any changes to the underlying hardware.}

\subsubsection{\textbf{\rbc{Customizing Feature Selection}}}\label{sec:py_eval_pythia_custom_feature}
\begin{figure}[!ht]
\centering
\includegraphics[width=5.75in]{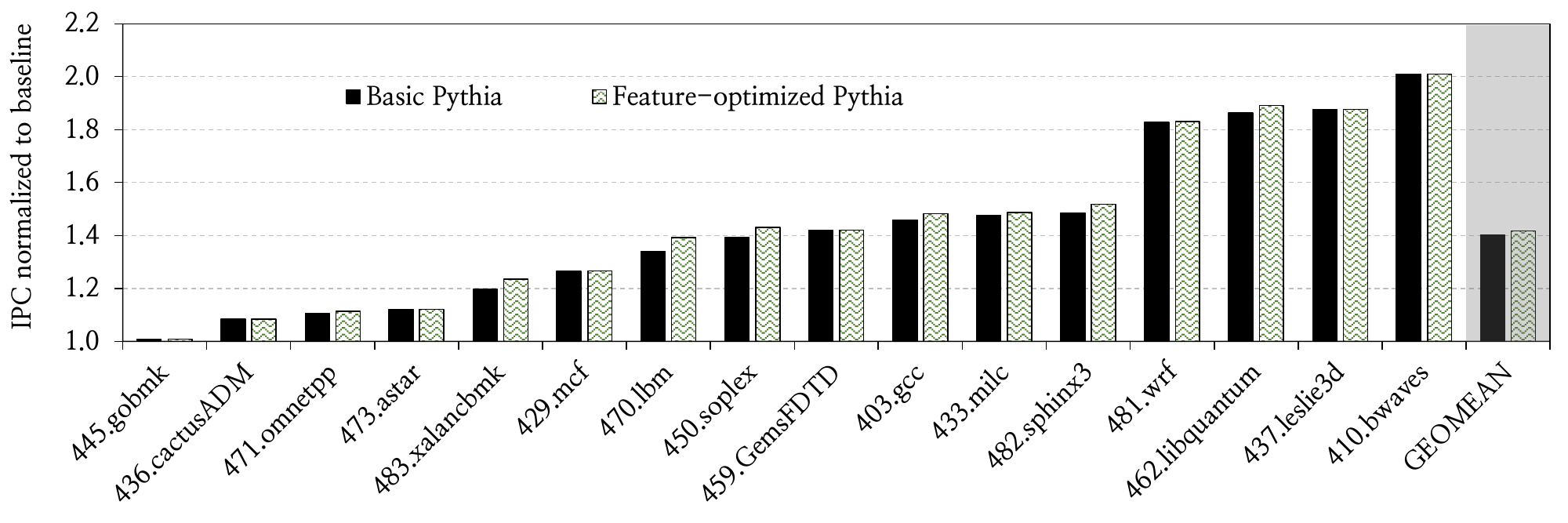}
\caption{Performance \rbc{of} \rbc{the basic} and \rbc{feature-optimized} Pythia \rbcb{on} \rbc{the} \texttt{SPEC CPU2006} suite.}
\label{fig:py_spec}
\end{figure}

To maximize the performance \rbc{benefits} of Pythia on \rbc{the} \texttt{SPEC CPU2006} workload suite, we run all one-combination and two-combination of program features from the initial set of $32$ supported features. For each workload, we fine-tune Pythia using the feature combination that provides the highest performance \rbc{benefit}. We call this the \emph{\rbc{feature-}optimized configuration} of Pythia for \texttt{SPEC CPU2006} suite. \Cref{fig:py_spec} shows the performance \rbc{benefits} of \rbcb{the basic} and optimized configurations of Pythia for all \texttt{SPEC CPU2006} workloads. The key takeaway is \rbc{that} by simply fine-tuning the program feature selection, Pythia delivers up to $5.1$\% ($1.5$\% on average) performance improvement on top of the \rbcb{basic} Pythia configuration.

\subsection{Performance Evaluation using DPC4 Traces}
\label{subsec:pythia_dpc4}

\begin{figure}[!ht]
\centering
\includegraphics[width=5.75in]{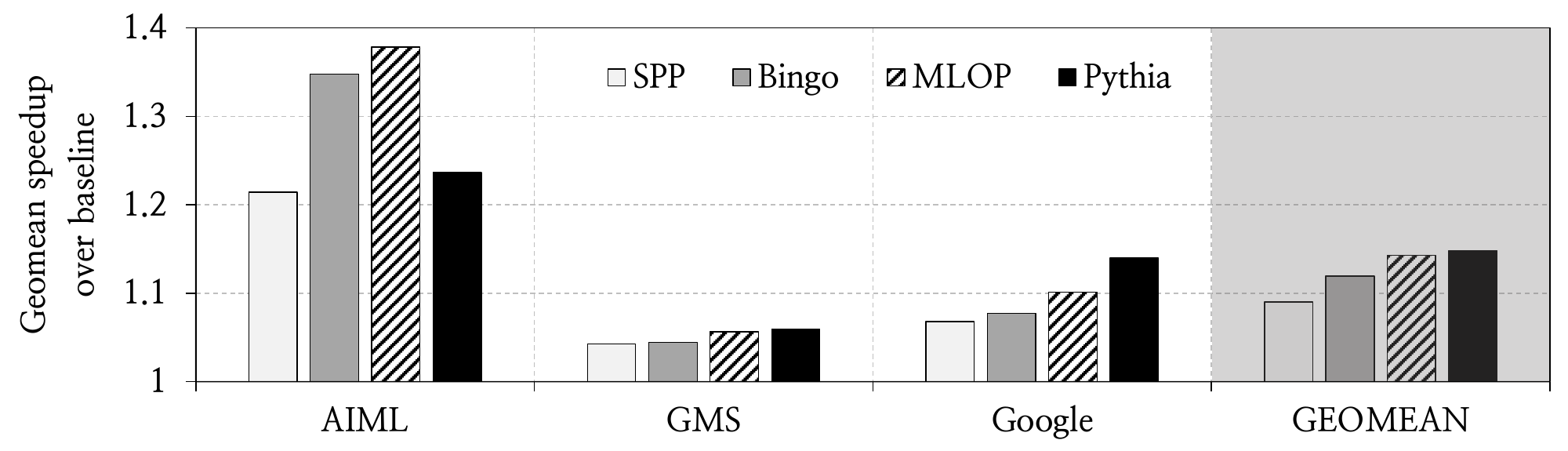}
\caption{Performance improvement in 483 single-core DPC4 traces.}
\label{fig:py_perf_dpc4}
\end{figure}

\Cref{fig:py_perf_dpc4} shows the geometric mean performance improvement of SPP, Bingo, MLOP, and Pythia over the no-prefetcher baseline across all $483$ single-core DPC4 workloads, categorized into AIML, GMS, and Google datacenter workload categories (see~\Cref{sec:dpc4_meth}). 
We make three key observations. 
First, Pythia achieves the highest overall geometric mean performance improvement across all $483$ DPC4 traces, outperforming \emph{all} prior prefetching mechanisms without \rbfor{requiring} any workload-specific tuning. 
Pythia improves performance by $14.8\%$ over the no-prefetching baseline, whereas SPP, Bingo, and MLOP \rbfiv{improve} performance by $9.1\%$, $11.9\%$, and $14.2\%$, respectively.
Second, Pythia delivers significant performance improvements in Google datacenter workload category.
Across these $359$ workload traces, Pythia outperforms SPP, Bingo, and MLOP on average by $7.2\%$, $6.3\%$, and $3.9\%$.
This \rbfor{result} strongly indicates Pythia's ability to adapt to complex and often irregular memory access behavior found in real production-scale workloads running on datacenters.
Third, Pythia underperforms MLOP and Bingo on AIML workloads. This is because these workloads exhibit pronounced streaming memory access pattern, where aggressive multi-degree spatial prefetching employed by Bingo and MLOP achieve higher gains. 
As Pythia’s learning-driven policy does not explicitly bias toward such highly regular streaming patterns, Pythia significantly underperforms Bingo and MLOP by $11.2\%$ and $14.2\%$, respectively.
This observation highlights a potential direction for future work: incorporating adaptive multi-degree prefetch control within Pythia’s RL framework to better capture highly regular streaming access patterns while also preserving its general adaptability.

\begin{figure}[!ht]
\centering
\includegraphics[width=5.75in]{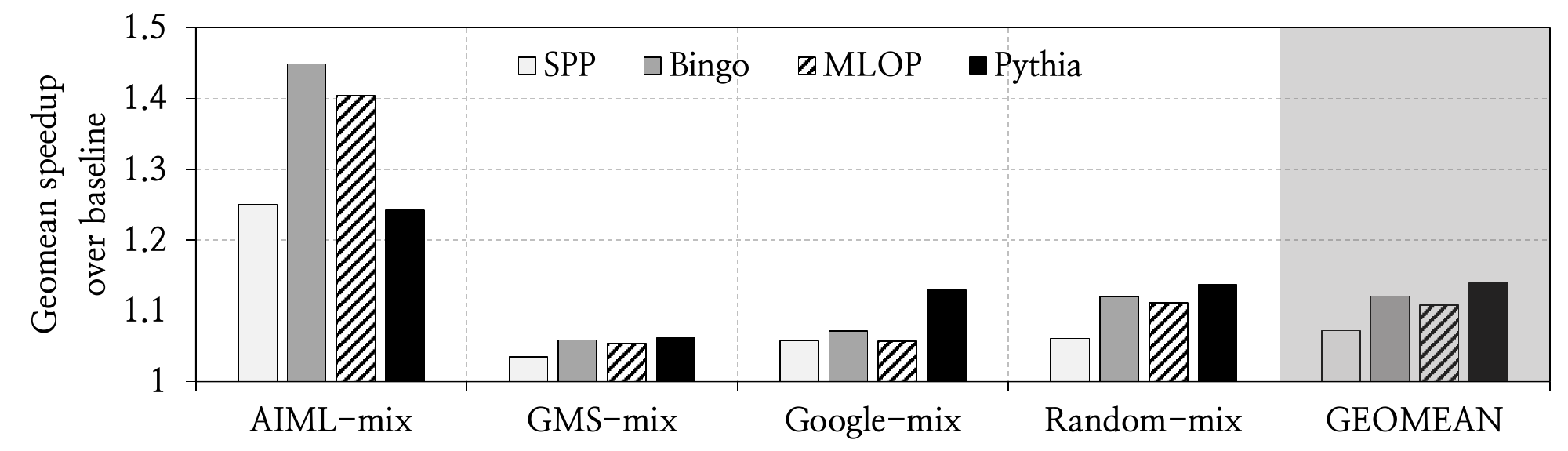}
\caption{Performance improvement in 966 four-core DPC4 traces.}
\label{fig:py_perf_dpc4_4c}
\end{figure}

\Cref{fig:py_perf_dpc4_4c} shows the category-wise geometric mean performance improvement of SPP, Bingo, MLOP, and Pythia across $966$ four-core DPC4 workload mixes (see~\Cref{sec:dpc4_meth}). 
We make three key observations. 
First, similar to the single-core evaluation \rbfiv{results}, Pythia achieves the highest overall geometric mean performance across all four-core mixes without \rbfor{requiring} any hyperparameter tuning or architectural modification for these trace mixes.
Pythia improves performance by $13.9\%$ over the no-prefetching baseline, whereas SPP, Bingo, and MLOP improves performance by $7.2\%$, $12.1\%$, and $10.8\%$, respectively.
Second, Pythia significantly outperforms SPP, Bingo, and MLOP by $7.2\%$, $5.9\%$, and $7.3\%$ on average in Google datacenter workload mixes. 
Third, similar to the single-core evaluation, Pythia significantly underperforms Bingo and MLOP (by $20.7\%$ and $16.2\%$, respectively) on AIML workloads mixes. 

Overall, these results provide strong empirical evidence that Pythia generalizes beyond its design-time workloads. 
Despite never observing these $483$ single-core and $966$ four-core workload traces during development and despite operating in a plug-and-play configuration without additional tuning, Pythia sustains state-of-the-art performance across both single-core and multi-core configurations.

\subsection{Overhead Analysis}\label{sec:py_eval_overhead}
\begin{sloppypar}
\rbc{To accurately estimate Pythia's chip area and power \rbc{overheads},} we implement \rbc{the full-blown} Pythia, including all fixed-point adders, multipliers, and the pipelined QVStore search operation (\Cref{sec:py_design_config_pipeline}), using \rbc{the} Chisel~\cite{chisel} \rbc{hardware design language (HDL). \rbc{We} extensively verify the functional correctness of the resultant register transfer logic (RTL) design \rbc{and} synthesize the RTL design using Synopsys Design Compiler~\cite{synopsys_dc} and 14-nm library from GlobalFoundries~\cite{global_foundries} to estimate Pythia's area and power overhead.}
Pythia \rbc{consumes} $0.33$ mm2 of area and $55.11$ mW of power in each core. 
\rbc{The QVStore component}
consumes $90.4$\% and $95.6$\% of the \rbc{total} area and power \rbc{of Pythia,} respectively. 
\rbc{With respect to the overall die area and power consumption of a}
$4$-core desktop-class Skylake processor with \rbc{the} \rbc{lowest} TDP budget~\cite{skylake-4c}, \rbc{and a $28$-core server-class Skylake processor with the highest TDP budget}, \rbc{Pythia (implemented in all cores) incurs \rbc{area \& power} overheads of only $1.03$\% \& $0.4$\%, \rbc{and $1.33$\% \& $0.75$\%,} respectively. 
We conclude that Pythia's performance benefits come at a \rbc{very modest} cost in area and power \rbc{overheads} across a variety of commercial processors.
}
\end{sloppypar}

\begin{table}[htbp]
  \centering
  \small
    \begin{tabular}{p{17em}C{4.2em}C{4.2em}}
    \multicolumn{3}{c}{\textbf{Pythia's area}: $0.33$ mm2/core; \textbf{Pythia's power}: $55.11$ mW/core\vspace{0.5em}} \\
    \toprule
    \textbf{Overhead \rbc{compared to real systems}} & \textbf{Area} & \textbf{Power} \\
    \midrule
    4-core Skylake D-2123IT, 60W TDP~\cite{skylake-4c} & 1.03\% & 0.37\% \\
    18-core Skylake 6150, 165W TDP~\cite{skylake-18c} & 1.24\% & 0.60\% \\
    28-core Skylake 8180M, 205W TDP~\cite{skylake-28c} & 1.33\% & 0.75\% \\
    \bottomrule
    \end{tabular}%
  \caption{Area and power \rbc{overhead} of Pythia.}
  \label{tab:py_overhead_analysis}%
\end{table}%

\sectionRB{Pythia: Summary}{Summary}{sec:py_conclusion}

\noindent We introduce Pythia, \rbc{the first} customizable prefetching framework that formulates prefetching as a reinforcement learning (RL) problem. Pythia autonomously learns to prefetch using multiple program features and system-level feedback \rbc{information} to predict memory accesses. 
Our extensive evaluations show that Pythia not only outperforms five \rbc{state-of-the-art} prefetchers but also provides robust performance \rbc{benefits} across a wide-range of \rbc{workloads} and system configurations.
\rbc{Pythia's benefits come with very modest area and power overheads.}

\subsection{Influence on the Research Community}
Pythia has been presented at the 54th IEEE/ACM International Symposium on Microarchitecture (MICRO) on October, 2021~\cite{pythia} and has been officially artifact evaluated with all three badges (i.e., available, functional, and reproducible).
We have made Pythia freely-downloadable from our GitHub repository~\cite{pythia_github} with all evaluated workload traces, scripts, and implementation code required to reproduce and extend it.

Since its release, \rbone{Pythia has influenced numerous subsequent works: both as a state-of-the-art baseline~\cite{mab,duong2024new,pmp,eris2022puppeteer,lin2024pars,micro_mama}, as well as a definitive reference for modeling architectural decision making using machine learning~\cite{singh2022sibyl,rakesh2025harmonia,chrome,chen2025gaze,gogineni2024swiftrl,long2023deep,zhang2022resemble,zhou2023efficient,pandey2023neurocool,yi2025artmem,alkassab2024deepref,huang2023rlop,Xue2025AugurST,Zeng2025ChartenOnlineRL,Xing2025ProactiveDP}}.
Most notably, a subsequent work, Micro-Armed Bandit (MAB)~\cite{mab}, extends Pythia by distilling its learning tables and demonstrates performance benefits similar to Pythia while reducing Pythia's overhead by two orders of magnitude.
Another follow-up work, Micro-MAMA~\cite{micro_mama}, improves upon Pythia and MAB further to demonstrate higher performance gains in multi-core processor configurations.
Pythia served as one of the baseline \rbfor{prefetchers} in the 4th Data Prefetching Championship (DPC4)~\cite{dpc4} and was independently verified by competing teams to provide the state-of-the-art performance gains among known prefetching techniques at that time.

\rbc{We believe \rbc{and hope} that Pythia \rbone{will continue to encourage} the \rbfor{the design of the} next generation data-driven autonomous prefetchers that automatically learn far-sighted prefetching policies by interacting with the system. Such prefetchers \rbc{can} not only improve performance under a wide variety of workloads and system configurations, but also reduce the system architect's burden in designing sophisticated \rbc{prefetching mechanisms}.}

\chapterRB{Perceptron-Based Off-Chip Load Prediction}{Accelerating Long-Latency Loads via Perceptron-Based Off-Chip Load Prediction}
\label{chap:hermes}

\noindent Load requests that miss the on-chip cache hierarchy and go to off-chip main memory often block instruction retirement from the reorder buffer (ROB) of modern out-of-order (OOO) processors, \rbc{preventing the processor} from allocating new instructions into the ROB~\cite{mutlu2003runahead,mutlu2003runahead2,mutlu2005techniques,hashemi2016continuous}, limiting performance.
To tolerate long memory latency, architects have primarily relied on two key latency-hiding techniques. 
First, they have significantly scaled up the size of on-chip caches in modern high-performance processor (e.g., each Intel Alder Lake core~\cite{goldencove_microarch} employs $4.3$MB on-chip cache (including L1, L2 and a per-core last-level cache (LLC) slice), which is $1.88\times$ larger than the \rbd{on-chip cache in the} previous-generation Skylake core~\cite{skylake}). 
Second, architects have designed and deployed increasingly sophisticated hardware prefetchers (likes of Pythia, discussed in \cref{chap:pythia}; or complex prefetchers in real processors ~\cite{goldencove,goldencove_microarch2}) that can \rbc{more effectively} predict the addresses of load requests in advance and fetch their corresponding data to on-chip caches before the program demands it, \rbd{thereby} completely or partially hiding the long off-chip load latency \rbc{for a fraction of off-chip loads}~\cite{goldencove,ppf,dspatch,pythia}.

\sectionRB{Hermes: Motivation and Goal}{Motivation and Goal}{sec:her_motivation}

\noindent Despite \rbc{these advances}, we observe two key trends \rbc{in} processor \rbc{design} that leave a significant performance improvement opportunity on the table: (1) a large fraction of load requests continues to go \rbd{off-chip} even in \rbd{the} presence of state-of-the-art prefetchers, and (2) an increasing fraction of the latency of an off-chip load request is spent accessing \rbc{the increasingly larger} on-chip caches.
 
\subsubsection{Large Fraction of Loads are Still Uncovered by State-of-the-Art Prefetchers}

\begin{figure}[!ht]
\centering
\includegraphics[width=5.75in]{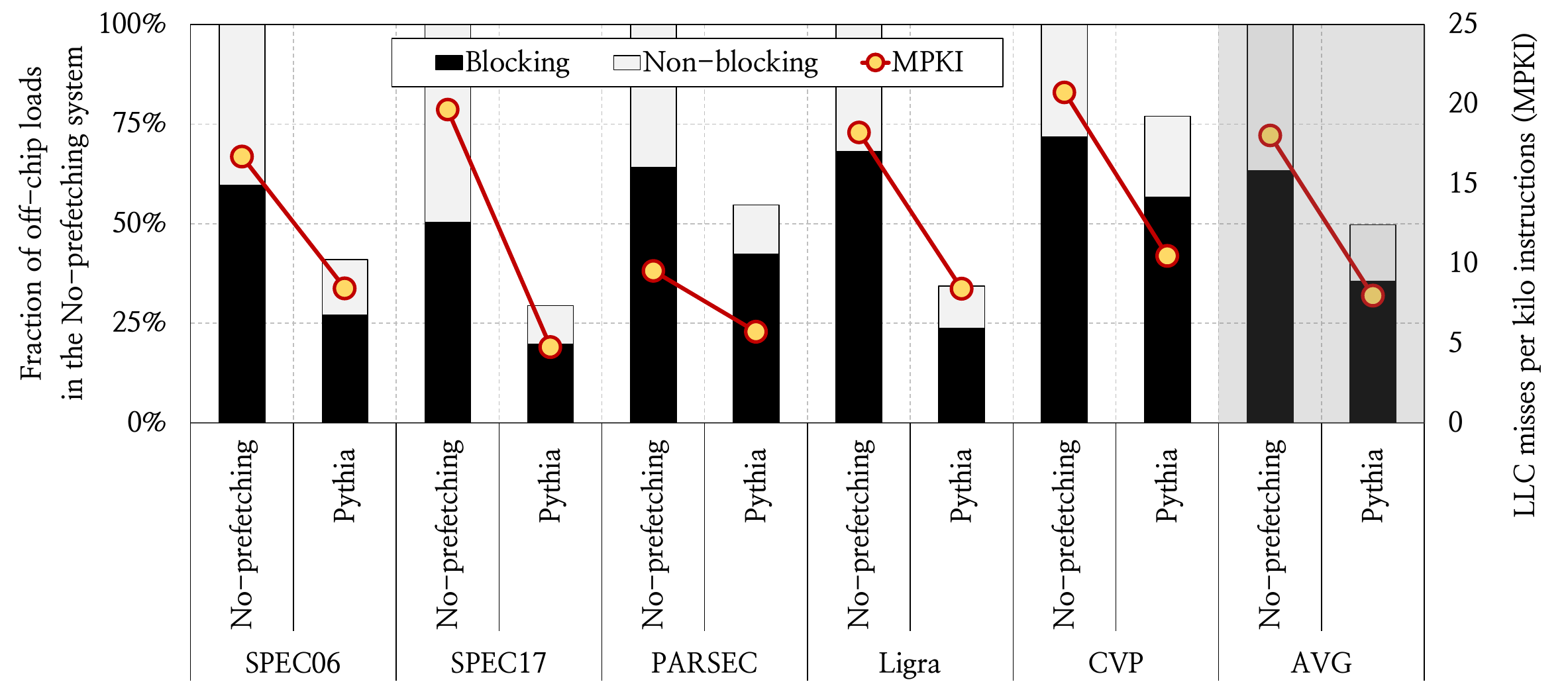}
\caption{The distribution of ROB-blocking and non-blocking load requests (on \rbe{the left} y-axis), \rbd{and LLC misses per kilo instructions} (on \rbe{the right} y-axis) in \rbd{the} absence and presence of \rbd{a state-of-the-art hardware} data prefetcher~\cite{pythia}.}
\label{fig:her_impact_of_pref}
\end{figure}

Over the past decades, researchers have proposed \rbc{many} hardware prefetching techniques that have consistently pushed the limits of performance improvement (e.g.,~\cite{stride,streamer,baer2,jouppi_prefetch,ampm,fdp,footprint,sms,sms_mod,spp,vldp,sandbox,bop,dol,dspatch,mlop,ppf,ipcp,pythia,litz2022crisp}). We observe that state-of-the-art prefetchers provide a \rbc{large} performance gain by accurately predicting future load addresses.
Yet, a large fraction of \rbc{off-chip} load requests cannot be predicted even by the most advanced prefetchers. 
These uncovered requests limit the \rbd{processor's} performance by blocking instruction retirement in \rbd{the} ROB. 
\Cref{fig:her_impact_of_pref} shows a stacked graph of total number off-chip load requests in a no-prefetching system and a system with the recently-proposed hardware \rbf{data} prefetcher Pythia~\cite{pythia}, normalized to the no-prefetching system, \rbe{across $110$ workload traces categorized into five workload categories}.\footnote{We select Pythia as the baseline prefetcher as it provides the highest prefetch coverage and performance benefit among the five contemporary prefetchers considered in this \rbfor{work} (see \Cref{sec:her_dmf_configuration} and \Cref{sec:her_prefetchers}). Nonetheless, \rbc{our qualitative} observation holds equally true for other prefetchers \rbc{considered in this work} (see~\Cref{sec:her_dmf_configuration}).} Each bar further categorizes load requests into two classes: loads that block instruction retirement from \rbc{the} ROB (called \emph{blocking}) and loads that do not (called \emph{non-blocking}).
\Cref{sec:her_methodology} discusses our evaluation methodology. 

We make two key observations from \Cref{fig:her_impact_of_pref}. 
First, on average, Pythia \rbc{accurately prefetches} nearly half of \rbc{all} off-chip load requests in the no-prefetching system, \rbd{thereby} improving \rbf{the} overall performance (not shown here; see~\Cref{sec:her_perf_1c}).
Second, the remaining half of the off-chip loads \rbc{are not prefetched even by} a sophisticated prefetcher like Pythia. $71.4\%$ of these \rbd{non-prefetched} off-chip loads block instruction retirement from \rbc{the} ROB, \rbc{significantly limiting performance}. \rbc{We} conclude that, \rbc{state-of-the-art prefetchers, while effective at improving performance,} still leave a significant performance \rbd{improvement} opportunity on the table.

\subsubsection{An Increasing Fraction of Off-Chip Load Latency is Spent in Accessing the On-Chip Cache Hierarchy}

We observe that the on-chip cache hierarchy has not only grown tremendously in size but also in design complexity (e.g., sliced last-level cache organization~\cite{beckmann2004managing,kim2002adaptive,hardavellas2009reactive}) in recent processors, \rbd{in order} to cater \rbc{to} workloads with \rbc{large} data \rbc{footprints}. A larger on-chip cache hierarchy, on \rbc{the} one hand, improves a core's performance by \rbc{preventing more} load requests from going off-chip. 
\rbd{On the other hand, all on-chip caches need to be accessed to determine if a load request should be sent off-chip. As a result, on-chip cache access latency significantly contributes to the total latency of an off-chip load. With increasing on-chip cache \rbe{sizes}, \rbe{and the complexity of the} cache hierarchy design and \rbf{the} on-chip network~\cite{wang2021stream,besta2018slim}, the on-chip cache access latency is increasing in processors~\cite{l3_lat_compare1,llc_lat3}.}
An analysis of the Intel \rbe{Alder Lake} core suggests that the
load-to-use latency of an LLC access
has increased to $14$~ns (which \rbe{is} equivalent to $55$ cycles for a core running at $4$~GHz)~\cite{llc_lat1,llc_lat2,llc_lat3}. 

\begin{figure}[!ht]
\centering
\includegraphics[width=5.75in]{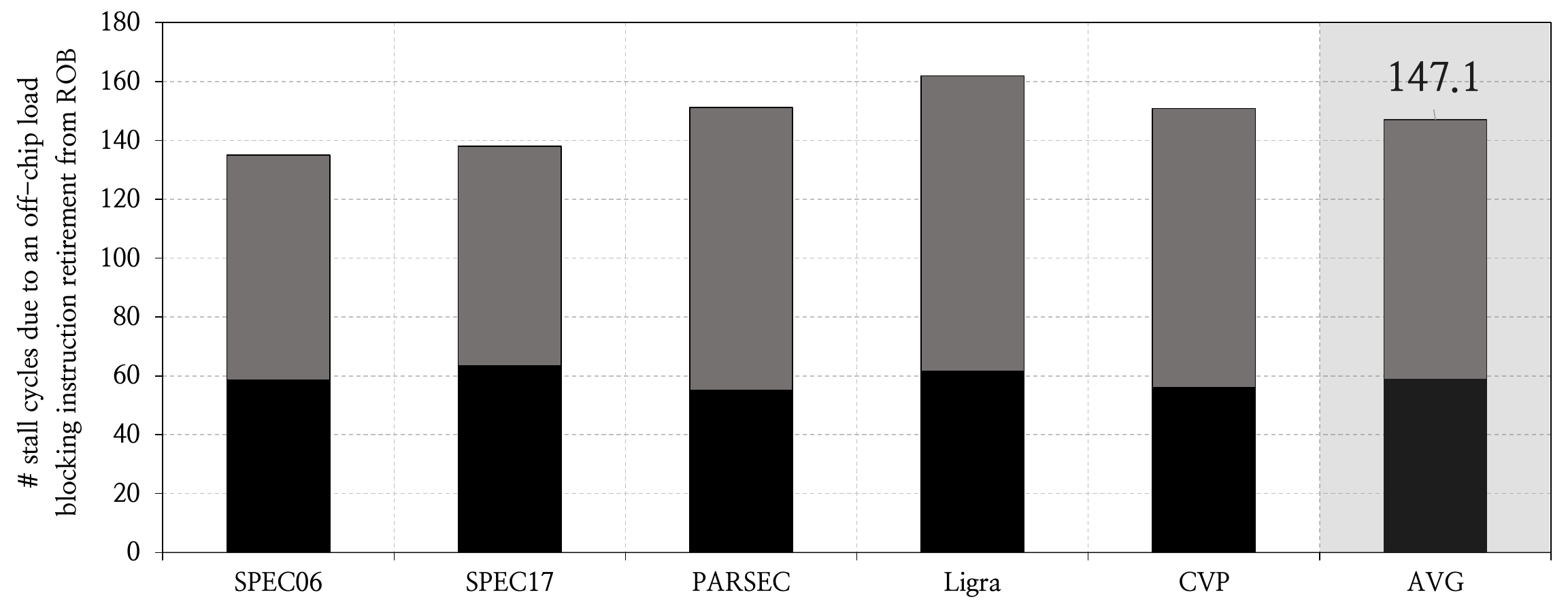}
\caption{\rbd{The average number of cycles a core stalls due to an off-chip load blocking any instruction from retiring from the ROB} across all workload categories. \rbc{The dark portion in each bar shows the cycles that can be \rbe{completely} \rbd{eliminated} by removing the on-chip cache access latency from an off-chip load's critical path.}}
\label{fig:her_load_lat}
\end{figure}

To demonstrate the effect of long on-chip cache access latency on the \rbd{total latency of an off-chip load}, \Cref{fig:her_load_lat} plots the 
average number of cycles a core stalls due to an off-chip load blocking any instruction from retiring from the ROB, averaged across each workload category in our baseline system with Pythia. 
\rbd{Each bar further shows the \rbe{average} number of cycles an off-chip load spends for accessing the on-chip cache hierarchy.}
Our simulation configuration faithfully models an Intel Alder Lake performance-core with \rbc{a large} ROB, large on-chip caches \rbc{and} publicly-reported cache access latencies (see \Cref{sec:her_methodology}). As \Cref{fig:her_load_lat} shows, an off-chip load stalls the core for \rbc{an average of} $147.1$ cycles. $40.1\%$ of these stall cycles (i.e., $58.9$ cycles) can be \rbe{completely} \emph{eliminated} by removing the on-chip cache access latency from \rbe{the off-chip load's} critical path.
We conclude that a large \rbc{and complex} on-chip cache hierarchy \rbd{is directly responsible for a large fraction of} the overall stall cycles \rbc{caused by} an off-chip load request. We envision that this problem will only get exacerbated with new \rbe{processor} designs \rbc{as on-chip caches \rbe{continue to} grow in size and complexity}~\cite{l3_lat_compare1}.

\subsection{Our Goal}

\textit{\textbf{Our goal}} is to improve processor performance by removing the on-chip cache access latency from the critical path of off-chip load requests.

\sectionRB{Hermes: Headroom and Challenges}{Hermes: Headroom and Challenges}{sec:her_hermes_headroom_challenges}

\begin{figure}[!h]
\centering
\includegraphics[width=5in]{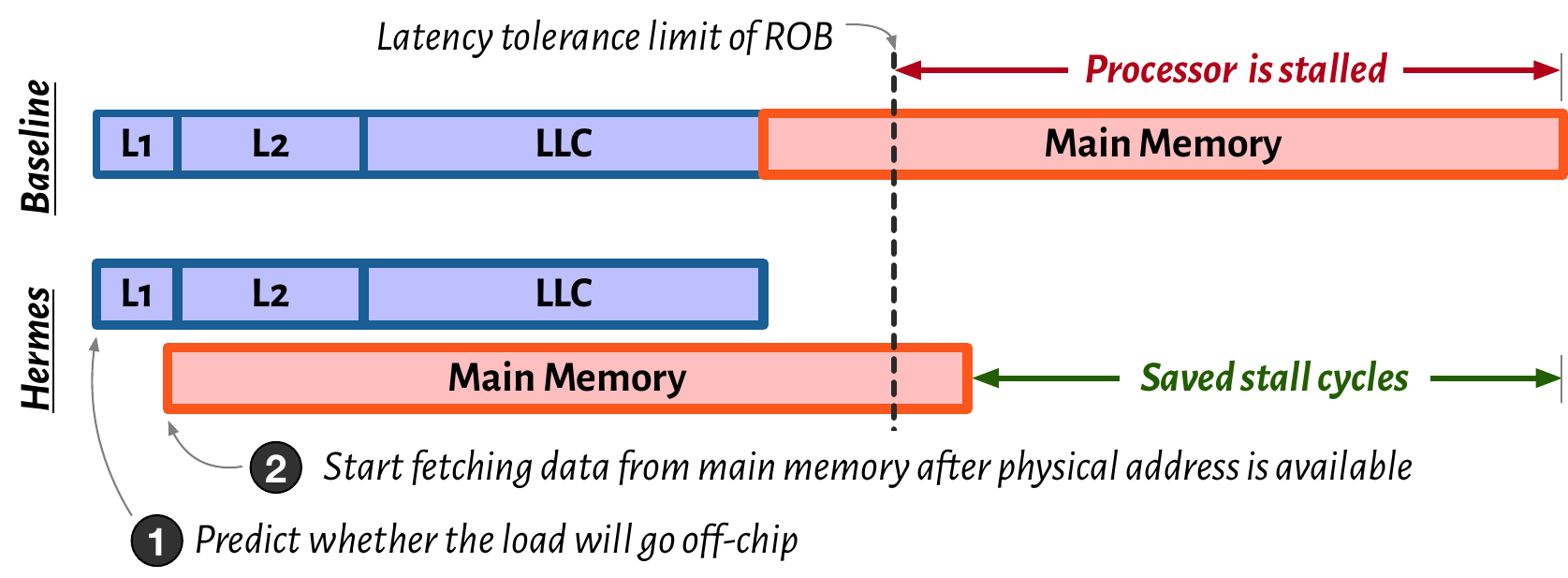}
\caption{Comparison of the \rbc{execution timeline} of an off-chip load request in a \rbe{conventional processor} and in Hermes.}
\label{fig:hermes_overall_idea}
\end{figure}

\noindent \rbd{To this end}, we propose \rbd{a new technique called} \textit{\textbf{Hermes}},
\rbc{whose} \textbf{\emph{key idea}} is \rbd{to predict which load requests might go off-chip and start fetching their corresponding data \emph{directly} from the main memory, while also concurrently accessing the cache hierarchy for such a load.}\footnote{Hence named after Hermes, the Olympian deity~\cite{hermes_wiki} who can quickly move between the realms of the divine (i.e., the \rbc{processor}) and the mortals (i.e., the main memory).}
\rbd{By doing so}, Hermes hides the on-chip cache access latency under the shadow of the main memory access latency (as \rbd{illustrated} in Fig.~\ref{fig:hermes_overall_idea}), \rbc{thereby} significantly reducing the overall latency of an off-chip load request.

\subsection{Headroom Analysis} \label{sec:her_headroom_study}

\begin{figure}[!ht]
\centering
\includegraphics[width=5.75in]{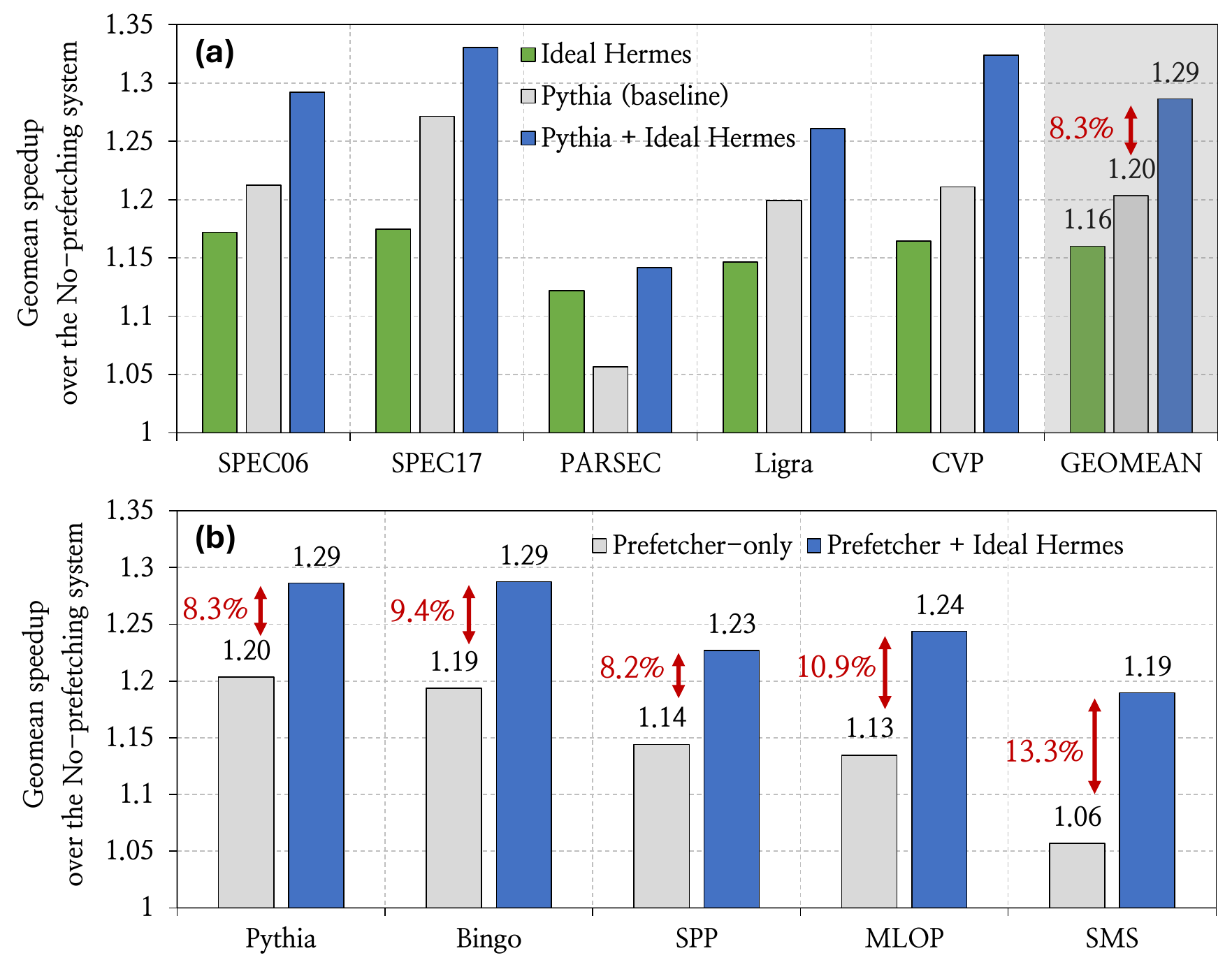}
\caption{(a) \rbd{\rbe{Speedup of} Ideal Hermes \rbd{by itself and when combined with} Pythia in single-core workloads}. (b) \rbe{Speedup of} Ideal Hermes \rbd{when combined with} four recently-proposed prefetchers: Bingo~\cite{bingo}, SPP~\cite{spp,ppf}, MLOP~\cite{mlop}, and SMS~\cite{sms}.}
\label{fig:her_Ideal_dmf_master}
\end{figure}

To understand the potential performance benefits of Hermes, we model an \emph{Ideal Hermes} system \rbd{in simulation} where we reduce the main memory access latency of \emph{every} off-chip load request by the \rbe{post-L1 on-chip cache hierarchy access latency (which includes L2 and LLC access, and interconnect latency)}. 
\rbd{In other words, in \rbe{the} Ideal Hermes system, we \rbe{(1)} magically and perfectly know if a load request would go off-chip after its physical address is available (i.e., after the translation lookaside buffer access, which happens in parallel with the L1 \rbf{data} cache access in modern processors~\cite{wood1986cache,cekleov1997virtual,patterson2016computer,basu2012reducing}), and \rbe{(2)} directly access the off-chip main memory for such a load, eliminating the non-L1-cache related on-chip cache \rbf{hierarchy} access latency from such a load’s total latency.}
\Cref{fig:her_Ideal_dmf_master}(a) shows \rbe{the speedup} of Ideal Hermes \rbd{by itself} and \rbd{when combined with} Pythia \rbe{normalized to the no-prefetching system in single-core workloads.} 
We make two key observations from \Cref{fig:her_Ideal_dmf_master}(a). 
First, Ideal Hermes \rbd{combined with} Pythia outperforms Pythia \rbd{alone} by $8.3\%$ on average across \rbd{all} workloads. 
Second, Ideal Hermes \rbd{by itself} provides nearly $80\%$ \rbd{of the} performance improvement \rbd{that} Pythia \rbd{provides}.
\Cref{fig:her_Ideal_dmf_master}(b) shows the \rbe{speedup} of Ideal Hermes \rbd{when combined with} four other recently-proposed high-performance prefetchers: Bingo~\cite{bingo}, SPP~\cite{spp} (with perceptron filter~\cite{ppf}), MLOP~\cite{mlop}, and SMS~\cite{sms}.
\rbd{Ideal Hermes improves performance by $9.4\%$, $8.2\%$, $10.9\%$, and $13.3\%$ on top of four state-of-the-art prefetchers Bingo, SPP, MLOP, and SMS, respectively.}
Based on these results, we conclude that Hermes has high potential performance benefit not only \rbd{when implemented alone} but also \rbd{when combined with} a wide variety of high-performance prefetchers.

\subsection{Key Challenges} \label{sec:her_key_challenge}

Even though Hermes has a significant potential \rbd{to improve} performance, Hermes's performance gain heavily depends on the accuracy (i.e., the fraction of predicted off-chip loads that actually go off-chip) \rbd{and the coverage (i.e., the fraction of off-chip loads that are successfully predicted)} of the off-chip load prediction. 
\rbe{A low-accuracy off-chip load predictor}
generates \rbd{useless} main memory requests, which incur both latency and bandwidth overheads, and causes interference to the useful requests in the main memory.
\rbe{A low-coverage predictor}
loses opportunity \rbe{to improve} performance.

\begin{figure}[!ht]
\centering
\includegraphics[width=5.75in]{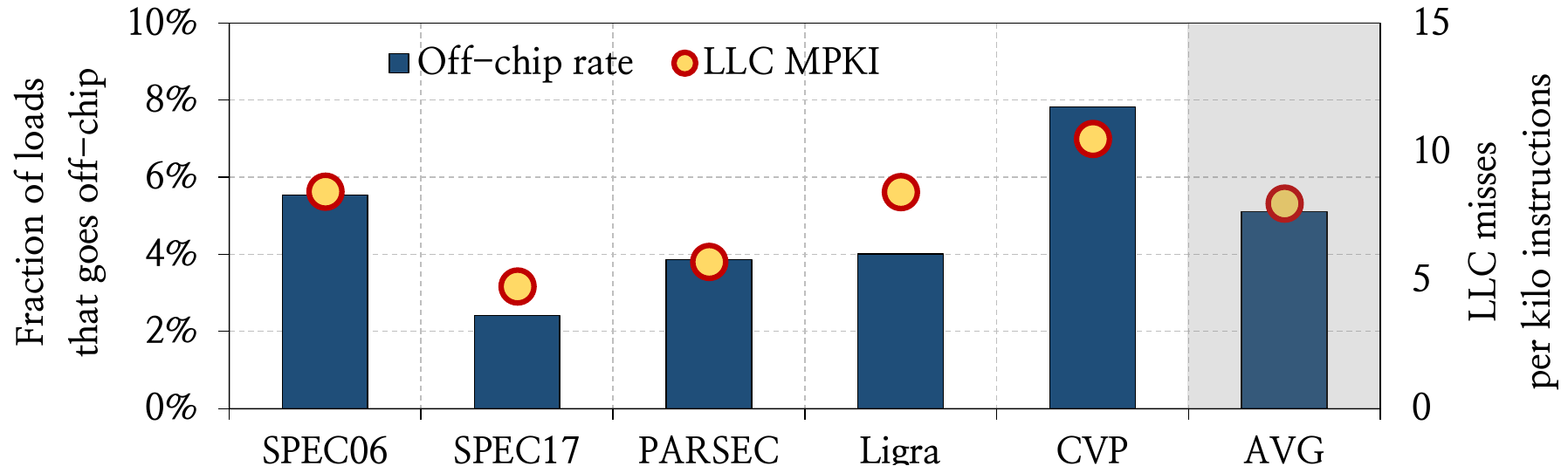}
\caption{\rbd{\rbe{Percentage of} loads that miss the LLC and goes off-chip (on \rbe{the left} y-axis) and the LLC MPKI (on \rbe{the right} y-axis) in \rbe{the} baseline system with Pythia.}}
\label{fig:her_off_chip_rate}
\end{figure}

We identify two key challenges in designing an off-chip \rbf{load} predictor with high accuracy and high coverage.
\rbd{First, only a small fraction of the total loads generated by a workload goes off-chip in presence of a sophisticated data prefetcher. As shown in~\ref{fig:her_off_chip_rate}, on average $7.9$ loads per kilo instructions miss the LLC and go off-chip in our baseline system with Pythia. However, these loads \rbe{constitute} only $5.1\%$ of the total loads generated by a workload.}
This small fraction of off-chip loads makes it difficult for an off-chip load predictor to accurately learn from the workload behavior to produce highly-accurate predictions.

Second, the off-chip predictability of a workload \rbd{can change} in \rbd{the} presence of modern sophisticated data prefetchers. This is because in \rbe{the} presence of a sophisticated  prefetcher, the likelihood of a load request going off-chip not only depends on the program behavior but also on the prefetcher's ability to successfully prefetch for the load.

In this work, we \rbd{overcome these two key challenges by} \rbe{designing} a new off-chip \rbf{load} prediction technique, called \pred, based on perceptron learning~\cite{mcculloch1943logical,rosenblatt1958perceptron,jimenez2002neural}.
\rbd{By learning to identify off-chip loads using multiple program features (e.g., sequence of program counters, byte offset of a load request, \rbe{page number of the load address}), \pred provides both higher accuracy and coverage than \rbi{a} prior cache hit-miss prediction \rbi{technique~\cite{yoaz1999speculation} and \rbi{higher accuracy than} another off-chip load prediction technique that we develop (see~\Cref{sec:her_dmf_configuration}),} in \rbf{the} presence of modern sophisticated prefetchers, without requiring large metadata storage overhead.}
With \rbd{small} changes to the existing on-chip datapath design, we demonstrate that Hermes with \pred significantly outperforms the baseline system with \rbf{a} state-of-the-art prefetcher \rbe{across} a wide range of workloads and system configurations.

\section{Hermes: Overview} \label{sec:her_design_overview}

\begin{figure}[!ht]
\centering
\includegraphics[width=4.5in]{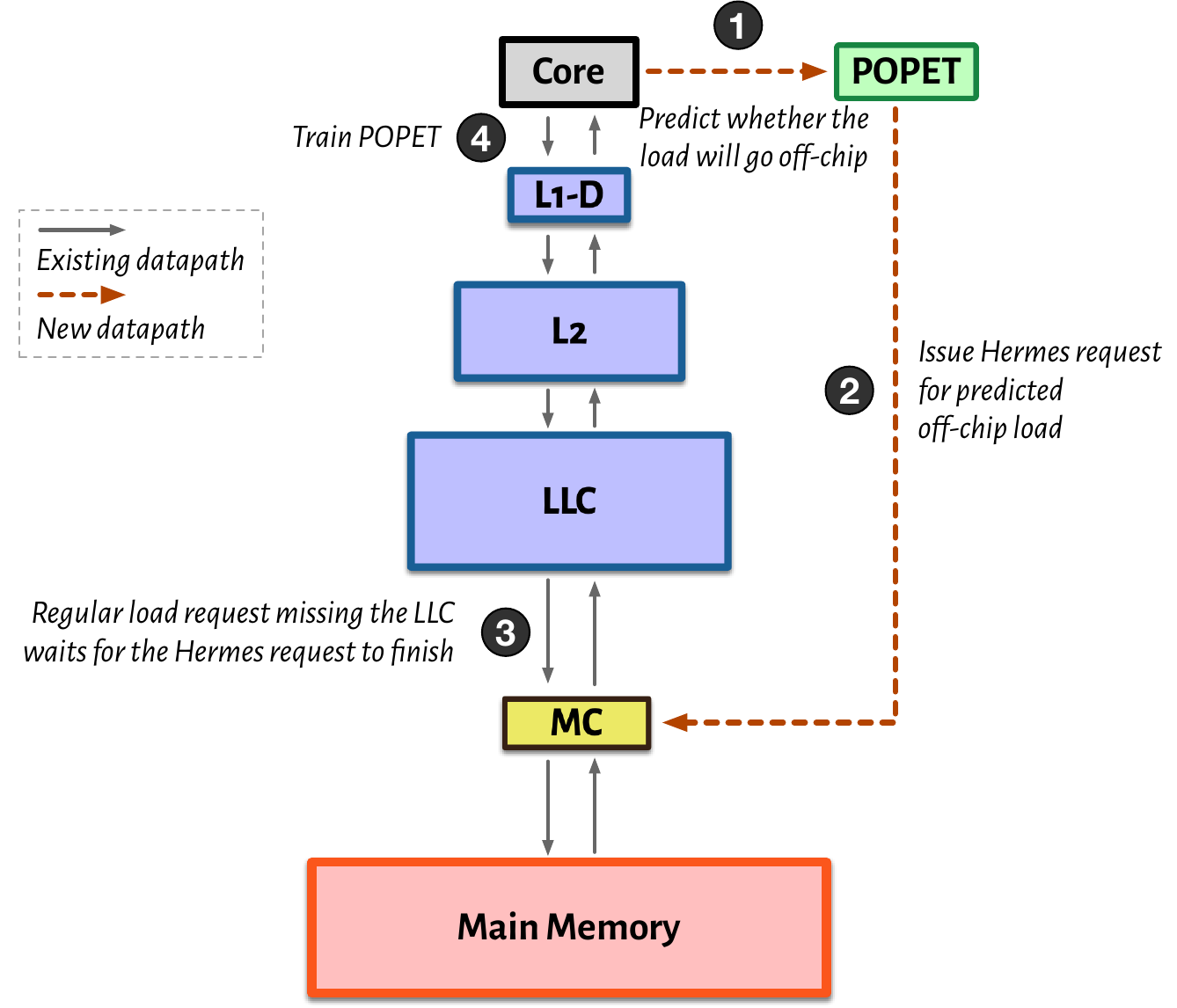}
\caption{Overview of Hermes.}
\label{fig:her_dmf_overview}
\end{figure}

\Cref{fig:her_dmf_overview} shows a high-level overview of Hermes. 
\pred is the key component of Hermes that is responsible for making highly-accurate off-chip load predictions.  
For every demand load request generated by the processor, \pred predicts whether or not the load request would go off-chip (\circled{1}). 
If the load is predicted to go off-chip, Hermes issues a speculative \rbd{memory} request \rbd{(called a \emph{Hermes request})} \emph{directly} to the main memory controller once the \rbd{load's} physical address is generated to start fetching the corresponding data from the main memory (\circled{2}). 
\rbe{This Hermes request is serviced by the main memory controller concurrently with the \emph{regular load request} (i.e., the load issued by the processor that generated the Hermes request) that accesses the on-chip cache hierarchy.}
If the prediction is correct, the \rbd{regular} load request to the same address
eventually misses the LLC and waits for  the ongoing \rbd{Hermes request} to finish, \rbd{thereby completely hiding the on-chip cache hierarchy access latency from the critical path of the correctly-predicted \rbe{off-chip} load} (\circled{3}).
\rbd{\rbe{If a} Hermes request \rbe{returns} from the main memory \rbe{but there has been no} regular load request to the same address, Hermes drops the request and does not fill the data into the cache hierarchy. \rbe{By doing so,} Hermes keeps the on-chip cache hierarchy fully coherent even in case of a misprediction.}
For every regular load request returning to the core, Hermes trains \pred based on whether or not this load has actually gone off-chip (\circled{4}).

\sectionRB{Hermes: Detailed Design}{Hermes: Detailed Design}{sec:her_detailed_design}

\noindent We first describe the design of \pred in \Cref{sec:her_ocp_design}, followed by the changes introduced by Hermes to the on-chip cache access datapath in \Cref{sec:her_data_path_design}.

\subsection{\pred Design} \label{sec:her_ocp_design}

The purpose of \pred is to accurately predict whether or not a load request generated by the processor will go off-chip. We model \pred using \rbe{the} multi-feature perceptron learning mechanism~\cite{rosenblatt1958perceptron, perceptron,jimenez2002neural,jimenez2003fast,jimenez2017multiperspective,teran2016perceptron,ppf,garza2019bit}, more specifically as a \emph{hashed-perceptron}~\cite{tarjan2005merging} model.
A hashed-perceptron model hashes multiple feature values to retrieve weights of each feature from small tables. If the sum of these weights exceeds a threshold, the model makes a positive prediction. Hashed-perceptron, as compared to other perceptron models, is lightweight and easy to implement in hardware. Prior works  successfully \rbe{apply} hashed-perceptron for various microarchitectural predictions, e.g., branch outcome~\cite{jimenez2002neural,jimenez2003fast,garza2019bit}, LLC reuse~\cite{jimenez2017multiperspective,teran2016perceptron}, prefetch usefulness~\cite{ppf}. This is the first work that applies hashed-perceptron \rbe{to} off-chip load prediction.

\subsubsection{Why is Perceptron Learning a Good Fit for Modeling Off-Chip Prediction?}
We choose to model \pred based on perceptron learning for two key reasons. First, by learning using multiple program features, perceptron learning can provide highly accurate predictions that could not be otherwise provided by simple history-based learning prediction (e.g., HMP~\cite{yoaz1999speculation}). Second, perceptron learning can be implemented with low storage overhead, without requiring any impractical metadata support (e.g., extending TLB~\cite{d2d,d2m} or in-memory metadata storage~\cite{lp}).

\subsubsection{\pred Design Overview}
\pred is organized as a collection of one-dimensional tables (\rbd{each} called a \emph{weight table}), where each table corresponds to a single program feature. Each table entry stores a \emph{weight} value, implemented using a $5$-bit saturating signed integer, that represents the correlation between the corresponding program feature value and the true outcome \rbd{(i.e., whether a given load actually went off-chip)}. A weight value saturated near the maximum (i.e., $+15$) or the minimum (i.e., $-16$) value represents a strong positive or negative correlation between the program feature value and the true outcome, respectively. A weight value closer to zero signifies a weak correlation. The weights are adjusted \rbd{during training (step \circled{4} in \Cref{fig:her_dmf_overview})} to \rbe{update} \pred's prediction with the true outcome. Each weight table is sized differently based on its corresponding program feature (see \Cref{table:her_overhead}).

\subsubsection{Making a Prediction using \pred} \label{sec:her_inference}

\begin{figure}[!h]
\centering
\includegraphics[width=5in]{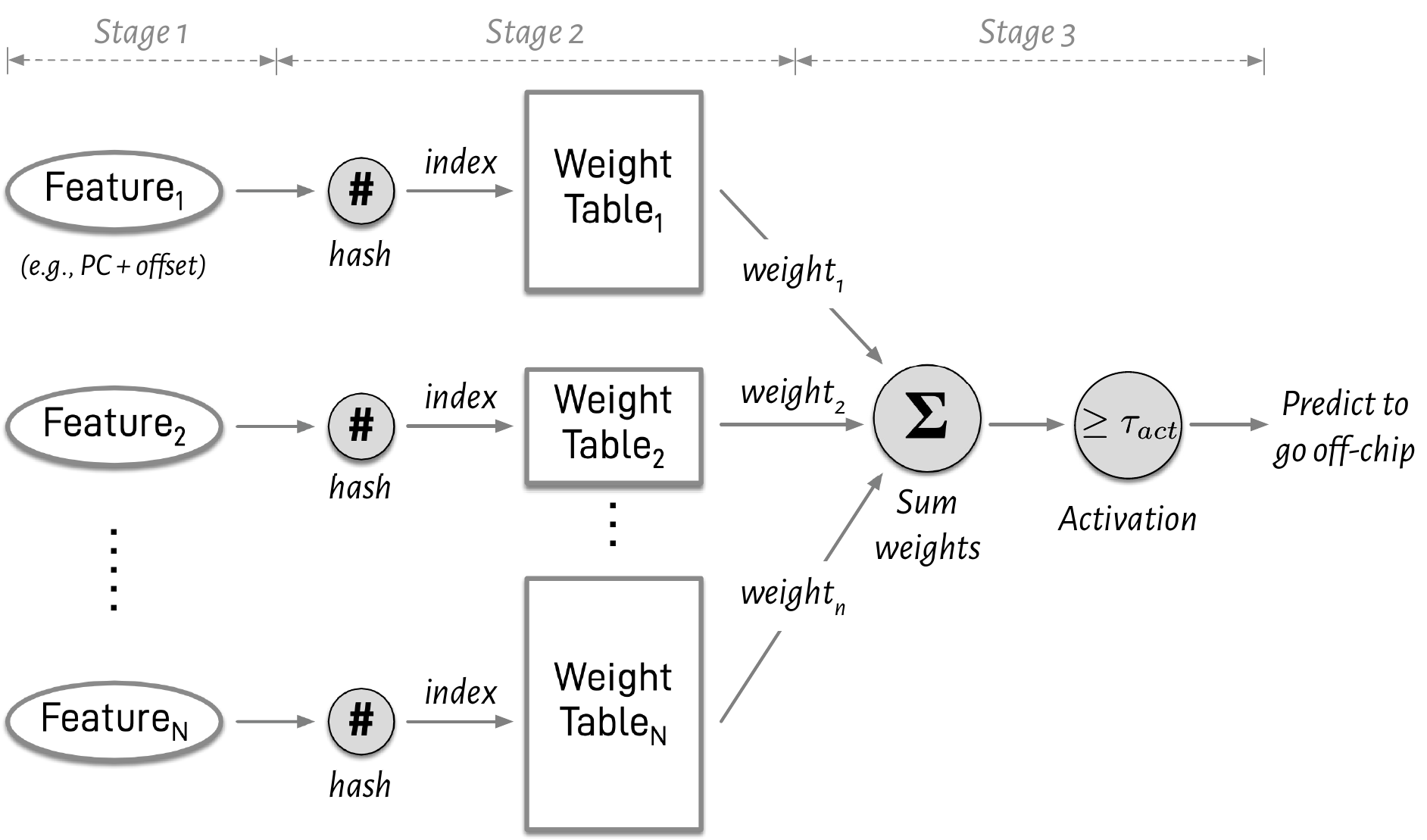}
\caption{Stages to make a prediction by \pred.}
\label{fig:her_perc_design}
\end{figure}

During load queue (LQ) allocation for a load generated by the core (step \circled{1} in \Cref{fig:her_dmf_overview}), \pred makes a binary prediction on whether or not the load request would go off-chip.
The prediction happens in three stages \rbd{as shown in \Cref{fig:her_perc_design}}.
In the first stage, \pred extracts a set of program features from the current load request and a history of prior requests (\Cref{sec:her_feature_selection} shows the list of program features used by \pred). 
In the second stage, each feature value is hashed and used as an index to retrieve a weight value from the weight table of the corresponding feature.
In the third stage, all weight values from individual features are accumulated to generate the \emph{cumulative perceptron weight} ($W_{\sigma}$).
If $W_{\sigma}$ exceeds a predefined threshold (called the \emph{activation threshold}, $\tau_{act}$), \pred
makes a positive prediction (\rbe{i.e., it predicts that} the current load request would go off-chip). Otherwise, \pred makes a negative prediction.
The hashed feature values, the cumulative perceptron weight $W_{\sigma}$, and the predicted outcome are stored in the LQ entry to be reused \rbe{to train} \pred when the load request returns to the processor core (step \circled{4} in \Cref{fig:her_dmf_overview}).

\subsubsection{Training \pred} \label{sec:her_training}

\pred training is invoked when a demand load request returns to the core and prepares to release its corresponding LQ entry (step \circled{4} in \Cref{fig:her_dmf_overview}). Every demand load that misses the LLC and goes to the main memory controller is marked as a \emph{true} off-chip load request. This true off-chip outcome, along with the predicted outcome stored in the LQ entry of the demand load, are used to appropriately train the feature weights of \pred. 
The training happens in two stages. 
In the first stage, the $W_{\sigma}$ (computed during prediction) is retrieved from the LQ entry. If $W_{\sigma}$ is neither positively nor negatively saturated (i.e., $W_{\sigma}$ lies within a negative and a positive training threshold, $T_N$ and $T_P$, respectively), the weight training is triggered. This saturation check prevents the individual feature weight values from getting over-saturated, \rbd{thereby} helping \pred to quickly adapt its learning to program phase changes. In the second stage, if the weight training is triggered, the weights for each individual program feature are retrieved from their corresponding weight table using the hashed feature indices stored in the LQ entry. If the true outcome is positive (meaning the load actually went off-chip), the weight value for each feature is incremented by one. If the true outcome is negative, the weight values are decremented by one. This simple weight update mechanism moves each individual feature weight \rbe{towards} the direction of the true outcome, thus gradually increasing the prediction accuracy.

\subsubsection{Automated Feature Selection} \label{sec:her_feature_selection}

The selection of the program features used to make the off-chip \rbf{load} prediction is critical to \pred's performance. 
A \rbd{carefully-crafted and selected} set of features can significantly improve the accuracy and the coverage of \pred. In this section we propose an automated, offline, performance-driven methodology to find a set of program features for \pred.

We initially select a set of $16$ individual program features using our domain expertise that can correlate well with \rbd{a load going off-chip}. \Cref{table:her_initial_feature_set} shows the initial feature set.

\begin{table}[htbp]
  \centering
  \small
    \begin{tabular}{m{20em}m{20em}}
    \thickhline
    \Tabval{\textbf{Features without control-flow information}} & \Tabval{\textbf{Features with control-flow information}} \\
    \hline
    \Tstrut 
    \begin{minipage}{20em}
      \vskip 2pt
      \begin{enumerate}[leftmargin=1.5em]
        \setlength{\itemsep}{1pt}
        \setlength{\parskip}{0pt}
        \setlength{\parsep}{0pt}
        \item Load virtual address
        \item Virtual page number
        \item Cacheline offset in page
        \item First access
        \item Cacheline offset + first access
        \item Byte offset in cacheline
        \item Word offset in cacheline
      \end{enumerate}
    \end{minipage} 
    \Bstrut & 
    \Tstrut 
    \begin{minipage}{20em}
      \vskip 2pt
      \begin{enumerate}[leftmargin=2em]
        \setcounter{enumi}{7}
        \setlength{\itemsep}{1pt}
        \setlength{\parskip}{0pt}
        \setlength{\parsep}{0pt}
        \item Load PC
        \item PC $\oplus$ load virtual address
        \item PC $\oplus$ virtual page number
        \item PC $\oplus$ cacheline offset
        \item PC + first access
        \item PC $\oplus$ byte offset
        \item PC $\oplus$ word offset
        \item Last-$4$ load PCs
        \item Last-$4$ PCs
      \end{enumerate}
    \end{minipage}
    \Bstrut \\
    \thickhline
    \end{tabular}%
  \caption{The initial \rbd{set} of program features used for automated feature selection. $\oplus$ represents a \rbd{bitwise} XOR operation.}
  \label{table:her_initial_feature_set}%
\end{table}%

The automated feature selection process happens offline during the design time of \pred. The process starts with the initial set of 16 individual program features and iteratively creates a list of \emph{feature sets}, each containing $n$ features, at every iteration $n$ in the following way. In the first iteration, we design \pred with each \rbe{of the} 16 initial program features and test its prediction accuracy in $10$ randomly-selected workload traces (called \emph{testing workloads}). We select the top-$10$ features that produce the highest prediction accuracy for the second iteration. In the second iteration, we create $160$ two-combination feature sets (meaning, each feature set contains two initial features from~\Cref{table:her_initial_feature_set}) by \rbe{combining} each \rbd{of the} $16$ initial features with each of the $10$ winning feature sets from the last iteration, and test the prediction accuracy on the testing workloads. We select the top-$10$ two-combination feature sets that produce the highest prediction accuracy for the third iteration. This iterative process repeats until the maximum prediction accuracy gets saturated \rbd{(i.e., the difference in accuracy of two successive iterations is less than $3\%$)}.\footnote{\rbe{For simplicity, our automated feature selection process optimizes for accuracy. A more comprehensive feature selection process can also include coverage or directly optimize for performance (i.e., execution time).}}
\Cref{table:her_ocp_config} shows the final list of program features selected by the automated feature selection process.

\begin{table}[htbp]
  \centering
  \small
    \begin{tabular}{m{9em}m{18em}}
    \thickhline
    \Tabval{\textit{\textbf{Selected features}}} & 
    \Tstrut 
    \begin{minipage}{18em}
      \vskip 4pt
      \begin{itemize}[leftmargin=1em]
        \setlength{\itemsep}{1pt}
        \setlength{\parskip}{0pt}
        \setlength{\parsep}{0pt}
        \item PC $\oplus$ cacheline offset
        \item PC $\oplus$ byte offset
        \item PC + first access
        \item Cacheline offset + first access
        \item Last-$4$ load PCs
      \end{itemize}
    \end{minipage} 
    \Bstrut \\
    \hline
    \Tabval{\textit{\textbf{Threshold values}}} & \Tabval{$\tau_{act} = -18$, $T_{N} = -35$, $T_{P} = 40$} \\
    \thickhline
    \end{tabular}%
  \caption{\pred configuration parameters.}
  \label{table:her_ocp_config}%
\end{table}%

\paraheading{Rationale \rbd{for Selected Features}.} 
Each selected feature correlates with the likelihood of observing an off-chip load request with a different program context information. \rbd{We explain the rationale for each selected feature below.}

\textbf{(1) PC $\oplus$ cacheline offset.} This feature is computed by XOR-ing the load PC value with the cacheline offset \rbd{of the load address} in the virtual page of the load request. 
The goal of this feature is to learn the likelihood of a load request going off-chip when a given load PC touches a certain cacheline offset in a virtual page.
The use of cacheline offset information, instead of load virtual address or virtual page number, enables this feature to apply the learning across different virtual pages.

\textbf{(2) PC $\oplus$ cacheline byte offset.} This feature is computed by XOR-ing the load PC with the byte offset of the load cacheline address. This feature is particularly useful in accurately predicting off-chip load requests when a program has a streaming access pattern over a linearly allocated data structure. For example, when a program streams through a large array of $4$B integers, every $16^{th}$ load (as a $64$B cacheline stores $16$ integers) generated by a load PC that is iterating over the array will go off-chip, and the remaining loads will hit in on-chip caches. In this case, this feature learns to identify only those loads \rbd{that have a byte offset of $0$ to go off-chip}.

\textbf{(3) PC + first access.} This feature is computed by left-shifting the load PC and adding the \emph{first access} hint at the most-significant bit position. \rbd{The first access hint is a binary value that represents whether or not a cacheline has been recently touched by the program}. The hint is computed using a small $64$-entry buffer (called the \emph{page buffer}) that tracks the demanded cachelines from last $64$ virtual pages. Each page buffer entry holds two \rbd{pieces of} information: a virtual page tag, and a $64$-bit bitmap, where each bit represents one cacheline in the virtual page. During every load request generation, \pred searches the page buffer with the virtual page number \rbd{of the load address}. \rbd{If a matching entry is found, \pred} uses the value of the bit corresponding to the cacheline offset \rbd{in the matching page buffer entry's bitmap} as the first access hint.
\rbd{If the bit is set (or unset), it signifies that the corresponding cacheline has (not) been recently accessed by the program. If the bit is unset, \pred sets the bit in the page buffer entry's bitmap.}
The first access hint provides a crude estimate of a cacheline's reuse in a short temporal window. However, it alone cannot determine the cacheline's residency in on-chip caches, as the memory footprint tracked by the page buffer is much smaller than the total cache size.

\textbf{(4) Cacheline offset + first access.} This feature is similar to the PC + first access feature, except that it learns the likelihood of a load request going off-chip when a given cacheline offset is recently touched by the program.

\textbf{(5) Last-4 load PCs.} This feature value is computed as a shifted-XOR of last four load PCs. \rbe{It} represents the execution path of a program and correlates it with the likelihood of observing an off-chip load request whenever the program follows the same execution path.

\subsubsection{Parameter Threshold Tuning} \label{sec:her_threshold_selection}

\rbd{\pred has three tunable parameters: negative and positive training thresholds ($T_{N}$ and $T_{P}$, respectively), and the activation threshold ($\tau_{act}$). Properly tuning the values of all these three parameters}
is also critical to \pred's performance, since both \pred's accuracy and coverage are sensitive to parameter values. 

We employ a three-step grid search technique to tune each of the three parameters separately. 
In the first stage, we uniformly sample values from \rbd{a parameter's} range. For example, $\tau_{act}$ can take values in the range $[-80,75]$.\footnote{\rbd{As \pred uses five program features (see~\Cref{sec:her_feature_selection}), the sum of all five weights (each represented by a 5-bit saturating signed integer as described in ~\Cref{sec:her_ocp_design}) can take a maximum and minimum value of $75$ and $-80$, respectively.}} 
We uniformly sample values from this range with a grid size of $5$. In the second stage, we run Hermes with the randomly-selected 10 test workloads (as mentioned in~\mbox{\Cref{sec:her_feature_selection}}) for each of the sampled values and pick the top-10 values that provide the highest performance gain. In the third stage, we run Hermes with all single-core workload traces using the selected 10 parameter values from the second stage. We finally select the value that provides the highest average performance gain.
\Cref{table:her_ocp_config} shows the \rbe{selected} threshold values of each parameter. 

\subsection{Hermes Datapath Design} \label{sec:her_data_path_design}

In this section, we describe the key changes introduced to the existing well-optimized on-chip cache access datapath to incorporate Hermes.
First, we show how the core issues \rbd{a Hermes request} directly to the main memory controller \rbe{if \pred predicts the load would go off-chip} and how a \rbd{regular} load request that misses the LLC waits for an ongoing \rbd{Hermes} request (see~\Cref{sec:her_spec_load_issue}). Second, we discuss \rbd{how the data fetched from main memory is properly sent back to the core in presence of Hermes} while maintaining cache coherence (see~\Cref{sec:her_returning_data_to_core}).

\subsubsection{Issuing a \rbd{Hermes} Request} \label{sec:her_spec_load_issue}

For every load request predicted to go off-chip, Hermes issues a Hermes request directly to the main memory controller (step~\circled{2} in \Cref{fig:her_dmf_overview}) once the load's physical address is generated.
The main memory controller enqueues the \rbd{Hermes request} in its read queue (RQ) and starts fetching the corresponding data from the main memory as dictated by its scheduling policy, \rbe{while the regular load request is concurrently accessing the on-chip cache hierarchy}. 
\rbd{If the off-chip prediction is correct}, the \rbd{regular} load request eventually misses the LLC and checks the main memory controller's RQ for any ongoing main memory access to the same load address (step~\circled{3}). If the address is found, the \rbd{regular} load request waits for the ongoing \rbd{Hermes} request to finish before sending the Hermes-fetched data back to the core.

Hermes's performance gain depends on the latency to directly issue \rbd{a Hermes} request to the main memory controller \rbd{(called \emph{Hermes request issue latency})}.
\rbe{Although} a \rbd{Hermes request} experiences a significantly shorter latency to arrive at the main memory controller than its corresponding \rbd{regular load request} \rbe{because a Hermes request bypasses the cache hierarchy and on-chip queueing delays}, \rbd{a Hermes request} nonetheless pays for a latency to route through the on-chip network. We model two variants of Hermes using an optimistic and a pessimistic estimate of \rbd{Hermes request} issue latency \rbe{to take into account a wide range of potential differences in on-chip interconnect designs} (see \Cref{sec:her_dmf_configuration}). \rbd{In~\Cref{sec:her_load_issue_latency}, we also evaluate Hermes with a wide range of Hermes request issue latencies (from $0$ cycle to $24$ cycles)} and show that Hermes consistently provides performance benefit even with the most pessimistic Hermes request issue latency.

\subsubsection{Returning Data to the Core} \label{sec:her_returning_data_to_core}

For every \rbd{Hermes request} returning from main memory, Hermes checks the RQ of the main memory controller and returns the fetched data back to the LLC if there is a \rbd{regular} load request already waiting for the same load address. If there is no \rbd{regular} load request waiting for the completed \rbd{Hermes} \rbd{request}, \rbd{Hermes drops the request and does \emph{not} fill the data into the cache hierarchy, \rbe{which} keeps the on-chip cache hierarchy \rbe{internally} coherent.}

\subsection{Storage Overhead} \label{sec:her_storage_overhead}

\Cref{table:her_overhead} shows the total storage overhead of Hermes. Hermes requires only $4$~KB of metadata storage \rbd{per processor core}. \pred 
consumes $3.2$~KB, whereas the metadata stored in LQ for \pred training consumes $0.8$~KB.

\begin{table}[htbp]
  \centering
  \small
    \begin{tabular}{|L{5em}||m{25em}||R{4em}|}
    \thickhline
    \Tabval{\textbf{Structure}} & \Tabval{\textbf{Description}} & \Tabval{\textbf{Size}} \\
    \hline
    \hline
    \Tabval{\textbf{\pred}} & 
    \Tstrut
    \begin{minipage}{25em}
      \vskip 6pt
      \begin{itemize}[leftmargin=1em]
        \setlength{\itemsep}{1pt}
        \setlength{\parskip}{0pt}
        \setlength{\parsep}{0pt}
        \item Perceptron weight tables
        \begin{itemize}[leftmargin=1em]
            \setlength{\itemsep}{1pt}
            \setlength{\parskip}{0pt}
            \setlength{\parsep}{0pt}
            \item PC $\oplus$ cacheline offset: $1024\times5b$
            \item PC $\oplus$ byte offset: $1024\times5b$
            \item PC + first access: $1024\times5b$
            \item Cacheline offset + first access: $128\times5b$
            \item Last-$4$ load PCs: $1024\times5b$
        \end{itemize}
        \item Page buffer: $64\times80b$
      \end{itemize}
    \end{minipage} 
    \Bstrut &
    \Tabval{\textbf{3.2 KB}} \\
    \hline
    \hline
    \Tabval{\textbf{LQ Metadata}} & \Tabval{Hashed PC: $128\times32b$; Last-4 PC: $128\times10b$; First access: $128\times1b$; perceptron weight: $128\times5b$; prediction: $128\times1b$} & \Tabval{\textbf{0.8 KB}} \\
    \hline
    \hline
    \Tabval{\textbf{Total}} &   & \Tabval{\textbf{4.0 KB}} \\
    \hline
    \end{tabular}%
  \caption{Storage overhead of Hermes.}
  \label{table:her_overhead}%
\end{table}%

\sectionRB{Hermes: Evaluation Methodology}{Methodology} {sec:her_methodology}

\noindent We use the ChampSim trace-driven simulator~\cite{champsim} to evaluate Hermes. We faithfully model the latest-generation Intel Alder Lake performance-core~\cite{goldencove} with its \rbd{large} ROB, large caches with publicly-reported on-chip cache access latencies~\cite{llc_lat1,llc_lat2,l3_lat_compare1}, and 
the state-of-the-art prefetcher Pythia~\cite{pythia} at the LLC. 
\Cref{table:her_sim_params} shows the key microarchitectural parameters. For single-core simulations, we \rbd{warm up} the core using $100$M instructions and simulate the next $500$M instructions. For multi-programmed simulations, we use $50$M and $100$M instructions from each workload for warmup and simulation, respectively. If a core finishes early, the workload is replayed \rbd{until} every core has finished executing at least $100$M instructions. 
\rbd{The source code of Hermes, along with all workload traces and scripts to reproduce \rbe{our} results are \rbe{freely} available at~\cite{hermes_github}.}

\begin{table}[!ht]
    \centering
    \small
    \begin{tabular}{L{4em}||L{32em}}
         \thickhline
         \Tabval{\textbf{Core}} & \Tabval{ 1 and 8 cores, 6-wide fetch/execute/commit, 512-entry ROB, 128/72-entry LQ/SQ, Perceptron branch predictor~\cite{perceptron} with 17-cycle misprediction penalty}\\
         \hline
         \Tabval{\textbf{L1/L2 Caches}} & \Tabval{Private, 48KB/1.25MB, 64B line, 12/20-way, 16/48 MSHRs, LRU, 5/15-cycle round-trip latency~\cite{llc_lat2}} \\
         \hline
         \Tabval{\textbf{LLC}} & \Tabval{3MB/core, 64B line, 12 way, 64 MSHRs/slice, SHiP~\cite{ship}, 55-cycle round-trip latency~\cite{llc_lat1,llc_lat2}, \textbf{Pythia} prefetcher~\cite{pythia}}\\
         \hline
         \Tabval{\textbf{Main Memory}} & \Tabval{\textbf{1C:} 1 channel, 1 rank per channel; \textbf{8C:} 4 channels, 2 ranks per channel; 8 banks per rank, DDR4-3200 MTPS, 64b data-bus per channel, 2KB row buffer per bank, tRCD=12.5ns, tRP=12.5ns, tCAS=12.5ns} \\
        \hline
        \Tabval{\textbf{Hermes}} & \Tabval{\textbf{Hermes-O/P}: $6$/$18$-cycle \rbd{Hermes request} issue latency}\\
        \thickhline
    \end{tabular}
    \caption{Simulated system parameters.}
    \label{table:her_sim_params}
\end{table}

\subsection{Workloads} \label{sec:her_workloads}

We evaluate Hermes using a wide range of memory-intensive workloads spanning \texttt{SPEC CPU2006}~\cite{spec2006}, \texttt{SPEC CPU2017}~\cite{spec2017}, \texttt{PARSEC}~\cite{parsec}, \texttt{Ligra} graph processing workload suite~\cite{ligra}, and commercial workloads from \rbd{the} 2nd data value prediction championship (\texttt{CVP}~\cite{cvp2}). For \texttt{SPEC CPU2006} and \texttt{SPEC CPU2017} workloads, we reuse the instruction traces provided by the 2nd and the 3rd data prefetching championships (DPC~\cite{dpc2,dpc3}). For \texttt{PARSEC} and \texttt{Ligra} workloads, we reuse the instruction traces open-sourced by Pythia~\cite{pythia}. 
The \texttt{CVP} workload traces are collected by the Qualcomm Datacenter Technologies and capture complex program behavior from various integer, floating-point, cryptographic, and server applications in the \rbd{field}. 
We only consider workload traces in our evaluation that have at least $3$ LLC misses per kilo instructions (MPKI) in the no-prefetching system. In total, we evaluate Hermes using $110$ single-core workload traces \rbd{from} $73$ workloads, which are summarized in~\Cref{table:her_workloads}.
For multi-programmed simulations, we create both homogeneous and heterogeneous trace mixes. For an eight-core homogeneous multi-programmed simulation, we run eight copies of each trace from our single-core trace list, one trace in each core. For heterogeneous multi-programmed simulation, we \emph{randomly} select any eight traces from our single-core trace list and run one trace in each core. In total, we evaluate Hermes using $110$ homogeneous and $110$ heterogeneous eight-core workloads.

\begin{table}[htbp]
  \centering
  \small
    \begin{tabular}{L{7em}L{5em}L{5em}m{15em}}
    \toprule
    \textbf{Suite} & \multicolumn{1}{l}{\textbf{\# Workloads}} & \multicolumn{1}{l}{\textbf{\# Traces}} & \textbf{Example Workloads} \\
    \midrule
    \Tabval{SPEC06} & \Tabval{14}    & \Tabval{22}    & \Tabval{gcc, mcf, cactusADM, lbm, ...} \\
    \Tabval{SPEC17} & \Tabval{11}    & \Tabval{23}    & \Tabval{gcc, mcf, pop2, fotonik3d, ...} \\
    \Tabval{PARSEC} & \Tabval{4}     & \Tabval{12}    & \Tabval{canneal, facesim, raytrace, ...} \\
    \Tabval{Ligra} & \Tabval{11}    & \Tabval{20}    & \Tabval{BFS, PageRank, Radii, ...} \\
    \Tabval{CVP} & \Tabval{33}     & \Tabval{33}    & \Tabval{integer, floating-point, server, ...} \\
    \bottomrule
    \end{tabular}%
  \caption{Workloads used for evaluation.}
  \label{table:her_workloads}%
\end{table}%

\subsection{Evaluated System Configurations} \label{sec:her_dmf_configuration}

For a comprehensive analysis, we compare Hermes with various off-chip load prediction mechanisms, as well as \rbd{in combination with} various recently proposed prefetchers. \Cref{table:her_overhead_comparison} compares the storage overhead of all evaluated mechanisms.

\textbf{\textit{(1) Various off-chip prediction mechanisms.}}
\rbg{We compare \pred against two cache hit/miss prediction techniques: (1) HMP, proposed by Yoaz et al.~\cite{yoaz1999speculation}, and (2) a simple cacheline tag-tracking based predictor, called \emph{TTP}, \rbh{which we design}.}
\rbh{HMP uses three predictors similar to a hybrid branch predictor: local~\cite{yeh1991two}, gshare~\cite{mcfarling1993combining,yeh1991two}, and gskew~\cite{michaud1997trading}, each of which individually \rbi{predicts} off-chip loads using a different \rbi{prediction} mechanism. For a given load, HMP consults each individual predictor and selects the majority prediction.}
\rbg{\rbh{We design} TTP \rbh{by taking inspiration from} prior cacheline address tracking-based mechanisms~\cite{loh_miss_map,lp,d2d}. \rbh{TTP} tracks partial tags of cacheline addresses that are likely to be present in the \rbh{entire on-chip} cache hierarchy in a separate metadata structure. For every cache fill (LLC eviction), the partial tag of the filled (evicted) cacheline address is inserted \rbi{into} (evicted \rbi{from}) TTP's metadata. To predict whether or not a given load would go off-chip, TTP searches the metadata \rbi{structure} with the partial tag of the load address. If the tag is not present in the metadata \rbi{structure}, TTP predicts the load would go off-chip. \rbh{We open-source TTP in our repository~\cite{hermes_github}.}}

\textbf{\textit{(2) Various data prefetchers.}}
We \rbd{evaluate} Hermes \rbd{combined with} five recently-proposed high-performance prefetching techniques: Pythia~\cite{pythia}, Bingo~\cite{bingo}, SPP~\cite{spp} (with perceptron filter~\cite{ppf}), MLOP~\cite{mlop}, and SMS~\cite{sms}. As mentioned in~\Cref{table:her_sim_params}, Pythia is incorporated in our baseline system.

\begin{table}[!ht]
    \centering
    \small 
    \begin{tabular}{m{23.6em}||R{5em}}
    \thickhline
    \Tabval{\textbf{HMP}~\cite{yoaz1999speculation} with local, gshare, and gskew predictors} & \Tabval{\textbf{11 KB}}\\
    \Tabval{\textbf{TTP} with a metadata budget similar to the L2 cache} & \Tabval{\textbf{1536 KB}} \\
    \hline
    \hline
    \Tabval{\textbf{Pythia}~\cite{pythia} with the same configuration in~\cite{pythia}} & \Tabval{\textbf{25.5 KB}} \\
    \Tabval{\textbf{Bingo}~\cite{bingo} with the same configuration in~\cite{bingo}} & \Tabval{\textbf{46 KB}} \\
    \Tabval{\textbf{SPP}~\cite{spp} with perceptron-based prefetch filter~\cite{ppf}} & \Tabval{\textbf{39.3 KB}} \\
    \Tabval{\textbf{MLOP}~\cite{mlop} with the same configuration in~\cite{mlop}} & \Tabval{\textbf{8 KB}} \\
    \Tabval{\textbf{SMS}~\cite{sms} with the same configuration in~\cite{sms}} & \Tabval{\textbf{20 KB}} \\
    \hline
    \hline
    \Tabval{\textit{\textbf{Hermes with \pred}}} \textit{(this work)} & \Tabval{\textbf{4 KB}} \\
    \thickhline
    \end{tabular}
    \caption{Storage overhead of all evaluated mechanisms.}
    \label{table:her_overhead_comparison}
\end{table}

We evaluate two variants of Hermes: \textbf{Hermes-O} and \textbf{Hermes-P}. These two variants differ only in Hermes request issue latency. \emph{Hermes-O} (i.e., the optimistic Hermes) and \emph{Hermes-P} (i.e., the pessimistic Hermes) use a request issue latency of $6$ cycles and $18$ cycles, respectively. Unless stated otherwise, \emph{Hermes} represents the optimistic variant \emph{Hermes-O}.

\sectionRB{Hermes: Evaluation}{Evaluation}{sec:her_her_evaluation}

\subsection{\pred Prediction Analysis} \label{sec:her_ocp_pred}

\subsubsection{Accuracy and Coverage of \pred} \label{sec:her_acc_cov_ocp}

\begin{figure}[!ht]
\centering
\includegraphics[width=5.75in]{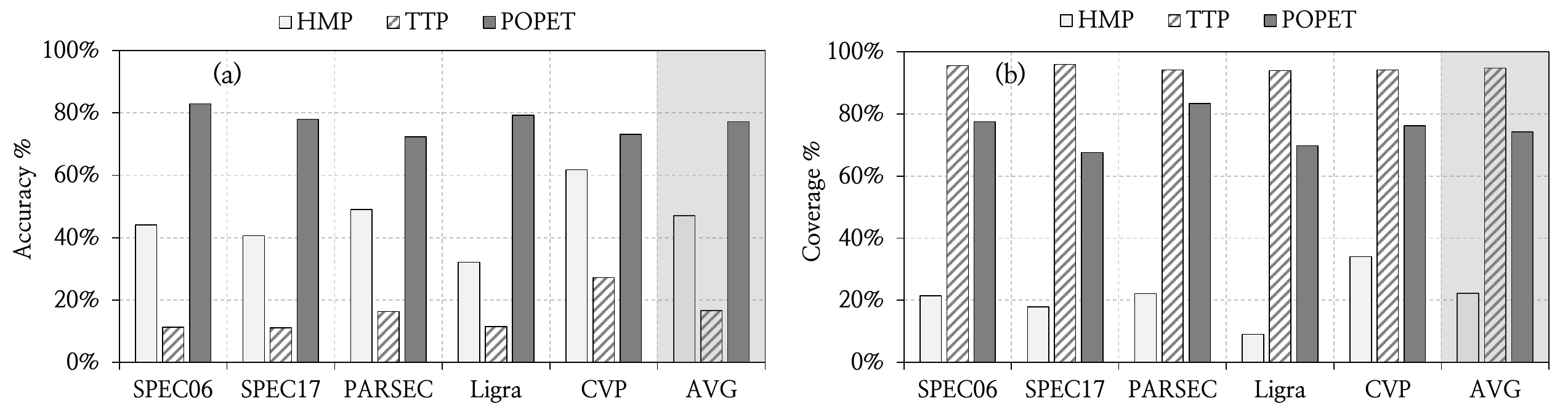}
\caption{Comparison of (a) accuracy and (b) coverage of \pred against those of HMP~\cite{yoaz1999speculation} and TTP.
}
\label{fig:her_ocp_acc_cov}
\end{figure}

\Cref{fig:her_ocp_acc_cov} shows the comparison of \rbh{\pred's} \rbe{off-chip load prediction} accuracy and coverage  against \rbh{those of} HMP and TTP in the baseline system. The key takeaway is that \pred has significantly higher accuracy \emph{and} coverage than HMP. \rbh{\pred provides} $77.1\%$ accuracy with $74.3\%$ coverage \rbd{on average across all single-core workloads}, whereas HMP \rbh{provides} $47\%$ accuracy with $22.3\%$ coverage. 
\rbg{TTP, with a metadata budget of $1.5$~MB, \rbh{provides the highest} coverage \rbh{($94.8\%$)} but with a significantly lower accuracy \rbh{($16.6\%$)}.}
\pred's superior accuracy \emph{and} coverage directly translates to performance benefits both in single-core and eight-core system configuration (see ~\Cref{sec:her_perf_1c_main} and~\Cref{sec:her_perf_mc}).

\subsubsection{\rbd{Effect of} Different \pred Features} \label{sec:her_acc_cov_ocp_contri}

\begin{figure}[!ht]
\centering
\includegraphics[width=5.75in]{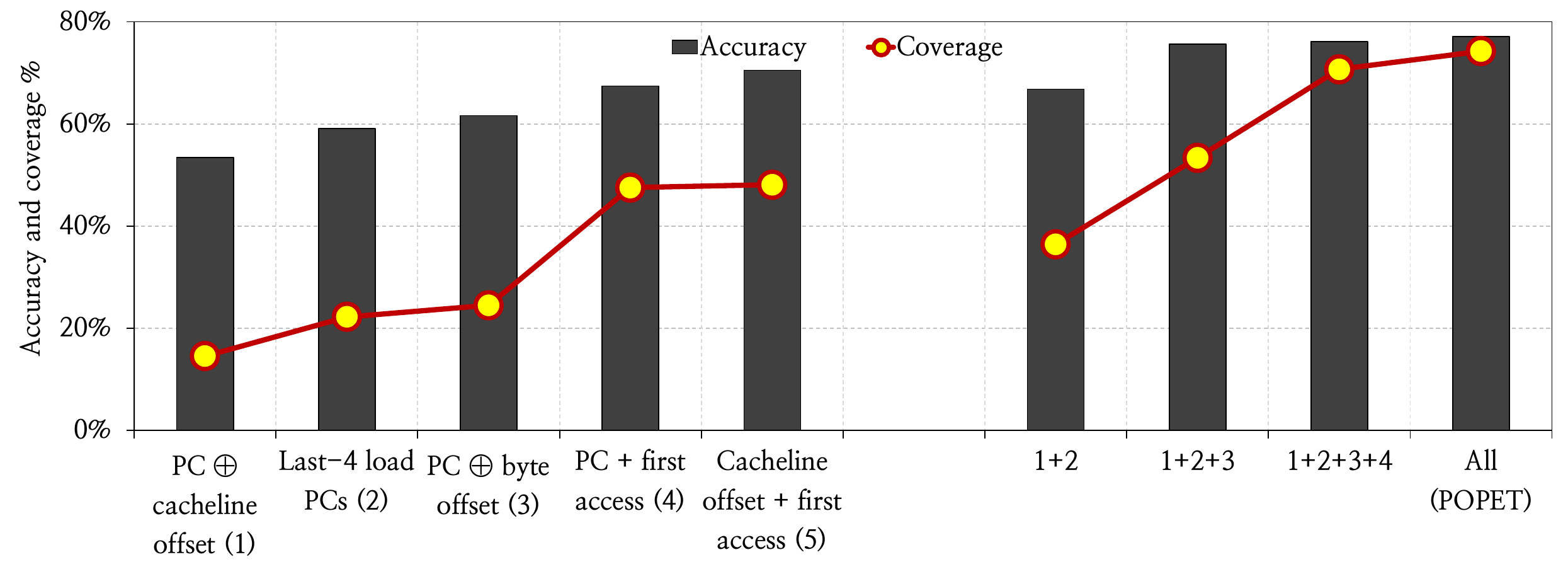}
\caption{The accuracy and coverage of \pred using each program feature individually and in various combinations.}
\label{fig:her_acc_cov_contri}
\end{figure}

\Cref{fig:her_acc_cov_contri} shows the accuracy and coverage of \pred using the five selected program features used individually and in \rbe{various} combinations. We make two key observations.
First, each program feature individually produces predictions with a wide range of accuracy and coverage.
The PC $\oplus$ cacheline offset feature produces the lowest-quality predictions with only $53.4\%$ accuracy and $14.5\%$ coverage,
whereas the cacheline offset + first access feature produces the highest-quality predictions with $70.6\%$ accuracy and $48.1\%$ coverage.
Second, by stacking multiple features together, the final \pred design achieves \emph{both} higher accuracy and coverage than \rbd{those provided by} any \rbd{single} individual program feature.
We conclude that \pred is capable of learning from multiple program features to achieve both higher off-chip load prediction accuracy and coverage than any individual program feature can provide.

\subsubsection{Usefulness of all features} \label{sec:her_feature_necessity}
\begin{figure}[!ht]
\centering
\includegraphics[width=5.75in]{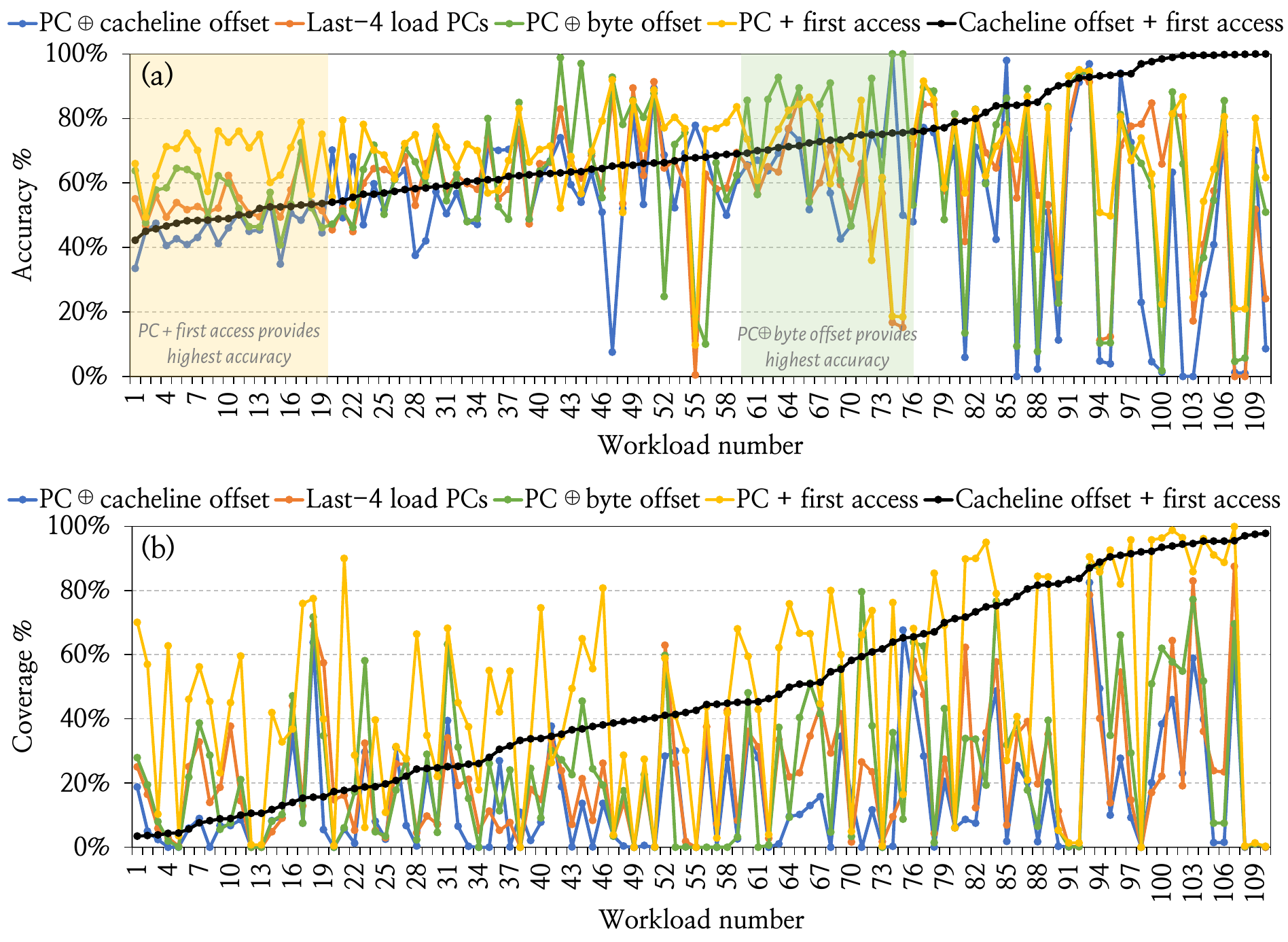}
\caption{Line graph of \pred's \rbd{(a) accuracy and (b) coverage} using each \rbd{of the} five program features individually across all 110 single-core workloads. 
No single feature can provide the best accuracy \rbd{or coverage} across all workloads.
}
\label{fig:her_acc_line_chart}
\end{figure}

To understand the \rbd{usefulness} of multi-feature learning, we analyze per-trace accuracy and coverage of \pred using each individual program feature. \Cref{fig:her_acc_line_chart}(a) shows the line graph of \pred's prediction accuracy with each \rbd{of the} five program features individually for all single-core workload traces. The traces are sorted in ascending order of \pred accuracy using the feature cacheline offset + first access, since this feature individually has the highest average accuracy (as shown in \Cref{fig:her_acc_cov_contri}(a)). The key takeaway from~\Cref{fig:her_acc_line_chart}(a) is that there is no \emph{single} program feature that individually provides the highest prediction accuracy across \emph{all} workloads. Out of $110$ workload traces, the features PC + first access, cacheline offset + first access, PC $\oplus$ byte offset, PC $\oplus$ cacheline offset, and last-4 load PCs provide the highest prediction accuracy in $47$, $29$, $20$, $9$, and $5$ workload traces, respectively. We observe similar variability in \pred's coverage, \rbd{as shown in \Cref{fig:her_acc_line_chart}(b)}, where no single program feature individually provides the highest coverage across \emph{all} workloads. This large variability of accuracy/coverage with different features in different workloads warrants learning using all features \emph{in unison} to provide higher accuracy \emph{and} coverage than any individual program feature across a \emph{wide range} of workloads.

\subsection{Single-Core Performance Analysis} \label{sec:her_perf_1c_main}

\subsubsection{Performance Improvement} \label{sec:her_perf_1c}

\begin{figure}[!ht]
\centering
\includegraphics[width=5.75in]{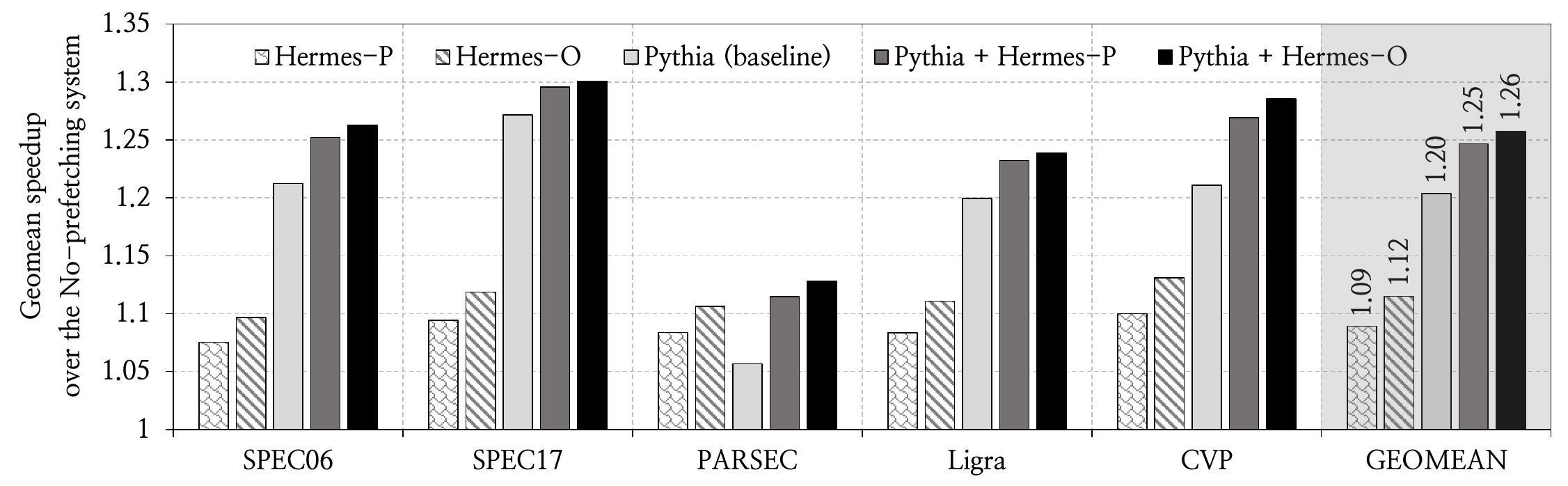}
\caption{\rbd{Speedup in single-core workloads.}}
\label{fig:her_perf_1c}
\end{figure}

\Cref{fig:her_perf_1c} shows \rbd{performance of} Hermes \rbd{(O and P)}, Pythia, and Hermes \rbd{combined with} Pythia \rbd{normalized to} the no-prefetching system in single-core workloads. We make three key observations. 
First, Hermes provides nearly half of the performance \rbd{benefit of} Pythia with only $\frac{1}{5}\times$ the storage overhead. On average, Hermes-O improves performance by $11.5\%$ over a no-prefetching system, whereas Pythia improves performance by $20.3\%$.
Second, Hermes-O \rbd{(Hermes-P) combined with} Pythia outperforms Pythia by $5.4\%$ ($4.3\%$). Third, Hermes \rbd{combined with} Pythia \emph{consistently} outperforms Pythia in \emph{every} workload category.

\begin{figure}[!ht]
\centering
\includegraphics[width=5.75in]{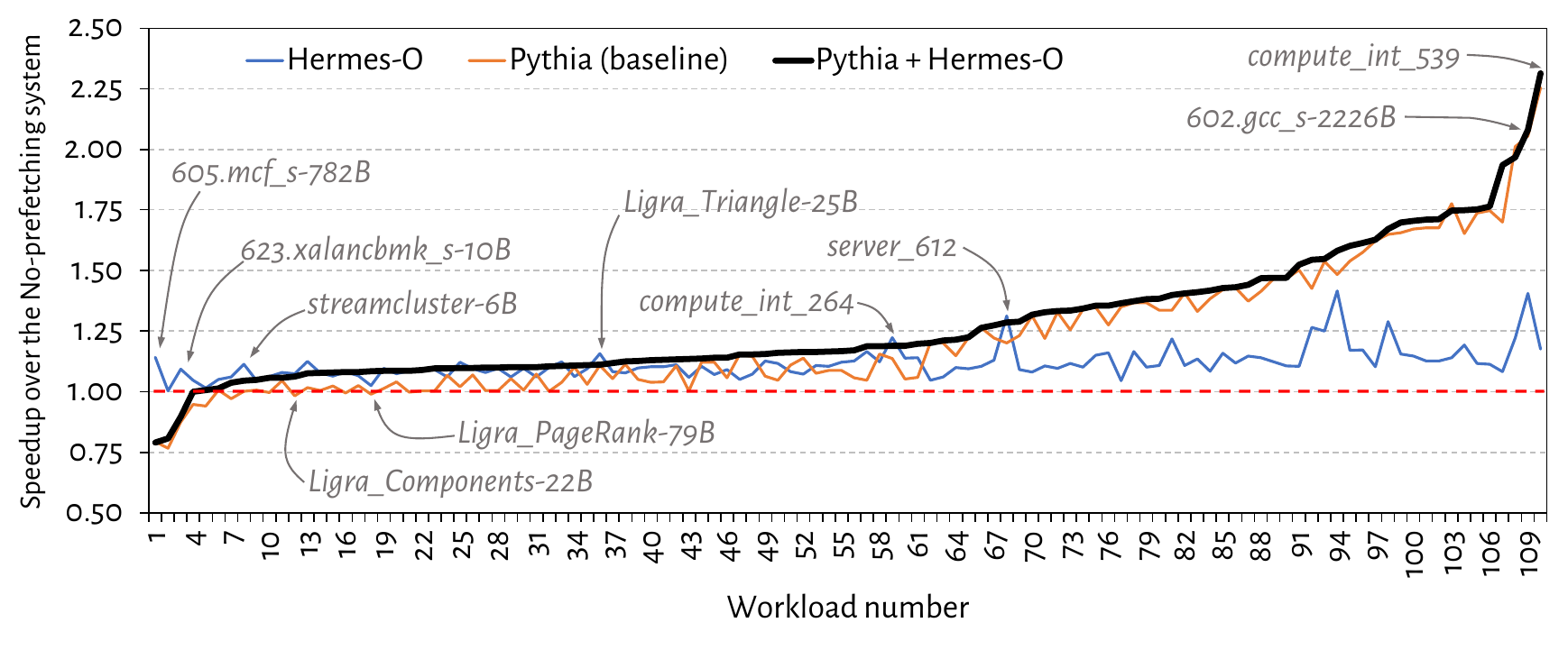}
\caption{Single-core performance of all 110 workloads.}
\label{fig:her_perf_1c_line}
\end{figure}

\begin{sloppypar}
To better understand Hermes's performance improvement, \Cref{fig:her_perf_1c_line} shows the performance line graph of Hermes, Pythia, and Hermes \rbd{combined with} Pythia for every single-core workload trace. The traces are sorted in ascending order of performance gains by Hermes \rbd{combined with Pythia} over the no-prefetching system. We make four key observations from \Cref{fig:her_perf_1c_line}.
First, \rbd{Hermes combined with Pythia} outperforms the no-prefetching system in all but three single-core workload traces. The \texttt{compute\_int\_539} and \texttt{605.mcf\_s-782B} \rbd{traces} \rbd{experience} the highest and the lowest \rbd{speedup} \rbd{($2.3\times$ and $0.8\times$, respectively)}.
Second, unlike Pythia, Hermes \emph{always} improves performance over the no-prefetching system in \emph{every} workload trace. 
Third, Hermes outperforms Pythia by $7.9\%$ on average in $51$ traces (e.g., \texttt{streamcluster-6B}, \texttt{Ligra\_PageRank-79B}). In the remaining $59$ traces, Pythia outperforms Hermes by $26\%$ on average.
Fourth, \rbd{Hermes combined with Pythia} consistently outperforms \emph{both} \rbd{Hermes and Pythia alone} in almost every workload trace.
\end{sloppypar}
Based on our performance results, we conclude that, Hermes provides significant and consistent performance \rbd{improvements} over a wide range of workloads both \rbd{by itself and when combined with} the state-of-the-art prefetcher \rbd{Pythia}.

\subsubsection{\rbd{Effect of the Off-chip Load Prediction Mechanism}} \label{sec:her_dmp_with_hmp_perf}

\begin{figure}[!ht]
\centering
\includegraphics[width=5.75in]{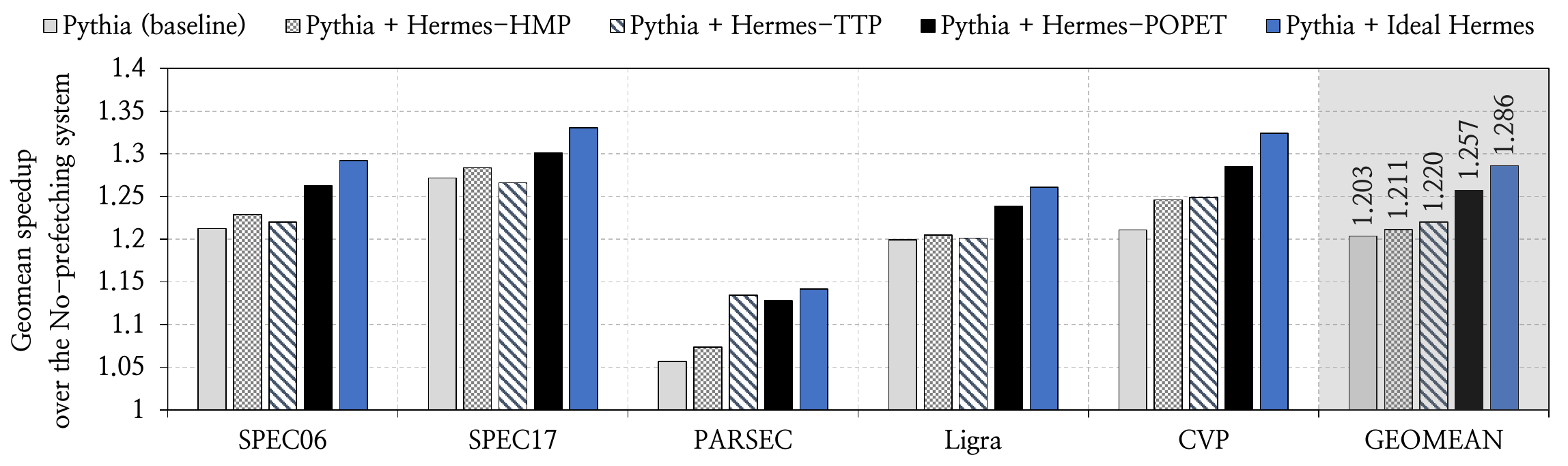}
\caption{\rbd{Speedup of Hermes with three off-chip load predictors (HMP, \rbg{TTP}, and \pred) and the Ideal Hermes.}}
\label{fig:her_perf_1c_with_hmp}
\end{figure}

\Cref{fig:her_perf_1c_with_hmp} shows the performance of Hermes with \pred, Hermes-HMP, \rbg{Hermes-TTP}, and the Ideal Hermes (see~\Cref{sec:her_headroom_study}) combined with Pythia normalized to the no-prefetching system in single-core workloads.
\rbe{We make two key observations.}
\rbe{First,} \rbd{Hermes with \pred} outperforms both Hermes-HMP and \rbg{Hermes-TTP}. 
\rbg{On average, \rbh{Hermes-HMP, Hermes-TTP}, and \rbd{Hermes with \pred} \rbd{combined with} Pythia provide \rbh{$0.8\%$, $1.7\%$}, and $5.4\%$ performance \rbh{improvement} \rbd{over} Pythia, respectively.}
\rbe{Second, Hermes-\pred provides nearly $90\%$ of the performance improvement provided by the Ideal Hermes that employs an ideal off-chip load predictor with $100\%$ accuracy and coverage.}
\rbd{We conclude that Hermes provides performance gains due to \rbg{both} \rbe{the} high \rbe{off-chip load prediction} accuracy and coverage of \pred. \rbe{Thus,} designing a good off-chip predictor is critical for Hermes to improve performance.}

\subsubsection{\rbd{Effect on Stall Cycles}}

\begin{figure}[!ht]
\centering
\includegraphics[width=5.75in]{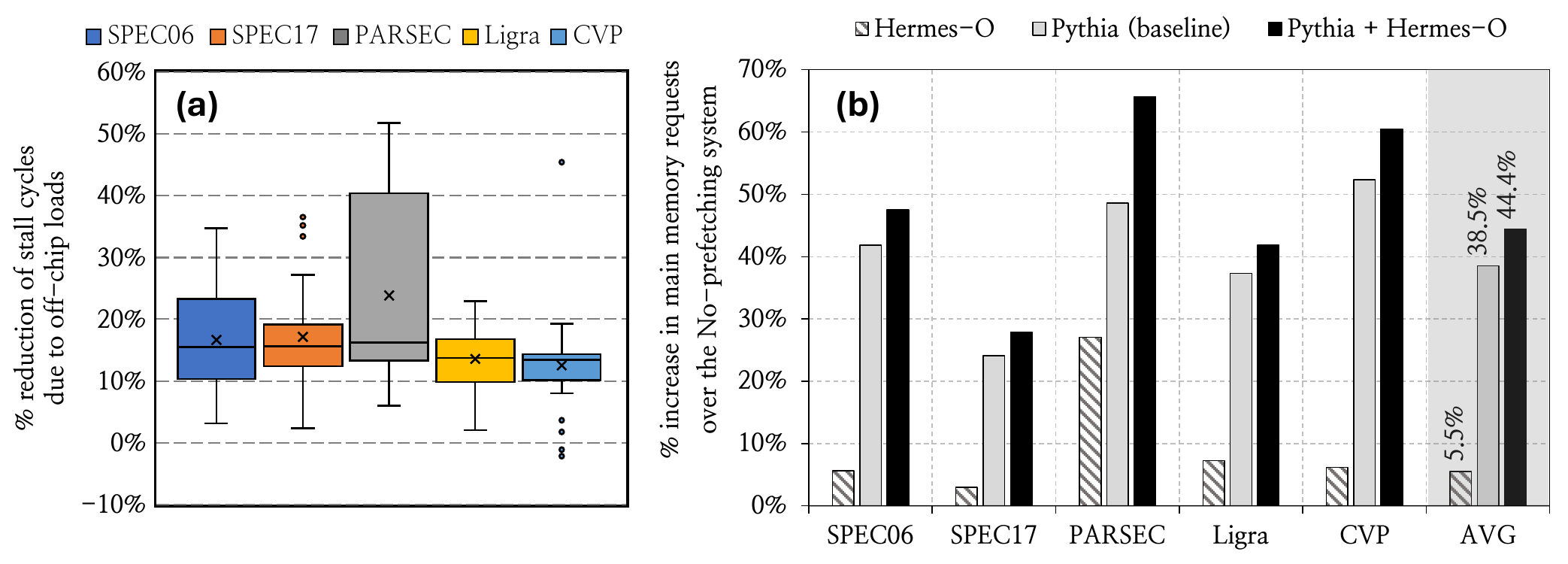}
\caption{(a) Reduction in stall cycles caused by off-chip loads. (b) Overhead in the main memory requests.}
\label{fig:her_system_impact}
\end{figure}

\Cref{fig:her_system_impact}(a) \rbd{plots the distribution of} the percentage reduction \rbd{in} stall cycles due to off-chip load requests in a system with Hermes over the baseline system \rbe{in single-core workloads} as a \rbe{box-and-whiskers} plot.\footnote{\rbd{Each box is lower-bounded by the first quartile (i.e., the middle value between the lowest value and the median value of the data points) and upper-bounded by the third quartile (i.e., the middle value between the median and the highest value of the data points). The inter-quartile range ($IQR$) is the distance between the first and the third quartile (i.e., the length of the box). Whiskers extend an additional $1.5\times IQR$ on the either side of the box. Any outlier values that falls outside the range of whiskers are marked by dots.}
The cross marked value within each box represents the mean.
}
The key observation is that Hermes reduces the stall cycles caused by off-chip loads by $16.2\%$ on average (up to $51.8\%$) across all workloads. \texttt{PARSEC} workloads experience the highest \rbd{average} stall cycle reduction of $23.8\%$. $90$ out of $110$ \rbd{workloads experience} at least $10\%$ stall cycle reduction. 
We conclude that Hermes considerably reduces the stall cycles due to off-chip load requests, which \rbd{leads to} performance \rbd{improvement}.

\subsubsection{Overhead in Main Memory Requests} \label{sec:her_bw_1c}

\Cref{fig:her_system_impact}(b) shows the percentage increase in main memory requests in Hermes, Pythia, and \rbd{Hermes combined with Pythia} over the no-prefetching system in all single-core workloads. We make two key observations. First, Hermes increases main memory requests by only $5.5\%$ \rbd{(on average)} over the no-prefetching system, whereas Pythia by $38.5\%$. This means that, every $1\%$ performance gain (see \Cref{fig:her_perf_1c}) comes at a cost of only $0.5\%$ increase in main memory requests in Hermes, whereas nearly $2\%$ increase in main memory requests in Pythia.
\rbd{We attribute this result} to the highly-accurate predictions made by \pred, as compared to less-accurate prefetch decisions made by Pythia. 
Second, Hermes \rbd{combined with} Pythia further increases main memory requests by only $5.9\%$ over Pythia. This means that, every $1\%$ performance benefit by Hermes on top of Pythia comes at a cost of only $1\%$ overhead in main memory requests.
We conclude that, Hermes, due to its underlying high-accuracy prediction mechanism, adds considerably lower overhead in main memory requests while providing significant performance improvement both \rbd{by itself and when combined with Pythia}.

\subsection{Eight-Core Performance Analysis} \label{sec:her_perf_mc}

\begin{figure}[!ht]
\centering
\includegraphics[width=5.75in]{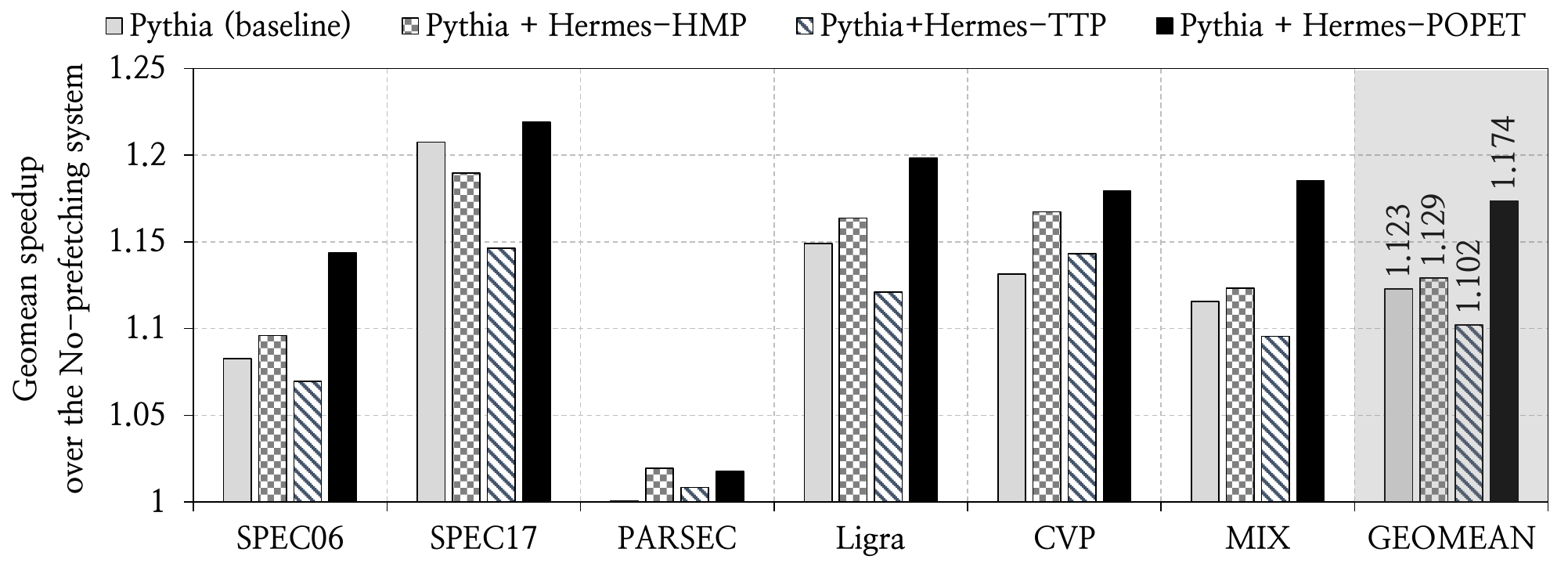}
\caption{\rbd{Speedup in} eight-core workloads.}
\label{fig:her_perf_8c}
\end{figure}

\rbd{\Cref{fig:her_perf_8c} shows \rbe{the} performance of Pythia, Hermes-HMP, \rbh{Hermes-TTP}, and Hermes-\pred combined with Pythia normalized to the no-prefetching system in all eight-core workloads.}
The key takeaway is that due to the highly-accurate predictions by \pred, Hermes-\pred combined with Pythia consistently outperforms Pythia in \emph{every} workload category. On average, Hermes-HMP, \rbg{Hermes-TTP}, and Hermes-\pred combined with Pythia \rbe{provide} $0.6\%$, \rbg{$-2.1\%$}, and $5.1\%$ \rbe{higher} performance on top of Pythia, respectively.
\rbh{Due to its inaccurate predictions, TTP generates many unnecessary main memory requests, which reduce the performance of Hermes-TTP combined with Pythia as compared to Pythia alone in the bandwidth-constrained four-core configuration.}
\rbe{We conclude that Hermes provides significant and consistent performance improvement in the bandwidth-constrained eight-core system due to its highly-accurate off-chip load prediction.}

\subsection{Performance Sensitivity Analysis} \label{sec:her_1c_sensitivity}

\subsubsection{\rbd{Effect of} Main Memory Bandwidth} \label{sec:her_dram_bw}

\begin{figure*}[!ht]
\centering
\includegraphics[width=5.75in]{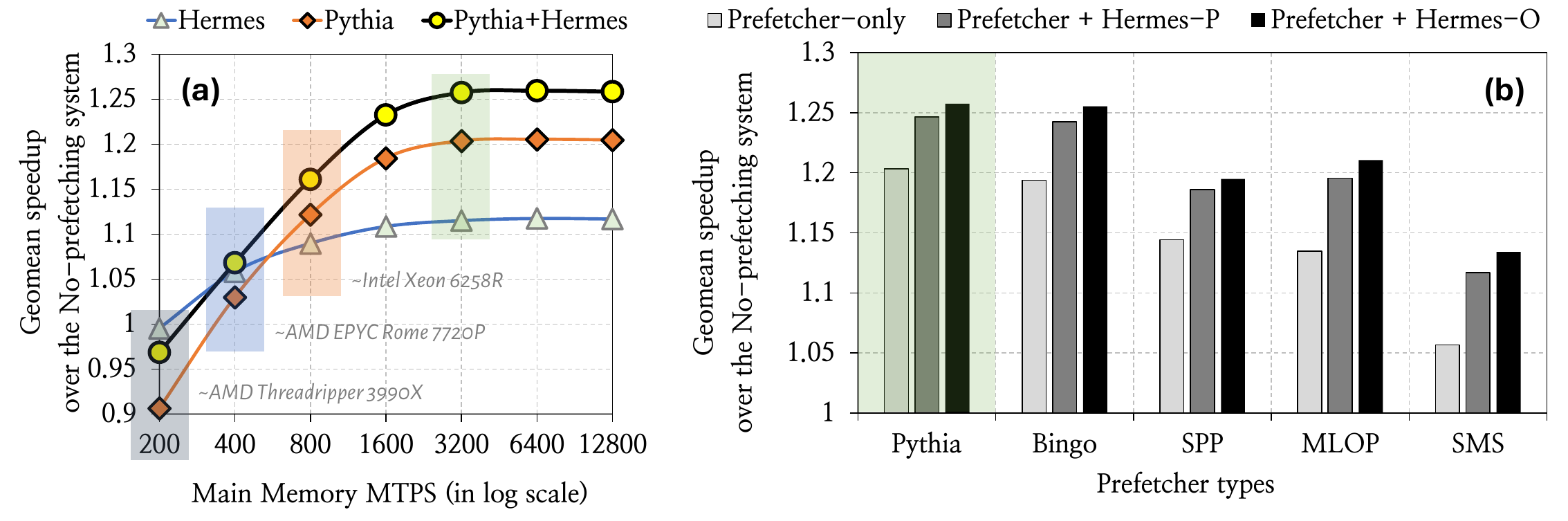}
\caption{Performance sensitivity to (a) main memory bandwidth and (b) baseline prefetcher. The baseline system configuration is highlighted in green. Other highlighted configurations closely match with various commercial processors~\cite{intel_xeon_gold,zen_epyc,zen_threadripper}.}
\label{fig:her_sensitivity_master}
\end{figure*}

\rbd{\Cref{fig:her_sensitivity_master}(a) shows the speedup of Hermes, Pythia, and Hermes combined with Pythia over the no-prefetching system in single-core workloads by scaling the main memory bandwidth.}
We make two key observations. 
First, Hermes \rbd{combined with} Pythia \emph{consistently} outperforms Pythia in \emph{every} main memory bandwidth configuration from $\frac{1}{16}\times$ to $4\times$ of the baseline system. 
\rbd{Hermes combined with Pythia} outperforms Pythia \rbe{alone} by $6.2\%$ and $5.5\%$ in \rbe{the} main memory bandwidth configuration \rbd{with} $200$ and $12800$ million transfers per second (MTPS), \rbe{respectively}.
Second, Hermes \rbe{by itself} \emph{outperforms} Pythia in highly-bandwidth-constrained configurations. This is \rbd{due} to the highly-accurate off-chip load predictions made by \pred, which incurs less main memory \rbd{bandwidth} overhead than the aggressive, less-accurate prefetching decisions made by Pythia.  
Hermes outperforms Pythia by $2.8\%$ and $8.9\%$ in $400$ and $200$ MTPS configurations, respectively.

\subsubsection{\rbd{Effect of the Baseline} Prefetcher} \label{sec:her_prefetchers}

We evaluate Hermes \rbd{combined with} four recently-proposed data prefetchers: Bingo~\cite{bingo}, SPP~\cite{spp} (with perceptron filter~\cite{ppf}), MLOP~\cite{mlop}, and SMS~\cite{sms}. For each experiment, we replace the baseline LLC prefetcher Pythia with a new prefetcher and measure the performance improvement of the prefetcher \rbd{by itself} and Hermes \rbd{combined with} the prefetcher. 
\Cref{fig:her_sensitivity_master}(b) \rbe{shows the performance of the baseline prefetcher, and Hermes-P/O combined with the baseline prefetcher, normalized to the no-prefetching system in single-core workloads}. The key takeaway is that Hermes \rbd{combined with} \emph{any} \rbd{baseline} prefetcher consistently outperforms \rbd{the baseline prefetcher by itself} for \rbd{all four evaluated} prefetching techniques. Hermes+prefetcher outperforms the prefetcher \rbd{alone} by $6.2\%$, $5.1\%$, $7.6\%$, and $7.7\%$, \rbd{for} Bingo, SPP, MLOP, and SMS as the \rbd{baseline} prefetcher.

\subsubsection{\rbd{Effect of \rbe{the} Hermes Request} Issue Latency} \label{sec:her_load_issue_latency}

\begin{figure*}[!ht]
\centering
\includegraphics[width=5.75in]{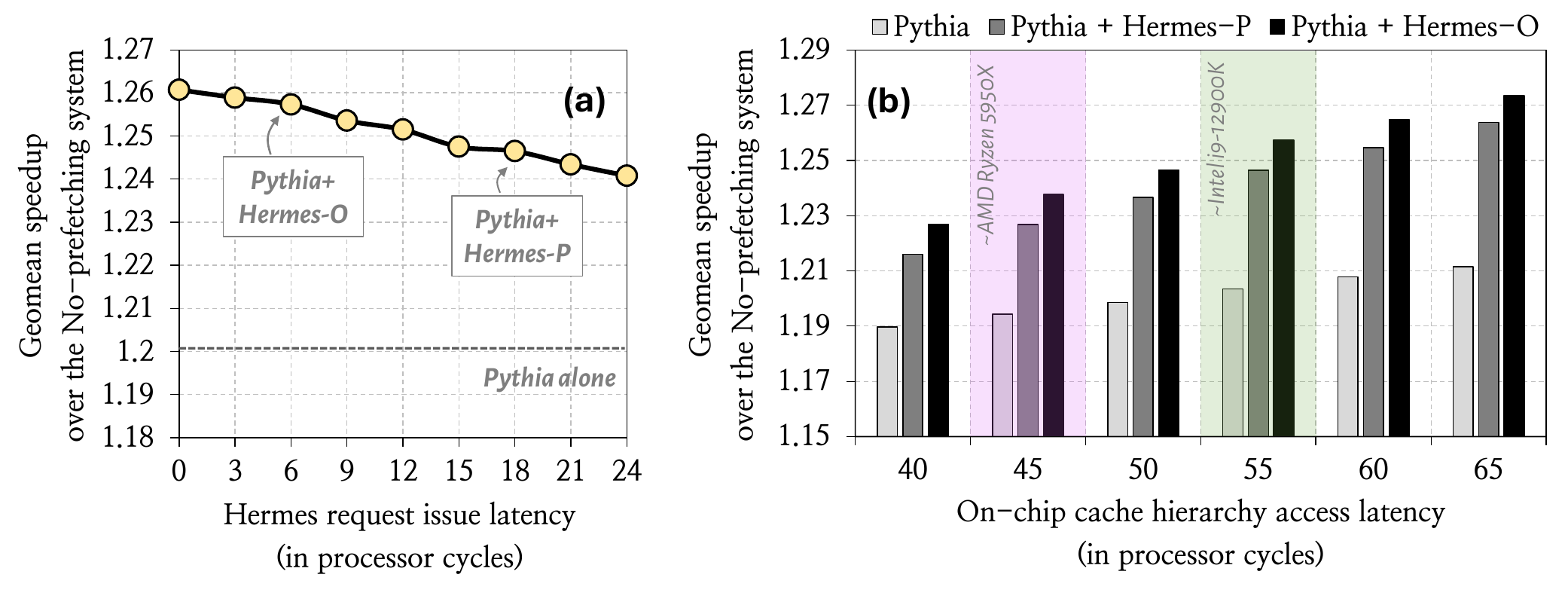}
\caption{Performance sensitivity to (a) \rbe{Hermes request issue latency} and (b) \rbe{on-chip cache hierarchy access latency}. The baseline system configuration is highlighted in green.}
\label{fig:her_sensitivity_master2}
\end{figure*}

To analyze the performance benefit of Hermes over a wide range of processor \rbd{designs} with \rbd{simple or} complex on-chip datapath, we perform a performance sensitivity study by varying the Hermes request issue latency. 
\rbd{\Cref{fig:her_sensitivity_master2}(a) shows the \rbd{performance} of Hermes \rbe{combined with} Pythia \rbe{normalized to the no-prefetching system} in single-core workloads \rbe{as} Hermes \rbe{request} issue latency \rbe{varies from $0$ cycles to $24$ cycles}.}
\rbd{The dashed-line represents the performance of Pythia \rbe{alone}}.
We make two key observations. 
First, the \rbd{speedup} \rbe{of} Hermes \rbd{combined with} Pythia decreases \rbe{as} the \rbd{Hermes request} issue latency \rbe{increases}.
Second, even with a pessimistic \rbd{Hermes request} issue latency \rbe{of $24$ cycles}, Hermes \rbd{combined with} Pythia outperforms  Pythia.
Pythia+Hermes outperforms Pythia by $5.7\%$ and $3.6\%$ with $0$-cycle and $24$-cycle \rbd{Hermes request} issue latency, respectively.

\subsubsection{\rbd{Effect of} \rbe{the} On-chip Cache Hierarchy Access Latency} \label{sec:her_llc_latency}

We evaluate Hermes \rbe{by} varying the on-chip cache hierarchy access latency. 
For each experiment, we keep the L1 and L2 cache access latencies unchanged and vary the LLC access latency from $25$-cycles to $50$-cycles, to mimic the access latencies of a wide range of sliced LLC designs with \rbd{simple or complex} on-chip networks. 
\rbd{\Cref{fig:her_sensitivity_master2}(b) shows the \rbe{performance} of Pythia, and Hermes (O and P) combined with Pythia, normalized to the no-prefetching system in single-core workloads.}
\rbd{We make two key observations.} 
\rbd{First,} Hermes \rbd{combined with} Pythia consistently outperforms Pythia \rbe{for} \emph{every} on-chip cache hierarchy latency. Hermes-O combined with Pythia outperforms Pythia alone by $3.6\%$ and $6.2\%$ in system with $40$-cycle and $65$-cycle on-chip cache hierarchy access latency, respectively.
\rbd{Second, the performance improvement by Hermes combined with Pythia increases \rbe{as the on-chip cache hierarchy access latency increases}. Thus, we \rbe{posit} that Hermes can provide even higher performance benefit in future processors with longer \rbe{on-chip} cache access \rbe{latencies}.}

\subsubsection{Effect of Reorder Buffer Size} \label{sec:her_perf_rob}

\begin{figure*}[!ht]
\centering
\includegraphics[width=5.75in]{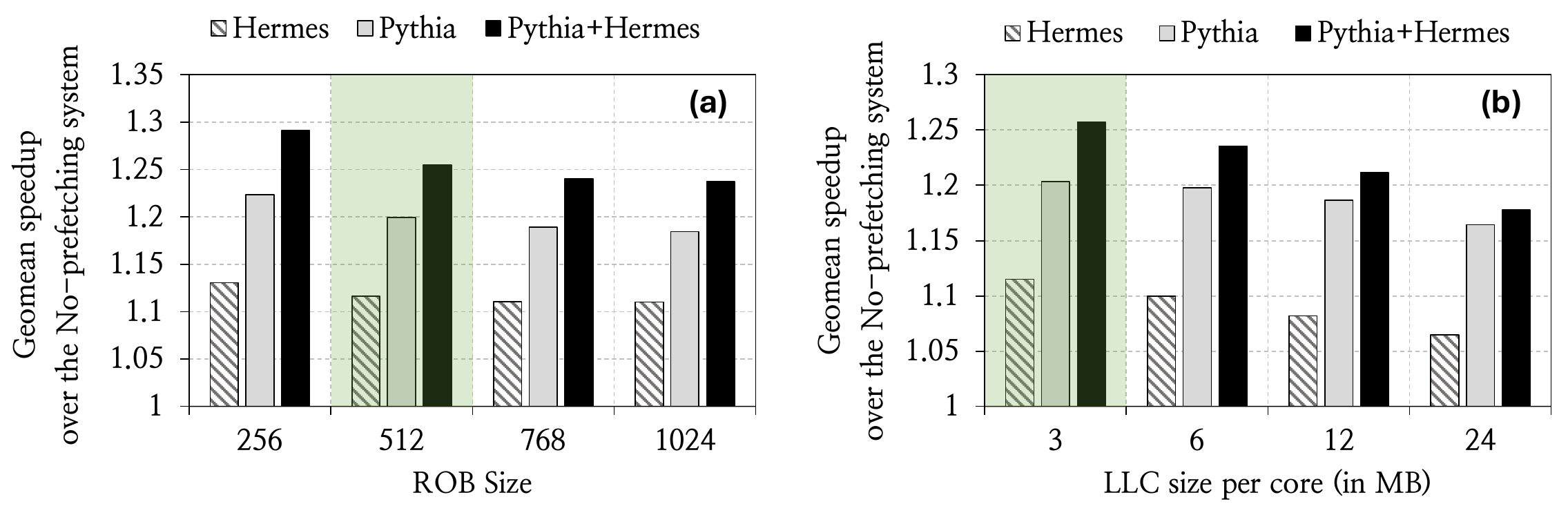}
\caption{Performance sensitivity to (a) reorder buffer size and (b) LLC size. The baseline system configuration is highlighted in green.}
\label{fig:her_sensitivity_master3}
\end{figure*}

\Cref{fig:her_sensitivity_master3}(a) shows the performance of Hermes, Pythia, and Hermes combined with Pythia normalized to the no-prefetching system in single-core workloads as the size of reorder buffer (ROB) varies from $256$ entries to $1024$ entries. The key takeaway is that Hermes combined with Pythia 
outperforms Pythia alone in every ROB size configuration. Pythia+Hermes outperforms Pythia by $6.7\%$ and $5.3\%$ in a system with $256$-entry and $1024$-entry ROB.

\subsubsection{Effect of LLC Size} \label{sec:her_perf_llc}

\Cref{fig:her_sensitivity_master3}(b) shows the performance of Hermes, Pythia, and Hermes combined with Pythia normalized to the no-prefetching system in single-core workloads as the per-core last-level cache (LLC) size varies from $3$~MB to $24$~MB. The key takeaway is that Hermes combined with Pythia outperforms Pythia alone in every LLC size configuration. 
Even in a system with a $12$~MB and $24$~MB LLC per core, Pythia+Hermes provides $2.5\%$ and $1.3\%$ performance benefit over Pythia alone.

\subsubsection{\rbd{Effect of} Perceptron Activation Threshold} \label{sec:her_act_thresh_sensi}

\begin{figure}[!ht]
\centering
\includegraphics[width=5.75in]{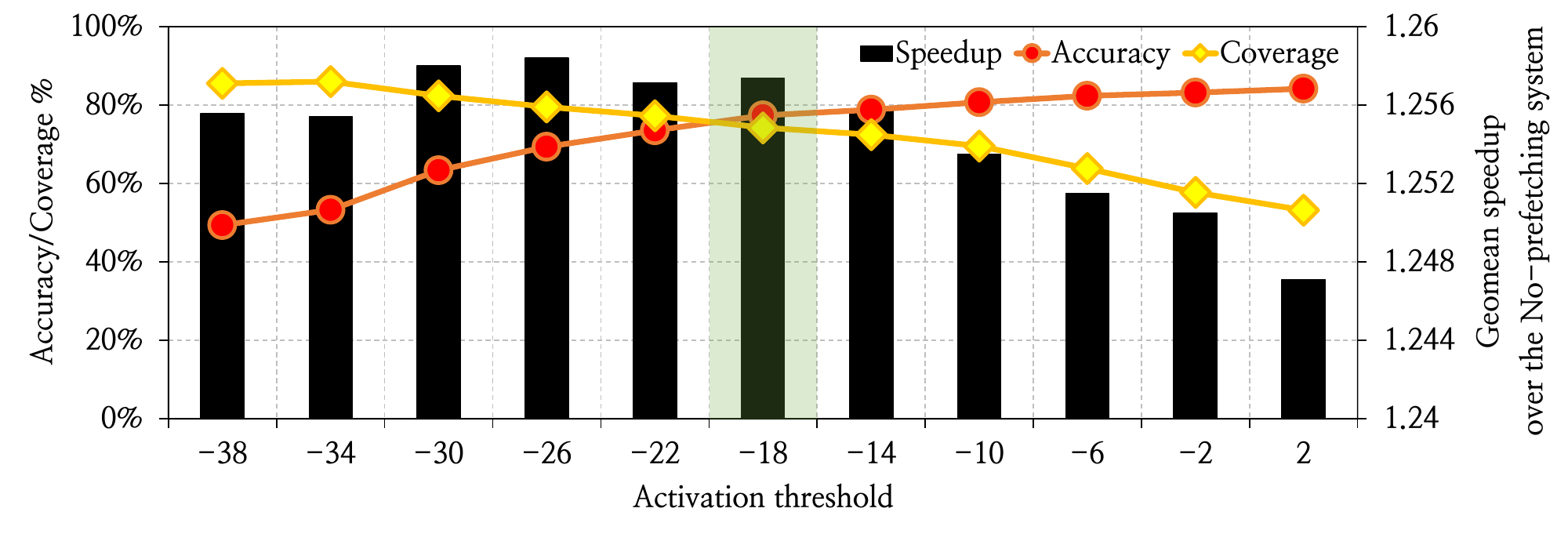}
\caption{\rbe{Effect of the activation threshold on \pred's accuracy and coverage (on the left y-axis) and Hermes's speedup (on the right y-axis) for all single-core workloads.}}
\label{fig:her_act_thresh_sensi}
\end{figure}

\rbe{We evaluate} the impact of the perceptron activation threshold ($\tau_{act}$) on \rbe{Hermes's performance} by varying $\tau_{act}$. \Cref{fig:her_act_thresh_sensi} shows \pred's accuracy and coverage (as line graphs on the \rbd{left} y-axis) and the \rbd{performance} of Hermes combined with Pythia over the no-prefetching system (as a bar graph on the \rbd{right} y-axis) across all single-core workloads \rbe{as $\tau_{act}$ varies from $-38$ to $2$}. 
The key takeaway from \Cref{fig:her_act_thresh_sensi} is that \pred's accuracy (coverage) increases (decreases) \rbe{as $\tau_{act}$ increases}. However, Hermes's performance \rbe{gain} peaks near $\tau_{act} = -26$, which favors higher coverage by trading off accuracy. As \pred's accuracy directly impacts Hermes's main memory request overhead (\rbe{and} hence its performance in bandwidth-constrained configurations), we set $\tau_{act} = -18$ in \pred. \rbd{Doing so} simultaneously optimizes both \pred's accuracy \emph{and} coverage.

\subsection{Effect of Prefetchers on Off-Chip Prediction}

\subsubsection{Effect on Prediction Accuracy and Coverage} \label{sec:her_acc_cov_variation_pref}

\begin{figure}[!ht]
\centering
\includegraphics[width=5.75in]{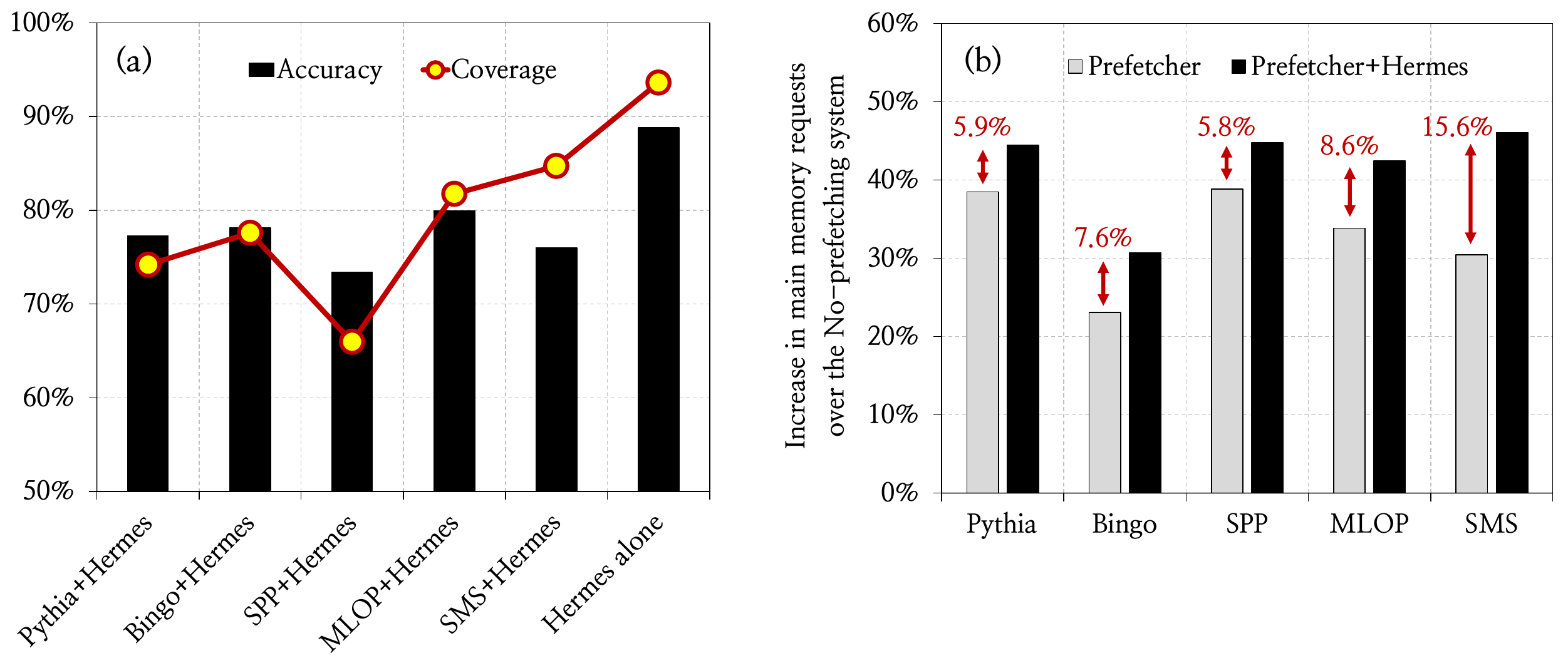}
\caption{(a) Variation of off-chip load prediction accuracy and coverage and (b) the increase in the main memory requests with different data prefetchers.
}
\label{fig:her_effects}
\end{figure}

\Cref{fig:her_effects}(a) shows the off-chip load prediction accuracy and coverage when Hermes is combined with different baseline data prefetchers. We make two key observations. First, \pred's accuracy and coverage varies widely based on the baseline data prefetcher. When combined with Pythia, Bingo, SPP, MLOP, and SMS, \pred provides accuracy of $77.3\%$, $78.1\%$, $73.4\%$, $79.9\%$, and $76.0\%$, while providing coverage of $74.2\%$, $77.6\%$, $65.9\%$, $81.7\%$, and $84.7\%$, respectively. Second, in a system without any baseline data prefetcher, \pred provides significantly higher accuracy ($88.9\%$) and coverage ($93.6\%$) than any configuration with a baseline prefetcher. This shows that, the prefetch requests generated by a sophisticated data prefetcher interfere with the off-chip load prediction. This is why \pred's accuracy and coverage increases in absence of a data prefetcher.

\subsubsection{Effect on Main Memory Requests} \label{sec:her_main_mem_overhead_pref}

\Cref{fig:her_effects}(b) shows the percentage increase in the main memory requests over the no-prefetching system by different types of data prefetchers alone, and in combination with Hermes in all single-core workloads. Combining Hermes with the baseline prefetcher increases the main memory request overhead by $5.9\%$, $7.6\%$, $5.9\%$, $8.6\%$, and $15.6\%$ for the baseline prefetchers Pythia, Bingo, SPP, MLOP, and SMS, respectively.

\subsection{Limit Study by Varying Hermes Request Issue Latency}
\label{sec:her_hermes_limit_study}

As we discuss earlier in~\Cref{sec:her_spec_load_issue}, Hermes's performance gain significantly depends on the Hermes request issue latency. 
We already conduct a bounded performance sensitivity study with Hermes request latency varying from $0$ cycle to $24$ cycles in~\Cref{sec:her_load_issue_latency}.
In this section, we conduct a performance limit study by pushing the range of the Hermes request issue latency even further to understand two key aspects: 
(1) under what range of Hermes request issue latency, Hermes's performance benefits become negligible, and 
(2) how much performance benefit we can expect from Hermes in a real system under a realistic on-chip network latency.

To accurately estimate the on-chip network latency in a real commercial processor, we first estimate the latency to access a banked-SRAM array of equal size to the L2/LLC of our baseline processor (which is modeled after the Intel Alder Lake processor; see~\Cref{sec:her_methodology}) using PCACTI~\cite{pcacti}, and then subtract the SRAM array access latency from the publicly-reported cache access latency (which inherently includes the on-chip network access latency).
Our evaluation yields an optimistic estimate of $2.01$ns (equivalent to $8$ processor cycles for our baseline processor clocked at 4GHz) and $2.65$ns (equivalent to $11$ processor cycles) latency for the L2- and LLC-sized SRAM arrays, respectively. This gives us a pessimistic estimate of $31$ cycles\footnote{Estimated latency spent on on-chip network for accessing L2 = reported L2 access latency (i.e., $10$ cycles) - estimated L2-sized SRAM access latency (i.e., $8$ cycles) = $2$ cycles; Estimated latency spent on on-chip network for accessing LLC = reported LLC access latency (i.e., $40$ cycles) - estimated LLC-sized SRAM access latency (i.e., $11$ cycles) = $29$ cycles;} for the total on-chip network latency for accessing L2 and LLC.

\begin{figure}[!ht]
\centering
\includegraphics[width=5.5in]{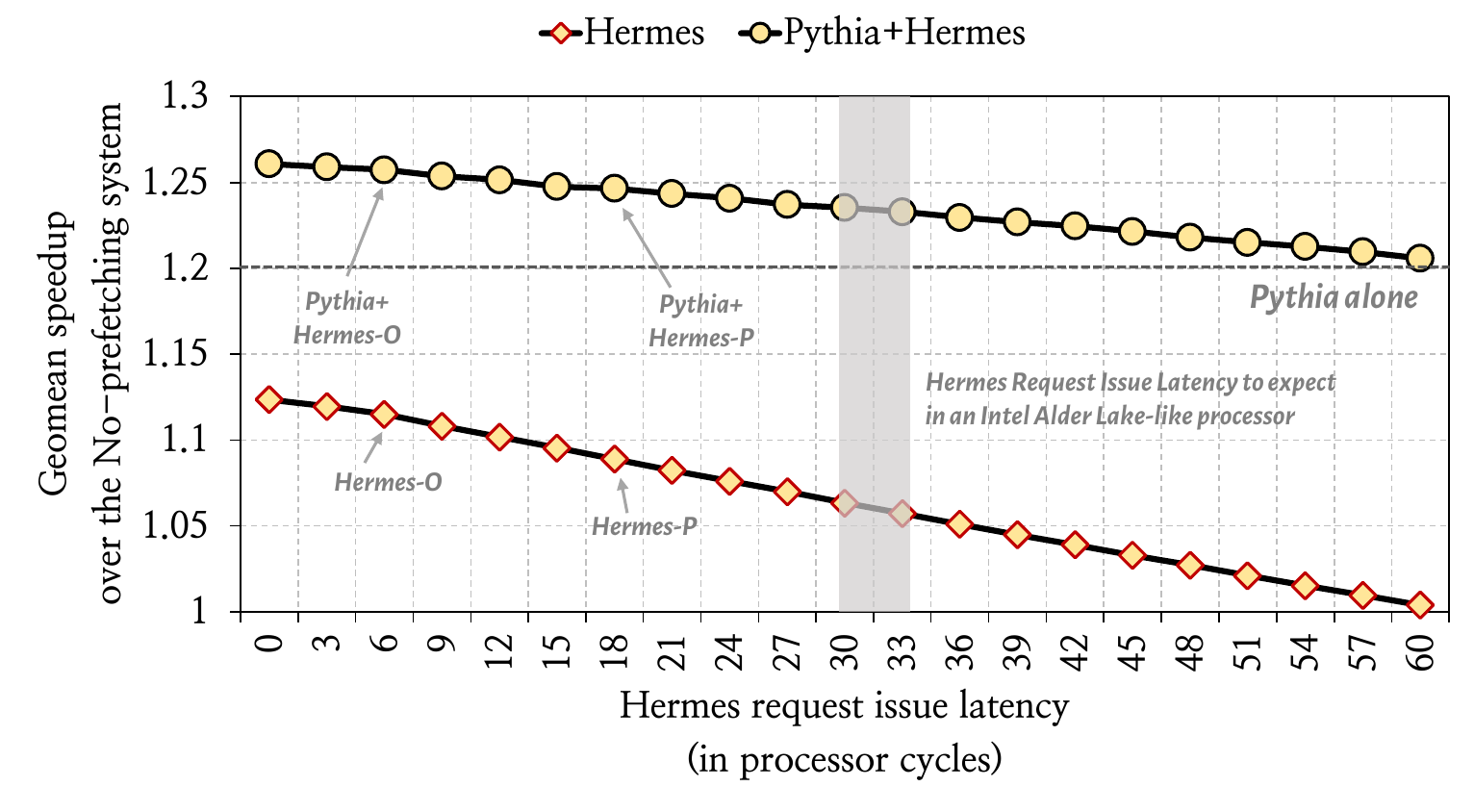}
\caption{Performance of Hermes and Pythia+Hermes while varying Hermes request issue latency.}
\label{fig:her_limit_study}
\end{figure}

\Cref{fig:her_limit_study} shows the geomean performance of Hermes and Pythia+Hermes under varying Hermes request issue latency. We make two key observations. First, Hermes's performance gain, both in standalone and in presence of Pythia, reduces for Hermes request issue latency of $51$ cycles or more. 
This is expected, as with such latency, a speculative Hermes request would arrive at the memory controller nearly at the same time as its corresponding regular load request,\footnote{The minor performance gains we observe even with Hermes request issue latencies higher than $51$ cycles is primarily stemming from hiding any latency caused by queuing delays during cache hierarchy traversal.} thus loosing the latency hiding opportunity. With $51$-cycle Hermes request issue latency, Hermes improves performance by only $2.1\%$ on average over a no-prefetching system, and Hermes combined with Pythia improves performance by only $1.5\%$ over Pythia-alone.
Second, even with our realistic $31$-cycle on-chip network latency, Hermes-alone provides a significant $6.3\%$ performance gain over no-prefetching system, and Hermes with Pythia provides $3.3\%$ performance over Pythia-alone.
Thus we conclude that Hermes would provide tangible performance benefit even on a real-system with complex on-chip network topology and latency.

\subsection{Performance Evaluation using DPC4 Traces}
\label{subsec:hermes_dpc4}

\begin{figure}[!ht]
\centering
\includegraphics[width=5.75in]{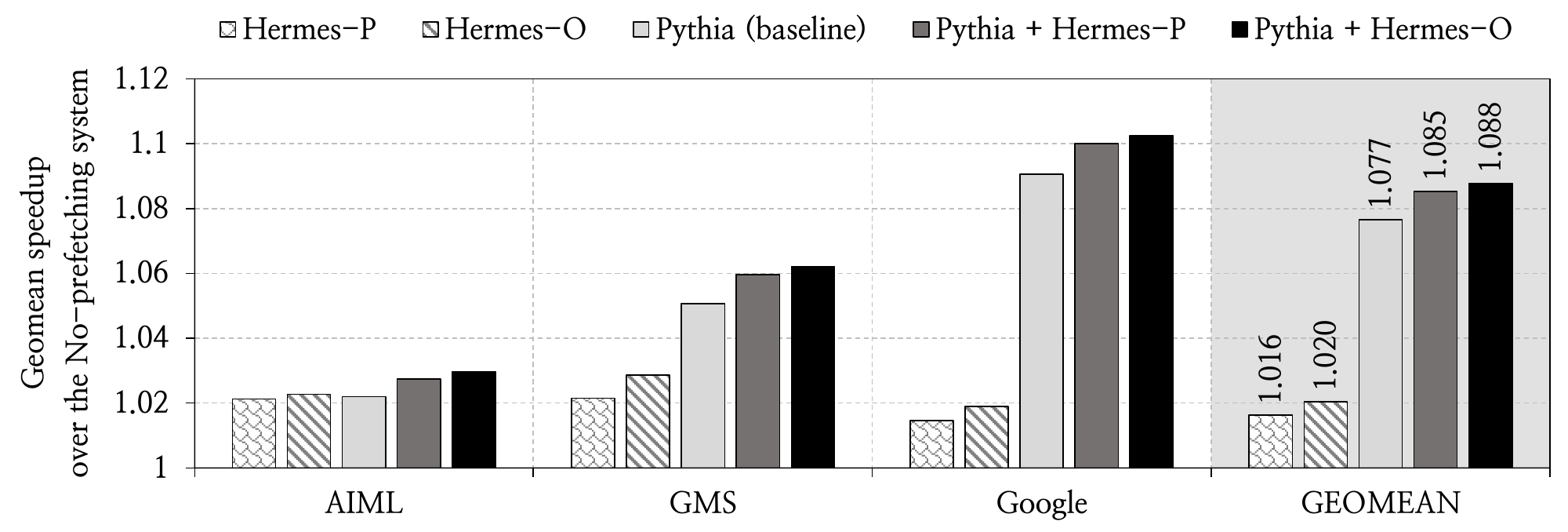}
\caption{Speedup in 483 single-core DPC4 workload traces.}
\label{fig:her_dpc4_overall}
\end{figure}

\Cref{fig:her_dpc4_overall} shows performance of Hermes (\rbfor{P and O}), Pythia, and Hermes combined with Pythia normalized to the no-prefetching system in $483$ single-core DPC4 workloads.
The key observation is that, without any additional tuning \rbfor{for} these traces, Hermes-O alone and combined with Pythia improve performance over a no-prefetching system by $2.0\%$ and $8.8\%$ on average, respectively, whereas Pythia alone provides performance gain of $7.7\%$.\footnote{Note that, Pythia's performance benefit shown here is substantially lower than that shown in~\Cref{subsec:pythia_dpc4}. This is because while~\Cref{subsec:pythia_dpc4} evaluates Pythia at the L2 cache (see~\Cref{sec:pythia_methodology}), here we evaluate Pythia at the LLC (see~\Cref{sec:her_methodology}). 
Relocating Pythia from the L2 cache to the LLC significantly reduces the geometric mean performance improvement for AIML and Google workloads.
Specifically, the performance improvement decreases from $23.6\%$ to $2.2\%$ for AIML workloads and from $14\%$ to $9.1\%$ for Google workloads.
In contrast, GMS workloads exhibit only a marginal change, with performance \rbfor{improvement} decreasing from $5.9\%$ to $5.1\%$.
}
In every workload category, Hermes provides performance benefit when applied alone and when combined with Pythia.

\begin{figure}[!ht]
\centering
\includegraphics[width=5.75in]{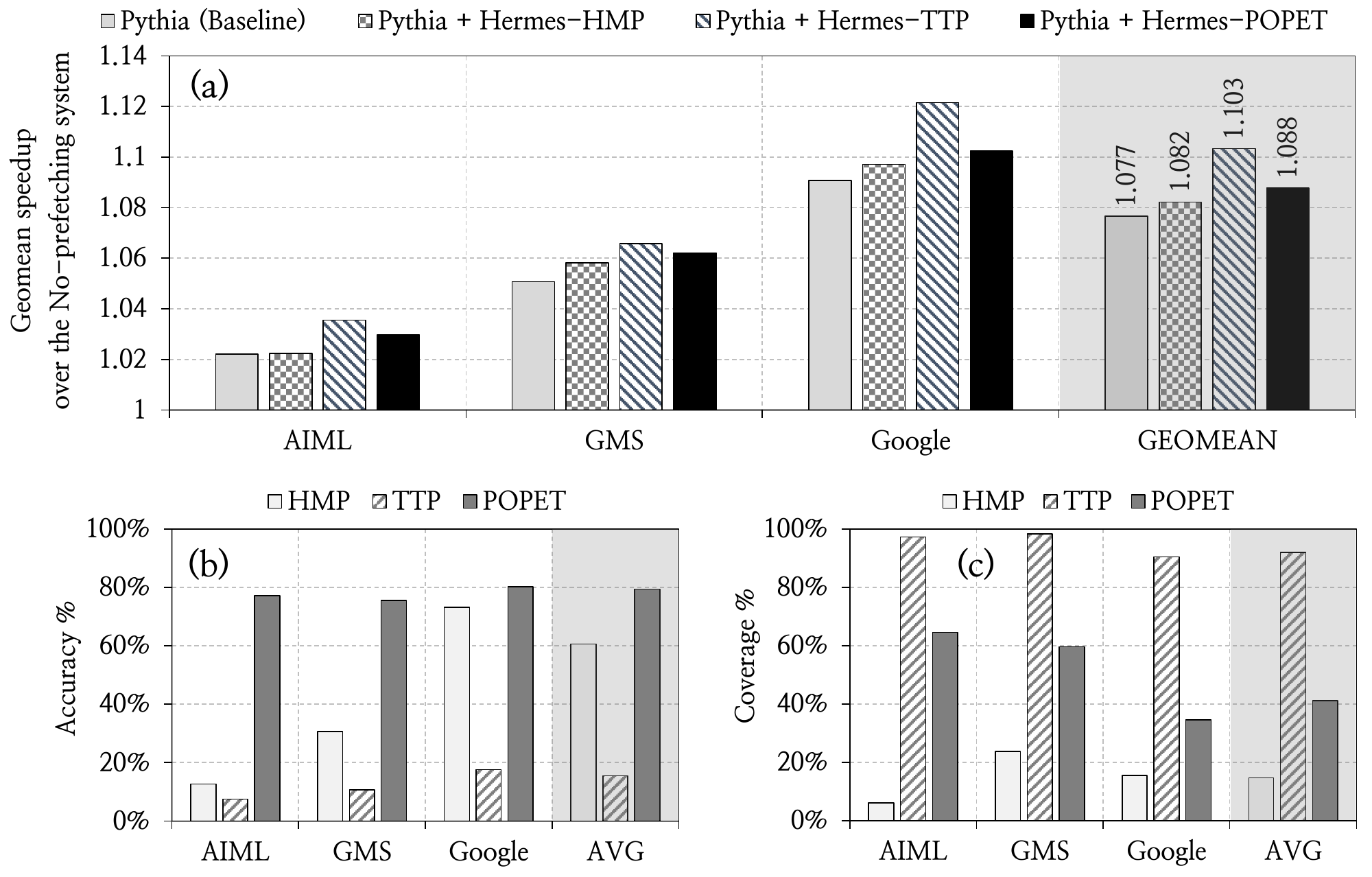}
\caption{(a) Speedup of Hermes with three different off-chip predictors, HMP, TTP, and POPET in single-core DPC4 workloads. Comparison of (b) accuracy and (c) coverage of POPET against those of HMP and TTP in single-core DPC4 workloads.
}
\label{fig:her_dpc4_ocp}
\end{figure}

\paraheading{Effect of the \rbfor{Off-Chip Load Prediction Mechanism}.}
\Cref{fig:her_dpc4_ocp}(a) shows the geomean performance of Pythia alone and Pythia combined with Hermes with three different off-chip prediction mechanisms: POPET, HMP, and TTP, across all $483$ DPC4 workloads.
We make two key observations.
First, Hermes with POPET consistently outperforms both Pythia alone and Pythia with Hermes-HMP in all workload categories.
\rbfor{Second, when combined with Pythia, Hermes-POPET underperforms Hermes-TTP.
To understand the reason behind POPET's under-performance,~\Cref{fig:her_dpc4_ocp}(b) and~\Cref{fig:her_dpc4_ocp}(c) report the off-chip prediction accuracy and coverage of HMP, TTP, and POPET, respectively.
Although POPET achieves a substantially higher prediction accuracy ($79.5\%$) than TTP ($15.5\%$), its coverage is significantly lower than that of TTP}.
\rbfor{POPET only covers $41.2\%$ of the true off-chip load requests, whereas TTP achieves a coverage of $92.1\%$}.
This limited coverage constrains POPET’s overall effectiveness and ultimately results in lower performance compared to TTP when combined with Pythia.

\begin{figure}[!ht]
\centering
\includegraphics[width=5.75in]{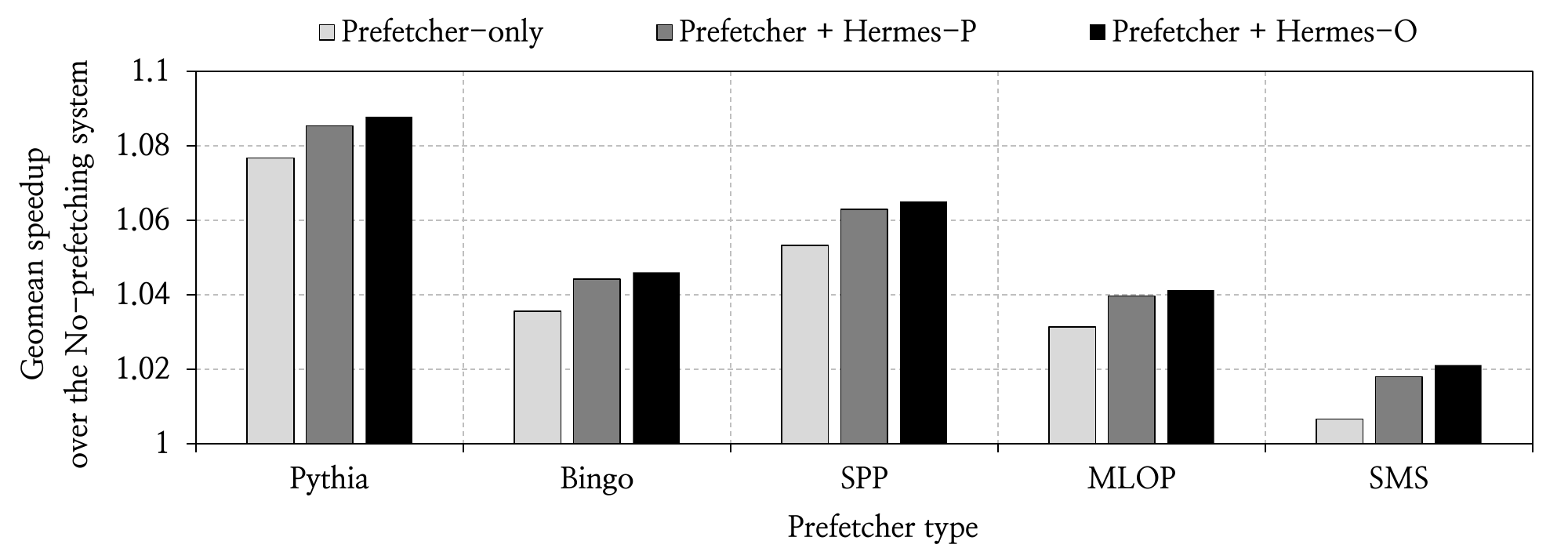}
\caption{Performance sensitivity to baseline prefetcher in single-core DPC4 workloads.}
\label{fig:her_dpc4_pref}
\end{figure}

\paraheading{Effect of the \rbfor{Baseline Prefetcher}.}
\Cref{fig:her_dpc4_pref} shows the geomean performance of the baseline prefetcher, and Hermes-P/O combined with the baseline prefetcher, normalized to the no-prefetching system across all $483$ single-core DPC4 workload traces.
The key takeaway is that Hermes combined with any baseline prefetcher consistently outperforms the baseline prefetcher by itself for all four evaluated prefetching techniques. Hermes+prefetcher outperforms the prefetcher alone by $1.1\%$, $1.0\%$, $1.2\%$, $1.0\%$, $1.5\%$, for Pythia, Bingo, SPP, MLOP, and SMS as the baseline prefetcher, respectively.

Overall, the performance results with DPC4 traces provide empirical evidence that Hermes generalizes beyond its design-time workloads. Despite never observing these traces during development and despite operating in a plug-and-play configuration without additional tuning, Hermes provides considerable, and often the state-of-the-art, performance gains in emerging workloads.

\subsection{Power Overhead Analysis} \label{sec:her_power_overhead}

\begin{figure}[!ht]
\centering
\includegraphics[width=5.75in]{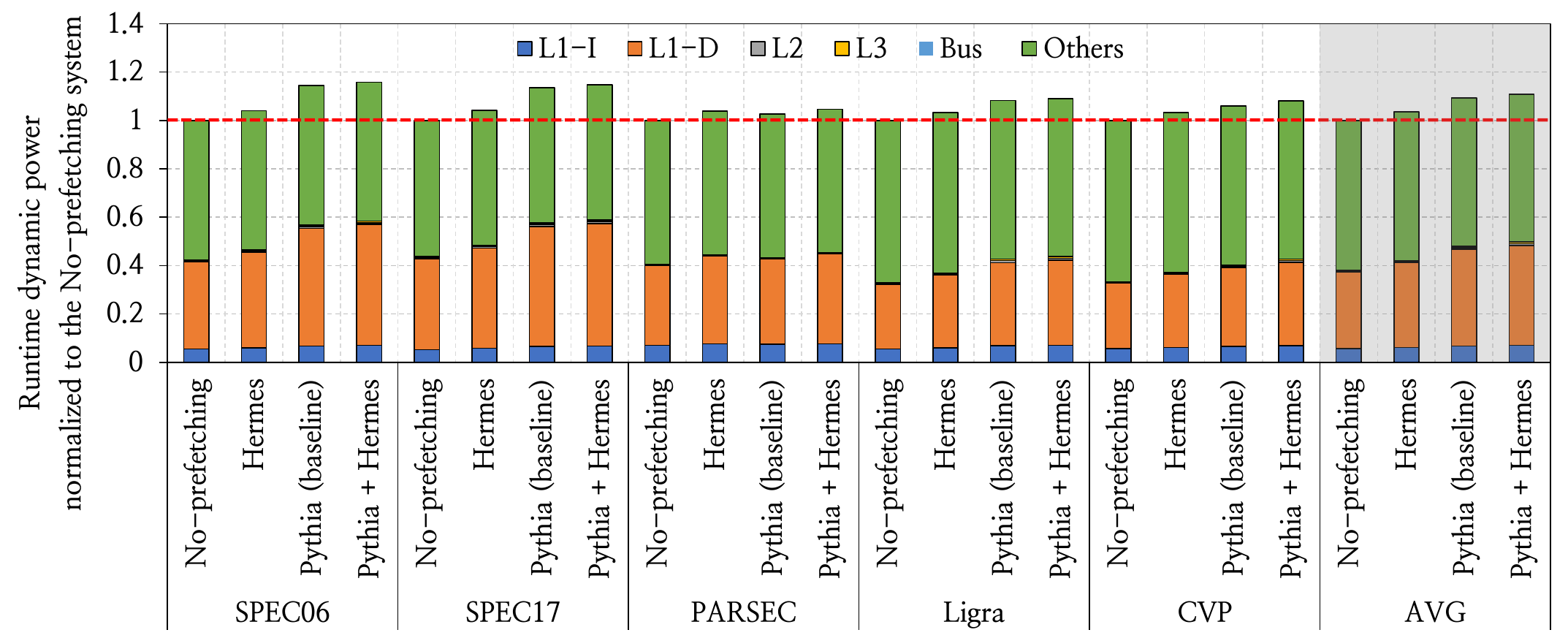}
\caption{\rbe{Processor power consumption of Hermes, Pythia, and Hermes combined with Pythia.}}
\label{fig:her_power_overhead}
\end{figure}

To accurately estimate Hermes's dynamic power consumption, we model our single-core configuration in McPAT~\mbox{\cite{mcpat}} and \rbd{compute \rbe{processor} power consumption} using statistics from performance simulations. \Cref{fig:her_power_overhead} shows the runtime dynamic power consumed by Hermes, Pythia, and \rbd{Hermes combined with Pythia}, normalized to the no-prefetching system \rbd{for all single-core workloads}. We make two key observations. First,  Hermes increases processor power consumption by only $3.6\%$ on average over the no-prefetching system, whereas Pythia increases power \rbe{consumption} by $8.7\%$. 
Second, Hermes \rbd{combined with} Pythia incurs only $1.5\%$ additional power overhead on top of Pythia. \rbe{We conclude} that Hermes incurs \rbd{only a} modest power overhead \rbd{and is more efficient than Pythia alone}.

\sectionRB{Hermes: Summary}{Summary}{sec:her_summary}

\noindent \rbe{We introduce Hermes, a technique that accelerates long-latency off-chip load requests by eliminating the on-chip cache hierarchy access latency from their critical path. To enable Hermes, we propose a perceptron-learning based off-chip load predictor (\pred) that accurately predicts which load requests might go off-chip.}
Our extensive evaluations using a wide range of workloads \rbd{and system configurations} show that Hermes provides significant performance benefits over a baseline system with \rbe{a} state-of-the-art prefetcher. 

\subsection{Influence on the Research Community}

Hermes has been presented at the 55th IEEE/ACM International Symposium on Microarchitecture (MICRO) on October, 2022~\cite{hermes} and was recognized with the \textit{\textbf{Best Paper Award}} at MICRO 2022~\cite{hermes_award}.
Hermes has been officially artifact evaluated with all three badges (i.e., available, functional, and reproducible).
We have made Hermes freely-downloadable from our GitHub repository~\cite{hermes_github} with all evaluated workload traces, scripts, and implementation code required to reproduce and extend it.

Since its release, Hermes has already influenced multiple subsequent works as a state-of-the-art baseline~\cite{jamet2024tlp,clip,krause2022hbpb,fang2025dhcm,wu2025concurrency,sato2024learning}. 
Most notably, one follow-up work, CLIP~\cite{clip}, extends Hermes' key observation of predicting only off-chip loads as \emph{performance-critical} to predicting any load that might block the retirement from reorder-buffer as critical loads. CLIP exploits this prediction to make more informed prefetch decision. 
Another notable follow-up work, PrefetchX~\cite{prefetchx}, \rbfor{has} revealed the existence of an previously-undocumented speculation mechanism already employed in Intel 3rd generation Xeon processors~\cite{intel_xeon_3rdgen} that is similar to Hermes in principle.

We believe and hope that Hermes' key observation and off-chip load prediction mechanism would continue inspiring future works to explore a multitude of other memory system optimizations.

\chapterRB{Prefetcher-OCP Coordination via Reinforcement Learning}{Synergizing Prefetching and Off-Chip Prediction via Online Reinforcement Learning}
\label{chap:athena}

\noindent In last two chapters, we introduce data-driven designs for prefetching and off-chip prediction - two speculative mechanisms that share the same goal of hiding memory access latency. 
While each improves performance individually, their simultaneous use often requires careful orchestration to avoid negating one another’s benefits.
In this chapter, we demonstrate how data-driven design enables autonomous coordination of prefetching and off-chip prediction, unlocking performance gains beyond what either mechanism can achieve alone.

\sectionRB{Athena: Motivation and Goal}{Motivation and Goal}{sec:ath_motivation}
\noindent Data prefetching and off-chip prediction are two key techniques used for hiding long memory access latency in high-performance processors.

\emph{Data prefetching} is a well-studied speculation technique that predicts addresses of memory requests and fetches their corresponding data into on-chip caches before the processor demands them. 
When accurate, prefetching improves performance by hiding the memory access latency. 
However, incorrect speculation can lead to significant memory bandwidth overhead and cache pollution~\cite{fdp, ebrahimi2009techniques}.
Prior works have shown that prefetchers often lose their performance benefits (and even severely degrade performance) in processors with limited memory bandwidth or cache capacity~\cite{fdp, clip, ebrahimi_paware, ebrahimi2009coordinated, ebrahimi2009techniques, pythia, memory_bank, lee2008prefetch, micro_mama}.

\emph{Off-chip prediction} is a more recently proposed speculation technique that predicts which memory requests would go off-chip and fetches their data directly from main memory~\cite{hermes, lp, krause2022hbpb}. 
Unlike a prefetcher that predicts \emph{full cacheline addresses} of future memory requests, an off-chip predictor (OCP) makes a \emph{binary} prediction on a memory request with \emph{known} cacheline address: speculating whether or not it will access the off-chip main memory.
This key difference allows OCP to often produce more accurate predictions than a prefetcher (see~\Cref{subsubsec:ath_complementary}). 
However, OCP can only hide on-chip cache access latency from the critical path of an off-chip memory request, offering lower timeliness than a prefetcher.

\subsection{Key Observations}

We make three key observations, highlighting room for performance improvement in processors that employ both OCP and prefetchers: 
(1) prefetcher and OCP often provide complementary performance benefits, especially in a bandwidth-constrained processor configuration, 
yet (2) naively combining these two techniques often fails to realize their full performance potential, 
and (3) existing microarchitectural policies are either not capable of coordinating OCP with multiple prefetchers, or leave substantial room for performance gain.

\subsubsection{\textbf{Off-chip Prediction and Prefetching Provide Complementary Performance Benefits}} \label{subsubsec:ath_complementary}
\begin{figure}[!ht]
    \centering
    \includegraphics[width=\columnwidth]{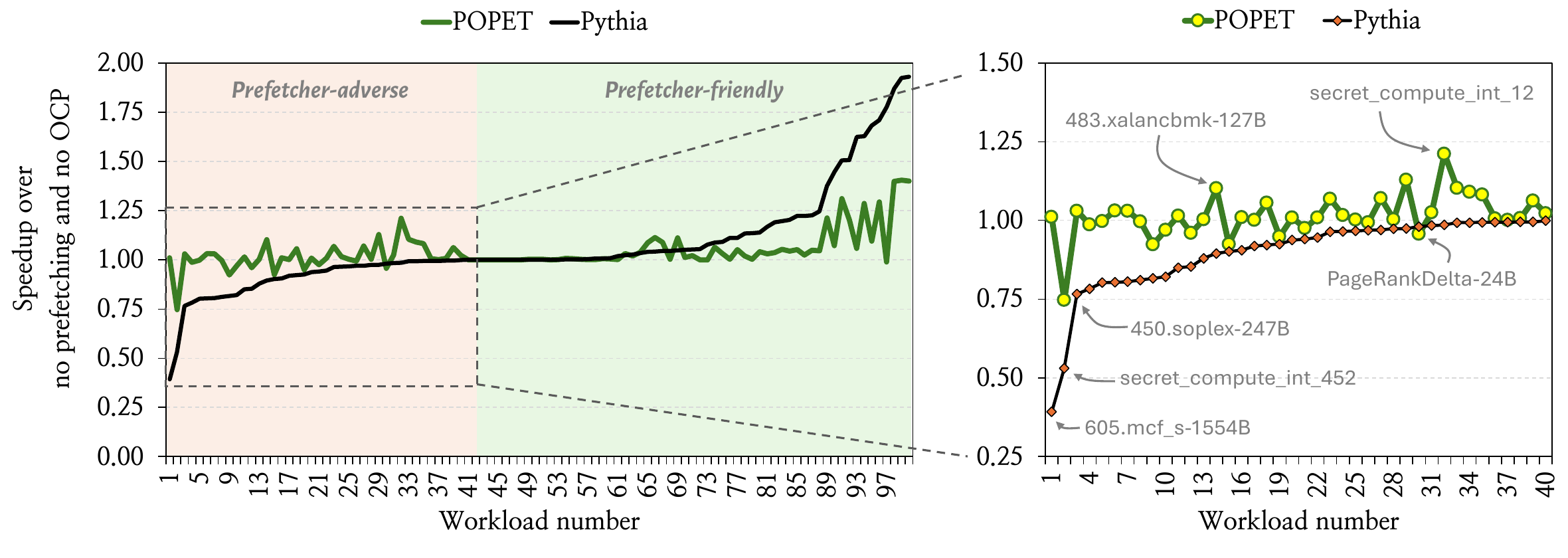}
    \caption{Performance line graph of a state-of-the-art off-chip predictor (OCP), POPET~\cite{hermes}, and a state-of-the-art prefetcher, Pythia~\cite{pythia}, across 100 workloads.}
\label{fig:ath_motiv_perf_line}
\end{figure}

\Cref{fig:ath_motiv_perf_line} shows the performance line graph of a state-of-the-art OCP, POPET~\cite{hermes}, against a state-of-the-art data prefetcher, Pythia~\cite{pythia}, deployed at the L2 cache (L2C) in a memory bandwidth-constrained single-core processor\footnote{We model the memory bandwidth-constrained processor with $3.2$ GB/s of main memory bandwidth (see~{\Cref{sec:ath_methodology}}). 
This configuration closely matches the per-core main memory bandwidth of many commercial datacenter-class processors, e.g., AMD EPYC 9754S~\cite{epyc_9754}, AmpereOne A192~{\cite{ampere_one}}, Amazon Graviton 3~{\cite{graviton3}}, and the Arm Neoverse V2 platform~\cite{bruce2023arm, neoverse2}. Nonetheless, our technique's benefits hold true for a wide range of memory bandwidth configurations, as shown in \Cref{sec:ath_eval_sen_bw}.} across $100$ workloads. 
The graph is sorted in increasing order of Pythia's speedup over the baseline system without a prefetcher or an OCP.

We make three key observations.
First, even though Pythia improves performance for the majority of workloads (highlighted in green), it also \emph{degrades} performance for a significant number of workloads ($40$ out of $100$; highlighted in red) even with its built-in bandwidth-aware throttling mechanism.
For ease of discussion, we call the workloads with performance improvement (or degradation) \emph{prefetcher-friendly} (\emph{prefetcher-adverse}). 
Second, in many prefetcher-adverse workloads, POPET \emph{improves} performance over the baseline.
Pythia \emph{degrades} performance by $11.6\%$ on average across all prefetcher-adverse workloads, whereas POPET \emph{improves} performance by $1.4\%$.
This contrast arises because, in prefetcher-adverse workloads, it is often easier to predict whether a memory request would go off-chip than to predict the full cacheline address of a future memory request.
For instance, in \texttt{483.xalancbmk-127B}, a workload known for its irregular memory access pattern, POPET predicts off-chip requests with $84.1\%$ accuracy, while Pythia generates prefetch requests with only $28.7\%$ accuracy.
As a result, POPET \emph{improves} performance by $10.3\%$, whereas Pythia \emph{degrades} performance by $10.5\%$.
Third, in prefetcher-friendly workloads, however, Pythia provides significantly higher performance benefits ($16.0\%$ on average) than POPET ($5.9\%$ on average). 
This is because, in these workloads, Pythia brings data to the cache well ahead of demand, hiding more memory access latency than only hiding the on-chip cache access latency by POPET.\footnote{Even though we demonstrate this observation using POPET and Pythia as the OCP and prefetcher, respectively, we observe this dichotomy across various prefetcher and OCP implementations. 
In~\Cref{sec:ath_evaluation}, we extend this observation to six prefetcher types~\cite{ipcp, navarro2022berti, pythia, spp, ppf, sms, mlop} and three OCP types~\cite{lp, hermes, yoaz1999speculation}.}

We conclude that OCP and prefetching often provide complementary performance benefits due to their fundamentally different forms of speculation.

\subsubsection{\textbf{Naively Combining OCP with Prefetching Often Fails to Realize Their Full Performance Potential}} \label{subsubsec:ath_naive_combo}
\begin{figure}[!ht]
    \centering
    \includegraphics[width=\columnwidth]{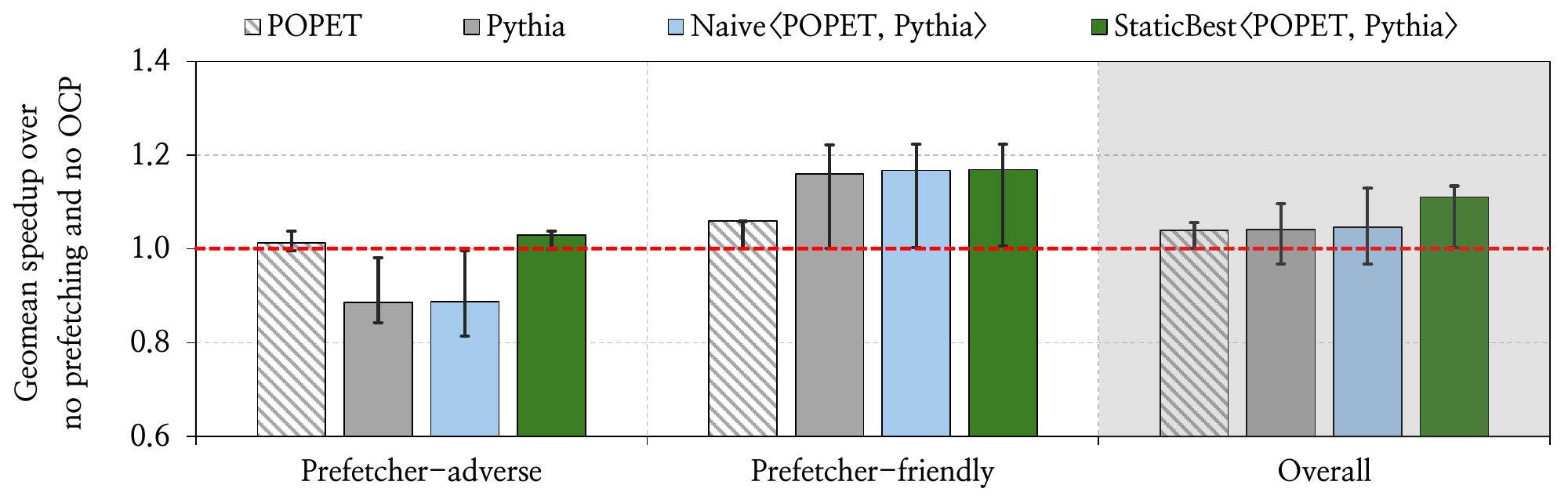}
    \caption{Geomean speedup of POPET, Pythia, Naive, and StaticBest combinations across all workloads.}
\label{fig:ath_motiv2}
\end{figure}

Although OCP and prefetching offer different tradeoffs for different workload categories, naively combining the two mechanisms often fails to realize their full performance potential together.
\Cref{fig:ath_motiv2} compares the performance of POPET and Pythia individually against two combinations of them: 
(1) \emph{Naive<POPET, Pythia>}, that simultaneously enables both POPET and Pythia without any coordination, 
and (2) \emph{StaticBest<POPET, Pythia>}, that retrospectively (i.e., using end-to-end workload execution results offline) selects the best-performing option for each workload among four possibilities: POPET-only, Pythia-only, both enabled, and both disabled.\footnote{While StaticBest estimates the performance headroom of an intelligent coordination policy, it is possible to further improve upon StaticBest by \emph{dynamically} identifying the best combination for \emph{each workload phase}, rather than the entire end-to-end workload.
In~\cref{sec:ath_eval_case_study} we show that, by adapting to each workload phase, our proposed technique can outperform StaticBest.} 
The error bar indicates the range between the first and third quartiles.

We make two key observations from~\Cref{fig:ath_motiv2}. 
First, even though Naive provides $4.7\%$ performance improvement over the baseline across \emph{all} workloads, Naive \emph{degrades} performance by $11.2\%$ in prefetcher-adverse workloads, effectively masking the performance improvement that POPET alone could have delivered otherwise (i.e., $1.2\%$ on average).
This shows that, even though off-chip prediction provides complementary performance benefits to prefetching, especially in prefetcher-adverse workloads, naively combining both techniques does not realize their full performance potential together.
Second, StaticBest combination provides \emph{consistent} performance benefits in \emph{both} prefetcher-adverse and prefetcher-friendly workloads and significantly outperforms the Naive combination (by $6.5\%$ on average) across all workloads.
These observations demonstrate the need to design an intelligent coordination mechanism between OCPs and prefetchers.

\subsubsection{\textbf{Existing Coordination Policies are Either Inflexible or Leave a Large Performance Potential Behind}} \label{subsubsec:ath_prior}

\begin{figure}[!ht]
    \centering
    \includegraphics[width=\columnwidth]{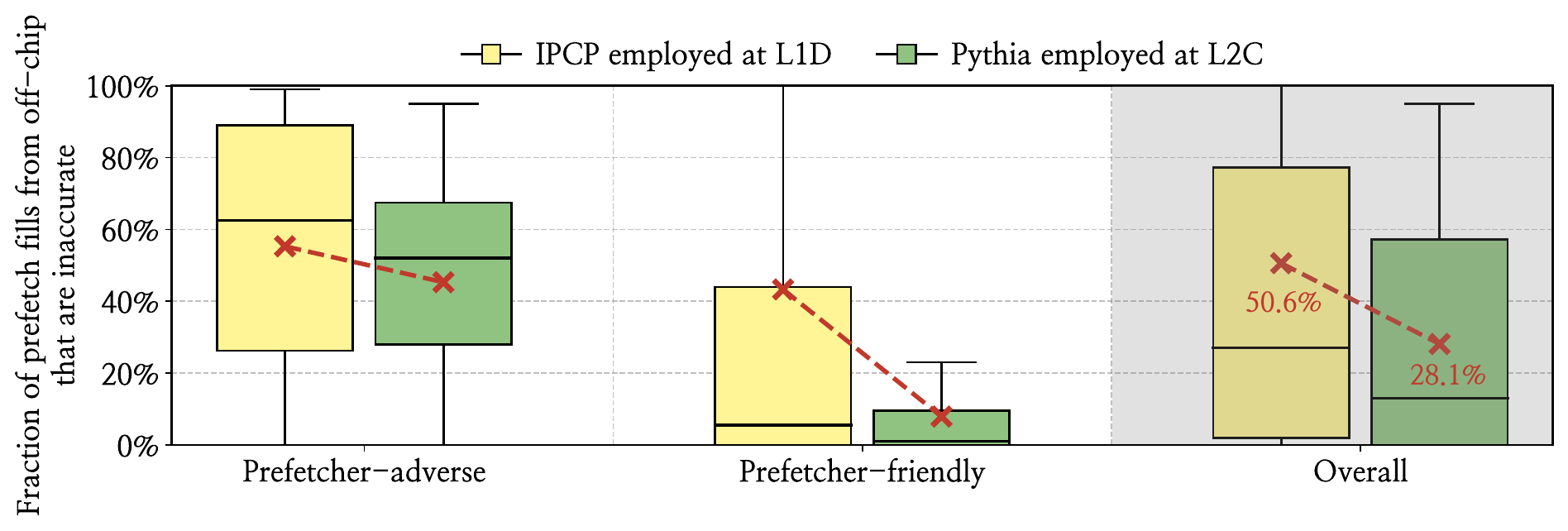}
    \caption{Fraction of prefetch fills from off-chip main memory that are inaccurate.}
\label{fig:ath_tlp_motiv}
\end{figure}

While researchers have proposed numerous techniques to control multiple prefetchers (e.g., ~\cite{ibm_power7, fdp, ebrahimi2009coordinated, ebrahimi2009techniques, eris2022puppeteer, mab, jalili2022managing, yang2024rl, yang2025reinforcement, US20080243268A1, US8583894B2, US9292447B2, US10073785B2, US8156287B2, US9645935B2, US8924651B2, US8892822B2, US9904624B1, US10331567B1,sridharan2017bandpass,panda2016spac,panda2016expert,panda2015caffeine,panda2014criticality,panda2013tcpt}), TLP~\cite{jamet2024tlp} is the \emph{only} prior technique that aims to control a prefetcher in the presence of an OCP.
TLP uses off-chip prediction as a hint to filter out prefetch requests to the L1 data cache (L1D), based on the empirical observation that prefetches filled from off-chip main memory into L1D are often inaccurate (i.e., the cacheline is not subsequently demanded during its cache residency)~\cite{jamet2024tlp}.
While TLP's observation is often effective for an L1D prefetcher, we observe that it may not hold true for prefetchers employed at higher (i.e., further away from the core) cache levels.
\Cref{fig:ath_tlp_motiv} shows the fraction of prefetch fills from the off-chip main memory that are inaccurate as a box-and-whisker plot.\footnote{Each box is lower- (upper-) bounded by the first (third) quartile. 
The box size represents the inter-quartile range (IQR). The whiskers extend to $1.5\times$ IQR range on each side. 
The cross-marked value represents the mean.}
We show the fraction for two state-of-the-art prefetchers, IPCP~\cite{ipcp} and Pythia, individually employed at two different cache levels. 
IPCP fills prefetch requests to L1D, whereas Pythia fills to L2C.
The key observation is that, while $50.6\%$ prefetch fills to L1D caused by IPCP are inaccurate, only $28.1\%$ of the prefetch fills to L2C caused by Pythia are inaccurate. 
In other words, an off-chip prefetch fill to L2C is \emph{nearly half as likely} to be inaccurate as an off-chip prefetch fill to L1D, while employing state-of-the-art prefetchers.
This fundamental limitation in TLP's key observation significantly limits its ability to coordinate OCPs with prefetchers that are placed beyond L1D (as we demonstrate in~\Cref{subsubsec:ath_L1D} and~\Cref{subsubsec:ath_L1DL2C}).

\begin{figure}[!ht]
    \centering
    \includegraphics[width=\columnwidth]{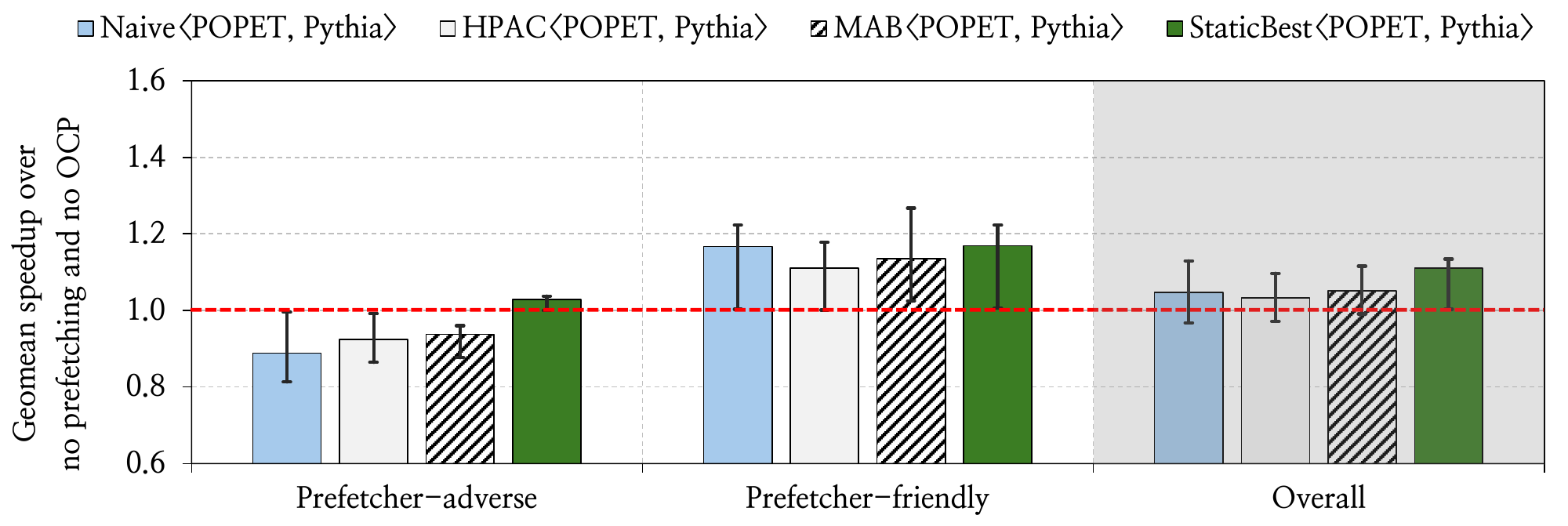}
    \caption{Geomean speedup of Naive, HPAC, MAB, and StaticBest combinations across all workloads.}
\label{fig:ath_motiv3}
\end{figure}

Besides TLP, other prior techniques focus solely on prefetcher control, without considering OCP. 
We \emph{extend} two best-performing prior techniques, heuristic-based HPAC~\cite{ebrahimi2009coordinated} and learning-based MAB~\cite{mab}, to coordinate an OCP and a prefetcher.
\Cref{fig:ath_motiv3} compares the performance of these techniques, coordinating POPET as the OCP and Pythia as the L2C prefetcher against Naive and StaticBest combinations across all workloads.\footnote{\Cref{fig:ath_motiv3} excludes TLP as its key observation does not reliably extend to coordinating an L2C prefetcher (Pythia) with an OCP, as shown earlier.}
We make two key observations.
First, in prefetcher-adverse workloads, both HPAC and MAB considerably mitigate the performance degradation of the Naive combination. However, neither policy matches the performance of the baseline (without prefetching or OCP), let alone harnesses the potential performance gains of StaticBest.
Second, in prefetcher-friendly workloads, these coordination techniques fall short of the Naive combination.
The heuristic-based HPAC falls short due to its reliance on the statically tuned thresholds that are optimized for average-case behavior across workloads. 
These fixed thresholds cannot adapt to per-workload or phase-specific characteristics, causing HPAC to make conservative coordination decisions even when prefetching is beneficial.
While MAB avoids such static thresholds, it still falls short as it makes decisions agnostic to any system-level features (e.g., prefetcher/OCP accuracy, prefetch-induced cache pollution).

We conclude that while there is a rich literature on prefetcher coordination techniques, the only OCP-aware prefetcher control mechanism (i.e., TLP) lacks flexibility (i.e., the ability to coordinate OCP with multiple prefetchers employed at various levels of the cache hierarchy), while other techniques (e.g., HPAC, MAB) leave significant performance potential behind.

\subsection{Our Goal}

\textbf{Our goal} in this work is to design a holistic framework that can autonomously coordinate off-chip prediction with \emph{multiple} prefetching techniques employed at various levels of the cache hierarchy by taking multiple system-level features into account, thereby delivering \emph{consistent} performance benefits, regardless of the underlying prefetcher-OCP combination, workload, and system configuration.

\section[Formulating Prefetcher-OCP Coordination using RL]{Formulating Prefetcher-OCP Coordination using Reinforcement Learning}

To this end, we formulate the dynamic coordination between prefetching and off-chip prediction as a reinforcement learning (RL) problem.
More specifically, we propose \emph{\textbf{Athena}},\footnote{Named after the Greek goddess of wisdom and strategic warfare~\cite{athena_wiki}.}  an RL-based agent that dynamically learns to synergize off-chip prediction with multiple prefetching techniques employed throughout the cache hierarchy of modern state-of-the-art processor by interacting with the processor and memory system.

\subsection{Why is RL a Good Fit for Prefetcher-OCP Coordination?}
RL is well-suited for coordinating data prefetchers and OCP due to the following three key advantages.

\paraheading{Less Reliance on Static Heuristics and Thresholds.}
Heuristic-based prefetcher coordination policies typically rely on statically defined thresholds~\cite{fdp, ebrahimi_fst, ebrahimi2009coordinated, ebrahimi2009techniques, ebrahimi_paware}. 
These thresholds are manually tuned and inherently inflexible, often resulting in suboptimal performance when workloads or system conditions change~\cite{rlmc, morse}.
Formulating prefetcher-OCP coordination as an RL problem allows a hardware architect to focus on \textit{what} performance targets the coordinator should achieve and \textit{which} system-level features might be useful, rather than spending time on manually devising fixed algorithms and/or thresholds that describe \textit{precisely how} the coordinator should achieve that target. 
This not only significantly reduces the human effort needed for a coordinator design, but also yields higher-performing coordination (as shown in~\Cref{sec:ath_evaluation}).

\paraheading{Online Feedback-Driven Learning.}
RL provides two key benefits over supervised learning methods (e.g., SVM~\cite{svm1,svm2}, decision tree~\cite{decision_tree1}) in formulating the prefetcher-OCP coordination problem. 
First, prefetcher-OCP coordination lacks a well-defined ground-truth label (e.g., which mechanism is beneficial to enable) for each system state. 
The optimal coordination decision depends on delayed, system-level performance outcomes that manifest only after executing an action under dynamic resource contention and changing workload behavior. 
As a result, generating labeled training data would require exhaustive offline exploration across workloads, phases, and system configurations, and such labels may not generalize as the underlying hardware, prefetchers, or OCPs change. 
Second, supervised learning models are inherently static once trained and cannot naturally adapt online to changing workload behavior without repeated retraining. 
In contrast, RL directly optimizes long-term performance using online reward feedback and continuously updates its policy, making it better suited to model prefetcher-OCP coordination.

\paraheading{Q-Value-Driven Prefetcher Aggressiveness Control.}
The Q-values learned by the RL agent provide a natural ranking over available actions, reflecting their expected utility. 
For actions involving prefetching (either standalone or combined with OCP), the corresponding Q-values \emph{implicitly} encode the agent's confidence in the prefetcher's effectiveness relative to alternative actions.
An RL agent can leverage these Q-values to also dynamically control prefetcher aggressiveness. 
Higher Q-values correspond to stronger confidence, prompting more aggressive prefetching, whereas lower Q-values imply uncertain benefits, resulting in more conservative prefetching.
Importantly, this Q-value-driven prefetcher aggressiveness control incurs no additional hardware overhead, since the learned Q-values jointly govern both prefetcher/OCP selection and prefetcher aggressiveness.

\sectionRB{Athena: Overview}{Athena: Overview}{sec:ath_overview}

\begin{figure}[!ht]
    \centering
    \includegraphics[width=\columnwidth]{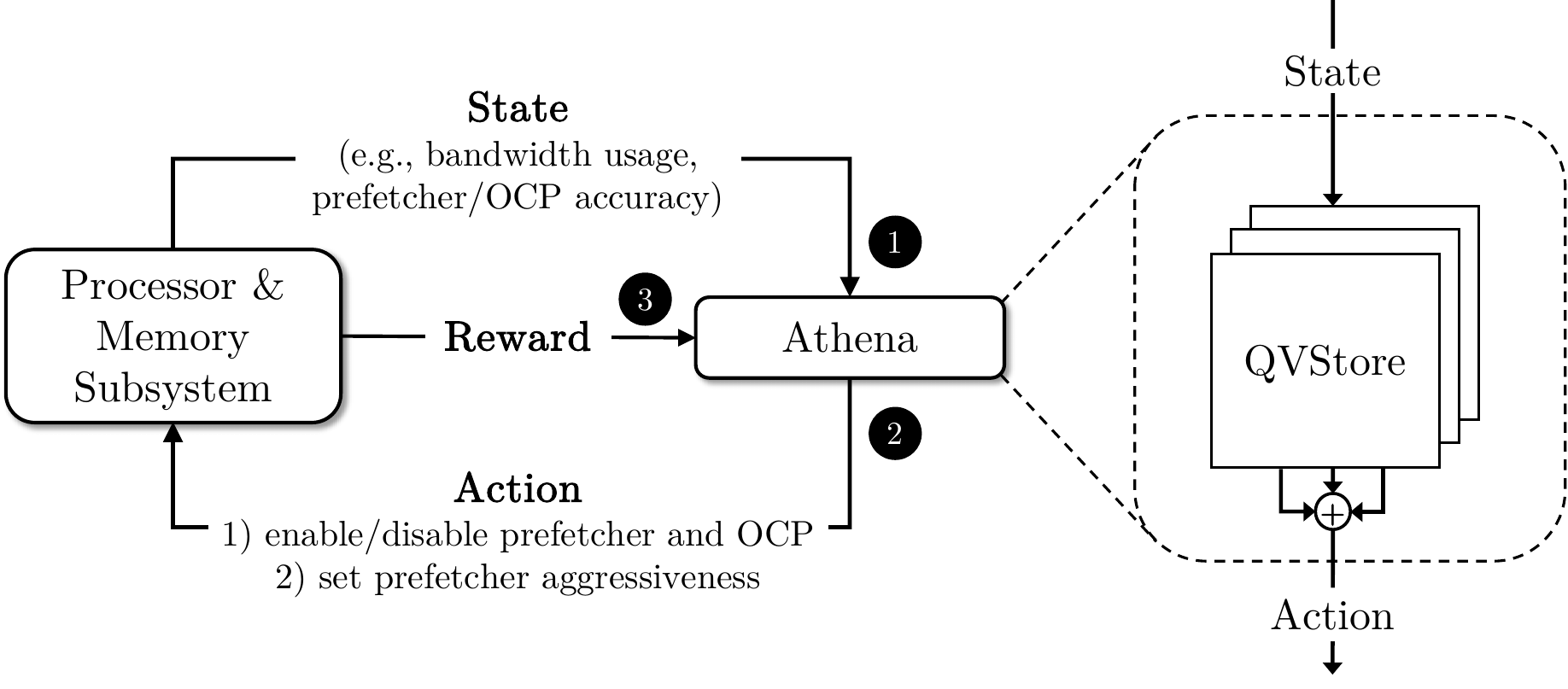}
    \caption{High-level overview of Athena as an RL agent.}
    \label{fig:athena_rl}
\end{figure}

\noindent Athena formulates the coordination of data prefetchers and the off-chip predictor as an RL problem, as illustrated in~\Cref{fig:athena_rl}. 
Here, Athena acts as an RL agent that continuously learns and adapts its prefetcher-OCP coordination policy by interacting with the processor and memory system.
As~\Cref{fig:athena_rl} shows, Athena comprises a key hardware structure, \emph{Q-Value Storage (QVStore)}, whose purpose is to store the Q-values of state-action pairs encountered during Athena's online operation.

Each timestep for Athena corresponds to a fixed-length epoch of workload execution (e.g., $N$ retired instructions). 
During an execution epoch, Athena observes and records various system-level features (e.g., prefetcher/OCP accuracy, memory bandwidth usage).
At the end of every epoch, Athena uses the recorded feature values as \emph{\textbf{state}} information to index into the QVStore (step \circled{1} in~\Cref{fig:athena_rl}) to select a coordination \emph{\textbf{action}}, i.e., whether to enable only prefetcher, only OCP, both mechanisms, or none of them (step \circled{2}).
If Athena decides to enable the prefetcher, it further determines the prefetcher aggressiveness based on the magnitude of the selected action's Q-value (see~\Cref{subsec:ath_action}). 
At the end of every epoch, Athena receives a numerical \emph{\textbf{reward}} that measures the change in multiple system-level metrics (e.g., the number of cycles taken to execute an epoch) to evaluate the impact of its actions on system performance (step \circled{3}). 
Athena uses this reward to autonomously and continuously learn a prefetcher-OCP coordination policy, that can adapt to diverse workloads and system configurations. 

\subsection{State} \label{subsec:ath_state}

We define the state as a vector of system-level features, where each feature encapsulates a distinct behavior of the memory subsystem (e.g., the accuracy of the prefetcher, the main memory bandwidth usage) observed during the current execution epoch.
Collectively, these features represent the runtime conditions relevant to making a prefetcher-OCP coordination decision.
While Athena can, in principle, learn a coordination policy using \emph{any} arbitrary set of features, increasing the state dimension rapidly increases the storage required to maintain the Q-values of all observed state-action pairs. 
To bound the storage overhead of Athena, we fix the state representation offline using a two-step process.
First, guided by domain knowledge, we identify a set of seven candidate system-level features that are expected to influence the effectiveness of prefetcher-OCP coordination.
\Cref{tab:ath_feature_list} summarizes each feature, its computation method, and the rationale behind its inclusion.
Second, we perform offline feature selection using automated design-space exploration (as described in~\Cref{subsec:ath_automated}) to determine the final subset of features that Athena uses to construct the state vector.

\begin{scriptsize}
\begin{table}[!ht]
    \centering
    \small
    \begin{tabular}{L{10em}||C{15em}|L{10em}}
        \thickhline
        \textbf{Feature} & \textbf{Measurement} & \textbf{Rationale} \\ 
        \thickhline
        \Tabval{\textbf{Prefetcher accuracy}} & 
        \Tstrut
        \begin{minipage}{15em}
            \vskip 6pt
            \centering
            $\dfrac{\text{\# demand hits}}{\text{\# prefetches issued}}$
        \end{minipage}
        \Bstrut & \Tabval{Effectiveness of prefetching} \\
        \hline 
        \Tabval{\textbf{OCP accuracy}} & 
        \Tstrut
        \begin{minipage}{15em}
            \vskip 6pt
            \centering
            $\dfrac{\text{\# off-chip demand hits}}{\text{\# off-chip predictions}}$
        \end{minipage}
        \Bstrut & \Tabval{Effectiveness of OCP} \\
        \hline
        \Tabval{\textbf{Bandwidth usage}} & 
        \Tstrut
        \begin{minipage}{15em}
            \vskip 6pt
            \centering
            $\dfrac{\text{current DRAM bandwidth}}{\text{max DRAM bandwidth}}$
        \end{minipage}
        \Bstrut & \Tabval{Memory bus pressure} \\
        \hline
        \Tabval{\textbf{Cache pollution}} & 
        \Tstrut
        \begin{minipage}{15em}
            \vskip 6pt
            \centering
            $\dfrac{\text{\# prefetch-evicted demand misses }}{\text{\# total demand misses}}$
        \end{minipage}
        \Bstrut & \Tabval{Interference caused by prefetches} \\
        \hline 
        \Tabval{\textbf{Prefetch bandwidth}} & 
        \Tstrut
        \begin{minipage}{15em}
            \vskip 6pt
            \centering
            $\dfrac{\text{\# prefetch requests to DRAM}}{\text{\# total DRAM requests}}$
        \end{minipage}
        \Bstrut & \Tabval{Prefetcher's share of memory traffic} \\
        \hline
        \Tabval{\textbf{OCP bandwidth}} & 
        \Tstrut
        \begin{minipage}{15em}
            \vskip 6pt
            \centering
            $\dfrac{\text{\# OCP requests to DRAM}}{\text{\# total DRAM requests}}$
        \end{minipage}
        \Bstrut & \Tabval{OCP's share of memory traffic} \\
        \hline
        \Tabval{\textbf{Demand bandwidth}} & 
        \Tstrut
        \begin{minipage}{15em}
            \vskip 6pt
            \centering
            $\dfrac{\text{\# demand requests to DRAM}}{\text{\# total DRAM requests}}$
        \end{minipage}
        \Bstrut & \Tabval{Demand's share of memory traffic} \\
        \thickhline
    \end{tabular}
    \caption{Candidate features considered for Athena's state representation.}
    \label{tab:ath_feature_list}
\end{table}
\end{scriptsize}

\subsection{Action} \label{subsec:ath_action}

Athena's action space consists of four coordination decisions, i.e., whether to enable (1) only prefetcher, (2) only OCP, (3) both mechanisms, or (4) none of them. 
These actions allow Athena to explicitly coordinate the prefetcher and the OCP at a coarse granularity, i.e., only enabling or disabling any given mechanism as a whole.
However, when Athena selects an action that enables prefetching, it further determines the prefetcher aggressiveness.
This aggressiveness is derived directly from the learned Q-values as a function of the \emph{relative difference} between the Q-value of the selected action and the average Q-value of the remaining actions.
The underlying rationale for this Q-value-driven prefetcher aggressiveness control is that the magnitude of the selected action's Q-value implicitly encodes Athena's confidence in the prefetcher's effectiveness.
A larger separation between the Q-value of the selected action (that enables the prefetcher, either without or with the OCP) and those of the alternative actions indicates stronger historical evidence that enabling prefetching is beneficial in the current state, thus warranting more aggressive prefetching.
On the other hand, a smaller separation reflects uncertainty in prefetcher effectiveness, prompting more conservative prefetching.
\Cref{algo:ath_pref_deg} formalizes this Q-value-based prefetcher aggressiveness control mechanism.\footnote{Here, we represent prefetcher aggressiveness using the prefetch degree, i.e., the number of prefetch requests issued per demand trigger. However, the proposed Q-value-driven aggressiveness control mechanism can also be applied to alternative aggressiveness definitions.}

\begin{algorithm}[!ht]
\small
\begin{algorithmic}[1]
\Procedure{SelectPrefetchDegree}{}
    \State $a^* \gets \arg\max_{a \in \mathcal{A}} Q(a)$
    \State $avgQ \gets$ average Q-value of all actions except $a^*$
    \State $\Delta Q \gets Q(a^*) - avgQ$ \Comment{compute the Q-value confidence}
    \State $r \gets \min(1,\, \Delta Q / \tau)$ \Comment{normalize confidence w.r.t. hyperparam $\tau$}
    \State $d \gets \lfloor r \cdot d_{\max} \rfloor$ \Comment{adjust prefetch degree based on normalized confidence}
    \State \Return $d$
\EndProcedure
\end{algorithmic}
\caption{Q-value-driven prefetcher aggressiveness control}
\label{algo:ath_pref_deg}
\end{algorithm}

First, Athena selects the action $a^*$ with the highest Q-value.
It then estimates the confidence of this decision by comparing $Q(a^*)$ against the average Q-value of all remaining actions.
The resulting confidence ratio $\Delta Q$ captures how strongly the selected action is preferred over the alternatives.
Using this confidence signal, Athena determines the final prefetch degree as a fraction of $d_{\max}$, where $d_{\max}$ denotes the number of prefetch requests that the underlying prefetcher can issue while operating at full aggressiveness.
If this confidence ratio exceeds a hyperparameter $\tau$, Athena enables the prefetcher at full aggressiveness, reflecting high confidence in the benefits of prefetching in the current system state.
Otherwise, Athena scales the prefetch degree proportionally to $\Delta  / \tau$, issuing fewer prefetch requests.

\subsection{Reward} \label{subsec:ath_reward}

The reward defines the optimization objective for Athena.
Prior works often use the change in instructions committed per cycle (IPC) as the \emph{only} system-level metric to train the RL agent~\cite{mab, jalili2022managing, micro_mama, eris2022puppeteer}.
However, a change in IPC may originate from two different sources: 
(1) the coordination actions taken by the agent, and 
(2) the inherent variations in workload behavior, that are independent of the agent's actions. 
As such, using IPC as the sole reward can be unreliable and may mislead the learned policy.

To address this limitation, Athena introduces a composite reward framework that explicitly separates the effects of Athena's action on the system from the inherent variations in workload behavior.
More specifically, we define the reward for Athena at a given timestep $t$ using two components:
(1) \emph{correlated reward} ($R_{t}^{corr}$), which encapsulates the effect of Athena's action on the system,
and (2) \emph{uncorrelated reward} ($R_{t}^{uncorr}$), which encapsulates the inherent change in program behavior.
The overall reward ($R_t$) is defined using these two component rewards as:
\begin{equation} 
    \label{eq:reward}
    R_t = R_{t}^{corr} - R_{t}^{uncorr}
\end{equation}

By subtracting the uncorrelated reward component from the correlated reward, Athena aims to isolate the performance impact that is causally attributable to its coordination actions from variations induced by inherent workload behavior.
This allows Athena to learn a higher-performing prefetcher-OCP coordination policy than it would have otherwise learned using a single, conflated reward signal (as we show in~\Cref{subsec:ath_eval_ablation}).

\paraheading{Correlated Reward.}
We define the correlated reward at a given timestep $t$ as a \emph{linear combination} of the changes in the constituent system-level metrics that are influenced by Athena's actions in two consecutive timesteps. 
Formally, the correlated reward, $R_{t}^{corr}$, is defined as:

\begin{equation} 
    \label{eq:corr_reward}
    R_{t}^{corr} = \sum_{i} \lambda_i \cdot \Delta M_{i,t}^{corr}
\end{equation}

Here, $\Delta M_{i,t}^{corr}$ denotes the change in the $i$-th correlated system-level metric observed between timesteps $(t-1)$ and $t$, and $\lambda_i$ is a hyperparameter that captures the relative weight of this metric to the overall reward.\footnote{The values of weight parameters are tuned offline using the automated design-space exploration (see~\Cref{subsub:hyp_tuning}).}

In principle, any system-level metric that is directly affected by Athena's actions (e.g., execution cycles, number of last-level cache misses) can be incorporated into the correlated reward.
In practice, we conduct an offline sensitivity analysis over a broad set of candidate metrics and select three metrics shown in~\Cref{tab:reward_decomp} as the constituents of the correlated reward that provide a stable and informative learning signal across diverse workloads and system configurations.

\paraheading{Uncorrelated Reward.}
We define the uncorrelated reward at a given timestep $t$ as a \emph{linear combination} of the changes in its constituent metrics that are largely independent of Athena's actions but are influenced by the inherent variations in workload behavior. 
Formally, the uncorrelated reward, $R_{t}^{uncorr}$, is defined as:

\begin{equation} 
    \label{eq:uncorr_reward}
    R_{t}^{uncorr} = \sum_{j} \lambda_j \cdot \Delta M_{j,t}^{uncorr}
\end{equation}

Here, $\Delta M_{j,t}^{uncorr}$ denotes the change in the $j$-th uncorrelated metric observed between timesteps $(t-1)$ and $t$, and $\lambda_j$ is a hyperparameter that captures its relative weight to the overall reward.
While any metric that is affected by changes in workload behavior can be incorporated into the uncorrelated reward, we select two metrics shown in~\Cref{tab:reward_decomp} as the constituents of the uncorrelated reward, based on offline sensitivity analysis.

\begin{table}[h]
    \centering
    \small
    \begin{tabular}{C{10em}||L{11em}|C{4em}}
        \thickhline
        \textbf{Reward Component} & \textbf{Constituent Metric} & \textbf{Weight} \\ 
        \thickhline
        \multirow{3}{*}{\textbf{\( R_{t}^{\text{corr}} \)}} 
        & \# Cycles & $\lambda_{\text{cycle}} $ \\
        & \# LLC misses & $\lambda_{\text{LLC}_\text{m}}$ \\
        & LLC miss latency & $\lambda_{\text{LLC}_\text{t}}$ \\ 
        \hline
        \multirow{2}{*}{\textbf{\( R_{t}^{\text{uncorr}} \)}} 
        & \# Load instructions & $\lambda_{\text{load}}$ \\ 
        & \# Mispredicted branches &$\lambda_{\text{MBr}}$ \\
        \thickhline
    \end{tabular}
    \caption{Constituent metrics of Athena's reward.}
    \label{tab:reward_decomp}
\end{table}

\sectionRB{Athena: Detailed Design}{Athena: Detailed Design}{sec:ath_implementation}

\subsection{QVStore Organization}\label{subsec:qvstore}
The QVStore maintains the Q-values of all state-action pairs encountered by Athena during online execution. 
Unlike prior RL-based approaches that rely on deep neural networks to approximate Q-values~\cite{jalili2022managing}, or operate without state information~\cite{mab, micro_mama}, Athena adopts a lightweight and hardware-friendly tabular organization for storing Q-values, tailored for low-latency access and online updates.

\begin{figure}[!ht] 
\centering \includegraphics[width=\columnwidth]{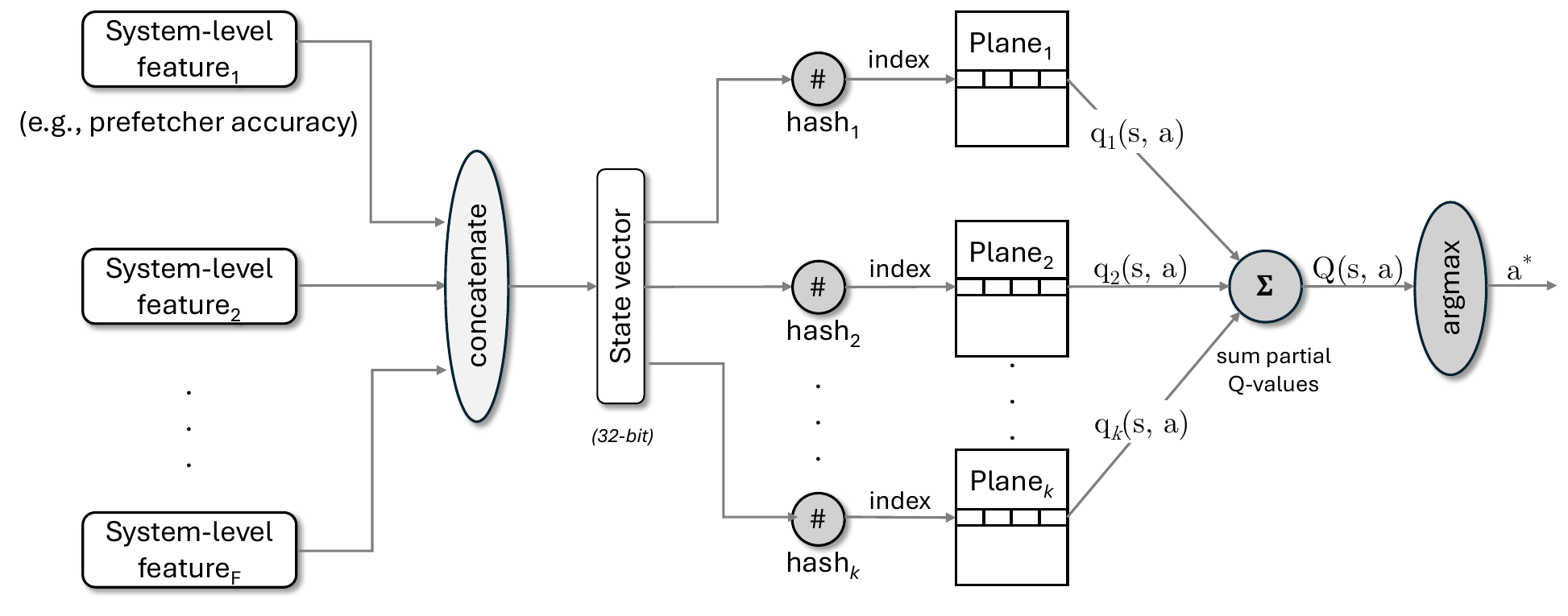}
\caption{Organization of QVStore.} 
\label{fig:at_mechanism} 
\end{figure}

\Cref{fig:at_mechanism} demonstrates the organization of QVStore and the procedure for retrieving the Q-value of a given state $S$ and action $A$.
As the number of Q-values that need to be stored for every possible state-action pair grows rapidly with 
(1) the number of constituent features in a state vector, 
and (2) the number of bits used to represent each feature, naively implementing the QVStore as a monolithic table quickly becomes impractical due to its storage overhead.
Such a design also incurs prohibitive access latency and power overhead, making it unsuitable for designing a timing-critical hardware RL agent.

To address these challenges, Athena organizes the QVStore as a partitioned structure comprising $k$ independent tables, each of which we call a \emph{plane}. 
Each plane stores a partial Q-value of a given state-action pair. 
This partitioned organization enables Athena to decouple the storage cost from the full combinatorial state space while simultaneously supporting fast, parallel access.

For a given state-action pair, Athena retrieves its corresponding Q-value from the partitioned QVStore in three stages, as shown in~\Cref{fig:at_mechanism}.
First, Athena constructs the state vector by concatenating all feature values.
Second, Athena applies $k$ distinct hash functions to the state vector, each producing an index to a plane to retrieve the corresponding partial Q-value \emph{in parallel}.
Third, Athena computes the final Q-value by summing the partial Q-values across all planes.
At the end of each epoch, Athena updates the Q-value using the SARSA update rule (see~\Cref{sec:background_rl}), applying the update independently to each plane.

The partitioned, multi-hash organization of QVStore provides two key benefits. 
First, hashing the same state into multiple planes strikes a balance between generalization and resolution: similar states are likely to collide in at least some planes, promoting value sharing, while dissimilar states are likely to be de-aliased through independent hashes.
Second, partitioning keeps each plane compact, enabling low-latency, energy-efficient parallel reads and updates.
Together, these properties substantially reduce storage overhead while supporting fast and scalable Q-value access and updates during Athena's operation.

\subsection{State Measurement}\label{subsec:state_measurement}

\subsubsection{\textbf{Prefetcher Accuracy}}
Athena employs a Bloom filter~\cite{bloom} to track prefetcher accuracy.
For every prefetch issued, the corresponding address is inserted into the Bloom filter.  
Upon each demand access, Athena queries the filter to determine whether the address was prefetched.  
Prefetch accuracy is computed as the ratio of demand accesses that hit in the filter over the number of issued prefetches. 
Athena resets the filter at the end of each epoch.

\subsubsection{\textbf{OCP Accuracy}}
Athena measures OCP accuracy using two simple counters.
When OCP predicts that a demand load will go off-chip, it issues a speculative request to the memory controller to start fetching the data directly from main memory. 
Athena tracks such predictions using a dedicated counter.
When a demand request misses all cache levels and arrives at the memory controller, it indicates that this request indeed went off-chip.
Athena increments another counter to track correct predictions.
OCP accuracy is computed as the ratio of correctly predicted off-chip accesses over the total number of off-chip predictions. 
Athena resets both counters at the end of each epoch.

\subsubsection{\textbf{Prefetch-Induced Cache Pollution}}
Athena uses a Bloom filter~\cite{bloom} to track prefetch-induced cache pollution at the last-level cache (LLC).
When a cache block is evicted from the LLC for a prefetch fill, Athena inserts the evicted address into the filter.  
If the address of a subsequent LLC miss hits the filter, Athena increments a dedicated counter.
Athena resets the filter and the counter at the end of each epoch. 
This method of measuring prefetch-induced cache pollution is similar to prior works~\cite{fdp, ebrahimi2009techniques}.

\subsection{Automated Design-Space Exploration} \label{subsec:ath_automated}
We employ automated design-space exploration (DSE) to select the optimal state features (see \Cref{subsec:ath_state}), reward weights (see~\Cref{subsec:ath_reward}), and hyperparameters ($\alpha$, $\gamma$, $\epsilon$, $\tau$, and epoch length).
To prevent overfitting, DSE is conducted on $20$ selected workloads, which are \emph{not included} in the final set of $100$ workloads.
We perform DSE using Cache Design~1 (CD1) in a single-core configuration, with POPET as the OCP and Pythia as the L2C prefetcher (see~\Cref{sec:ath_meth_sys_config}).
The configuration that achieves the best performance across the $20$ selected workloads is then applied \textit{unaltered} to the full set of $100$ evaluation workloads, to all other cache designs (see~\Cref{sec:ath_eval_sc}), OCPs (see~\Cref{sec:ath_eval_sen_cd1}), prefetchers (see~\Cref{sec:ath_eval_sen_cd4}), and multi-core experiments (see~\Cref{sec:ath_eval_mc}).

\subsubsection{\textbf{Feature Selection}}
We derive the program features through an iterative process. 
Starting from the initial set of seven candidate features (see \cref{subsec:ath_state}), we begin with the feature that yields the highest standalone performance gain. 
In each iteration, we include the feature that provides the greatest additional performance improvement while retaining the previously selected features.
We observe diminishing performance gains after the fourth iteration. 
Consequently, we fix the feature set to the following four: prefetcher accuracy, OCP accuracy, bandwidth usage, and prefetch-induced cache pollution, as summarized in \Cref{table:config}.\\

\begin{table}[!ht]
  \centering
  \label{table:config}
  \small
  \begin{tabular}{L{8em}||L{26em}}
    \thickhline
    \Tabval{\textbf{Category}} & \Tabval{\textbf{Final Values}} \\
    \thickhline
    \Tabval{\textbf{\emph{Selected Features}}} & \Tabval{(1) prefetcher accuracy, (2) OCP accuracy, (3) bandwidth usage, (4) prefetch-induced cache pollution} \\
    \hline
    \Tabval{\textbf{\emph{Reward Weights}}} & \Tabval{$\lambda_{\text{cycle}} = 1.6$, $\lambda_{\text{LLC}_\text{m}} = 0.0$, $\lambda_{\text{LLC}_\text{t}} = 0.0$, $\lambda_{\text{load}} = 0.6$, $\lambda_{\text{MBr}} = 1.0$} \\
    \hline
    \Tabval{\textbf{\emph{Hyperparameters}}} & \Tabval{$\alpha = 0.6$, $\gamma = 0.6$, $\epsilon = 0.0$, $\tau = 0.12$, Epoch length ($N$) = $2K$ instructions} \\
    \thickhline
  \end{tabular}
  \caption{Final Athena configuration derived through automated design-space exploration.}
\end{table}

\subsubsection{\textbf{Reward and Hyperparameter Tuning}} \label{subsub:hyp_tuning}
We use grid search~\cite{grid_search1,grid_search2} to tune reward weights (see~\Cref{subsec:ath_reward}) and hyperparameters (see~\Cref{sec:background_rl}). 
For each reward weight and hyperparameter, we define a search range and discretize it into equally spaced grid points.
The learning rate \(\alpha\), the discount factor \(\gamma\), and the exploration rate \(\epsilon\) are searched over $[0, 1]$ in steps of $0.1$. 
Reward weights \(\lambda_{i}\) and \(\lambda_{j}\) are searched over $[0, 2]$ in steps of $0.2$.
The final reward weights and hyperparameters are shown in \Cref{table:config}.

\subsection{Overhead Analysis}\label{subsec:storageoverhead}

\subsubsection{\textbf{Storage Overhead}} \label{subsec:overhead_storage}
\Cref{table:ath_overhead} summarizes the storage overhead of Athena.
The QVStore is organized into eight planes, each containing $64$ rows and $4$ columns (one per action). Each entry stores an $8$-bit Q-value. 
To size the Bloom filter used for prefetcher accuracy tracking, we experimentally observe an average of $49$ prefetch requests per epoch (i.e., $2K$ retired instructions), with a standard deviation (SD) of $50$. 
We therefore size the Bloom filter for prefetcher accuracy tracking at $4096$ bits, which yields a false positive rate of $1\%$ when accommodating three SDs above the average (i.e., $199$ requests).
Similarly, we observe an average of $62$ LLC evictions per epoch, with an SD of $58$. 
Thus, we size the Bloom filter for prefetch-induced cache pollution tracking at $4096$ bits, providing a false positive rate of $1\%$ when inserting three SDs more evictions than the average (i.e., $236$ evictions).
\\

\begin{table}[!ht]
  \centering
  \label{table:ath_overhead}
  \small
  \begin{tabular}{l||l||r}
    \thickhline
    \textbf{Structure} & \textbf{Description} & \textbf{Size} \\
    \thickhline
    \multirow{2}[1]{*}{\textbf{QVStore}} & \# planes = $8$,  \# rows = $64$ & \multirow{2}[1]{*}{{2~KB}} \\
    & \# columns = $4$, entry size = $8$ bits & \\
    \hline
    \textbf{Accuracy Tracker} & $4096$-bit Bloom filter, $2$ hashes & {0.5~KB} \\
    \hline
    \textbf{Pollution Tracker} & $4096$-bit Bloom filter, $2$ hashes & {0.5~KB} \\
    \thickhline
    \textbf{Total} & & \textbf{3~KB} \\
    \thickhline
  \end{tabular}
  \caption{Storage overhead of Athena.}
\end{table}

\subsubsection{\textbf{Latency Overhead}} \label{subsec:overhead_latency}
At the end of every epoch, Athena needs to compute the overall reward from its constituent partial rewards and update the QVStore, both of which incur considerable computation overhead. 
To account for such computations, we model Athena with a delayed QVStore update latency of $50$ cycles (i.e., the QVStore is updated $50$ cycles after the end of an epoch).
We also sweep the update latency and observe that Athena's performance benefit is not sensitive to the update latency. 
This is because Athena needs to query the updated QVStore only at the end of the current epoch (i.e., $2K$ retired instructions), which takes considerably longer than even the pessimistic estimate of the update latency.

\sectionRB{Athena: Evaluation Methodology}{Methodology}{sec:ath_methodology}

\noindent We evaluate Athena using the ChampSim trace-driven simulator~\cite{champsim}.
We faithfully model an Intel Golden Cove-like microarchitecture~\cite{goldencove}, including its large reorder buffer (ROB), multi-level cache hierarchy, and publicly reported on-chip cache access latencies~\cite{llc_lat1, llc_lat2, l3_lat_compare1}. 
Table~\ref{table:athena_sim_params} summarizes the key microarchitectural parameters.
The source code of Athena is freely available at~\cite{athena_github}.

\begin{scriptsize}
    \begin{table}[!ht]
    \centering
    \small
    \begin{tabular}{L{4.5em}||L{34em}}
        \thickhline
        \hline
        \Tabval{\textbf{Core}} & \Tabval{6-wide fetch/issue/commit, 512-entry ROB, 128-entry LQ, 72-entry SQ, perceptron branch predictor~\cite{perceptron}, 17-cycle misprediction penalty} \\
        \hline
        \Tabval{\textbf{L1I/D}} & \Tabval{Private, 48KB, 64B line, 12-way, 16 MSHRs, LRU, 4/5-cycle round-trip latency} \\
        \hline
        \Tabval{\textbf{L2C}} & \Tabval{Private, 1.25MB, 64B line, 20-way, 48 MSHRs, LRU, 15-cycle round-trip latency~\cite{llc_lat2}} \\
        \hline
        \Tabval{\textbf{LLC}} &  \Tabval{Shared, 3MB/core, 64B line, 12-way, 64 MSHRs/slice, SHiP~\cite{ship}, 55-cycle round-trip latency~\cite{llc_lat1, llc_lat2}} \\
        \hline
        \Tabval{\textbf{Main Memory}} & \Tabval{1 rank per channel, 8 banks per rank, 64-bit data bus, 2KB row buffer, $t_{\text{RCD}}$ = $t_{\text{RP}}$ = $t_{\text{CAS}}$ = 12.5ns; DDR4 with 3.2 GB/s per core}\\
        \thickhline
    \end{tabular}
    \caption{Simulated system parameters.}
    \label{table:athena_sim_params}
\end{table}
\end{scriptsize}

\subsection{Workloads} \label{sec:ath_meth_workloads}

We evaluate Athena using a diverse set of workload traces spanning \texttt{SPEC CPU 2006}~\cite{spec2006}, \texttt{SPEC CPU 2017}~\cite{spec2017}, \texttt{PARSEC}~\cite{parsec}, \texttt{Ligra}~\cite{ligra}, and real-world commercial workloads from the first Value Prediction Championship (\texttt{CVP}~\cite{cvp1}). 
We only consider workloads in our evaluation that have at least $3$ LLC misses per kilo instructions (MPKI) in the no-prefetching and no-OCP system.
In total, we use $100$ memory-intensive single-core workload traces, as summarized in Table~\ref{table:athena_workloads}.
\texttt{SPEC CPU 2006} and \texttt{SPEC CPU 2017} workloads are collectively referred to as \texttt{SPEC}. All the workload traces used in our evaluation are freely available online~\cite{athena_github}.

\begin{table}[htbp]
  \centering
  \small
    \begin{tabular}{L{7em}L{6em}L{25em}}
    \toprule
    \Tabval{\textbf{Suite}} & \Tabval{\textbf{\# Workloads}} & \Tabval{\textbf{Example Workloads}} \\
    \toprule
    \Tabval{SPEC CPU06} & \Tabval{29} & \Tabval{astar, leslie3d, libquantum, milc, omnetpp, sphinx3, soplex  ...} \\
    \Tabval{SPEC CPU17} & \Tabval{20} & \Tabval{bwaves, cactuBSSN, fotonik3d, gcc, lbm, mcf, xalancbmk ...} \\
    \Tabval{Ligra} & \Tabval{13} & \Tabval{BC, BFS, BFSCC, PageRankDelta, Radii, ...} \\
    \Tabval{PARSEC} & \Tabval{13} & \Tabval{canneal, facesim, fluidanimate, raytrace, streamcluster ...} \\
    \Tabval{CVP} & \Tabval{25}  & \Tabval{integer, floating point, ...} \\
    \bottomrule
    \end{tabular}%
  \caption{Workloads used for evaluation.}
  \label{table:athena_workloads}%
\end{table}%

For multi-core evaluation, we construct $90$ four-core and $90$ eight-core workload mixes, each comprising three categories.
(1) $30$ \emph{prefetcher-adverse mixes}: each workload is randomly selected from the prefetcher-adverse workloads.
(2) $30$ \emph{prefetcher-friendly mixes}: each workload is randomly selected from the prefetcher-friendly workloads.
(3) $30$ \emph{random mixes}: workloads are drawn uniformly at random from the entire set of $100$ workloads.

For all single-core simulations, we perform a warm-up of $100$ million (M) instructions. 
\texttt{SPEC} workloads are simulated for $500$M, \texttt{PARSEC}, \texttt{Ligra}, and \texttt{CVP} workloads for $150$M instructions.
For multi-core simulations, each core performs a warm-up of $10$M instructions, followed by simulating $50$M instructions.
The workloads are replayed as needed to ensure all cores reach the required number of simulated instructions.

\subsection{Evaluated Prior Prefetcher Control Policies} 
\label{sec:ath_meth_sys_config_pol}

We compare Athena against three prior prefetcher control policies: 
TLP~\cite{jamet2024tlp}, HPAC~\cite{ebrahimi2009coordinated}, and MAB~\cite{mab}.

{\paraheading{Two Level Perceptron}}
(TLP)~\cite{jamet2024tlp} explicitly incorporates OCP into its coordination framework and combines it with prefetch filtering~\cite{jamet2024tlp} at the L1D. We adopt the same features, prediction thresholds ($\tau_{\text{low}}$, $\tau_{\text{high}}$), and filtering threshold ($\tau_{\text{pref}}$) as specified in~\cite{jamet2024tlp}.

{\paraheading{Hierarchical Prefetcher Aggressiveness Control}}
(HPAC)~\cite{ebrahimi2009coordinated} compares various system-level feature values against static thresholds to make prefetch control decisions. Although not designed with OCP in mind, we adapt HPAC to coordinate prefetchers and OCP. We use three system-level features for local aggressiveness control: prefetcher accuracy, OCP accuracy, and main-memory bandwidth usage.
We use the bandwidth needed by each core as the feature for global aggressiveness control.
For each feature, we tune its static threshold via extensive grid search using the same set of tuning workloads that we use to tune Athena (see~\cref{subsec:ath_automated}).

{\paraheading{Micro-Armed Bandit}}\label{subsubsec:mab}
(MAB)~\cite{mab} uses a multi-armed bandit algorithm~\cite{multi_arm1,multi_arm2} for its decision-making. 
Although originally not designed for OCP, we adapt MAB to coordinate OCP with prefetchers. 
MAB selects whether to enable the prefetcher and OCP based on previous reward feedback derived from the system's IPC.
Thus, our implementation of MAB uses four (eight) arms while coordinating one OCP in the presence of one (two) prefetcher(s).
We find the best-performing hyperparameters via grid search~\cite{grid_search1,grid_search2} using the same set of tuning workloads that we use to tune Athena (see~\cref{subsec:ath_automated}).

\subsection{Evaluated Cache Designs} \label{sec:ath_meth_sys_config}
To demonstrate Athena's adaptability, we evaluate Athena across diverse cache designs (CDs), as summarized in~\Cref{tab:ath_design_comparison}. 
All these cache designs include an OCP alongside the three-level cache hierarchy and differ only in the number and placement of prefetchers, closely mimicking cache hierarchy designs found in commercial processors.    
For each cache design, we identify suitable prior approaches as comparison points. HPAC and MAB can be adapted to all four cache designs to coordinate OCP with prefetchers. 
In contrast, TLP, by design, is restricted to cache designs involving an L1D prefetcher (i.e., CD2 and CD4; see~\Cref{subsubsec:ath_prior}).
Unless stated otherwise, CD1 serves as the default cache configuration.  

\begin{table}[htbp]
\centering
\small
    \begin{tabular}{C{7em}ll}
    \toprule
    \Tabval{\textbf{Cache Design}} & \Tabval{\textbf{Description}} & \Tabval{\textbf{Comparison Points}} \\
    \toprule
    \Tabval{\textbf{CD1}} & \Tabval{OCP + 1 L2C prefetcher} & \Tabval{HPAC, MAB} \\
    \Tabval{\textbf{CD2}} & \Tabval{OCP + 1 L1D prefetcher} & \Tabval{HPAC, MAB, TLP} \\
    \Tabval{\textbf{CD3}} & \Tabval{OCP + 2 L2C prefetchers} & \Tabval{HPAC, MAB} \\
    \Tabval{\textbf{CD4}} & \Tabval{OCP + 1 L1D + 1 L2C prefetcher} & \Tabval{HPAC, MAB, TLP} \\
    \bottomrule
    \end{tabular}
\caption{Evaluated cache designs (CD) and corresponding comparison points.}
\label{tab:ath_design_comparison}
\end{table}

\subsection{Evaluated Data Prefetchers} \label{sec:ath_meth_sys_config_pref}

We assess Athena's flexibility by integrating six prefetchers, i.e., IPCP~\cite{ipcp}, Berti~\cite{navarro2022berti}, Pythia~\cite{pythia}, SPP~\cite{spp} with perceptron-based prefetch filter (PPF)~\cite{ppf}, SMS~\cite{sms}, and MLOP~\cite{mlop}, at various levels of the cache hierarchy.
IPCP and Berti are evaluated at L1D and are trained using all memory requests looking up the L1D. 
Pythia, SPP+PPF, MLOP, and SMS operate at L2C and are trained using all memory requests looking up the L2C. 
All prefetchers prefetch in the physical address space.
Unless stated otherwise, we use IPCP as the default L1D prefetcher, and Pythia as the default L2C prefetcher.

\subsection{Evaluated Off-Chip Predictors} \label{sec:ath_meth_sys_config_ocp}

We also evaluate Athena across three OCPs: POPET~\cite{hermes}, HMP~\cite{yoaz1999speculation}, and TTP~\cite{lp, hermes}.
POPET uses a hashed-perceptron network with five program features to make accurate off-chip predictions. We evaluate the exact POPET configuration presented in~\cite{hermes}.
HMP combines three prediction techniques analogous to hybrid branch prediction: local~\cite{yeh1991two, yeh1992alternative}, gshare~\cite{mcfarling1993combining}, and gskew~\cite{michaud1997trading}. 
TTP, as introduced by~\cite{lp, hermes}, predicts off-chip loads by tracking cacheline tags. 
We evaluate the exact TTP configuration open-sourced by~\cite{hermes}.
Similar to prior work~\cite{hermes}, all speculative load requests issued by the OCPs incur a $6$-cycle latency before reaching the memory controller.
Unless stated otherwise, we use POPET as the default OCP.
Table~\ref{table:ath_eval_sys} summarizes the storage overhead of all evaluated mechanisms.

\begin{table}[htbp]
  \centering
  \small
    \begin{tabular}{|L{1em}||L{0.8em}|L{24em}||R{4em}|}
    \hline
    \multirow{6}[2]{*}{\begin{sideways}\textbf{Prefetchers}\end{sideways}} 
      & \multirow{2}[1]{*}{\begin{sideways}\textbf{L1D}\end{sideways}} 
          & \Tstrut \textbf{IPCP}~\cite{ipcp}, as an L1D-only prefetcher & \textbf{0.7 KB} \\ \cline{3-4}
      &   & \Tstrut \textbf{Berti}, as configured in~\cite{navarro2022berti} & \textbf{2.55 KB} \\ \cline{2-4}
      & \multirow{4}[2]{*}{\begin{sideways}\textbf{L2C}\end{sideways}}
          & \Tstrut \textbf{Pythia}, as configured in~\cite{pythia} & \textbf{25.5 KB} \\ \cline{3-4}
      &   & \Tstrut \textbf{SPP+PPF}, as configured in~\cite{spp, ppf} & \textbf{39.3 KB} \\ \cline{3-4}
      &   & \Tstrut \textbf{MLOP}, as configured in~\cite{mlop} & \textbf{8 KB} \\ \cline{3-4}
      &   & \Tstrut \textbf{SMS}, as configured in~\cite{sms} & \textbf{20 KB} \\
    \hline
    \addlinespace[0.5em]
    \hline
    \multirow{3}[2]{*}{\begin{sideways}\textbf{OCPs}\end{sideways}} 
      & \multicolumn{2}{l||}{\Tstrut \textbf{POPET}~\cite{hermes}, with 5 features} & \textbf{4 KB} \\ \cline{2-4}
      & \multicolumn{2}{l||}{\Tstrut \textbf{HMP}~\cite{yoaz1999speculation}, with 3 component predictors} & \textbf{11 KB} \\ \cline{2-4}
      & \multicolumn{2}{l||}{\Tstrut \textbf{TTP}~\cite{lp}, with metadata budget \textasciitilde L2 cache size} & \textbf{1536 KB} \\
    \hline
    \addlinespace[0.5em]
    \hline
    \multirow{4}[1]{*}{\begin{sideways}\textbf{Policies}\end{sideways}} 
      & \multicolumn{2}{l||}{\Tstrut \textbf{TLP}, as in~\cite{jamet2024tlp}} & \textbf{6.98 KB} \\ \cline{2-4}
      & \multicolumn{2}{l||}{\Tstrut \textbf{HPAC}~\cite{ebrahimi2009coordinated}, adapted for OCP} & \textbf{0.5 KB} \\ \cline{2-4}
      & \multicolumn{2}{l||}{\Tstrut \textbf{MAB}~\cite{mab}, adapted for OCP} & \textbf{0.1 KB} \\ \cline{2-4}
      & \multicolumn{2}{l||}{\Tstrut \emph{\textbf{Athena (this work)}}} & \textbf{3 KB} \\
    \hline
    \end{tabular}
    \caption{Storage overhead of all evaluated mechanisms.}
    \label{table:ath_eval_sys}
\end{table}

\sectionRB{Athena: Evaluation}{Evaluation}{sec:ath_evaluation}

\subsection{Single-Core Evaluation Overview} \label{sec:ath_eval_sc}

\subsubsection{\textbf{CD1: OCP with One L2C Prefetcher}}\label{subsubsec:ath_L2C}
\begin{figure}[!ht]
    \centering
    \includegraphics[width=\columnwidth]{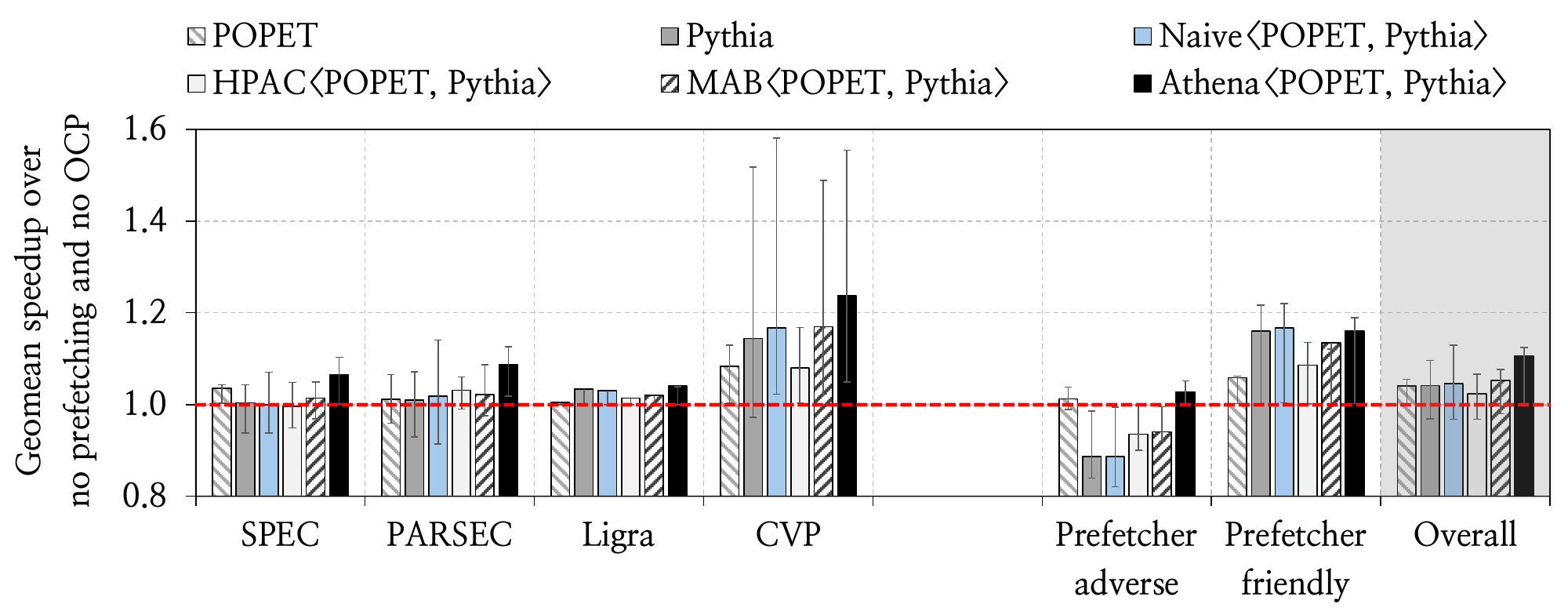}
    \caption{Speedup in cache design 1 (CD1).}
\label{fig:cd1}
\end{figure}

\Cref{fig:cd1} shows the performance improvement of Naive, HPAC, MAB, and Athena when coordinating POPET as the OCP and Pythia as the L2C prefetcher.
We make two key observations.
First, in prefetcher-adverse workloads, Naive degrades performance by $11.1\%$ compared to the baseline with no prefetching or OCP.
This degradation arises because Pythia negatively impacts performance in these workloads, undermining POPET's gains. 
In contrast, Athena dynamically identifies that POPET is beneficial and improves performance by $14.0\%$ over Naive, even surpassing POPET's standalone performance.
Second, although Naive harms performance in prefetcher-adverse workloads, it significantly improves performance by $16.7\%$ in prefetcher-friendly workloads. 
Athena dynamically determines that enabling both POPET and Pythia is advantageous in these workloads, thus closely matching Naive's performance.
Overall, Athena outperforms Naive, HPAC, and MAB by $5.7\%$, $7.9\%$, and $5.0\%$, respectively, across \emph{all} $100$ workloads, demonstrating \emph{consistent} performance improvements.

\begin{figure}[!ht]
    \centering
    \includegraphics[width=\columnwidth]{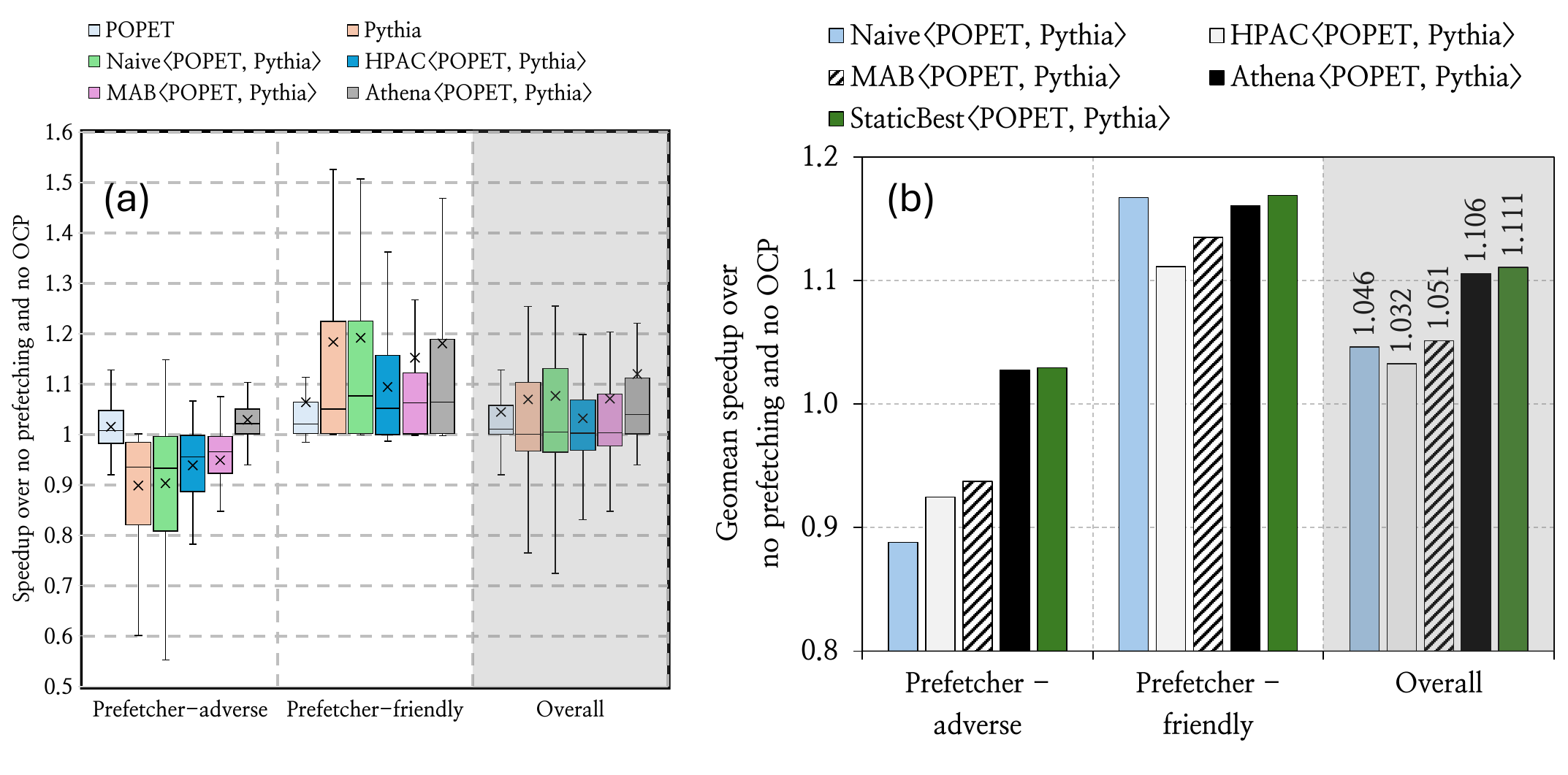}
    \caption{(a) Workload category-wise performance analysis in CD1. (b) Performance comparison with StaticBest in CD1.}
\label{fig:cd1_perf_deepdive}
\end{figure}

\paraheading{Workload Category-Wise Performance Analysis.}
To further analyze Athena's performance gains in CD1,~\Cref{fig:cd1_perf_deepdive}(a) shows the workload category-wise speedups as a box-and-whisker plot.
We make three key observations.
First, for prefetcher-adverse workloads, Athena substantially raises the lower quartile as well as the lower-end whisker relative to Naive, HPAC, and MAB.
This implies that Athena improves performance broadly across all prefetcher-adverse workloads, rather than merely alleviating a small number of extreme slowdowns.
Second, for prefetcher-friendly workloads, Athena increases both the upper quartile and the upper-end whisker compared to HPAC and MAB, demonstrating consistent performance improvements over these policies across all prefetcher-friendly workloads.
Third, when considering all workloads together, Athena improves mean performance relative to HPAC and MAB while simultaneously elevating both the lower and upper quartiles. 
This result indicates that Athena delivers robust and consistent performance gains across a broad spectrum of workload behaviors.

\paraheading{Performance Comparison with StaticBest.}
\Cref{fig:cd1_perf_deepdive}(b) compares the performance of Athena against the StaticBest combination (see~\Cref{subsubsec:ath_naive_combo}). 
The key takeaway is that, by dynamically coordinating POPET and Pythia, Athena provides similar performance gains as the StaticBest combination for both prefetcher-adverse and prefetcher-friendly workload categories. 
On average, Athena improves performance by $10.3\%$ over the baseline with no prefetching or OCP, whereas StaticBest improves performance by $11.1\%$.

\begin{figure}[!ht]
    \centering
    \includegraphics[width=\columnwidth]{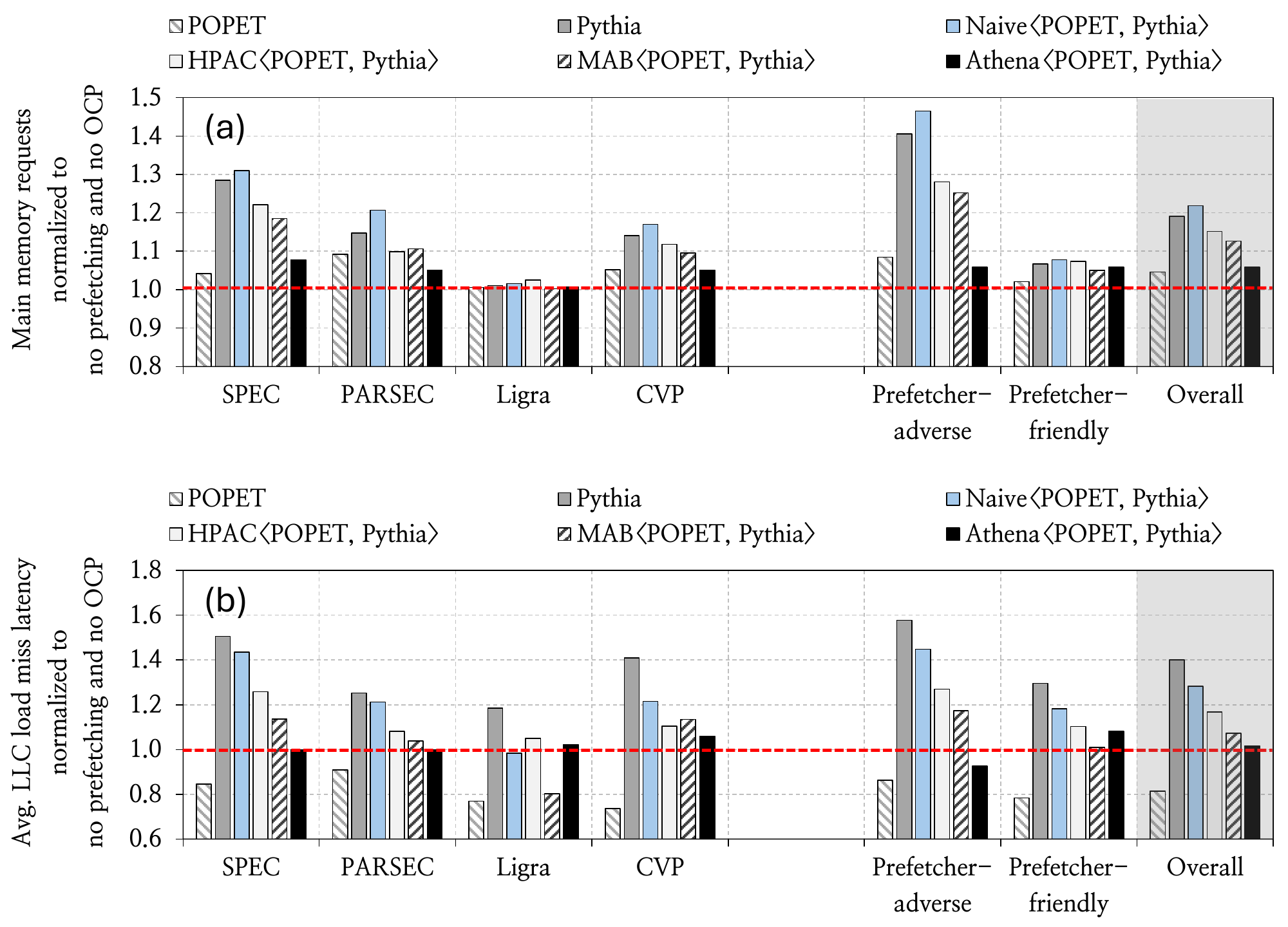}
    \caption{Comparison of (a) the number of main memory requests and (b) average last-level cache (LLC) load miss latency.}
\label{fig:cd1_dram_llc}
\end{figure}

\paraheading{Effect of Athena on Main Memory Requests.}
\Cref{fig:cd1_dram_llc}(a) shows the number of main memory requests issued by POPET-alone, Pythia-alone, Naive, HPAC, MAB, and Athena across all $100$ workloads in CD1.
We make two key observations.
First, in prefetcher-adverse workloads, 
Pythia-alone significantly increases the main memory requests due to its poor prefetch accuracy.
Naively combining POPET with Pythia further increases the main memory requests (by $46.5\%$ on top of the baseline system without any prefetcher or OCP).
This overhead in main memory requests severely harms the overall performance (see~\cref{subsubsec:ath_L2C}).
Athena, by dynamically coordinating POPET and Pythia, substantially reduces the main memory request overhead (only $5.9\%$ over the baseline).
Second, across all workloads, Naive, HPAC, and MAB increase the main memory requests by $21.9\%$, $15.2\%$, and $12.7\%$, respectively, whereas Athena increases them by only $5.8\%$ over the baseline without any prefetcher or OCP.

\paraheading{Effect of Athena on LLC Load Miss Latency.}
\Cref{fig:cd1_dram_llc}(b) shows the average last-level cache (LLC) load miss latency, normalized to the baseline system without prefetching or OCP, across all $100$ workloads. 
We make three key observations. 
First, naively enabling both Pythia and POPET increases average LLC miss latency, particularly for prefetcher-adverse workloads, indicating that uncoordinated speculation can exacerbate memory-system contention and interference. 
Second, HPAC and MAB partially reduce this latency overhead, but still fall short of the baseline.
Third, Athena consistently reduces the LLC miss latency overhead across all workload categories by dynamically coordinating Pythia and POPET.
Overall, Naive, HPAC, and MAB increase the average LLC load miss latency by $28.3\%$, $16.7\%$, and $7.3\%$, respectively, whereas Athena increases it by only $1.7\%$ over the baseline without any prefetcher or OCP.

\subsubsection{\textbf{CD2: OCP with One L1D Prefetcher}}\label{subsubsec:ath_L1D}

\begin{figure}[!ht]
    \centering
    \includegraphics[width=\columnwidth]{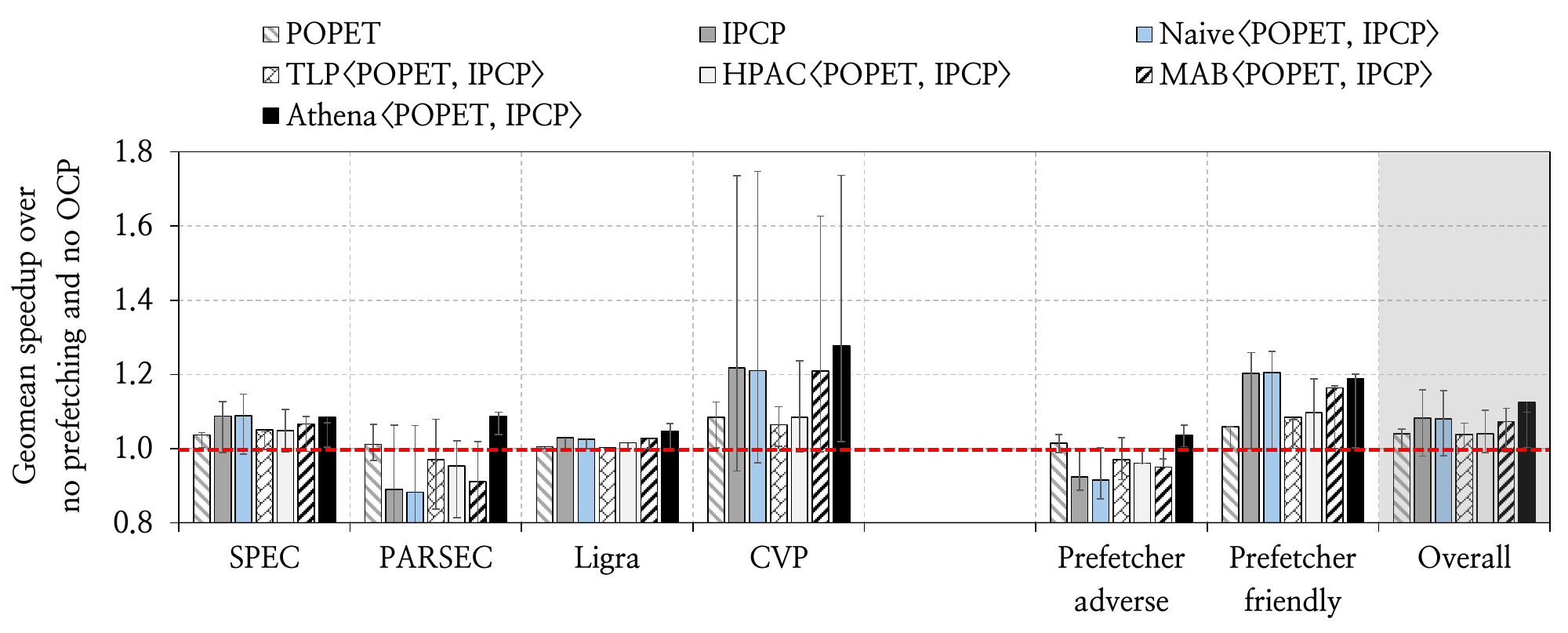}
    \caption{Speedup in cache design 2 (CD2).}
\label{fig:cd2}
\end{figure}

\Cref{fig:cd2} shows the performance improvement of Naive, TLP, HPAC, MAB, and Athena when coordinating POPET as the OCP and IPCP as the L1D prefetcher.
We make two observations. 
First, TLP outperforms Naive by $5.5\%$ in prefetcher-adverse workloads by filtering out prefetches that are predicted to go off-chip.
However, this filtering strategy hurts performance in prefetcher-friendly workloads where prefetch requests are indeed helpful, causing TLP to underperform Naive by $12.0\%$.
Second, Athena, by dynamically learning using multiple system-level features, outperforms TLP in both prefetcher-adverse and prefetcher-friendly workloads by $6.5\%$ and $10.4\%$, respectively, highlighting its robust and \emph{consistent} performance improvements.
Overall, Athena outperforms Naive, TLP, HPAC, and MAB on average by $4.5\%$, $8.7\%$, $8.4\%$, and $5.2\%$, respectively.

\subsubsection{\textbf{CD3: OCP with Two L2C Prefetchers}} \label{subsubsec:ath_2L2C}

\begin{figure}[!ht]
    \centering
    \includegraphics[width=\columnwidth]{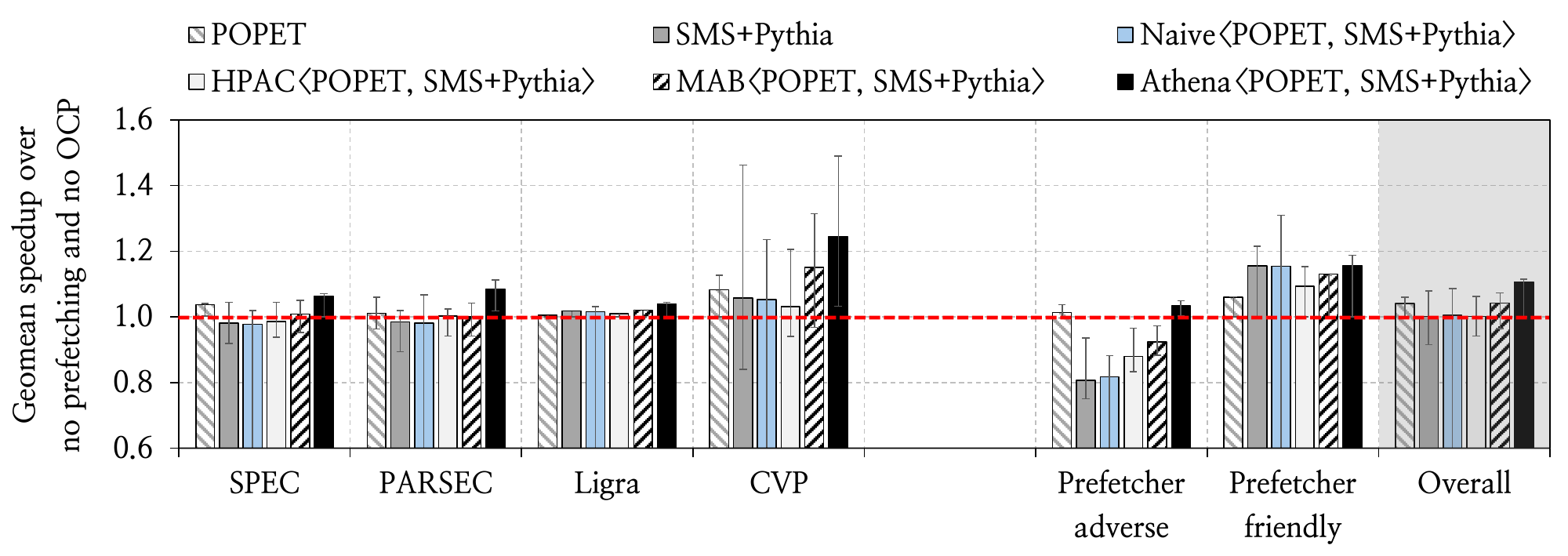}
    \caption{Speedup in cache design 3 (CD3).}
\label{fig:cd3}
\end{figure}

\Cref{fig:cd3} shows the performance improvement of Naive, HPAC, MAB, and Athena when coordinating POPET as the OCP, along with SMS and Pythia as two L2C prefetchers.
We observe that, in prefetcher-adverse workloads, HPAC and MAB only partially alleviate Naive's performance degradation, failing to match the baseline with no prefetching or OCP.
Athena, in contrast, achieves a $3.2\%$ improvement over the baseline, surpassing POPET's standalone performance.
In prefetcher-friendly workloads, Athena matches Naive's performance.
Overall, Athena outperforms Naive, HPAC, and MAB on average by $10.1\%$, $10.4\%$, and $6.4\%$, respectively, underscoring Athena's effectiveness irrespective of the cache design.

\subsubsection{\textbf{CD4: OCP with One L1D and One L2C Prefetcher}} \label{subsubsec:ath_L1DL2C}

\begin{figure}[!ht]
    \centering
    \includegraphics[width=\columnwidth]{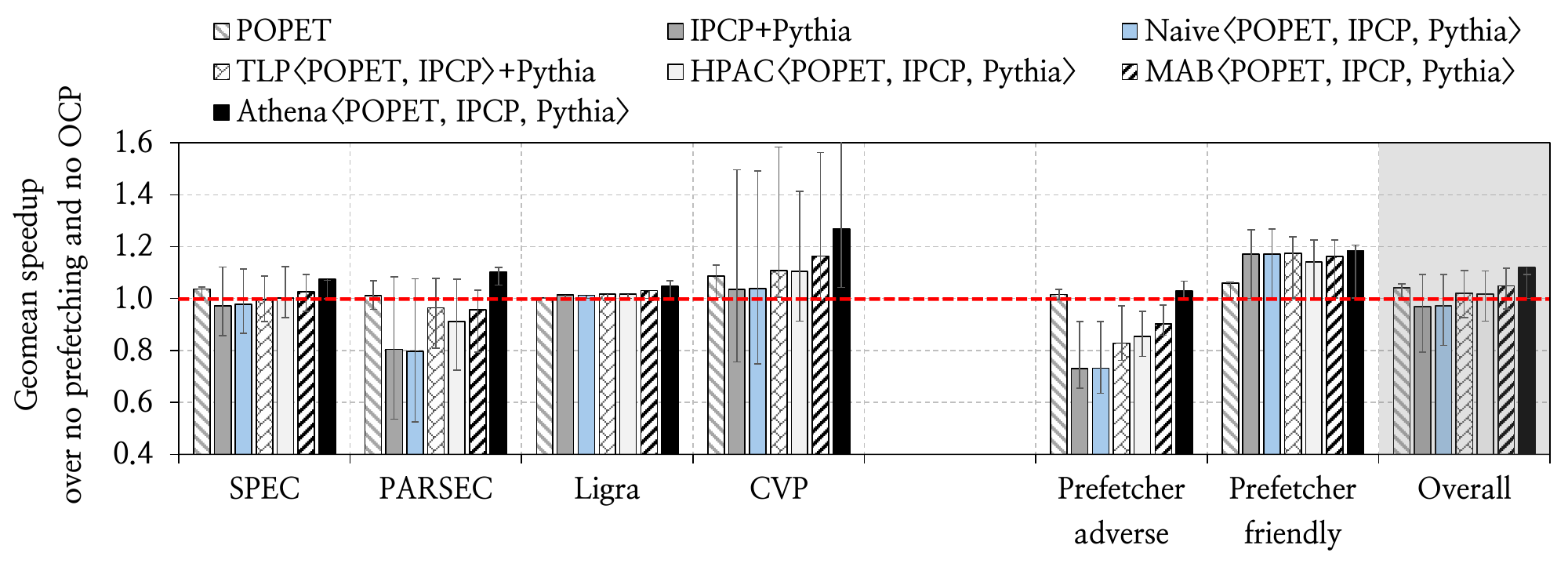}
    \caption{Speedup in cache design 4 (CD4).}
\label{fig:cd4}
\end{figure}

\Cref{fig:cd4} shows the performance improvement of Naive, TLP, HPAC, MAB, and Athena when coordinating POPET as the OCP, IPCP as the L1D prefetcher, and Pythia as the L2C prefetcher.
We make two key observations. 
First, in prefetcher-adverse workloads, enabling both prefetchers and the OCP without coordination results in a severe $26.8\%$ performance degradation, the worst among all evaluated cache designs.
TLP, due to its lack of control over the L2C prefetcher, fails to throttle harmful L2C prefetch requests, resulting in a performance degradation of $16.7\%$. 
Athena, benefiting from its flexibility, effectively coordinates prefetchers across two cache levels, significantly outperforming TLP by $19.9\%$.
Second, in prefetcher-friendly workloads, TLP closely matches Naive's performance since it inherently has no control over the L2C prefetcher.
In contrast, Athena dynamically determines that prefetching is beneficial, thereby providing higher performance compared to TLP.
Overall, Athena outperforms Naive, TLP, HPAC, and MAB on average by $14.9\%$, $9.9\%$, $10.3\%$, and $7.0\%$, respectively. 

Based on our extensive evaluation, we conclude that Athena \emph{consistently} outperforms prior coordination mechanisms (e.g., TLP, HPAC, and MAB) across diverse cache designs that employ OCP with multiple prefetchers at different cache levels.

\subsection{Performance Sensitivity Analysis in CD1} \label{sec:ath_eval_sen_cd1}

While~\Cref{sec:ath_eval_sc} demonstrates the benefits of Athena across various cache designs, this section further demonstrates Athena's adaptability by fixing the cache design to CD1 and varying the underlying L2C prefetcher and OCP type.

\subsubsection{\textbf{Effect of L2C Prefetcher Type}}
\begin{figure}[!ht]
    \centering
    \includegraphics[width=\textwidth]{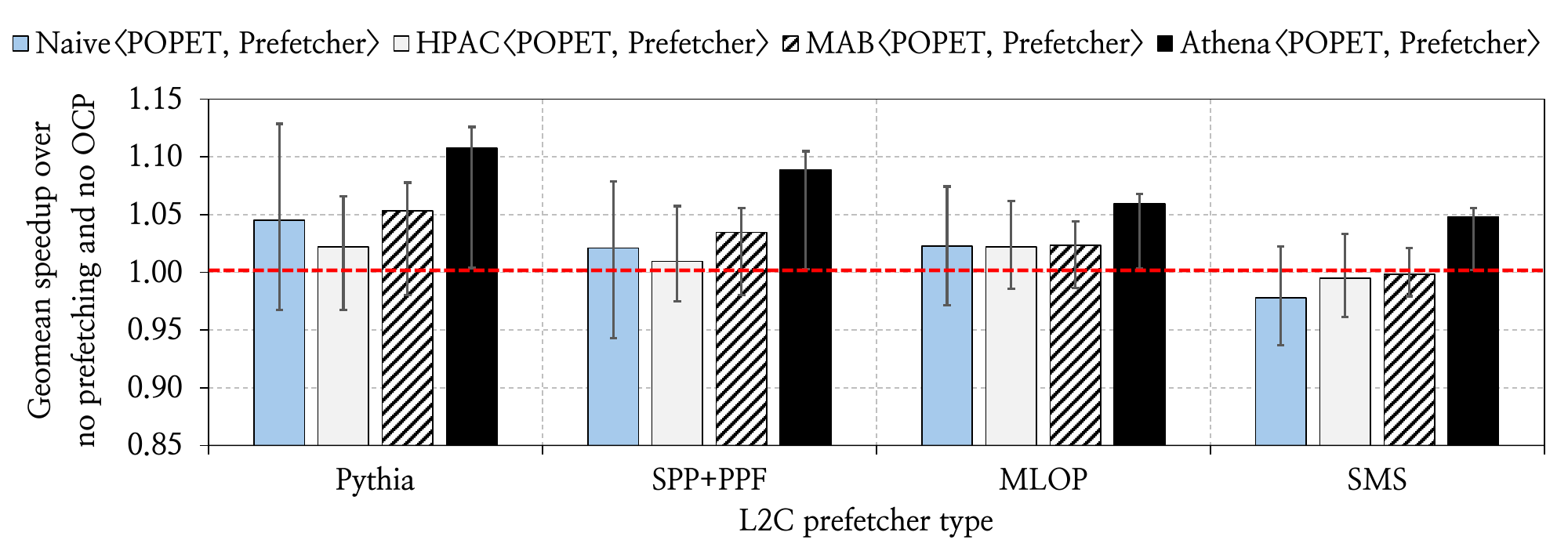}
    \caption{Performance sensitivity to underlying prefetching mechanism at L2C in CD1.}
\label{fig:cd1_sen_1}
\end{figure}

\Cref{fig:cd1_sen_1} shows the performance improvement of Naive, HPAC, MAB, and Athena across all workloads while using POPET as the OCP, and varying the underlying L2C prefetcher.
The key observation is that, by autonomously learning using system-level features and telemetry information, Athena \emph{consistently} outperforms Naive, HPAC, and MAB for \emph{every} prefetcher type, without requiring any changes to its configuration. 
On average, Athena outperforms the next-best-performing MAB by $5.0$\%, $5.4$\%, $3.6$\%, and $5.0$\%, when coordinating POPET with four types of L2C prefetchers, Pythia, SPP+PPF, MLOP, and SMS, respectively.
We conclude that Athena is able to adapt and provide consistent performance benefits across diverse prefetcher types.

\subsubsection{\textbf{Effect of Off-Chip Predictor Type}} \label{sec:ath_eval_sen_ocp}

\begin{figure}[!ht]
    \centering
    \includegraphics[width=\textwidth]{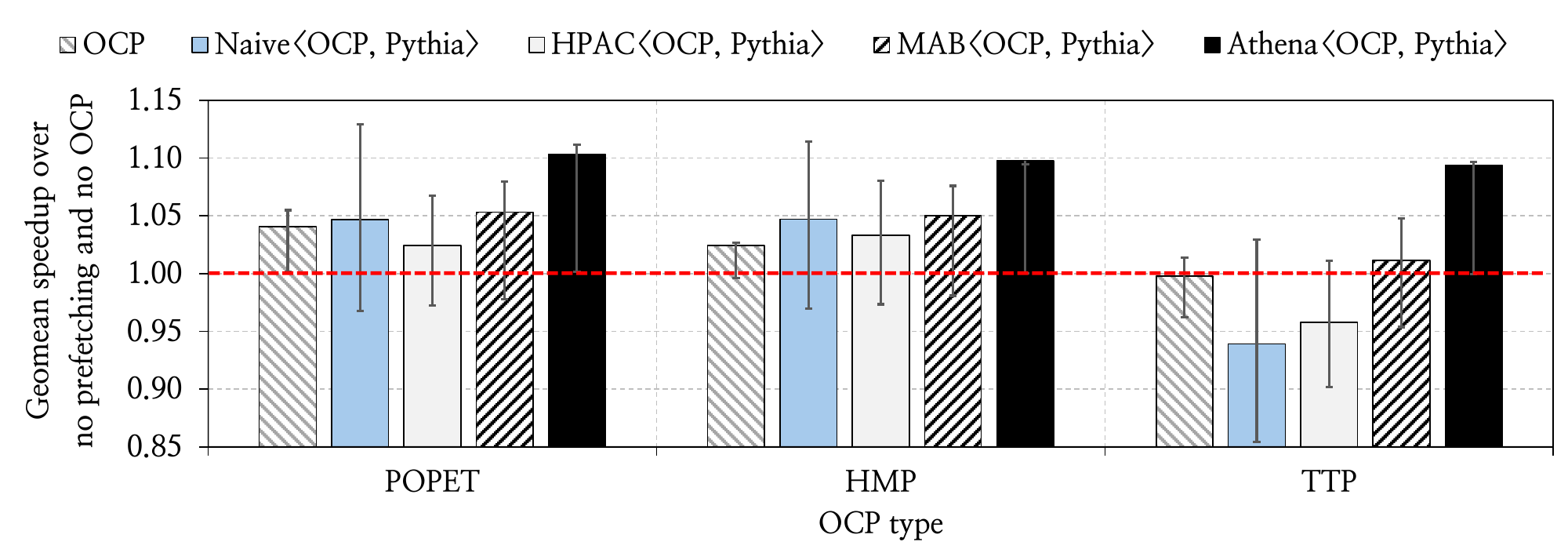}
    \caption{Performance sensitivity to off-chip prediction mechanisms in CD1.}
\label{fig:cd1_sen_2}
\end{figure}

\Cref{fig:cd1_sen_2} shows the performance improvement of Naive, HPAC, MAB, and Athena while using Pythia as the L2C prefetcher and varying the underlying OCP.
The key takeaway is that Athena \emph{consistently} outperforms Naive, HPAC, and MAB for \emph{every} OCP type. 
On average, Athena outperforms the next-best-performing MAB by $5.0\%$, $4.7\%$, and $8.2\%$, when employing POPET, HMP, and TTP as the underlying OCP, respectively.
We conclude that Athena is able to adapt and provide consistent performance benefits across diverse OCP types.

\subsubsection{\textbf{Effect of OCP Request Issue Latency}} \label{subsubsec:ath_ocp_latency}

For each load request predicted to go off-chip, an OCP issues a speculative memory request (we call it an \emph{OCP request}) directly to the main memory controller as soon as the physical address of the load becomes available. 
While an OCP request experiences substantially lower latency than a regular demand load, it still incurs a latency to traverse the on-chip network.
To faithfully evaluate OCP under a wide range of on-chip network designs, we vary the latency to directly issue an OCP request to the main memory controller (we call this \emph{OCP request issue latency}) from $6$ to $30$ cycles, in line with prior work~\cite{hermes}.

\begin{figure}[!ht]
    \centering
    \includegraphics[width=\textwidth]{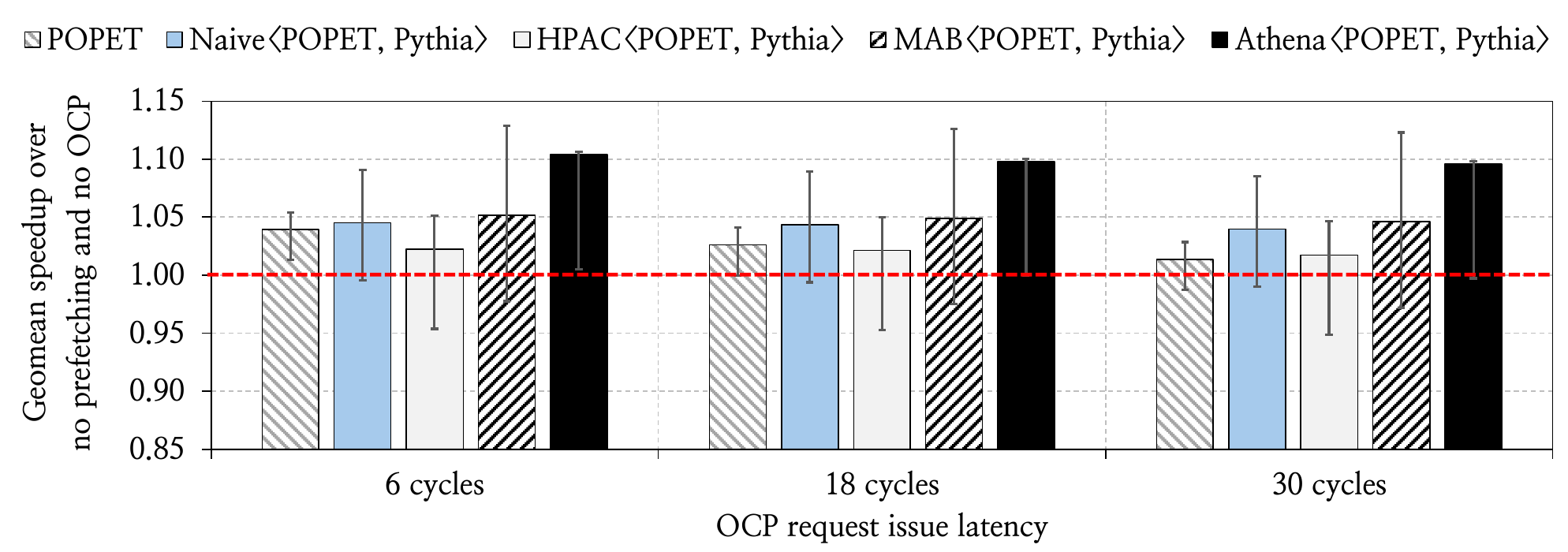}
    \caption{Performance sensitivity off-chip predicted request issue latency in CD1.}
\label{fig:cd1_sen_3}
\end{figure}

\Cref{fig:cd1_sen_3} shows the performance improvement of Naive, HPAC, MAB, and Athena across all workloads when coordinating Pythia as the L2C prefetcher, POPET as the OCP, and varying the OCP request issue latency. 
We make three key observations.
First, POPET's performance gains decrease by $2.5\%$ as the OCP request issue latency increases from $6$ to $30$ cycles, consistent with prior work~\cite{hermes}.
Second, although the overall benefit of OCP reduces with higher request latency, Athena's performance decreases by only $0.8\%$, demonstrating robust adaptability to varying request delays.
Third, Athena \emph{consistently} outperforms Naive, HPAC, and MAB for all evaluated OCP request issue latencies.
We conclude that Athena is able to adapt to diverse system configurations with variations in on-chip network design.

\subsection{Performance Sensitivity Analysis in CD4} \label{sec:ath_eval_sen_cd4}

This section further demonstrates Athena's adaptability in CD4, which employs one prefetcher each at L1D and L2C, by varying the L1D prefetcher type and main memory bandwidth.

\subsubsection{\textbf{Effect of L1D Prefetcher Type}} \label{subsubsec:ath_eva_sen_l1d}

\begin{figure}[!ht]
    \centering
    \includegraphics[width=\columnwidth]{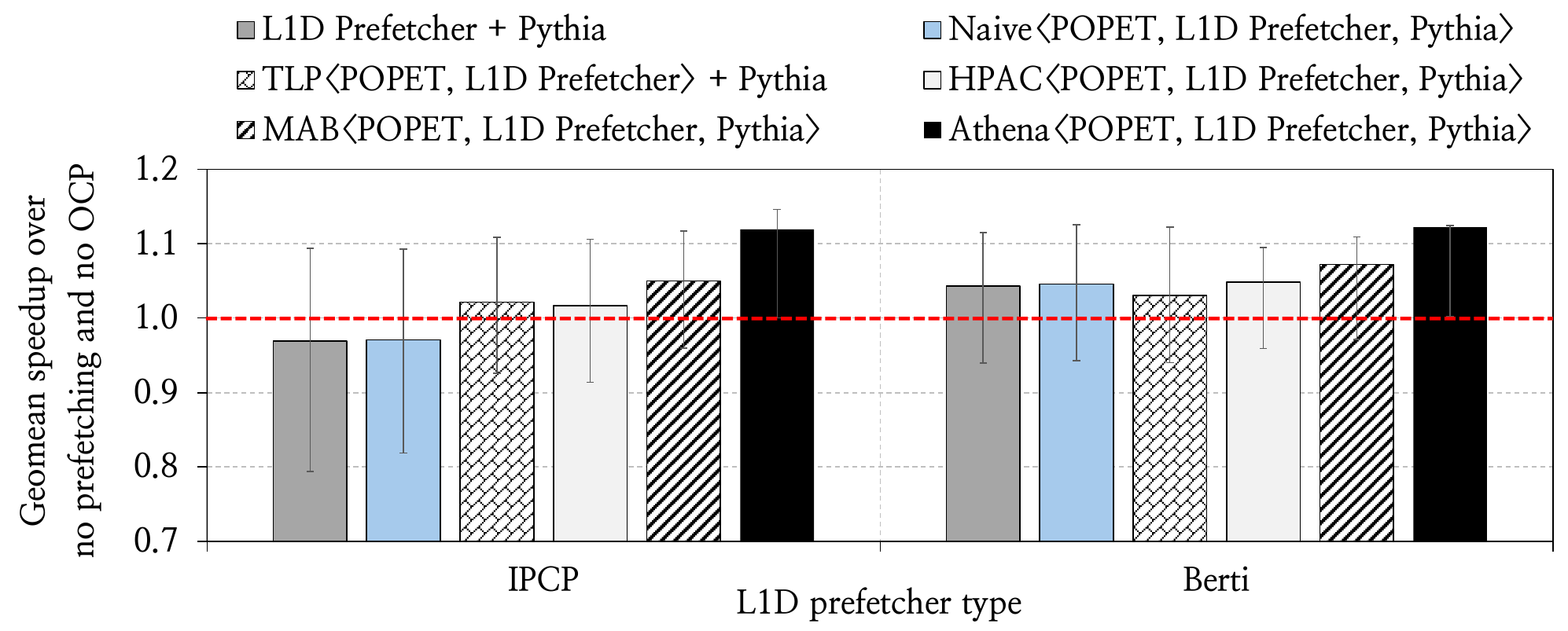}
    \caption{Performance sensitivity to prefetching mechanism at L1D in CD4.}
\label{fig:cd4_sen_1}
\end{figure}

\Cref{fig:cd4_sen_1} shows the performance improvement of Naive, TLP, HPAC, MAB, and Athena across all workloads while varying the underlying L1D prefetcher, but keeping POPET as the OCP and Pythia as the L2C prefetcher. 
We make two key observations.
First, Berti, due to its higher prefetch accuracy, provides a higher performance gain than IPCP.
Berti improves performance by $4.3\%$ on average over the baseline without any prefetcher or OCP, whereas IPCP degrades performance by $3.1\%$. 
Second, Athena \emph{consistently} improves performance, both over the baseline and prior coordination techniques, irrespective of the L1D prefetcher type.
Athena outperforms the next-best-performing MAB by $7.0\%$ and $5.0\%$ on average, while coordinating IPCP and Berti at L1D, respectively.

\subsubsection{\textbf{Effect of Main Memory Bandwidth}} \label{sec:ath_eval_sen_bw}

\begin{figure}[!ht]
    \centering
    \includegraphics[width=\columnwidth]{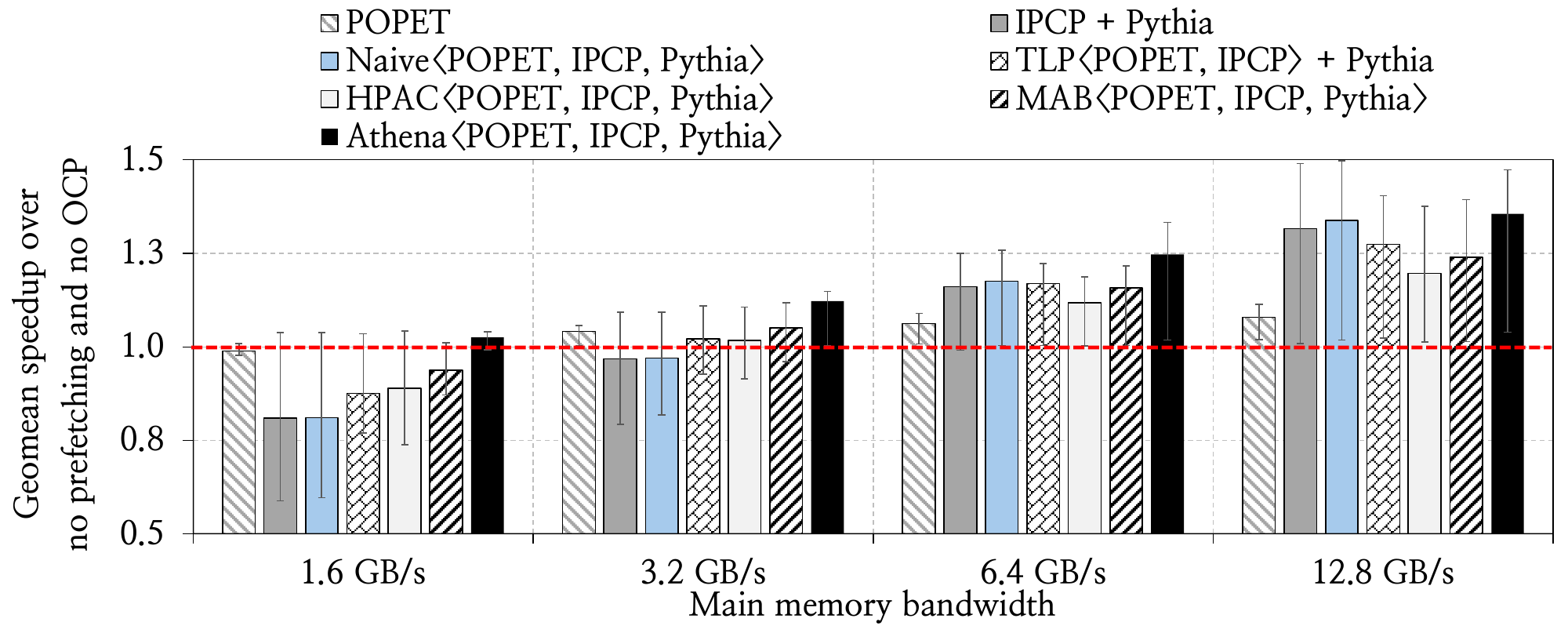}
    \caption{Performance sensitivity to main memory bandwidth in CD4.}
\label{fig:cd4_sen_2}
\end{figure}

\Cref{fig:cd4_sen_2} shows the performance improvement of Naive, TLP, HPAC, MAB, and Athena over the baseline with no prefetcher or OCP across all workloads, while varying the main memory bandwidth (measured in gigabytes per second (GB/s)).

We make three key observations.
First, while Naive significantly improves performance over the baseline in a system with ample main memory bandwidth, Naive's performance gain, which is largely dominated by the prefetchers, significantly deteriorates when the system has limited main memory bandwidth, akin to datacenter-class processors~\cite{epyc_9754, ampere_one, graviton3, bruce2023arm, neoverse2}.
For example, Naive \emph{improves} performance by $33.5\%$ on average in the system with $12.8$~GB/s main memory bandwidth, while it \emph{degrades} performance by $18.9\%$ in the system with $1.6$~GB/s main memory bandwidth.
Second, OCP alone also hurts performance in severely bandwidth-limited configurations, despite its highly accurate predictions. 
POPET degrades performance by $1.1\%$ on average over the baseline in the system with $1.6$~GB/s main memory bandwidth.
This indicates that neither prefetching nor off-chip prediction is beneficial for performance in severely bandwidth-constrained configurations.
Third, by autonomously learning using system-level features such as bandwidth usage, Athena \emph{consistently} outperforms Naive, TLP, HPAC, and MAB across \emph{all} bandwidth configurations. 
Athena's benefit is more prominent in bandwidth-constrained configurations since no static combination (i.e., POPET-alone, Pythia-alone, naive combination of POPET and Pythia, or none) is consistently good across all phases of all workloads.
However, in a system with ample bandwidth, Naive often yields good performance, and Athena correctly identifies this combination as the best-performing.
Overall, Athena outperforms Naive (MAB) by $21.4\%$ ($8.8\%$) and $1.7\%$ ($11.6\%$) in $1.6$~GB/s and $12.8$~GB/s bandwidth configurations, respectively.

\subsection{Multi-Core Evaluation Overview} \label{sec:ath_eval_mc}

\subsubsection{\textbf{Four-Core Performance Analysis}}\label{subsubsec:ath_4c}

\begin{figure}[!ht]
    \centering
    \includegraphics[width=\columnwidth]{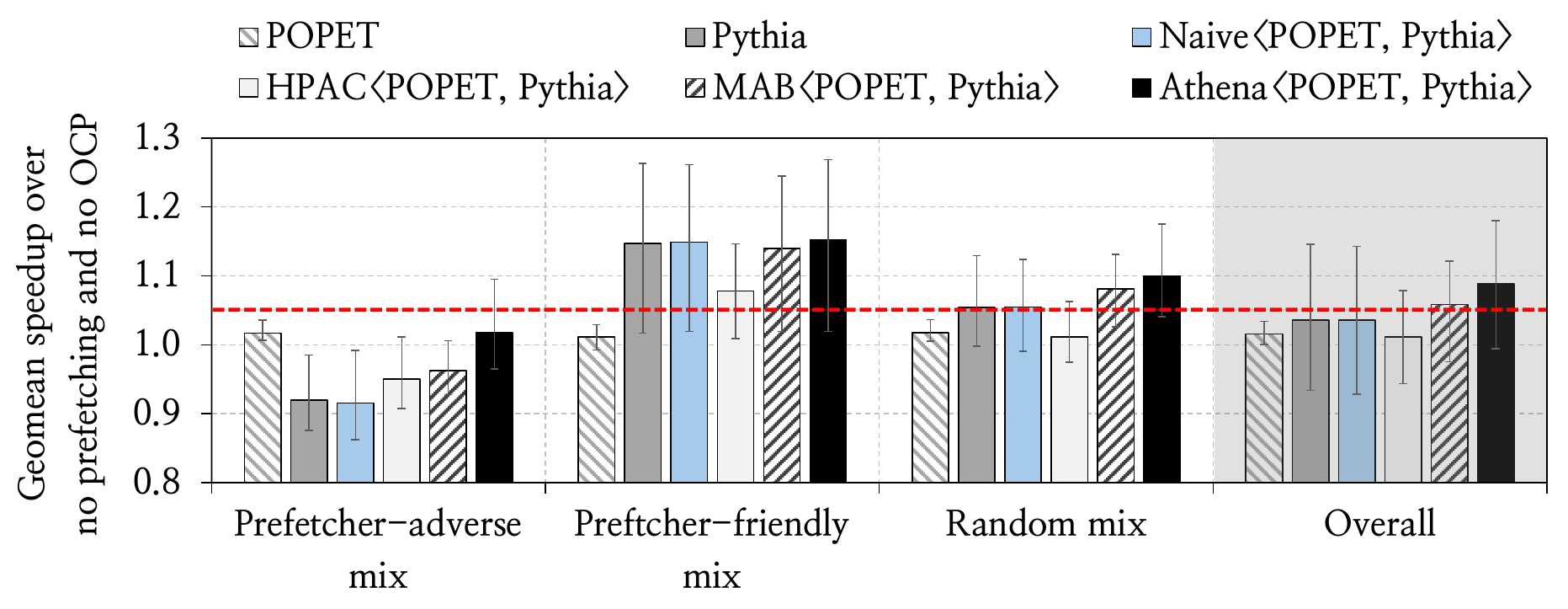}
    \caption{Speedup in four-core workloads.}
    \label{fig:ath_perf_4c}
\end{figure}

\Cref{fig:ath_perf_4c} shows the performance improvement of Naive, HPAC, MAB, and Athena when coordinating POPET as the OCP and Pythia as the L2C prefetcher in four-core workloads.

We make three key observations. 
First, across all workload mixes, Athena outperforms Naive, HPAC, and MAB by $5.3\%$, $7.7\%$, and $3.0\%$, respectively, despite utilizing hyperparameters exclusively tuned for single-core workloads (i.e., hyperparameters derived from the automated DSE described in~\Cref{subsec:ath_automated} are applied directly without alteration). 
Second, Athena \emph{consistently} outperforms \emph{all} prior coordination mechanisms in \emph{every} workload mix category.
Athena's performance gains over Naive, HPAC, and MAB are largest in prefetcher-adverse mixes, where no static configuration (i.e., POPET-alone, Pythia-alone, naive combination of POPET and Pythia, or none) is consistently good across all workload mixes. 
In contrast, Athena's relative benefit is smallest in prefetcher-friendly mixes, where the Naive combination is often beneficial for performance.
More specifically, Athena outperforms Naive (MAB) by $10.1\%$ ($5.5\%$), $0.4\%$ ($1.2\%$), and $4.5\%$ ($1.9\%$) on average in prefetcher-adverse, prefetcher-friendly, and random workload mixes, respectively. 
These results highlight Athena's ability to adaptively select effective coordination policies across diverse multi-core workload compositions.

\subsubsection{\textbf{Eight-Core Performance Analysis}} \label{subsubsec:ath_8c}

\begin{figure}[!ht]
    \centering
    \includegraphics[width=\columnwidth]{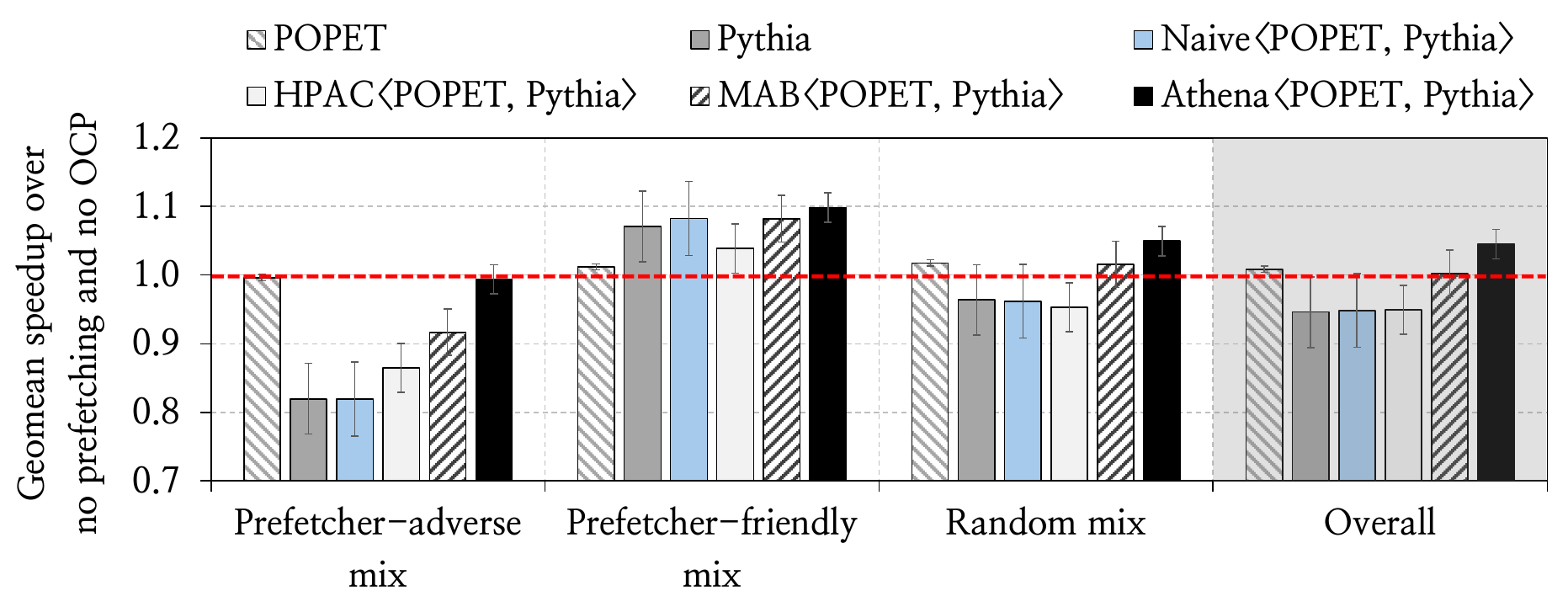}
    \caption{Speedup in eight-core workloads.}
    \label{fig:ath_perf_8c}
\end{figure}

\Cref{fig:ath_perf_8c} shows the performance improvement of Naive, HPAC, MAB, and Athena when coordinating POPET as the OCP and Pythia as the L2C prefetcher in eight-core workloads.
We highlight two key observations.
First, similar to four-core mixes, Athena consistently outperforms Naive, HPAC, and MAB by $9.7\%$, $9.6\%$, and $4.3\%$, respectively, across all eight-core mixes, despite using hyperparameters that are exclusively tuned for single-core workloads.
Second, Athena \emph{consistently} outperforms \emph{all} prior coordination mechanisms in \emph{every} workload mix category.

The multi-core results demonstrate that Athena surpasses existing coordination policies in multi-core settings, without requiring workload-specific tuning. We believe Athena could improve even further by tuning it specifically for multi-core.

\subsection{Understanding Athena} \label{subsec:ath_deepdive_athenad}

\subsubsection{Understanding Athena's Decision-Making using a Case Study} \label{sec:ath_eval_case_study}

\begin{figure}[!ht]
    \centering
    \includegraphics[width=\columnwidth]{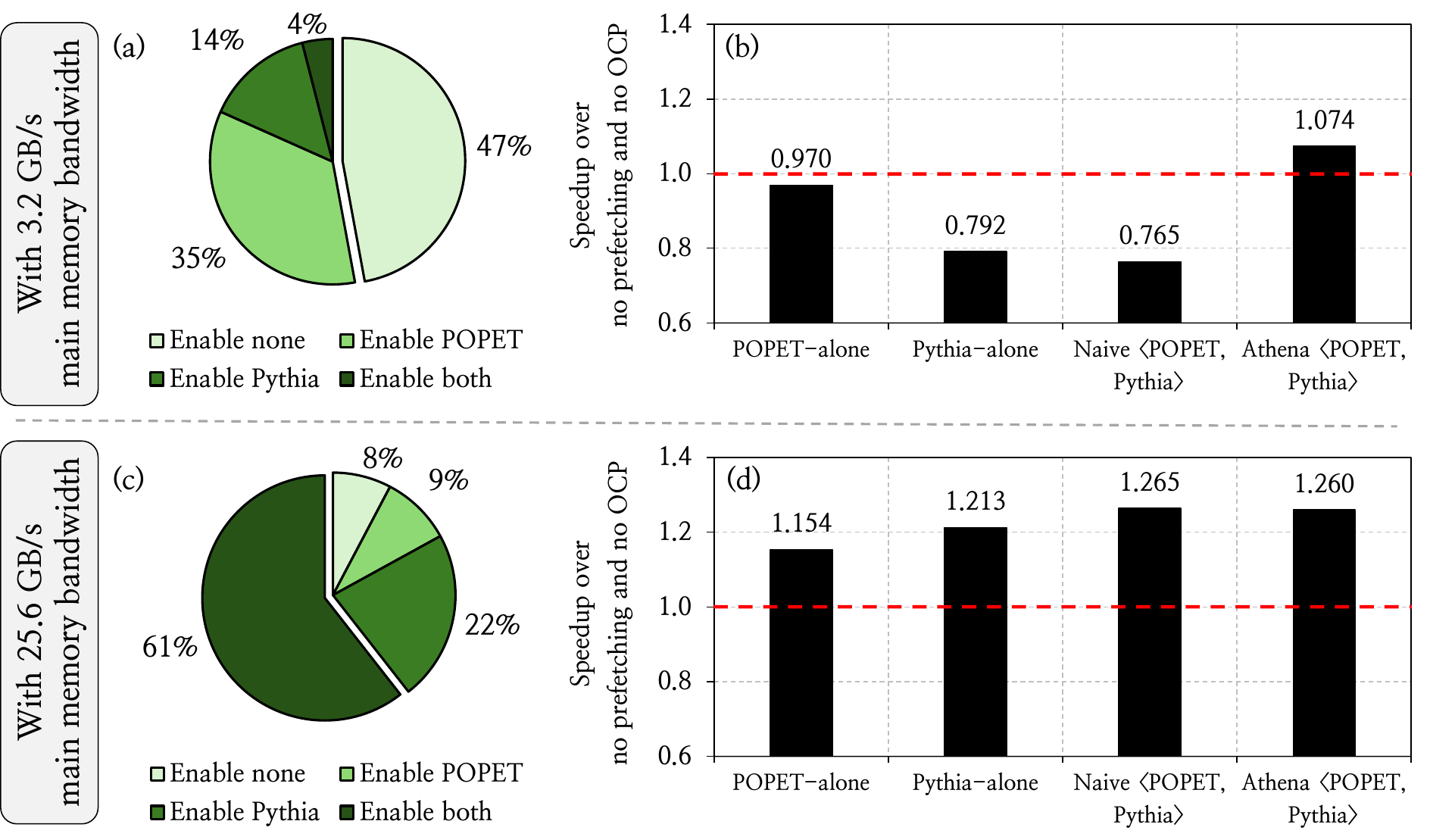}
    \caption{Distribution of Athena's action in coordinating Pythia and POPET and speedup of different Pythia-POPET combinations in \texttt{compute\_fp\_78} workload from \texttt{CVP} suite, while varying memory bandwidth: 3.2~GB/s and 25.6~GB/s.} 
    \label{fig:athena_deepdive1}
\end{figure}

To provide deeper insights into Athena's decision-making process, we analyze its actions in coordinating POPET as the OCP and Pythia as the L2C prefetcher in a representative workload, \texttt{compute\_fp\_78}, from the \texttt{CVP} suite. 
As~\Cref{fig:athena_deepdive1}(a) shows, in the system with $3.2$~GB/s main memory bandwidth, Athena disables both POPET and Pythia, or enables only POPET in $47\%$ and $35\%$ of its actions, respectively.
It enables only Pythia or both mechanisms in only $14\%$ and $4\%$ of its actions.
To find the rationale behind this action distribution, we independently evaluate the performance of three static combinations: POPET-alone, Pythia-alone, and Naive, for this workload. 
\Cref{fig:athena_deepdive1}(b) shows both Pythia-alone and Naive substantially degrade performance in the $3.2$~GB/s bandwidth configuration. 
While POPET-alone also incurs performance degradation, it does so less severely.
By selectively enabling POPET in specific epochs and disabling both mechanisms in most others, Athena effectively \emph{outperforms} all three static combinations.

When we evaluate the same workload in the system with $25.6$~GB/s main memory bandwidth, we observe that the action distribution significantly changes, and Athena strongly favors enabling \emph{both} Pythia and POPET.
As~\Cref{fig:athena_deepdive1}(c) shows, Athena enables both Pythia and POPET in $61\%$ of its actions.
The performance graph in~\Cref{fig:athena_deepdive1}(d) shows that, unlike in the $3.2$~GB/s configuration, both Pythia-alone and Naive significantly \emph{improve} performance.
We conclude that Athena is not only dynamically learning to find the best coordination, but also adapting to system configuration changes.

\subsubsection{Understanding the Sources of Athena's Performance Gains via Ablation Study} \label{subsec:ath_eval_ablation}

\begin{figure}[!ht]
\centering
\includegraphics[width=\linewidth]{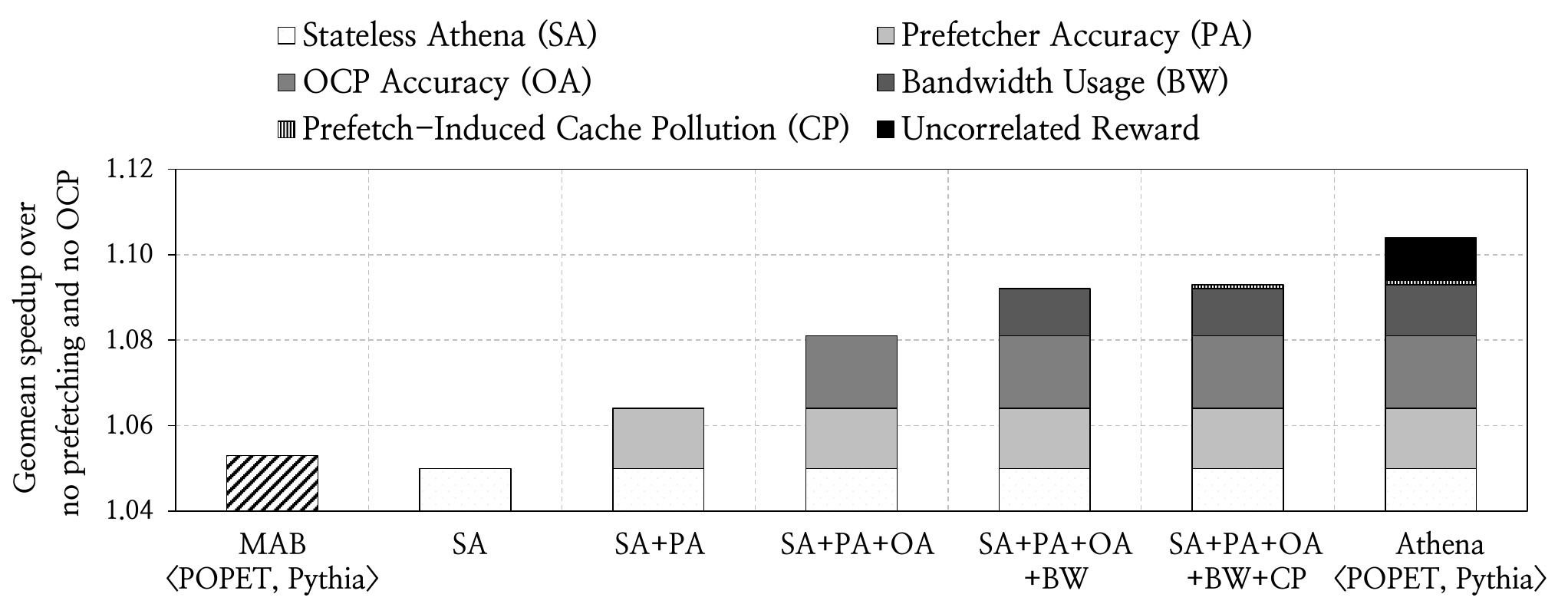}
\caption{Contribution of individual state features and reward components to Athena's geomean performance \edit[3]{across all 100 workloads}.}
\label{fig:ablation}
\end{figure}

To better understand the source of Athena's performance gains, we conduct an ablation study to evaluate the contribution of each state feature and reward component to overall performance. 
We begin with a version of Athena that uses no state information and employs only IPC as the correlated reward (see~\ref{subsec:ath_reward}). 
We call this configuration \emph{Stateless Athena}.
We then progressively introduce each state feature (i.e., prefetcher accuracy, OCP accuracy, bandwidth utilization, and prefetch-induced cache pollution), and finally include the uncorrelated reward \edit[3]{(see~\ref{subsec:ath_reward})}.
\Cref{fig:ablation} illustrates the effect of each state feature and reward component on Athena's geomean performance. 
Each bar represents Athena's performance up to and including the added state feature or reward component.

We make three key observations from~\Cref{fig:ablation}.
First, the stateless Athena, which operates without any state information, performs slightly worse than MAB, consistent with prior work~\cite{mab}. This result stems from the difference in exploration strategies: MAB employs Discounted Upper Confidence Bound (DUCB), whereas Athena uses $\epsilon$-greedy. In the stateless configuration, $\epsilon$-greedy selects random actions uniformly, with a non-decaying exploration rate, leading to slightly lower efficiency.
Second, incorporating prefetcher accuracy, OCP accuracy, bandwidth utilization, and prefetch-induced cache pollution progressively improves performance by $1.4\%$, $1.7\%$, $0.8\%$, and $0.1\%$, respectively, relative to the preceding configuration. However, as discussed in \Cref{subsec:ath_automated}, adding additional features beyond prefetch-induced cache pollution yields diminishing returns. Hence, we limit Athena's state to four features.
Finally, adding the uncorrelated reward further improves performance by $1.0\%$, highlighting that the uncorrelated reward component significantly helps improve performance by isolating the true impact of Athena's actions from inherent variations in the workload.

\subsection{Athena for Prefetcher-Only Management}\label{subsubsec:athena_pref_only}

\begin{figure}[!ht]
    \centering
    \includegraphics[width=\columnwidth]{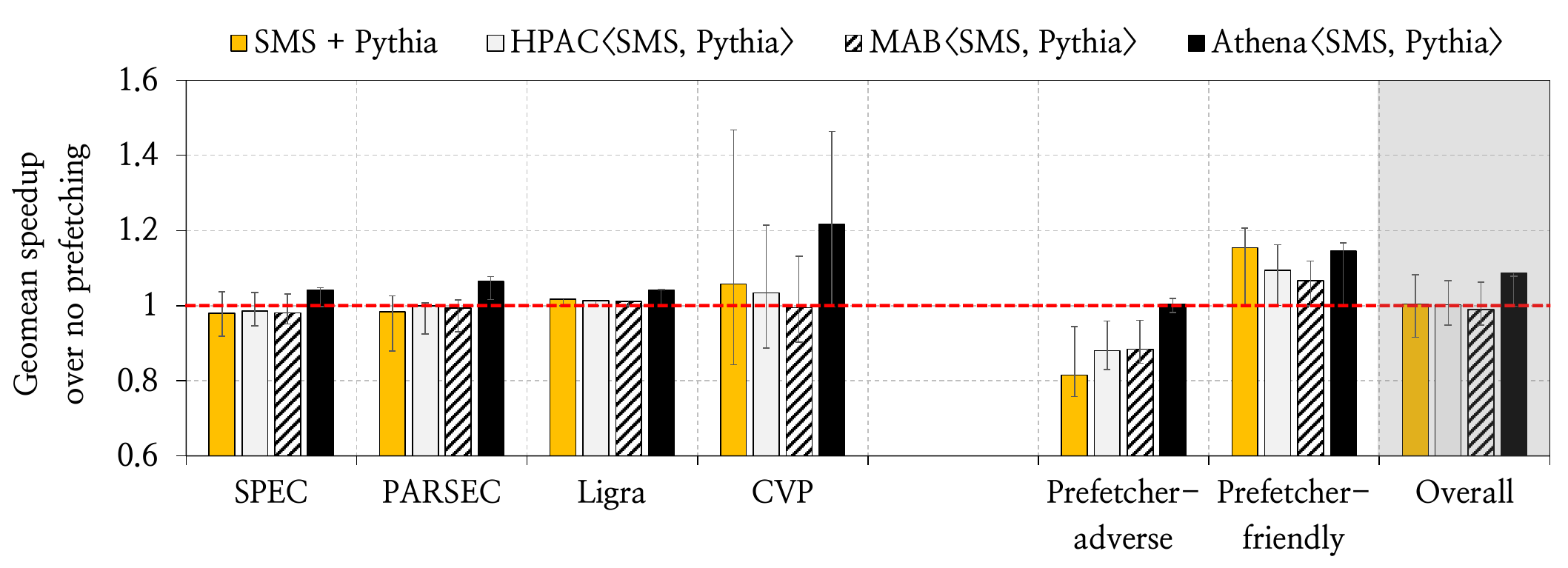}
    \caption{Geomean performance of Athena coordinating two L2C prefetchers without OCP.}
    \label{fig:generalizability}
\end{figure}

To evaluate the generality of Athena and its applicability beyond OCP-enabled systems, we conduct a generalizability study by comparing Athena against HPAC and MAB in a cache hierarchy without an OCP. 
Specifically, we evaluate Athena using a configuration that employs SMS and Pythia at the L2C (similar to CD3 in~\Cref{subsubsec:ath_2L2C}, but without OCP).
\Cref{fig:generalizability} shows the geomean performance of Naive, HPAC, MAB, and Athena, when coordinating SMS and Pythia as the L2C prefetchers.

We make two key observations.
First, in prefetcher-adverse workloads, both HPAC and MAB fail to adequately throttle prefetching, leading to performance degradation below the baseline with no prefetching. In contrast, Athena effectively mitigates these slowdowns, maintaining performance close to the baseline. However, in the absence of an OCP, which could otherwise act as a complementary mechanism, Athena only prevents performance loss as opposed to improving performance, as observed in \Cref{subsubsec:ath_2L2C}.
Second, in prefetcher-friendly workloads, Athena consistently outperforms HPAC and MAB by $5.1\%$ and $7.8\%$, respectively. 
Overall, Athena achieves $7.6\%$ and $8.8\%$ higher performance than HPAC and MAB, respectively.
We conclude that Athena generalizes across system configurations with multiple prefetchers and maintains its adaptability even when the alternative mechanism, OCP, is absent.

\subsection{Performance Evaluation using DPC4 Traces}
\label{subsec:athena_dpc4}

\begin{figure}[!ht]
    \centering
    \includegraphics[width=\columnwidth]{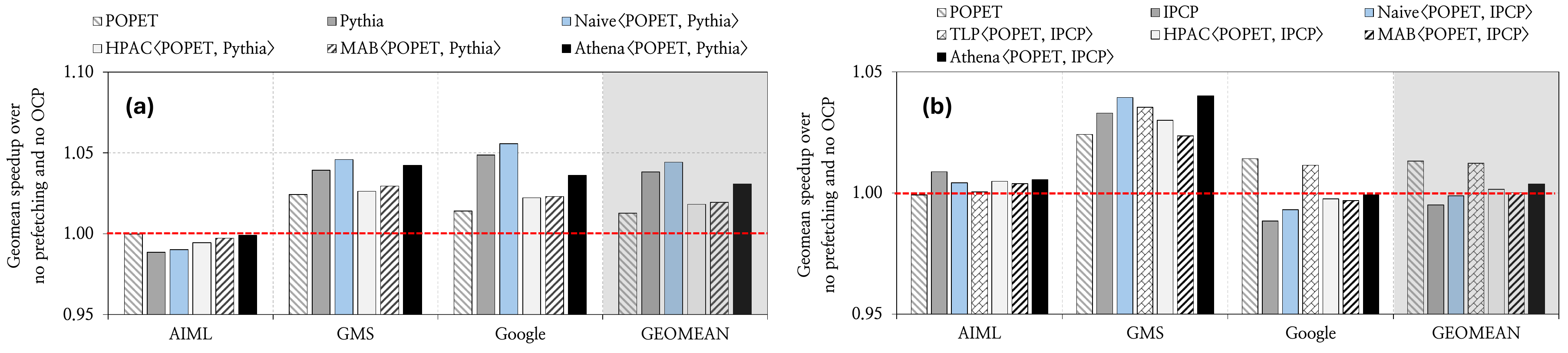}
    \caption{Speedup in (a) CD1 and (b) CD2 across 483 DPC4 traces.}
    \label{fig:ath_dpc4_cd12}
\end{figure}

\Cref{fig:ath_dpc4_cd12}(a) shows the geomean performance improvement of Naive, HPAC, MAB, and Athena when coordinating POPET as the OCP and Pythia as the
L2C prefetcher across all $483$ DPC4 workloads. The key observation is that while Athena provides the best performance gain among three coordination policies considered in this work, Athena still underperforms the naive combination of POPET and Pythia.
\rbfor{This underperformance is primarily due to the suboptimal coordination actions taken during the state-action space exploration by Athena (see~\Cref{subsubsec:rl_policy}).}
On average, Athena improves performance by $3.1\%$ on average over the baseline without any prefetcher or OCP, whereas Naive, HPAC, and MAB improve performance by $4.4\%$, $1.8\%$, and $1.9\%$.

\Cref{fig:ath_dpc4_cd12}(b) shows the geomean performance improvement of Naive, TLP, HPAC, MAB, and Athena when coordinating POPET as the OCP and IPCP as the
L1D prefetcher across all DPC4 workloads. Unlike CD1, in this case, Athena provides better performance gains than Naive, HPAC, and MAB, but falls short to TLP.
On average, Athena improves performance by $0.4\%$ on average over the baseline without any prefetcher or OCP, whereas Naive degrades performance by $0.2\%$ and TLP, HPAC, and MAB improve performance by $1.2\%$, $0.1\%$, and $0.01\%$.

\begin{figure}[!ht]
    \centering
    \includegraphics[width=\columnwidth]{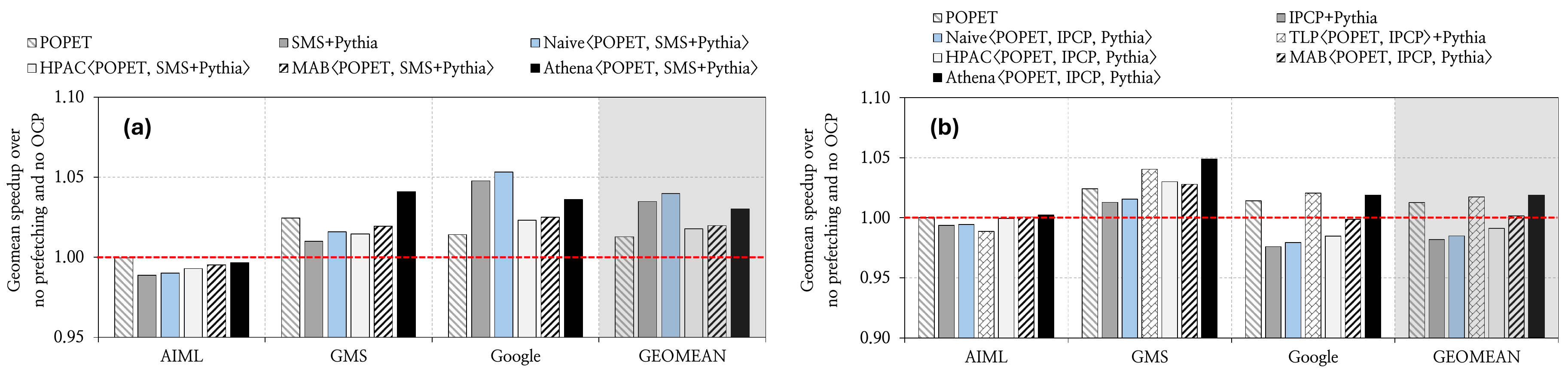}
    \caption{Speedup in (a) CD3 and (b) CD4 across 483 DPC4 traces.}
    \label{fig:ath_dpc4_cd34}
\end{figure}

\Cref{fig:ath_dpc4_cd34}(a) shows the geomean performance improvement of Naive, HPAC, MAB, and Athena when coordinating POPET as the OCP, along with
SMS and Pythia as two L2C prefetchers across all DPC4 workloads. Similar to CD1, here Athena provides the best performance gain among three coordination policies considered in this work, yet underperforms the naive combination.
On average, Athena improves performance by $3.0\%$ on average over the baseline without any prefetcher or OCP, whereas Naive, HPAC, and MAB improve performance by $4.0\%$, $1.8\%$, and $2.0\%$.

\Cref{fig:ath_dpc4_cd34}(b) shows the geomean performance improvement of Naive, HPAC, MAB, and Athena when coordinating POPET as the OCP, IPCP as the L1D prefetcher, and Pythia as the L2C prefetcher across all DPC4 workloads. Unlike all other cache designs, here in CD4 Athena provides the best performance gain among all coordination policies.
On average, Athena improves performance by $1.9\%$ on average over the baseline without any prefetcher or OCP, whereas Naive and HPAC degrade performance by $1.5\%$ and $0.9\%$, and TLP and MAB improve performance by $1.7\%$ and $0.2\%$, respectively.

Overall, these results suggest that while Athena is able to outperform prior coordination mechanisms in many cache designs without any additional fine-tuning for DPC4 traces, it can still be improved further for higher performance benefit.

\sectionRB{Athena: Summary}{Summary}{sec:ath_summary}

\noindent We introduce Athena, a reinforcement learning (RL)-based policy that coordinates data prefetchers and off-chip predictor (OCP) by autonomously learning from system behavior. 
Athena measures multiple system-level metrics (e.g., prefetcher/OCP accuracy, memory bandwidth usage) and uses them as state information to take an action: enabling the prefetcher and/or OCP and adjusting the prefetcher aggressiveness. 
Athena introduces a holistic reward framework that disentangles events correlated to its own actions (e.g., improvement in IPC) from the events that are uncorrelated to its actions (e.g., change in mispredicted branch instructions). 
This allows Athena to autonomously learn a coordination policy by isolating the true impact of its actions from inherent variations in the workload.
Our extensive evaluation 
shows that Athena \emph{consistently} outperforms a naive prefetcher-OCP combination, heuristic-based HPAC, and learning-based TLP and MAB
across a
wide range of system configurations with various combinations
of underlying prefetchers at various cache levels, OCPs, and
main memory bandwidth, while incurring only modest storage
overhead.

\subsection{Influence on the Research Community}

Athena has been presented at the 32nd International Symposium on High-Performance Computer Architecture (HPCA) on February, 2026~\cite{athena}. 
Athena has been officially artifact evaluated with all three badges (i.e., available, functional, and reproducible) and has \rbfor{been} recognized with the \textit{\textbf{Distinguished Artifact Award}} at HPCA 2026~\cite{athena_award}. 
We have made Athena freely-downloadable from our GitHub repository~\cite{hermes_github} with all evaluated workload traces, scripts, and implementation code required to reproduce and extend it.

We hope that Athena and its novel reward policy would influence future works on data-driven coordination policy design. Such policies would not only improve system performance and efficiency under a wide range of configurations, but \rbfor{would} also reduce an architect's burden in designing sophisticated control policies.

\chapterRB{Safely Eliminating Load Instruction Execution}{Improving Performance and Power Efficiency by Safely Eliminating Load Instruction Execution}
\label{chap:constable}

\noindent In the preceding chapters, we demonstrated how \emph{data-driven} prefetching (Pythia), off-chip prediction (Hermes), and their synergistic orchestration (Athena) can effectively \emph{hide} long memory access latency.
In this chapter, we shift our focus from latency-hiding mechanisms to \emph{latency-tolerance} mechanisms employed within the processor core.
We quantitatively demonstrate how conventional latency-tolerance mechanism often fail to realize their potential due to their inability to fully exploit the underlying data characteristics.
We introduce a new \emph{data-aware} technique that exploits the repetitive characteristics of load instructions, thereby unlocking performance and power efficiency benefit 
beyond what state-of-the-art mechanisms provide.

\section{Brief Background}

Extracting high instruction-level parallelism (ILP)~\cite{ilp,ilp2} is essential in providing high single-thread and multi-thread performance in modern processors~\cite{hill2008amdahl,suleman2009accelerating,joao2012bottleneck}. Unfortunately, ILP often gets limited by \emph{data dependence} and \emph{resource dependence} between instructions~\cite{tjaden1970detection,patt1985hps,jouppi1989available,smith1989limits,austin1995zero}.\footnote{
ILP also gets limited by frequent \emph{control dependence}~\cite{tjaden1970detection,patt1985hps,patt1985critical},
which is outside the scope of this work.
} 
Data dependence limits ILP due to data \rbd{flow (communications)} between instructions, whereas resource dependence (also called structural \rbd{dependence}) limits ILP due to \rbd{contention for} limited hardware resources in the system (e.g., execution unit, load port).

Load instructions are \rbd{a major} source of ILP limitation in modern workloads due to both data and resource dependence~\cite{austin1995zero}. 
Load instructions typically have longer latency than most non-memory instructions since they perform multiple component operations (i.e., address computation and data fetch) in a single instruction. This exacerbates stalls due to load data dependence, thus limiting ILP.
Load instructions also use several hard-to-scale pipeline resources (e.g., reservation station (RS) entry, ports to access address generation unit (AGU) and L1 data cache), which often cause resource dependence in the pipeline, thus limiting ILP.

Researchers have proposed numerous techniques to mitigate load data dependence by tolerating load instruction latency. \emph{Load Value Prediction} (LVP) and \emph{Memory Renaming} (MRN) are two such key techniques that mitigate load data dependence by speculatively executing load data-dependent instructions using a predicted load value.
Chapter~\ref{subsec:dep_pred} provides an extensive literature review of LVP and MRN.

\section{Motivation and Goal}

Even though LVP and MRN provide performance benefit by breaking load data dependence, the predicted load gets executed nonetheless to verify the speculated load value, which takes scarce and hard-to-scale hardware resources that otherwise could have been utilized for executing other load instructions. In other words, LVP and MRN provide performance benefits by mitigating load data dependence, but they \emph{do not} mitigate load resource dependence.

\begin{figure}[!ht]
\centering
\includegraphics[width=5in]{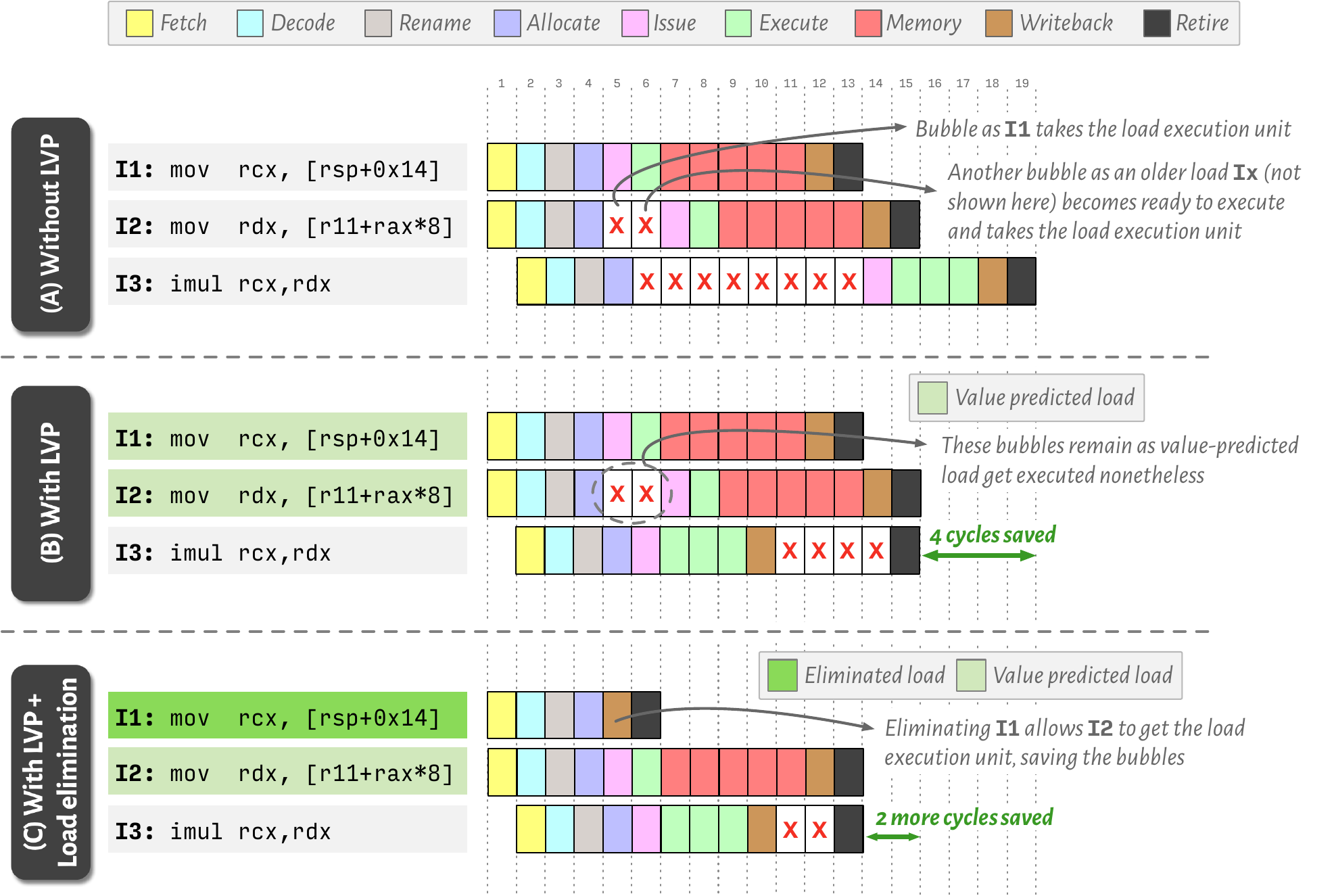}
\caption{
Execution timeline of a code \rbd{example} in a processor (a) without a load value predictor (LVP), (b) with LVP, and (c) with LVP and load elimination.
}
\label{fig:cst_elim_timeline}
\end{figure}

To illustrate how LVP provides performance benefit by mitigating data dependence,\footnote{Since LVP and MRN \rbd{work} on conceptually similar principles, we use LVP for this discussion without loss of generality.} yet the benefit may get limited by resource dependence, \Cref{fig:cst_elim_timeline}(a) and (b) show the execution timeline of a \rbd{code example} in a processor without and with LVP, respectively. 
For simplicity, we assume that the OOO processor has fetch, issue, and retire bandwidth of two instructions, and one load execution unit (comprised of an AGU and a load port).
We also assume a perfect LVP.
As \Cref{fig:cst_elim_timeline}(a) shows, $I_1$ gets issued to the load execution unit in cycle-$5$, thus stalling $I_2$. 
In cycle-$6$, an older load instruction $I_x$ (not shown in the figure) becomes ready to execute and gets issued to the load execution unit, thus stalling $I_2$ even further.
Stalls like these, where a load instruction gets delayed due to limited hardware resources (i.e., resource dependence), frequently occur in \rbd{a} modern high-performance processor with deeper and wider pipeline, as we quantitatively show in \Cref{sec:cst_headroom_resoruce_hazard}. 
These stalls get exacerbated further in \rbd{the} presence of performance-enhancement techniques like simultaneous multithreading (SMT)~\cite{smt}, where a hardware resource may get shared across SMT threads.

\rbd{When} LVP \rbd{is employed}, as shown in \Cref{fig:cst_elim_timeline}(b), both loads $I_1$ and $I_2$ get value-predicted and the data-dependent instruction $I_3$ retires 4 cycles earlier than \rbd{in} the processor without LVP.
However, since both $I_1$ and $I_2$ need to get executed to verify their respective predicted values, $I_2$ still experiences stalls in cycle-$5$ and $6$ due to resource dependence.
If we can safely eliminate \rbd{the execution of} $I_1$ while breaking its data dependence, as shown in \Cref{fig:cst_elim_timeline}(c), \rbd{we can enable} $I_2$ to get issued to the load execution unit in cycle-$5$, which provides an additional $2$ cycles savings on top of the processor with LVP.\footnote{\rbd{Similarly, $I_2$ can potentially be eliminated as well, providing further savings in execution time (not shown in \Cref{fig:cst_elim_timeline})}.}

\rbd{We conclude} that LVP and MRN may improve performance by mitigating load data dependence, but they leave performance improvement opportunity by not mitigating load resource dependence.

\subsection{Our Goal}
\textbf{Our goal} in this work is to improve ILP by mitigating \emph{both} load data dependence and resource dependence. 
To this end, we propose a \rbc{lightweight}, purely-microarchitectural technique \rbc{called} \textbf{\emph{Constable}}, which safely eliminates \rbc{the entire execution of a load instruction} (i.e., \rbc{both} load address \rbc{computation} and \rbc{data fetch from memory hierarchy}).

\sectionRB{Constable: Performance Headroom}{Performance Headroom of Constable}{sec:cst_headroom}

\noindent To understand the performance headroom of Constable, we first study the static load instructions that repeatedly fetch the same value from the same load address \emph{across the entire workload trace}. 
We call such a load \emph{\rbe{global-stable}}. 
Essentially, a \rbe{global-stable} load is a prime candidate for elimination since both load address computation and \rbc{data fetch} operations of its execution produce \rbd{the same} result across all dynamic instances of the instruction. 
We then quantify the resource dependence on \rbe{global-stable} loads \rbd{(\Cref{sec:cst_headroom_resoruce_hazard})},
\rbd{and} the performance benefit of \rbc{ideally} eliminating all \rbe{global-stable} load execution \rbd{(\Cref{sec:cst_headroom_perf_headroom})}.

\begin{figure*}[!ht]
\centering
\includegraphics[width=5in]{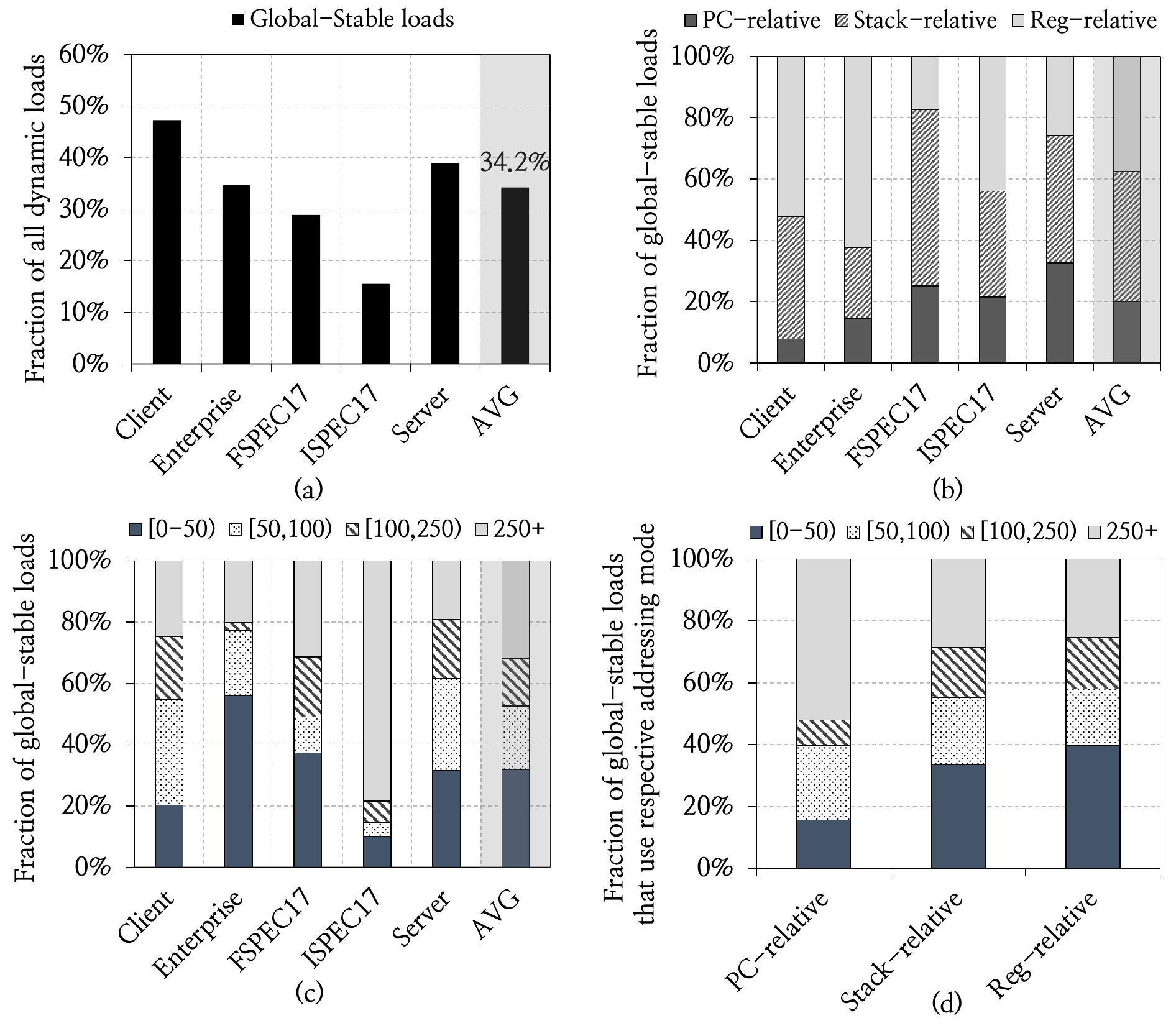}
\caption{
(a) Fraction of dynamic loads that are \rbe{global-stable}. Distribution of \rbe{global-stable} loads by their (b) addressing mode and (c) inter-occurrence distance. (d) Distribution of inter-occurrence distance of \rbe{global-stable} loads from each addressing mode.
}
\label{fig:cst_stable_load_characterization}
\end{figure*}

\subsection{\rbe{Global-Stable} Loads in Real Workloads} \label{sec:cst_headroom_stable_loads_why}

Intuitively, \rbe{global-stable} load instructions would be hard to find in real workloads since such instructions should already be optimized by the compiler. However, we observe that a significant fraction of load instructions in real workloads are \rbe{global-stable} \rbd{even after aggressive compiler optimizations applied}. \Cref{fig:cst_stable_load_characterization} shows the fraction of \rbc{dynamic} load instructions that are \rbe{global-stable} on average across $90$ workloads divided into five categories. \Cref{sec:cst_constable_methodology} discusses our evaluation methodology.
We make two key observations. First, $34.2\%$ of all dynamic loads are \rbe{global-stable}. Second, the fraction of \rbe{global-stable} loads are much higher in \texttt{Client}, \texttt{Enterprise}, and \texttt{Server} workloads as compared to SPEC CPU 2017 workloads (i.e., \texttt{ISPEC17} and \texttt{FSPEC17} categories). \rbd{We} conclude that \rbe{global-stable} load instructions are relatively abundant in real workloads.

\subsubsection{Characterization of \rbe{Global-Stable} Loads} \label{sec:cst_headroom_stable_load_charac}
To understand the source of the \rbe{global-stable} loads in workloads (e.g., accessing variables in global scope, memory accesses in a tight loop),
we further characterize these loads by their addressing mode and the inter-occurrence distance (i.e., the number of instructions between two successive dynamic instances of the same \rbe{global-stable} load instruction).
\mbox{\Cref{fig:cst_stable_load_characterization}}(b) shows the breakdown of \rbe{global-stable} loads based on their addressing mode. The key takeaway is that \rbe{global-stable} loads use various different addressing modes. On average, $20\%$, $42.6\%$, and $37.4\%$ of all \rbe{global-stable} loads use PC-relative (e.g., loads that access variables in the global scope), stack-relative (i.e., loads that access stack segment using RSP or RBP as their only source register), and register-relative (i.e., loads that use \rbd{other} general-purpose architectural registers as their source) addressing. 
\mbox{\Cref{fig:cst_stable_load_characterization}}(c) shows the breakdown of \rbe{global-stable} loads based on their inter-occurrence distance. The key takeaway is that \rbe{global-stable} loads have a bimodal inter-occurrence distance distribution. $31.9\%$ of \rbe{global-stable} loads reoccur within $50$ instructions (e.g., loads in a tight loop) on average, whereas $31.8\%$ loads reoccur more than $250$ instructions away (e.g., accessing a global-scope variable across function calls). 
\Cref{fig:cst_stable_load_characterization}(d) further shows the distribution of inter-occurrence distance of \rbe{global-stable} loads from each addressing mode. 
As we can see, \rbe{global-stable} loads that use PC-relative addressing have long inter-occurrence distance ($52\%$ of these loads have inter-occurrence distance of $250$ or more instructions), whereas \rbe{global-stable} loads that use register-relative addressing have short inter-occurrence distance ($39.6\%$ of these loads have inter-occurrence distance of less than $50$ instructions).

\rbd{We} conclude with three key takeaways. 
First, \rbe{global-stable} load instructions pose diverse characteristics, both in addressing mode and inter-occurrence distance.
Second, the inter-occurrence distance of \rbe{global-stable} loads changes significantly depending on their addressing mode.
Third, an effective load elimination technique should capture elimination opportunities across both short and long inter-occurrence distances.

\subsection{Why Do \rbe{Global-Stable} Loads Exist?} \label{sec:cst_headroom_workload_analysis}
\begin{figure*}[!ht]
\centering
\includegraphics[width=\columnwidth]{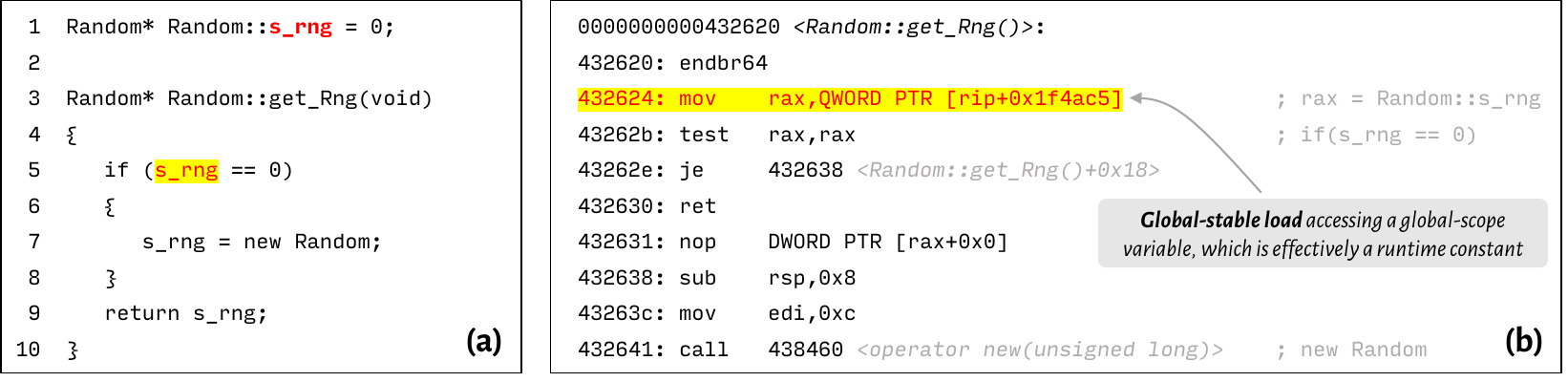}
\caption{
Code \rbd{example} and disassembly from \texttt{541.leela\_r} of SPEC CPU 2017 suite. The highlighted load instructions are \rbe{global-stable}.
}
\label{fig:cst_stable_load_example_leela}
\end{figure*}

To understand why \rbd{a} compiler \rbd{with aggressive optimization} \rbd{fails to avoid} \rbe{global-stable} load instructions, we use a custom-made binary instrumentation tool\footnote{We call this tool \emph{Load Inspector}, which is freely available at \url{https://github.com/CMU-SAFARI/Load-Inspector}.} to analyze the disassembly of workload binaries compiled with full optimization \rbd{using} a state-of-the-art off-the-shelf compiler.
\Cref{fig:cst_stable_load_example_leela}(a) and (b) show a code \rbd{example} from \texttt{541.leela\_r} from SPEC CPU 2017~\cite{spec2017} benchmark suite and its disassembly, respectively. The workload is compiled using the latest GNU g++-13.2 compiler~\cite{gcc_132} at full optimization (i.e., using -O3 flag~\cite{gcc_o3}) for x86-64 instruction set architecture to produce the most optimized binary.  
The highlighted load instruction in \Cref{fig:cst_stable_load_example_leela} fetches the object pointer \texttt{s\_rng} from memory.
Since \texttt{s\_rng} gets initialized only once at the beginning of the workload, the pointer variable effectively acts as a runtime constant and thus the highlighted load instruction is \rbe{global-stable}. 
The compiler could not eliminate this load instruction since it cannot reserve an architectural register across the global scope of the program to be reused for accessing the \texttt{s\_rng} pointer.

\begin{figure*}[!ht]
\centering
\includegraphics[width=\columnwidth]{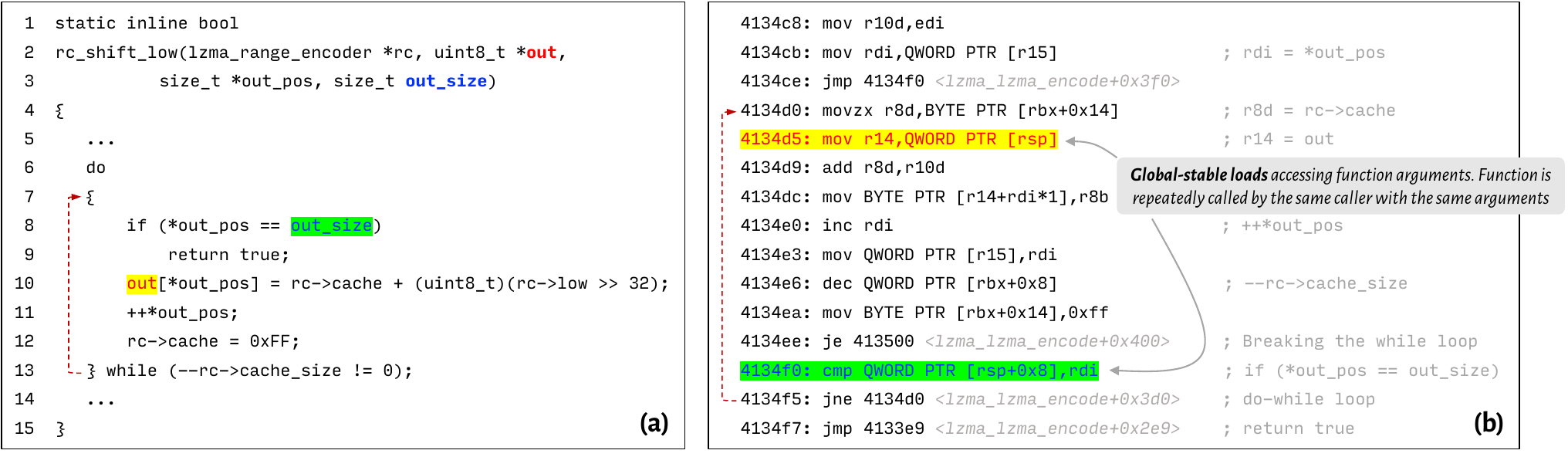}
\caption{
Code \rbd{example} and disassembly from \texttt{557.xz\_r} of SPEC CPU 2017 suite. The highlighted load instructions are \rbe{global-stable}.
}
\label{fig:cst_stable_load_example_xz}
\end{figure*}

\Cref{fig:cst_stable_load_example_xz}(c) and (d) show two more examples of \rbe{global-stable} load instructions \rbd{in} a code \rbd{example} from \texttt{557.xz\_r} of SPEC CPU 2017 suite and its disassembly, compiled \rbd{in} the same way as the previous workload. 
\rbd{Each} highlighted load instruction accesses an argument variable to the function \texttt{rc\_shift\_low} that does not change during the function invocation.
Since the function is repeatedly called using the same \rbd{arguments} from the same caller function throughout the workload trace, both the load instructions act as \rbe{global-stable} loads. 
However, as the function \texttt{rc\_shift\_low} gets \emph{inlined} within the body of its caller function (not shown here), the compiler could not allocate architectural registers to store and reuse these variables due to register pressure~\cite{reg_spill}.

We \rbd{conclude} that a state-of-the-art off-the-shelf compiler often fails to optimize \rbe{global-stable} loads due to various empirically-observed reasons, such as accessing runtime constants and local variables of inline functions, \rbd{combined with a limited number of architectural registers}.

\subsection{Can Increasing Architectural Registers Eliminate \rbe{Global-Stable} Loads at Compile Time?} \label{sec:cst_arxiv_incr_arch_reg}

Increasing architectural registers enables a compiler to exploit additional registers to capture data reuse, which otherwise would have been reused via memory. Thus increasing architectural registers typically reduces the number of load and store instructions in a program.
To understand the effect of increasing architectural registers on \rbe{global-stable} load instructions, we compile all C/C++-based workloads from SPEC CPU 2017 rate suite~\cite{spec2017} without and with Intel APX extension~\cite{intel_apx}, that doubles the number of architectural registers in x86-64 ISA from 16 to 32, using Clang 18.1.3~\cite{clang_18}.
We run these workloads using the test input and profile their \emph{end-to-end execution} using the Load Inspector tool (see \Cref{sec:cst_headroom_stable_loads_why}) to observe (1) the reduction in dynamic loads caused by APX, and (2) the fraction of all dynamic loads that are \rbe{global-stable} in workloads without and with APX.

\begin{figure}[!ht] 
\centering
\includegraphics[width=5.75in]{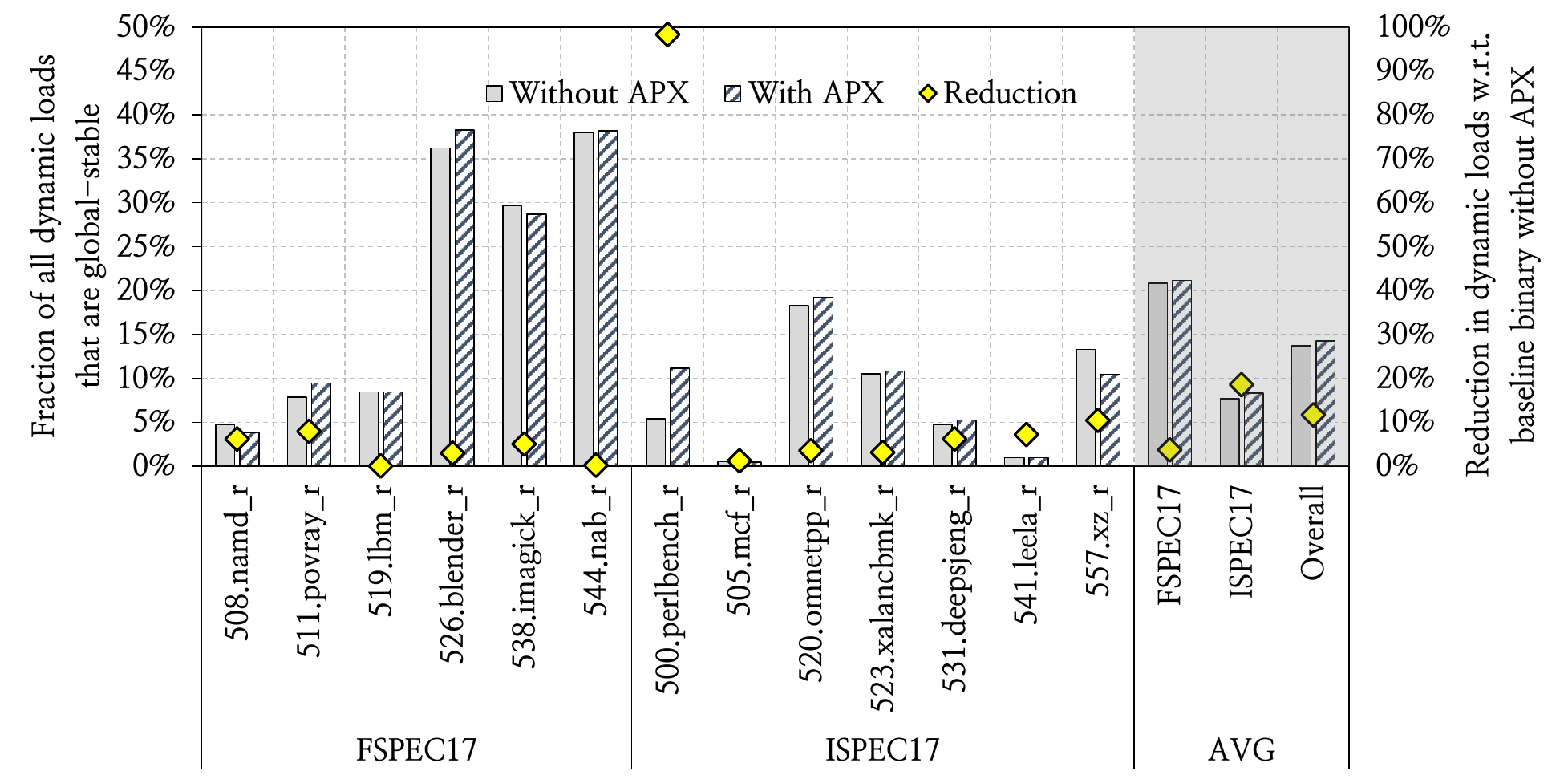}
\caption{Fraction of all dynamic loads that are \rbe{global-stable} in workloads compiled without and with APX (on the left y-axis) and the reduction in dynamic loads by APX (on the right y-axis).}
\label{fig:cst_apx_no_apx}
\end{figure}

\Cref{fig:cst_apx_no_apx} shows the fraction of all dynamic loads that are \rbe{global-stable} in each workload without and with APX (as bar graph on the left y-axis) and the reduction in dynamic loads with APX extension (as markers on the right y-axis) for all C/C++-based workloads from SPEC CPU 2017 suite.\footnote{\texttt{502.gcc\_r}, \texttt{510.parest\_r} and \texttt{525.x264\_r} are omitted from this study due to failed compilation using Clang.}

We make two key observations from \Cref{fig:cst_apx_no_apx}.
First, the fraction of dynamic loads that are \rbe{global-stable} (i.e., the elimination opportunities for Constable) is much higher than the reduction in dynamic loads by doubling the number of architectural registers. APX reduces the number of dynamic loads by $11.7\%$ on average. \texttt{500.perlbench\_r} is an outlier that observes a reduction of $98.3\%$ of loads. Without it, APX reduces the dynamic loads by only $4.5\%$ on average. On the other hand, $13.7\%$ and $14.2\%$ of all dynamic loads on average are \rbe{global-stable} in workloads without and with APX, respectively.\footnote{Note that, the \rbe{global-stable} load fraction reported here is slightly lower than that reported in \Cref{fig:cst_stable_load_characterization}(a). This is because, the earlier study in \Cref{fig:cst_stable_load_characterization}(a) uses the representative sections of the workloads (see~\Cref{sec:cst_methodology_workloads}) to limit the simulation overhead, while this study instruments each workload \emph{end-to-end}.}
The difference is more prominent for \texttt{FSPEC17} workloads, where APX reduces dynamic loads by only $3.7\%$, whereas $20.8\%$ and $21.2\%$ of dynamic loads are \rbe{global-stable} in without and with APX, respectively.
Second, the fraction of dynamic loads that are \rbe{global-stable} is nearly the same in workloads without and with APX. \texttt{500.perlbench\_r} and \texttt{557.xz\_r} are only two workloads that show more than $3\%$ absolute change in the \rbe{global-stable} load fraction.
This shows that the elimination opportunities for Constable is largely orthogonal to the benefits of increasing architectural registers.

\begin{figure}[!ht] 
\centering
\includegraphics[width=5.75in]{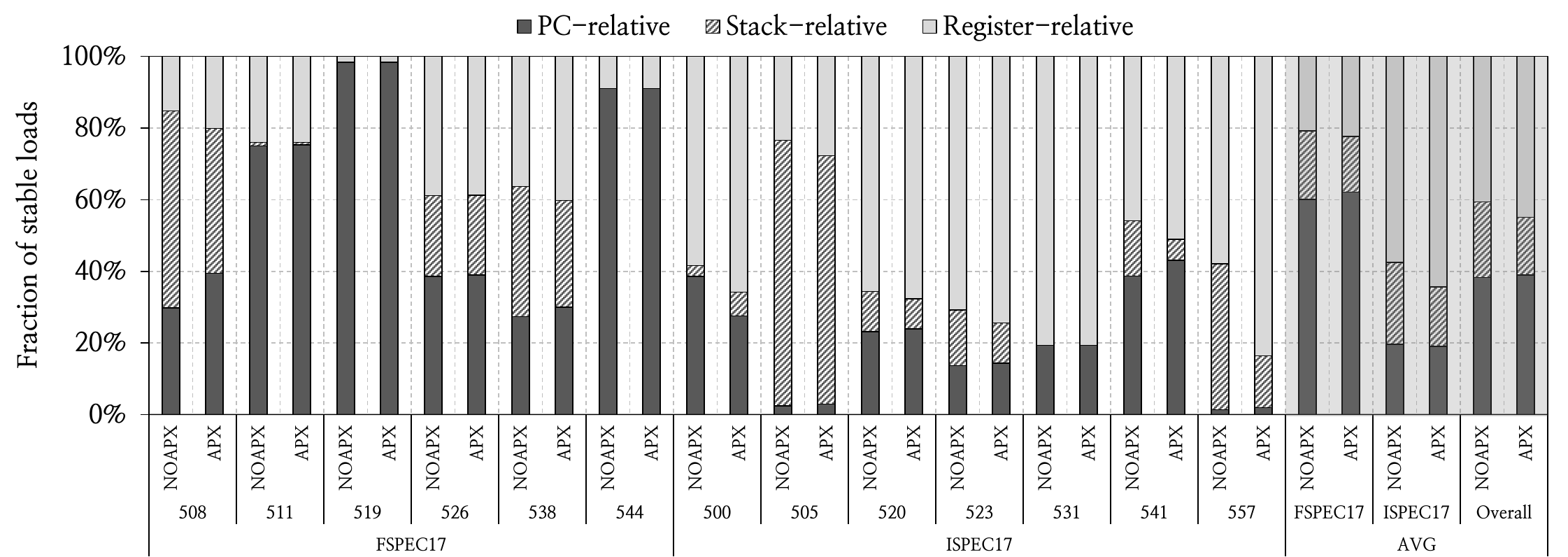}
\caption{Distribution of \rbe{global-stable} loads by their addressing modes in workloads without and with APX. Each number on the x-axis corresponds to the respective workload from SPEC CPU 2017 suite.}
\label{fig:cst_apx_no_apx_addr_mode}
\end{figure}

To further analyze the change in characteristics of \rbe{global-stable} loads in presence of APX, we break down the \rbe{global-stable} loads based on their addressing modes in workloads both without and with APX in \Cref{fig:cst_apx_no_apx_addr_mode}.
We make two key observations from this figure.
First, the fraction of stack-relative \rbe{global-stable} loads reduces in presence of APX. On average, $21.1\%$ and $16\%$ of all \rbe{global-stable} loads use stack-relative addressing in workloads without and with APX, respectively. This is expected, since increasing architectural registers predominantly reduces stack loads.
Second, the fraction of PC-relative \rbe{global-stable} loads stays nearly the same in presence of APX ($38.3\%$ without APX as compared to $38.9\%$ with APX). This shows that doubling architectural registers alone cannot eliminate all memory accesses to global-scope variables which are effectively runtime constant.

Based on these results, we conclude that the two load elimination techniques - at compile time by increasing architectural registers and at runtime by Constable - are largely \emph{orthogonal} to each other. Thus, Constable would likely be equally-performant and power-efficient in presence of increased architectural registers, as it is with the current set of architectural registers.

\subsection{Resource Dependence on \rbe{Global-Stable} Loads} \label{sec:cst_headroom_resoruce_hazard}

\begin{figure}[!ht] 
\centering
\includegraphics[width=5.75in]{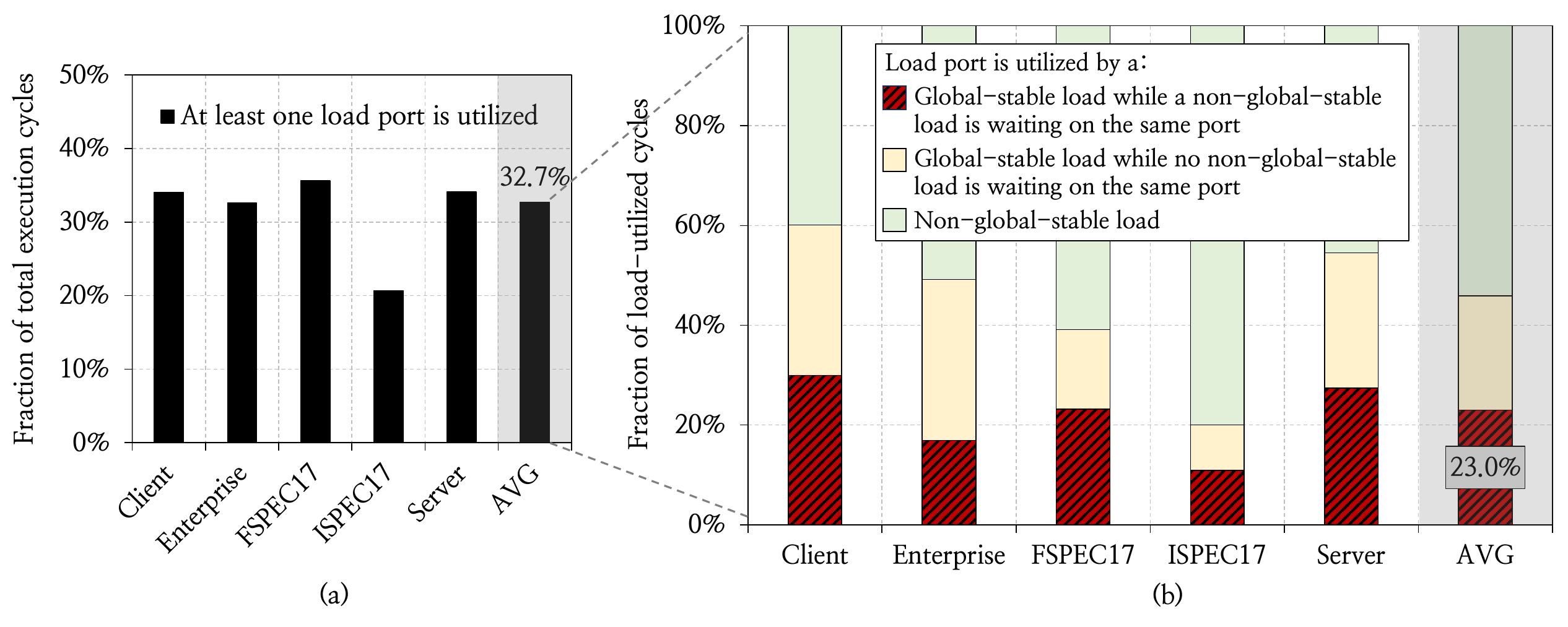}
\caption{(a) Fraction of total execution cycles where at least one load port is utilized (we call such cycles \emph{load-utilized}). (b) Categorization of load-utilized cycles based on whether or not a \rbe{global-stable} load utilizes a load port.}
\label{fig:cst_load_port_util}
\end{figure}

To quantify the loss of ILP due to resource dependence stemming from \rbe{global-stable} loads, we analyze the utilization of load ports. 
\rbd{Other} hardware resources that are used during a load execution (e.g., RS entry and AGU port) may also cause resource dependence, but we omit them here due to brevity.
\mbox{\Cref{fig:cst_load_port_util}}(a) shows the fraction of total execution cycles where at least one load port is utilized (we call such cycles \emph{load-utilized}), in our baseline processor\footnote{Our baseline processor has an issue width of six instructions per cycle with three AGU and \rbd{three} load ports, which support a maximum throughput of three loads per cycle.} augmented with a state-of-the-art load value predictor EVES~\cite{eves}.
As we can see, on average $32.7\%$ of the total execution cycles are load-utilized. 
\mbox{\Cref{fig:cst_load_port_util}}(b) further categorizes the load-utilized cycles of each workload category based on whether or not a \rbe{global-stable} load utilizes a load port.
As we can see, for $23.0\%$ of all load-utilized cycles, a \rbe{global-stable} load takes a load port for its execution, while a non-\rbe{global-stable} load (i.e., a static load instruction that does not fetch the same value from the same load addresses across all dynamic instances) is waiting to be scheduled on the same port.
Had the execution of \rbe{global-stable} loads be eliminated, the non-\rbe{global-stable} loads could have been scheduled faster, which in turn would provide performance benefit.
Thus, we conclude that \rbe{global-stable} load instructions causes significant resource dependence, which can be mitigated by eliminating their execution altogether.

\subsection{Performance Headroom}
\label{sec:cst_headroom_perf_headroom}

\begin{figure}[!ht]
\centering
\includegraphics[width=5.75in]{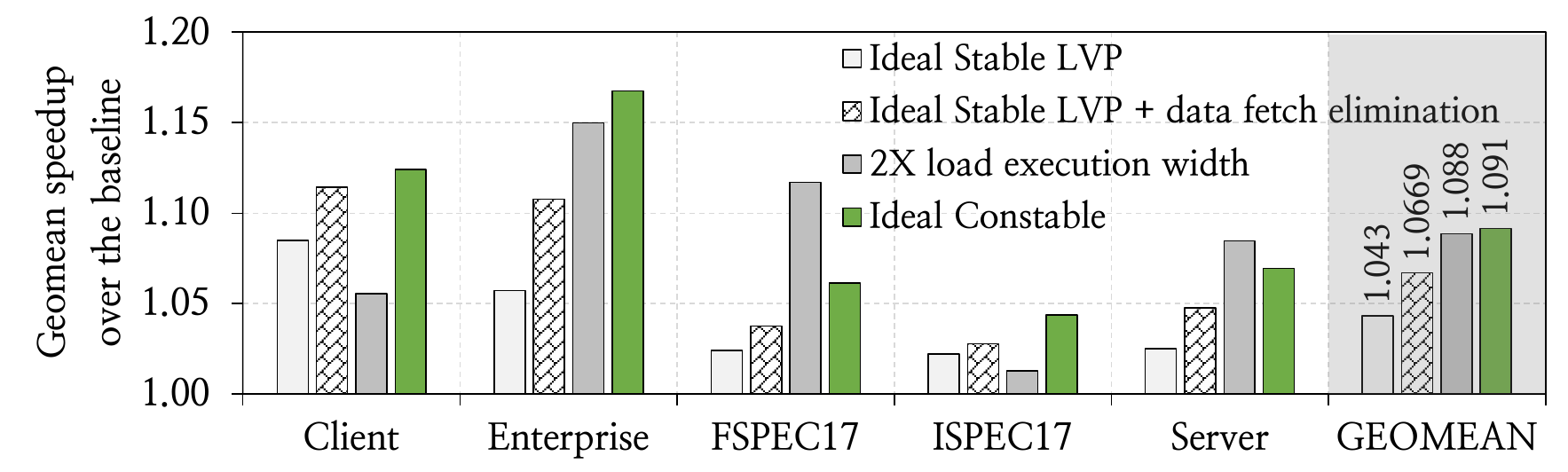}
\caption{Speedup of Ideal Constable against \rbd{Ideal Stable LVP} and a processor with $2\times$ load execution width of the baseline.}
\label{fig:cst_perf_headroom}
\end{figure}

To measure the performance headroom of eliminating \rbe{global-stable} loads, we model an \emph{Ideal Constable} configuration that identifies \rbd{\emph{all}} \rbe{global-stable} load instructions offline and eliminates both component operations of \rbd{their} execution (i.e., load address \rbc{computation} and data \rbc{fetch}).
We also compare Ideal Constable's performance against three other configurations: 
(1) \emph{Ideal \rbd{Stable} LVP}, where \emph{\rbd{all} \rbe{global-stable}} load instructions identified offline are perfectly value predicted, \rbd{and} they are are also executed to verify the predictions, 
(2) \emph{Ideal \rbd{Stable} LVP with data fetch elimination}, where all \rbe{global-stable} load instructions are perfectly value predicted, \rbd{and} the value-predicted loads are executed \rbd{only until the end of} address generation, and 
(3) \emph{$2\times$ load execution width} configuration, where the number of load execution units are doubled \rbd{over} the baseline.
\Cref{fig:cst_perf_headroom} shows the speedup of each configuration over baseline.

We make four key observations from \Cref{fig:cst_perf_headroom}. 
First, Ideal Constable provides $9.1\%$ performance improvement on average over the baseline. This shows that eliminating the execution of \rbe{global-stable} loads has high performance headroom. 
Second, Ideal Constable significantly outperforms \rbd{Ideal Stable LVP} ($4.3\%$ on average). This shows that mitigating both data and resource dependence (as done by Ideal Constable) has higher performance potential than only mitigating data dependence (as done by \rbd{Ideal Stable LVP}).
Third, \rbd{Ideal Stable LVP} with data fetch elimination outperforms \rbd{Ideal Stable LVP} ($6.7\%$ on average), yet it falls short to the Ideal Constable. This shows that eliminating \emph{both} the address computation and data \rbc{fetch} operations of a load execution has higher performance potential than just eliminating the data \rbc{fetch}.
Fourth, Ideal Constable even \rbd{slightly} outperforms $2\times$ load execution width configuration, which incurs significantly higher area and power overhead.
\rbd{We} conclude that Constable has significant potential performance benefit by mitigating \emph{both} load data and resource dependence.

\section{Constable: Key Insight} \label{sec:cst_key_idea}

\noindent Constable is based on the \textbf{key insight} that a dynamic instance $I_2$ of a static load instruction $I$ is bound to fetch the same value from the same memory location as the previous dynamic instance $I_1$ of the same static load instruction if the following two conditions are satisfied:
\begin{itemize}
    \item \textbf{Condition~1}: None of the source registers of $I$ has been written between the occurrences of $I_1$ and $I_2$.
    \item \textbf{Condition~2}: No store or snoop request has arrived to the memory address of $I_1$ between the occurrences of $I_1$ and $I_2$. 
\end{itemize}

Satisfying Condition~1 ensures that $I_2$ would have the same load address as $I_1$, and thus the address computation operation of $I_2$ can be safely eliminated.
Satisfying Condition~2 ensures that $I_2$ would fetch the same value from the memory as $I_1$, and thus the data \rbc{fetch operation} of $I_2$ can be safely eliminated.

Constable exploits this observation to operate in two key steps.
First, Constable dynamically identifies load instructions that have repeatedly fetched the same value from the same load address. We call such loads \emph{likely-stable}.\footnote{\rbc{Hence the name Constable, that \emph{polices} the likely-stable loads to safely eliminate them~\cite{constable_wiki}}.}
Second, when \rbc{Constable gains enough confidence that a given load instruction is \rbd{likely}-stable}, 
Constable tracks modifications to \rbc{the} source architectural registers \rbc{of the load instruction} and its memory location via two small \rbd{hardware} structures.
Constable \rbc{eliminates} the execution of \rbc{all future instances} of the likely-stable load and \rbd{breaks} the load data dependence using the last-fetched value - until there is a write to \rbd{the} source registers or a store or snoop request to \rbd{the} load address.

\sectionRB{Constable: Microarchitecture Design}{Constable: Microarchitecture Design}{sec:cst_const_design}

\subsection{Design Overview}

\begin{figure}[!ht] 
\centering
\includegraphics[width=5in]{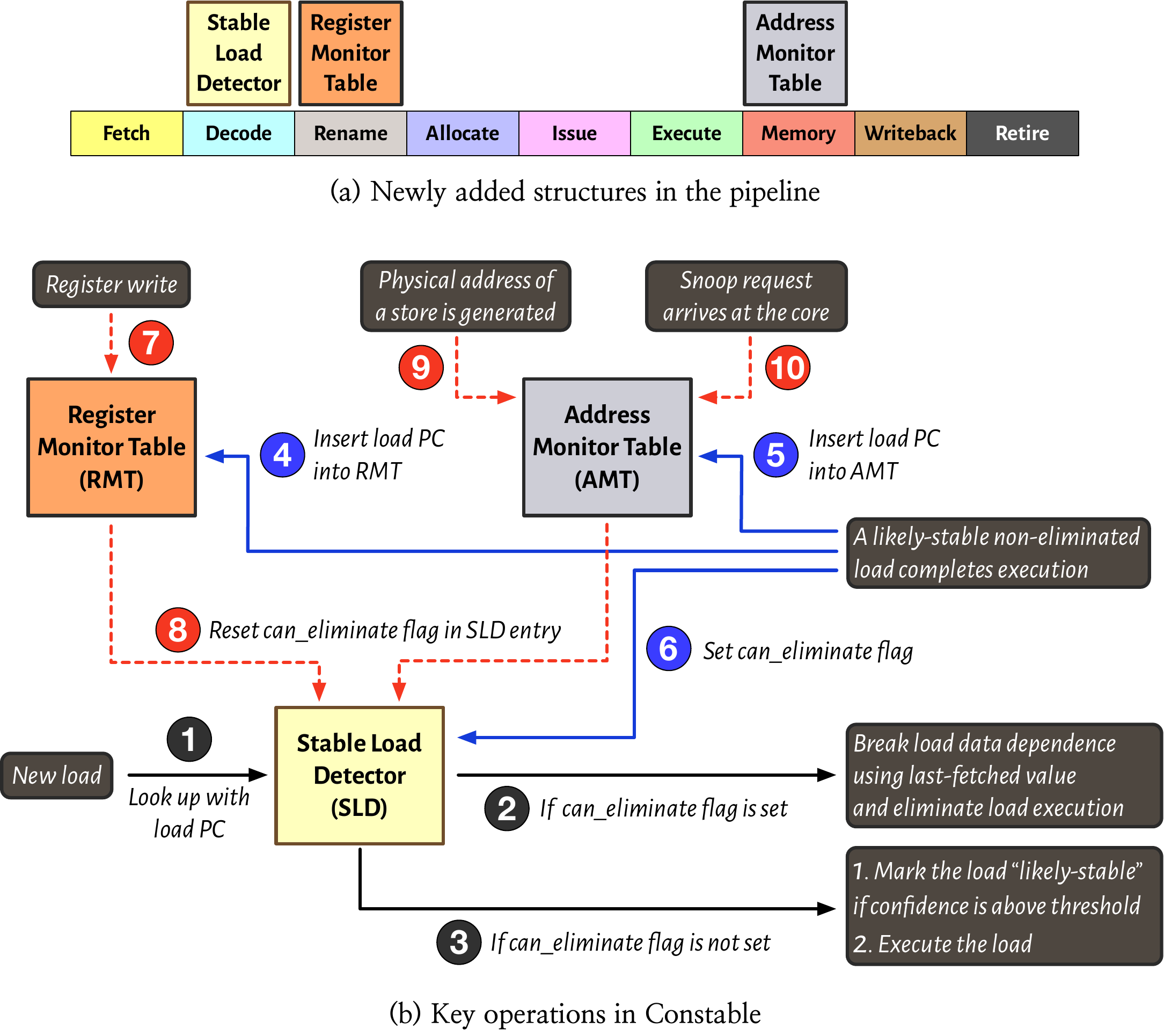}
\caption{Overview of Constable.}
\label{fig:cst_constable_overview}
\end{figure}

\Cref{fig:cst_constable_overview} shows a high-level overview of Constable. Constable is comprised of three \rbd{main} hardware structures:

\paraheading{Stable Load Detector (SLD).}
SLD is a program counter (PC)-indexed table that serves three key purposes. 
First, SLD identifies whether or not a given load instruction is likely-stable by \rbd{analyzing} its past dynamic instances. 
Second, SLD \rbd{decides} whether or not the execution of a load instruction can be eliminated.
Third, SLD provides the last-computed load address and the last-fetched data of a given \rbd{likely-stable} load instruction.

\paraheading{Register Monitor Table (RMT).} 
RMT is an architectural-register-indexed table whose key purpose is to monitor modifications to architectural registers and \rbd{avoid} eliminating a load instruction when its source architectural register gets modified. 
Each RMT entry stores a list of load PCs that are currently getting eliminated that use the corresponding architectural register as their source. 
In the rename stage, every instruction looks up RMT using its destination architectural register and resets the elimination status of any load PC from the corresponding RMT entry in SLD to ensure that any future instances of that load instruction will not be eliminated. In essence, RMT \rbd{enforces} the Condition~1 for eliminating a load instruction (\Cref{sec:cst_key_idea}).

\paraheading{Address Monitor Table (AMT).}
AMT is a physical-address-indexed table whose key purpose is to monitor modifications in the memory and \rbd{avoid} eliminating a load instruction when the memory location from which it fetches the data gets modified.
Each AMT entry stores a list of load PCs that are currently getting eliminated that access the corresponding physical memory address. 
Every store or snoop request looks up AMT using its physical address and resets the elimination status of any load PC from the corresponding AMT entry in SLD to ensure that any subsequent instances of that load will not be eliminated further. In essence, AMT \rbd{enforces} the Condition~2 for eliminating a load instruction (\Cref{sec:cst_key_idea}).

\subsection{Identifying Likely-Stable Loads} \label{sec:cst_design_identifying_likey_stable_loads}

SLD employs a confidence-based learning mechanism to identify likely-stable load instructions based on the execution \rbd{outcomes} of their past dynamic instances. Each SLD entry stores four key pieces of information: (1) last-computed load address, (2) last-fetched value, (3) a $5$-bit \emph{stability confidence \rbd{level}} and (4) a \texttt{can\_eliminate} flag that represents whether or not an instance of this load instruction can be eliminated. 
When a non-eliminated load instruction completes execution in the writeback stage, Constable checks the SLD using the load PC to compare the last-computed load address and last-fetched value with the current load address and value. If both the address and value match, Constable increments the stability \rbd{confidence level} by one; otherwise, it halves the confidence. If the stability \rbd{confidence level} surpasses a threshold (set to $30$ in our evaluation), Constable identifies subsequent load instances from the same PC as likely-stable.

\subsection{Eliminating Load Execution} \label{sec:cst_design_break_dep}

During the rename stage, a load instruction first checks the SLD using the load PC (\circled{1} in \Cref{fig:cst_constable_overview}). 
If the \texttt{can\_eliminate} flag is set in the corresponding SLD entry, Constable breaks the load data dependence using the last-fetched value stored in the SLD entry and eliminates its execution (\circled{2}). 
If the \texttt{can\_eliminate} flag is not set, Constable checks the stability \rbd{confidence level} stored in the SLD entry. If the \rbd{confidence level} is above threshold, Constable marks the load instruction as likely-stable and executes it normally as the baseline (\circled{3}). 
Only a load instruction marked as likely-stable can set the \texttt{can\_eliminate} flag during the writeback stage of its execution (see \Cref{sec:cst_design_update_load_complete}).

\paraheading{Microarchitecture for Breaking Load Data Dependence.}
Breaking load data dependence requires supplying the load value to all dependent in-flight instructions. Prior works on LVP achieve this by writing the value to the physical register file (PRF) or to a separate value table~\cite{rami_ap}. Since writing to PRF either requires adding expensive write ports to PRF~\cite{perais2014eole,perais2015bebop,orosa2018avpp} or a latency-sensitive arbitration of the existing write ports~\cite{perais2021ssr,rami_ap}, Constable implements load data dependence breaking using a small extra register file (only $32$ entries), called xPRF, which is dedicated to hold the values of the in-flight eliminated load instructions.\footnote{
We implement Constable using xPRF as having a small PRF to break data dependence has been shown to be more area- and energy-efficient than adding new write ports to the existing PRF~\cite{rami_ap}. However, Constable can also be implemented by adding or arbitrating PRF write ports. For a fair evaluation, we also implement the LVP and MRN techniques considered in this work using xPRF. As such, xPRF is not considered as an additional structure for Constable.
}

If SLD decides to eliminate the load execution, Constable stores the last-fetched value provided by SLD in an available xPRF register and converts the load instruction into a three-operand register move instruction: the source \rbd{is} the xPRF register, the destination \rbd{is} the destination architectural register of the load, and the third operand \rbd{is} the last-computed load address provided by the SLD. If there is no available xPRF register, Constable does not eliminate the load and executes it normally as the baseline. 
We observe this happens rarely \rbd{(only in $0.2\%$ of the instances)} in our evaluation with a $32$-entry xPRF. 
In the rename stage, the converted register move instruction simply maps its destination register to the source xPRF register to complete its execution (similar to move elimination~\cite{trace_cache,cont_opt}). 
\rbd{Doing so} enables the dependents of the converted register move instruction to get scheduled by reading the xPRF register value. 
In the allocation stage, the converted register move instruction allocates a reorder buffer (ROB) entry and a load buffer (LB) entry. 
The address field in the LB entry gets updated with the last-computed load address embedded within the move instruction as the third operand. 
This address field in the LB entry is later required to correctly disambiguate the eliminated load from the in-flight stores~\cite{mem_disambig} as discussed in \Cref{sec:cst_design_in_flight_stores}. 
Since the execution of the converted register move instruction has already been completed in the rename stage, the instruction bypasses the remaining pipeline stages \rbd{and resources} directly \rbd{to retirement} based on the in-order retirement logic.

\subsection{Updating Constable Structures}

\subsubsection{\textbf{Updates When a Likely-Stable Non-Eliminated Load Finishes Execution}} \label{sec:cst_design_update_load_complete}
During the writeback stage of the pipeline, when a likely-stable yet not eliminated load finishes its execution, Constable updates its structures to eliminate subsequent instances of the same load instruction. This happens in three steps. 
First, Constable looks up RMT with its source architectural registers. For each source register, Constable inserts the load PC into the corresponding RMT entry (\bluecircled{4}).
Second, Constable looks up AMT with the physical address of the load instruction. If the load address is found, Constable inserts the load PC into the corresponding AMT entry (\bluecircled{5}).
If the address is not found, Constable inserts a new AMT entry for the load address and inserts the load PC into the new AMT entry. 
Third, Constable looks up SLD with the load PC and sets the \texttt{can\_eliminate} flag of the corresponding entry (\bluecircled{6}). Setting the \texttt{can\_eliminate} flag allows Constable to eliminate the execution of subsequent instances of the same load instruction.

\subsubsection{\textbf{Updates during Register Renaming}}
In the rename stage, Constable checks the destination architectural register of every instruction and updates its structures to \rbd{avoid} eliminating subsequent instances of any load instruction that uses the destination register as its source.
This happens in two steps.
First, Constable looks up RMT with the architectural destination register of every instruction (\redcircled{7}).
If there is any load PC in the corresponding RMT entry, Constable looks up the SLD using each load PC and resets the \texttt{can\_eliminate} flag in the corresponding entry in SLD (\redcircled{8}).

\subsubsection{\textbf{Updates on a Store Instruction}} \label{sec:cst_design_update_store}
When the address of a store instruction gets generated, Constable updates its structures to \rbd{avoid} eliminating subsequent instances of any load instruction that fetches data from the same memory address as the store. 
This happens in two steps.
First, Constable looks up AMT using the physical store address (\redcircled{9}). 
If the address is found in AMT, Constable looks up SLD using each load PC in the AMT entry and resets the \texttt{can\_eliminate} flag from the corresponding entry in SLD (\redcircled{8}).
Second, after resetting \texttt{can\_eliminate} flag for all load PCs in the AMT entry, Constable evicts the AMT entry.

\subsubsection{\textbf{Updates on a Snoop Request}} \label{sec:cst_design_update_snoop}
To safely eliminate loads in multi-core systems, Constable monitors snoop requests coming to the core and updates its structures to \rbd{avoid} eliminating subsequent instances of any load instruction that fetches data from the same memory address as the snoop.
Constable handles \rbd{a} snoop \rbd{request} in a similar way as a store request. 
When a snoop request arrives at the core, Constable looks up AMT using the snoop address (\redcircled{10}). 
If the address is found, Constable looks up SLD using each load PC in the AMT entry and resets the \texttt{can\_eliminate} flag from the corresponding entry in SLD (\redcircled{8}).
Finally, Constable evicts the AMT entry.

\subsection{Disambiguating Eliminated Loads from In-Flight Stores} \label{sec:cst_design_in_flight_stores}
When a store instruction computes its address, Constable accesses AMT and resets the \texttt{can\_eliminate} flag for all load instructions accessing the same memory location (see \Cref{sec:cst_design_update_store}). 
This prevents Constable from eliminating any \emph{subsequent occurrences} of those load instructions. 
However, in a processor that aggressively issues loads out-of-order~\cite{mem_disambig,franklin1996arb,moshovos1997dynamic}, there may be eliminated loads in the pipeline that are younger than the store instruction and whose addresses match with the store address. 
We observe that this happens \rbd{rarely (see~\Cref{sec:cst_arxiv_ext_eval_pipe_flush} in the extended version~\cite{constable_extended})} since Constable considers a load instruction to be eligible for elimination only if it meets the stability \rbd{confidence level} threshold.
In such infrequent cases, Constable exploits the existing memory disambiguation logic~\cite{mem_disambig,data_preload} that matches the store address with the address of every load in the LB. If a violation is caught, Constable flushes the pipeline and \rbd{re-executes all younger instructions, including the incorrectly-eliminated load} (see the example in \Cref{sec:cst_design_working_example}).

\subsection{Maintaining Coherence in Multi-Core Systems} \label{sec:cst_design_multi_core_coherence}

Constable relies on monitoring snoop requests for tracking modifications in the memory by other processor cores to safely eliminate loads in a multi-core system. However, monitoring snoop requests poses the following two key challenges.

\subsubsection{Loss of Elimination Opportunity due to Clean Evictions}
In a multi-core system with a directory-based coherence protocol~\cite{censier1978new}, when a cacheline gets evicted from a core-private cache, the core-valid bit (CV-bit) corresponding that core \rbd{(i.e., the \emph{own core})} gets reset in the directory entry of that cacheline~\cite{scalable_directory,directory_evaluation,scd,gupta1992reducing}. Since resetting CV-bit prevents the directory from sending any further snoop request to that cacheline to the core, on every core-private cache eviction, Constable needs to \rbd{avoid} eliminating any load instruction that accesses the evicted cacheline. This poses two key drawbacks. First, if the evicted cacheline is clean (e.g., eviction due to limited cache capacity or cache conflict), Constable loses elimination opportunity \rbd{(we quantify the impact of such elimination opportunity loss in~\Cref{sec:cst_arxiv_eval_clean_eviction} in~\cite{constable_extended})}. Second, for every core-private cache eviction, Constable needs to look up and invalidate the corresponding AMT entry, which increases design complexity. 
To address these drawbacks, we propose to \emph{pin} the own core's CV-bit of a cacheline that is accessed by an \rbd{eliminated} load instruction.
When the memory request of a likely-stable yet not eliminated load returns from the cache hierarchy,
Constable pins the own core's CV-bit in the directory entry of that cacheline.
Pinning CV-bit ensures that (1) the coherence protocol would send any snoop request to that cacheline to the own core, even if the cacheline gets clean-evicted from the core-private cache, and (2) Constable does not need to look up AMT on every core-private cache eviction.
The CV-bit is reset as soon as a snoop request is delivered to the core, as per the normal directory-based coherence protocol.

\subsubsection{Tracking Snoop Requests \rbd{at} Cacheline Address Granularity}
Unlike a store instruction that contains a full memory address, a snoop request contains a cacheline address. 
Thus, to support AMT lookup using a snoop address, Constable indexes AMT using physical addresses 
at cacheline granularity. 
\rbd{This} may \rbd{cause loss of} elimination opportunities due to false address collisions (e.g., a store to a cacheline may reset the \texttt{can\_eliminate} flag of a load instruction that accesses different bytes of the same cacheline accessed by the store).
However, we find that the performance impact of such elimination opportunity loss is negligible. 
Constable with a cacheline-address-indexed AMT \rbd{has only} $0.4\%$ \rbd{lower average} performance than a \rbd{Constable with} full-address-indexed AMT.
This is primarily because the compiler tends to lay out memory addresses accessed by likely-stable load instructions together (e.g., a group of function arguments laid out in the same cacheline of stack memory segment), which reduces \rbd{the overhead of false address collisions}.

\subsection{Other Design Decisions} \label{sec:cst_design_design_decisions}

\subsubsection{\textbf{Architecting SLD}} \label{sec:cst_design_design_decisions_sld}
Designing SLD with sufficient read/write ports is crucial for realizing Constable's performance \rbd{benefits}.
Constable reads SLD for every load instruction to identify likely-stable loads in the rename stage (\mbox{\circled{1}} in~\Cref{fig:cst_constable_overview}). Thus SLD needs to support the read bandwidth of the expected number of load instructions in a group of instructions getting renamed together in every cycle (we call this a \emph{rename group}).\footnote{We model a six-wide rename architecture (see \Cref{sec:cst_methodology_perf}).} We observe that a rename group contains $1.93$ loads on average across all workloads, and $98.3\%$ of all rename groups have less than or equal to two loads. 
\rbd{Thus,} we model SLD with three read ports. If there are more than three loads in a rename group, we stall the rename stage until Constable finishes SLD lookup for every load in that group.

\begin{figure}[!ht] 
\centering
\includegraphics[width=5.75in]{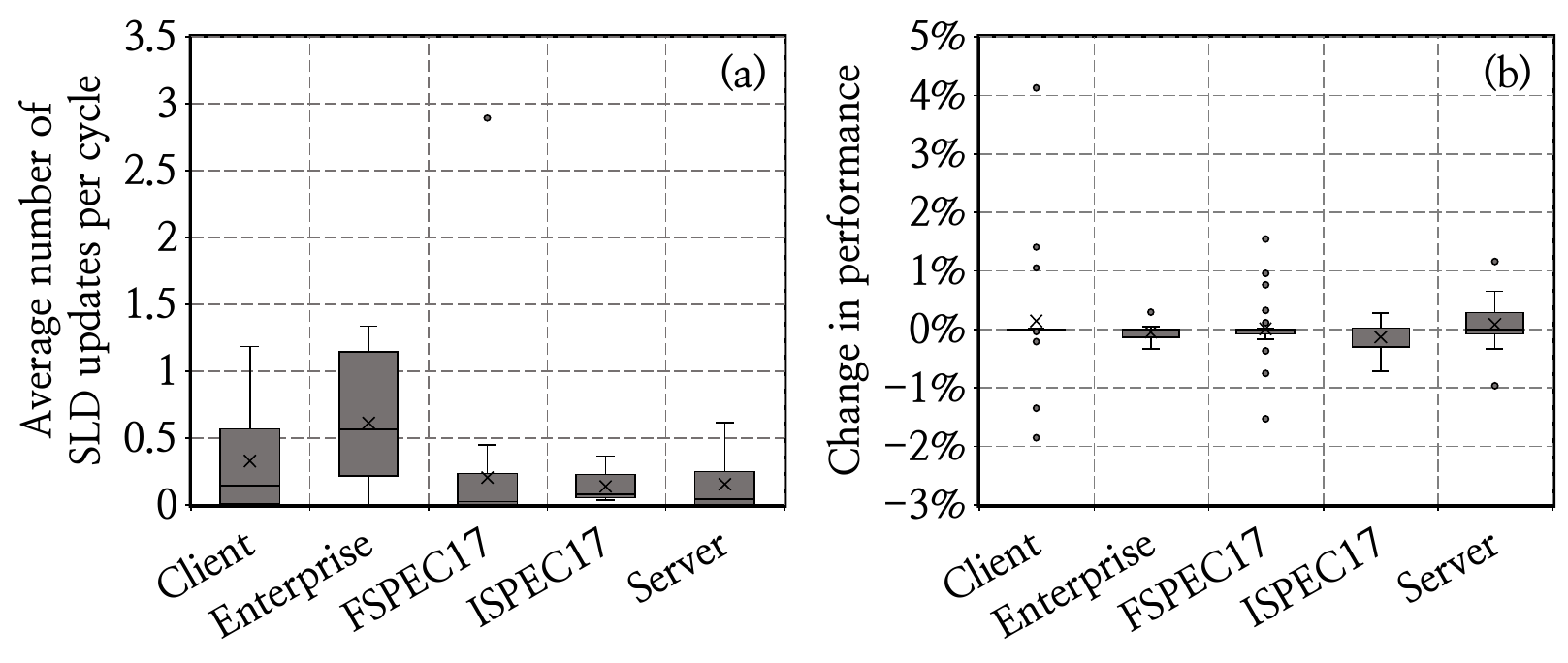}
\caption{(a) \rbd{Average} number of SLD updates per cycle during rename stage. (b) \rbd{Change} in performance when Constable's structures are updated only by correct path instructions vs. all instructions without \rbd{updating} \rbd{Constable's} structures on branch misprediction recovery.}
\label{fig:cst_design_decisions}
\end{figure}

Constable may need to update the \texttt{can\_eliminate} flag in SLD on every RMT update, which happens for each instruction in a rename group (\redcircled{7} and \redcircled{8}). Since each RMT entry may contain a list of likely-stable load PCs, the expected number of SLD updates per cycle can vary in a large range. 
\Cref{fig:cst_design_decisions}(a) shows the average number of observed SLD updates per cycle for every workload as a box-and-whiskers plot.\footnote{
Each box is lower- (upper-) bounded by the first (third) quartile. The box size represents the inter-quartile range (IQR). The whiskers extend to 1.5$\times$IQR range on each side, and the cross-marked values in the box show the mean.
}
As we can see, we observe only $0.28$ SLD updates per cycle on average across all workloads. $98.23\%$ of all cycles on average across all workloads have two or \rbd{fewer} SLD updates. This is because, at any point in time, only a small fraction of all load PCs \rbd{($14.7\%$ on average)} satisfies the stability \rbd{confidence level} threshold \rbd{in order} to be tracked by RMT entries. Thus, we model SLD with two write ports. If there are more than two SLD updates in a cycle, we stall the rename stage until Constable finishes SLD update for every load instruction in that rename group.

\subsubsection{\textbf{Handling Wrong Path Execution}} \label{sec:cst_design_design_decisions_wrong_path}
In presence of branch prediction, Constable's structures may get updated by wrong path instructions (especially, steps~\redcircled{7} and~\redcircled{8} in~\Cref{fig:cst_constable_overview}). This may result in an unnecessary loss of elimination opportunity, unless the structures are restored on a branch misprediction recovery. 
To understand the need for restoring Constable's structures, we measure the change in performance of Constable when its structures are updated only by the instructions on the correct path against when they are updated by all instructions without \rbd{an update} mechanism on branch misprediction recovery, and show it as a box-and-whiskers plot in~\Cref{fig:cst_design_decisions}(b). 
The key observation is that $82$ out of $90$ workloads show less than $1\%$ absolute change in performance, while the average performance change is only $0.2\%$. Thus, we model Constable without \rbd{any update} mechanism for its structures on a branch misprediction recovery.

\subsubsection{\textbf{Handling Changes in Physical Address Mapping}}
AMT monitors memory locations accessed by all \rbd{eliminated} load instructions in physical address space. 
This poses a challenge: when the physical memory mapping changes, the physical memory address tracked by an AMT entry may not be associated with the corresponding eliminated load anymore. In that case, to \rbd{avoid} incorrectly eliminating load execution, 
Constable resets the \texttt{can\_eliminate} flag of all SLD entries and invalidates all RMT and AMT entries when the physical memory mapping changes (e.g., context switch).

\subsection{An Illustrative Example} \label{sec:cst_design_working_example}
\begin{figure*}[!ht]
\centering
\includegraphics[width=\columnwidth]{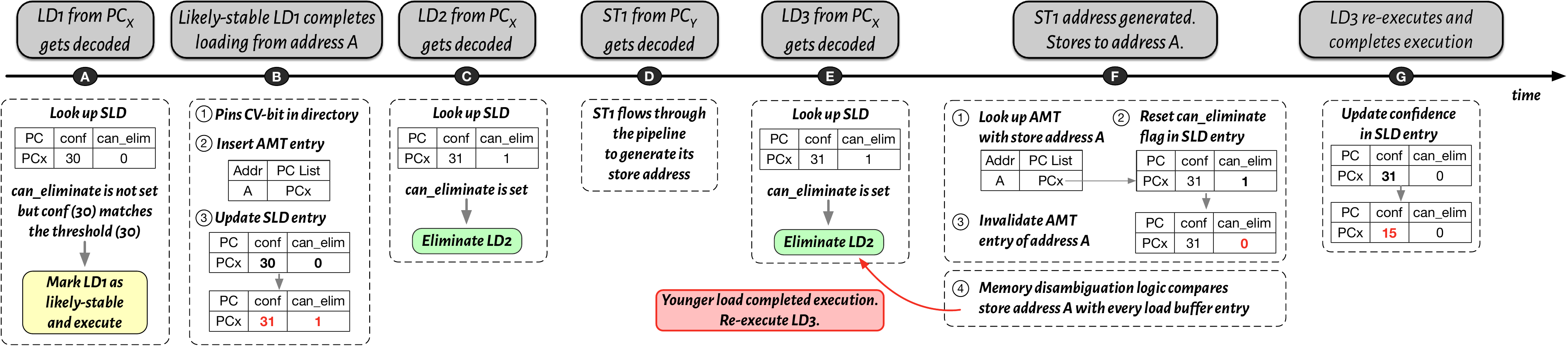}
\caption{An illustrative example of \rbd{Constable's operation}.}
\label{fig:cst_working_example}
\end{figure*}

To put it together, \Cref{fig:cst_working_example} illustrates an example of \rbd{Constable's operation}. For this example, we consider that the loads $LD_1$, $LD_2$, and $LD_3$ are three dynamic instances of the static load instruction $LD$ with a PC value $PC_x$ and the source registers of $LD$ do not get modified between $LD_1$ and $LD_3$.
We also assume the stability \rbd{confidence level} threshold is set to $30$. 

When $LD_1$ gets decoded (\circled{A} in \Cref{fig:cst_working_example}), Constable checks SLD and finds that the stability \rbd{confidence level} of $PC_x$ matches the threshold, yet the \texttt{can\_eliminate} flag is not set.
In this case, Constable marks $LD_1$ as likely-stable and executes it normally. 
When $LD_1$ finishes its execution (\circled{B}), Constable (1) pins the CV-bit corresponding to the own core in the coherence directory entry of address A, (2) updates AMT and RMT (not shown here), and (3) increments the stability \rbd{confidence level} in SLD. Since $LD_1$ is marked as likely-stable, Constable sets the \texttt{can\_eliminate} flag in SLD entry.
When $LD_2$ gets decoded (\circled{C}), Constable eliminates executing $LD_2$ since the \texttt{can\_eliminate} flag is set.
Now a store $ST_1$ from a different $PC_y$ gets decoded (\circled{D}). This store instruction would ultimately modify the memory address touched by $LD$. 
However, before $ST_1$ could generate its store address, $LD_3$, which is younger than $ST_1$ in program order, gets decoded (\circled{E}) and Constable \emph{incorrectly} eliminates its execution since the \texttt{can\_eliminate} flag is set. 
When $ST_1$ finally generates its store address (\circled{F}), Constable resets the \texttt{can\_eliminate} flag to prevent eliminating subsequent instances of $LD$ and evicts the AMT entry. 
However, the existing memory disambiguation logic probes the load buffer with the store address and finds out that a younger load $LD_3$ has been incorrectly completed. 
As a result, the memory disambiguation logic aborts and re-executes $LD_3$ \rbd{(and all instructions younger than $LD_3$ that are not shown here)}. \rbd{When $LD_3$ completes its execution} (\circled{G}), \rbd{it} halves the stability \rbd{confidence level} counter. 

\subsection{Storage Overhead} \label{sec:cst_design_storage_overhead}
\Cref{table:constable_storage_overhead} shows the storage overhead of Constable. 
Constable requires only $12.4$ KB storage \rbd{per core of the processor (see~\Cref{sec:cst_methodology_perf})}.

\begin{table}[htbp]
  \centering
  \small
    \begin{tabular}{L{4.5em}||m{28em}||R{4em}}
    \thickhline
    \Tabval{\textbf{Structure}} & \Tabval{\textbf{Description}} & \Tabval{\textbf{Size}}\\
    \thickhline
    \Tabval{\textbf{SLD}} &
    \Tstrut
    \begin{minipage}{28em}
      \small
      \vskip 6pt
      \begin{itemize}[leftmargin=1em]
        \setlength{\itemsep}{1pt}
        \setlength{\parskip}{0pt}
        \setlength{\parsep}{0pt}
        \item \# entries: $512$ ($32$ sets $\times$ $16$ ways)
        \item Entry size: tag ($24b$) + addr ($32b$) + val ($64b$) + \rbd{confidence level} ($5b$) + can\_eliminate flag ($1b$)
      \end{itemize}
    \end{minipage} 
    \Bstrut &
    \Tabval{\textbf{7.9~KB}} \\
    \hline
    \Tabval{\textbf{RMT}} &
    \Tstrut
    \begin{minipage}{28em}
      \small
      \vskip 6pt
      \begin{itemize}[leftmargin=1em]
        \setlength{\itemsep}{1pt}
        \setlength{\parskip}{0pt}
        \setlength{\parsep}{0pt}
        \item $16$ load PCs for each stack registers (RSP and RBP)
        \item $8$ load PCs for each remaining $14$ architectural registers in x86-64
      \end{itemize}
    \end{minipage}
    \Bstrut &
    \Tabval{\textbf{0.4~KB}} \\
    \hline
    \Tabval{\textbf{AMT}} &
    \Tstrut
    \begin{minipage}{28em}
      \small
      \vskip 6pt
      \begin{itemize}[leftmargin=1em]
        \setlength{\itemsep}{1pt}
        \setlength{\parskip}{0pt}
        \setlength{\parsep}{0pt}
        \item \# entries: 256 ($32$ sets $\times$ $8$ ways)
        \item Entry size: physical address tag ($32b$) + \# hashed load PCs ($4\times24b$)
      \end{itemize}
    \end{minipage} 
    \Bstrut &
    \Tabval{\textbf{4.0~KB}} \\
    \thickhline
    \Tabval{\textbf{Total}} & & \Tabval{\textbf{12.4~KB}} \\
    \thickhline
    \end{tabular}%
  \caption{Storage overhead of Constable.}
  \label{table:constable_storage_overhead}
\end{table}%

\sectionRB{Constable: Evaluation Methodology}{Methodology}{sec:cst_constable_methodology}

\subsection{Performance Modeling} \label{sec:cst_methodology_perf}
We evaluate Constable using \rbd{an} in-house, cycle-accurate, industry-grade simulator that simultaneously runs both functional and microarchitectural simulation on a workload. 
We faithfully model a $6$-wide out-of-order processor core configured similar to the Intel Golden Cove~\cite{goldencove,goldencove_microarch,goldencove_microarch2,spr1} as our baseline. 
\Cref{table:cst_const_sim_params} shows the key microarchitectural parameters.
\rbd{We include} MRN and various dynamic optimizations in the rename stage of the baseline processor, as highlighted in bold.
For a comprehensive analysis, we evaluate Constable and other competing mechanisms on the baseline system, both without SMT (called \emph{noSMT}) and with $2$-way SMT (called \emph{SMT2}). 
For noSMT \rbd{configuration}, all hardware resources inside core are fully available to the single running software context.
For SMT2 \rbd{configuration}, resources inside core (including Constable) are either statically-partitioned or dynamically-shared between both software contexts~\cite{amd_smt}.
Unless stated otherwise, all reported results are from noSMT simulations.

\begin{table}[htbp]
  \centering
  \small
    \begin{tabular}{L{8em}||L{30em}}
    \thickhline
    \Tabval{\textbf{Basic}} & \Tabval{x86-64 core clocked at 3.2~GHz with 2-way SMT support} \\
    \hline
    \Tabval{\textbf{Fetch \& Decode}} & \Tabval{8-wide fetch, TAGE/ITTAGE branch predictors~\cite{seznec2006case}, 20-cycle misprediction penalty, 32KB 8-way L1-I cache, 4K-entry 8-way micro-op cache, 6-wide decode, 144-entry IDQ, loop-stream detector~\cite{lsd}} \\
    \hline
    \Tabval{\textbf{Rename}} & \Tabval{6-wide, 288 integer, 220 512-bit and 320 256-bit physical registers, \textbf{Memory Renaming~\cite{mrn}, zero elimination~\cite{trace_cache}, move elimination~\cite{trace_cache,cont_opt}, constant folding~\cite{trace_cache,cont_opt}, branch folding~\cite{branch_folding}}} \\
    \hline
    \Tabval{\textbf{Allocate}} & \Tabval{512-entry ROB, 240-entry LB, 112-entry SB, 248-entry RS} \\
    \hline
    \Tabval{\textbf{Issue \& Retire}} & \Tabval{6-wide issue to 12 execution ports; 5, 3, 2, and 2 ports for ALU, load, store-address, and store-data execution. Port 0, 1, and 5 are used for vector instructions. Aggressive out-of-order load scheduling with memory dependence prediction~\cite{store_set,moshovos1997streamlining}, 6-wide retire} \\
    \hline
    \Tabval{\textbf{Caches}} & \Tabval{\textbf{L1-D}: 48KB, 12-way, 5-cycle latency, LRU, PC-based stride prefetcher~\cite{stride}; \textbf{L2}: 2MB, 16-way, 12-cycle round-trip latency, LRU, stride + streamer~\cite{streamer} + SPP~\cite{spp}; \textbf{LLC}: 3MB, 12-way, 50-cycle data round trip latency~\cite{goldencove_microarch,hermes}, dead-block-aware replacement policy~\cite{sdbp}, streamer, MESIF~\cite{mesif} protocol} \\
    \hline
    \Tabval{\textbf{Memory}} & \Tabval{4 channels, 2 ranks/channel, 8 banks/rank, 2KB row-buffer/rank, 64b bus/channel, DDR4, tCAS=22ns, tRCD=22ns, tRP=22ns, tRAS=56ns} \\
    \thickhline
    \end{tabular}%
  \caption{Simulation parameters. IDQ: Instruction Decode Queue, SB: Store Buffer.}
  \label{table:cst_const_sim_params}%
\end{table}%

\subsection{Power Modeling} \label{sec:cst_methodology_power}
We use \rbd{an} in-house, RTL-validated power model to measure the dynamic power consumption \rbd{of} the core. 
We report the overall core power consumption by breaking it down into four key units: (1) front end (FE), (2) out-of-order (OOO), (3) non-memory execution unit (EU), and (4) memory execution unit (MEU). We further break down the OOO power into three sub-units: RS, Register Alias Table (RAT), and ROB. We also break down the MEU power into two sub-units: L1-D cache, and data translation look-aside buffer (DTLB).
To faithfully model the power consumption of Constable, we estimate the read/write access energy and leakage power of Constable's structures using CACTI 7.0~\cite{cacti} $22$nm library. We scale the estimates to $14$nm technology using~\cite{tech_scaling} to make the estimates compatible with our core power model. \Cref{table:cst_constable_area_power} shows the access energy, leakage power, and the area estimate of Constable's structures. We report RMT and SLD power in the RAT component and AMT power in the L1-D component of the core power model.

\begin{table}[htbp]
  \centering
  \small
    \begin{tabular}{m{6.2em}||C{5.5em}C{5.5em}C{6em}C{6em}C{3em}}
    \thickhline
    \Tstrut \textbf{Component}\Bstrut & \Tabval{\textbf{Port count}} & \Tstrut \textbf{Read access energy (pJ)}\Bstrut & \Tstrut \textbf{Write access energy (pJ)}\Bstrut & \Tstrut \textbf{Leakage power (mW)}\Bstrut & \Tstrut \textbf{Area (mm2)}\Bstrut\\
    \thickhline
    \Tstrut \textbf{SLD} \Bstrut & \Tabval{$3$R/$2$W}  & \Tstrut$10.76$\Bstrut & \Tstrut$16.70$\Bstrut & \Tstrut$1.02$\Bstrut  & \Tstrut$0.211$\Bstrut \\
    \Tstrut \textbf{RMT} \Bstrut & \Tabval{$2$R/$6$W} & \Tstrut$0.15$\Bstrut  & \Tstrut$0.20$\Bstrut  & \Tstrut$0.31$\Bstrut  & \Tstrut$0.004$\Bstrut \\
    \Tstrut \textbf{AMT} \Bstrut & \Tabval{$1$R/$1$W} & \Tstrut$1.58$\Bstrut  & \Tstrut$4.22$\Bstrut  & \Tstrut$0.74$\Bstrut  & \Tstrut$0.017$\Bstrut \\
    \thickhline
    \end{tabular}%
  \caption{Access energy, leakage power, and area \rbd{estimates} of Constable's structures in 14nm technology. R: read port, W: write port.}
  \label{table:cst_constable_area_power}%
\end{table}%

\subsection{Workloads} \label{sec:cst_methodology_workloads}

We evaluate Constable using $90$ workload traces that span across a diverse set of $58$ workloads. Our workload suite contains all benchmarks from the SPEC CPU 2017 suite~\cite{spec2017}, and many well-known \texttt{Client}, \texttt{Enterprise}, and \texttt{Server} workloads.
Each trace contains a snapshot of the processor and the memory state (1) to drive both the functional and microarchitecture simulation \rbd{models}, and (2) to faithfully simulate wrong-path execution.
Each trace is carefully selected to be representative of the overall workload.
\Cref{table:cst_constable_workloads} summarizes the complete list of workloads.

\begin{table}[htbp]
  \centering
  \small
    \begin{tabular}{L{5em}L{6em}L{5em}L{18em}}
    \thickhline
    \Tabval{\textbf{Suite}} & \Tabval{\textbf{\#Workloads}} & \Tabval{\textbf{\#Traces}} & \Tabval{\textbf{Example Workloads}} \\
    \thickhline
    \Tabval{\textbf{Client}} & \Tabval{16}    & \Tabval{22}    & \Tabval{DaCapo~\cite{dacapo}, SYSmark~\cite{sysmark}, TabletMark~\cite{tabletmark}, JetStream2~\cite{jetstream}}\\
    \hline
    \Tabval{\textbf{Enterprise}} & \Tabval{9}    & \Tabval{14}    & \Tabval{SPECjEnterprise~\cite{specjenterprise}, SPECjbb~\cite{specjbb}, LAMMPS~\cite{lammps}} \\
    \hline
    \Tabval{\textbf{FSPEC17}} & \Tabval{13}     & \Tabval{29}    & \Tabval{All from SPECrate FP 2017~\cite{spec2017}} \\
    \hline
    \Tabval{\textbf{ISPEC17}}  & \Tabval{10}    & \Tabval{11}    & \Tabval{All from SPECrate Integer 2017~\cite{spec2017}} \\
    \hline
    \Tabval{\textbf{Server}}    & \Tabval{10}    & \Tabval{14}    & \Tabval{Hadoop~\cite{hadoop}, Linpack~\cite{linpack}, Snort~\cite{snort}, BigBench~\cite{bigbench}} \\ 
    \thickhline
    \end{tabular}%
  \caption{Workloads used for evaluation.}
  \label{table:cst_constable_workloads}%
\end{table}%

\subsection{Evaluated Mechanisms}

For a comprehensive analysis, we evaluate Constable standalone and in combination with three prior works: (1) a state-of-the-art load value predictor EVES~\cite{eves}, (2) early load address resolution (ELAR)~\cite{elar}, and (3) register file prefetching (RFP)~\cite{rfp}.
For EVES, we use the optimized implementation that won the first championship value prediction (CVP-1) in \rbd{the} $32$~KB storage budget track~\cite{cvp1}.
For ELAR, we follow the same microarchitecture design as proposed in~\cite{elar}.
For RFP, we sweep and select the configuration parameter values that provide the highest performance benefit over the baseline. EVES, RFP, and Constable apply their optimizations to load instructions with data size up to $64$ bits. \Cref{table:cst_constable_mech_overhead} shows the \rbd{overheads} of all evaluated mechanisms. 

\begin{table}[htbp]
  \centering
  \small
    \begin{tabular}{L{28em}||R{10em}}
    \thickhline
    \Tstrut \textbf{Mechanism} \Bstrut & \Tstrut \textbf{Overhead} \Bstrut \\
    \thickhline
    \Tstrut \textbf{EVES}~\cite{eves} with the same configuration as the winner of CVP-1 $32$~KB storage budget track~\cite{cvp1} \Bstrut & \Tstrut \textbf{32~KB} \Bstrut\\
    \hline
    \Tstrut \textbf{ELAR}~\cite{elar} with an additional adder in decode stage and a new copy of the ESP register \Bstrut & \Tstrut \textbf{Adder + 64-b register} \Bstrut \\
    \hline
    \Tstrut \textbf{RFP}~\cite{rfp} with 2K-entry prefetch table, 64-entry page address table, and 128-entry RFP-Inflight table \Bstrut & \Tstrut \textbf{12.4~KB} \Bstrut\\
    \hline
    \hline
    \Tstrut \emph{\textbf{Constable} (this work)} \Bstrut & \Tstrut \textbf{15.4~KB} \Bstrut \\
    \thickhline
    \end{tabular}%
  \caption{Overhead of all evaluated mechanisms.}
  \label{table:cst_constable_mech_overhead}%
\end{table}%

\subsection{Functional Verification of Constable} \label{sec:cst_methodology_func_veri}

Since Constable completely eliminates a load instruction in microarchitectural simulation, we cannot verify the functional correctness of Constable in the same way as LVP or MRN techniques. \rbd{To} functionally verify Constable, we enforce a \emph{golden check} at the \rbd{retirement} stage of every load instruction.
The golden check matches the load address and the load data from the functional simulation model with those from the microarchitectural simulation model. In case of a mismatch, the golden check aborts the simulation. We extensively verify the functional correctness of Constable using a broader set of $3400$ traces and ensure that no single trace fails the simulation.

\sectionRB{Constable: Evaluation}{Evaluation}{sec:cst_const_evaluation}

\subsection{Performance Improvement Analysis} \label{sec:cst_const_eval_perf}

\subsubsection{\textbf{noSMT \rbd{Configuration}}} \label{sec:cst_eval_perf_no_smt}
\begin{figure}[!ht] 
\centering
\includegraphics[width=5.75in]{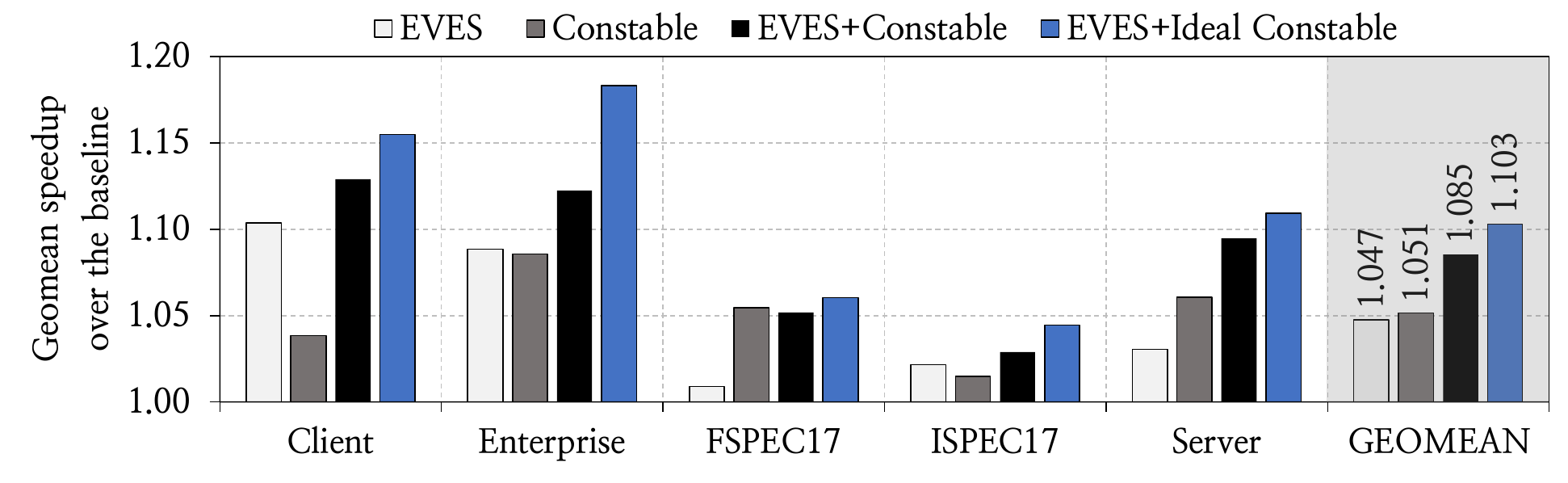}
\caption{Speedup over the baseline \rbd{(noSMT)}.}
\label{fig:cst_const_perf_main}
\end{figure}

\Cref{fig:cst_const_perf_main} shows the geomean performance of EVES, Constable, and Constable and Ideal Constable (\Cref{sec:cst_headroom_perf_headroom}) combined with EVES normalized to the baseline for each workload category. We make three key observations.
First, Constable alone improves performance by $5.1\%$ on average over the baseline, which is similar to the performance improvement of EVES ($4.7\%$ on average) while incurring $\frac{1}{2}\times$ of \rbd{EVES'} storage overhead.
Second, when combined with EVES, Constable improves performance by $8.5\%$ on average over the baseline, which is $3.7\%$ \rbd{higher} than EVES alone.
Third, when combined with EVES, Constable provides $82.9\%$ of the performance improvement provided by Ideal Constable.

\begin{figure}[!ht] 
\centering
\includegraphics[width=5.75in]{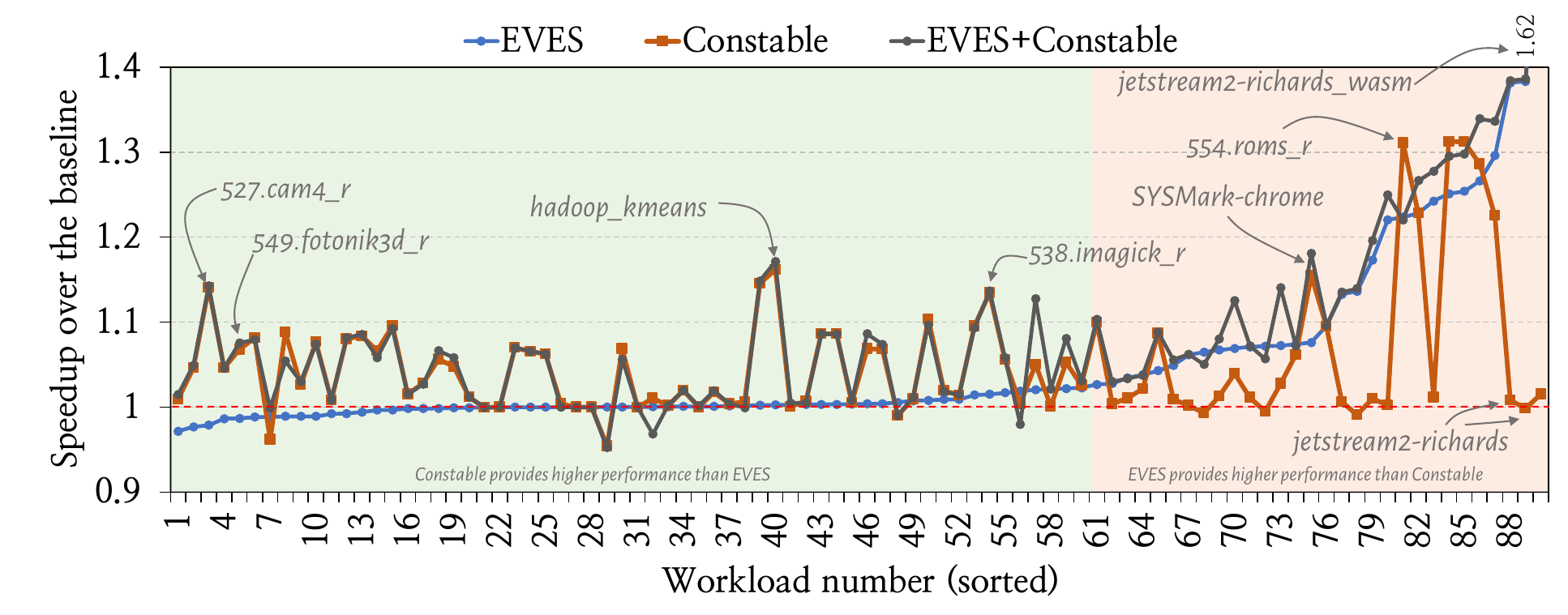}
\caption{\rbd{Speedup} of all workloads \rbd{(noSMT)}.}
\label{fig:cst_perf_line}
\end{figure}

\paraheading{Per-Workload Performance.} To better understand Constable's performance improvement, \Cref{fig:cst_perf_line} shows the performance line graph of EVES, Constable, and Constable combined with EVES for every workload. \rbd{Workloads} are sorted in ascending order of the performance gain of EVES over the baseline. We make three key observations.
First, Constable outperforms EVES by $4.9\%$ on average in $60$ \rbd{of the} $90$ workloads (highlighted in green). In the remaining 30 workloads (highlighted in red), EVES outperforms Constable by $9.2\%$ on average.
Second, Constable combined with EVES consistently outperforms both EVES and Constable alone in every workload.

\begin{figure}[!ht] 
\centering
\includegraphics[width=5.75in]{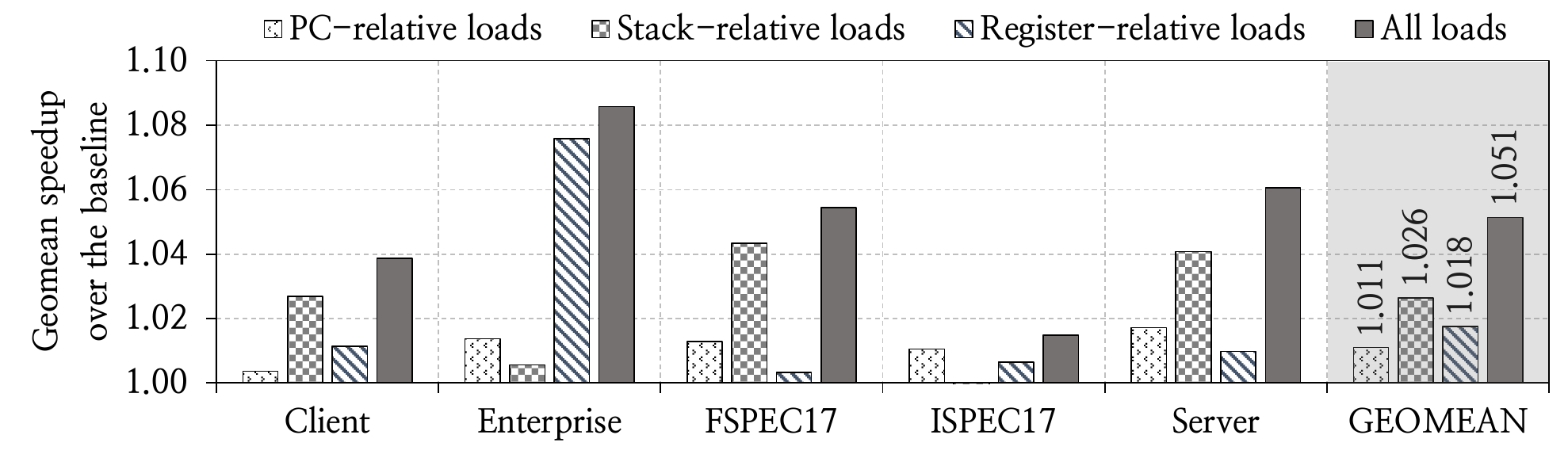}
\caption{Speedup of Constable by eliminating execution of only PC-relative, stack-relative, and register-relative loads.}
\label{fig:cst_perf_breakdown}
\end{figure}

\paraheading{Load Category-Wise Performance.} 
To understand the performance \rbd{benefits} contributed by different load categories, \mbox{\Cref{fig:cst_perf_breakdown}} compares the geomean performance of Constable when it eliminates only PC-relative, stack-relative, and register-relative loads with that of a full-blown Constable. 
The key takeaway is that each three individual types of loads contribute towards Constable's overall performance benefit. Eliminating only PC-relative, stack-relative, and register-relative loads provide a performance improvement of $1.1\%$, $2.6\%$, and $1.8\%$, respectively, which nearly get added up to a $5.1\%$ improvement by the full-blown Constable.

Based on these results, we conclude that (1) Constable provides a significant performance benefit over a wide range of workloads both by itself and when combined with a state-of-the-art load value predictor EVES, and (2) Constable's performance benefit comes from eliminating all types of loads.

\subsubsection{\textbf{SMT2 \rbd{Configuration}}} \label{sec:cst_eval_perf_smt}

\begin{figure}[!ht] 
\centering
\includegraphics[width=5.75in]{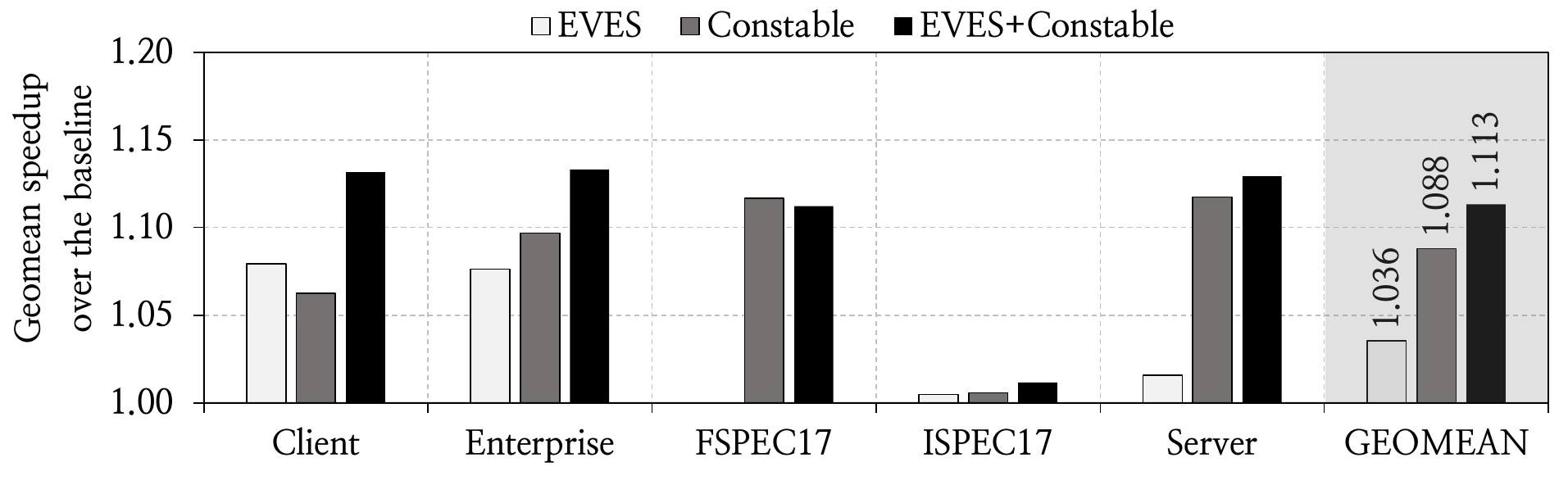}
\caption{Speedup over the baseline (SMT2).}
\label{fig:cst_const_perf_main_smt}
\end{figure}

\Cref{fig:cst_const_perf_main_smt} shows the geomean performance of EVES, Constable, and Constable combined with EVES normalized to the baseline. We make two key observations.
First, unlike \rbd{in} noSMT \rbd{configuration}, Constable significantly outperforms EVES in the baseline \rbd{with} SMT2. Constable alone improves performance by $8.8\%$ on average over the baseline, whereas EVES alone improves performance by $3.6\%$. This is because, unlike EVES, Constable's load elimination fundamentally reduces utilization of load execution resources, which face increased contention in presence of SMT.
Second, combining Constable with EVES continues to provide additional performance benefit than EVES alone. Constable with EVES improves performance by $11.3\%$ on average over the baseline in SMT2.
\rbd{We} conclude that Constable provides even more performance benefit in presence of SMT as compared to non-SMT system, \rbd{due to high resource contention in SMT systems}.

\subsection{Performance Comparison with Prior Works} \label{sec:cst_eval_prior_work}

\begin{figure}[!ht] 
\centering
\includegraphics[width=5.75in]{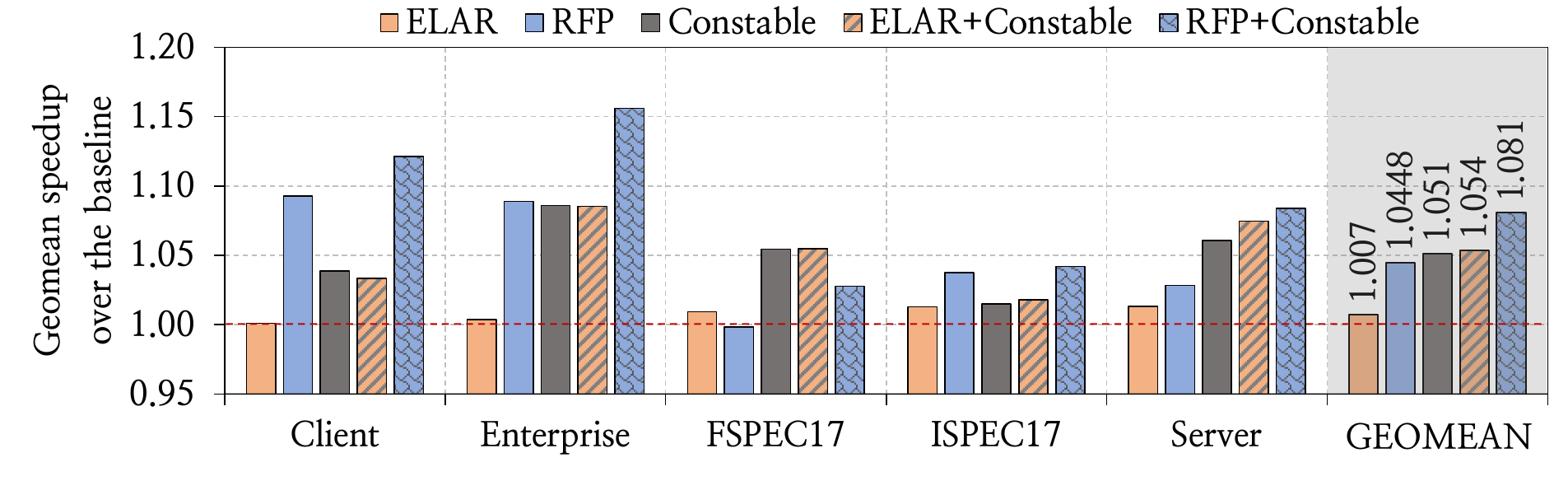}
\caption{Speedup of Constable \rbd{over} ELAR and RFP.}
\label{fig:cst_prior_work}
\end{figure}

\Cref{fig:cst_prior_work} shows the geomean performance of Constable standalone and when combined with ELAR and RFP in the baseline. We make two key observations.
First, Constable alone outperforms both ELAR and RFP. ELAR, RFP, and Constable \rbd{improve} performance on average by $0.74\%$, $4.4\%$ and $5.1\%$, respectively over the baseline. ELAR provides relatively small performance benefit over baseline as our baseline already implements constant folding~\cite{trace_cache,cont_opt}, which can track stack register modifications in the form of $RSP \leftarrow RSP \pm immediate$ before the execution stage. 
Second, when combined with ELAR and RFP, Constable provides \rbd{more} performance benefit than ELAR and RFP alone, respectively. This shows that Constable can be applied \rbd{along with} these proposals to provide even more performance benefit.

\subsection{Loads \rbd{Eliminated by Constable}} \label{sec:cst_evaluation_load_coverage}

\begin{figure}[!ht] 
\centering
\includegraphics[width=5.75in]{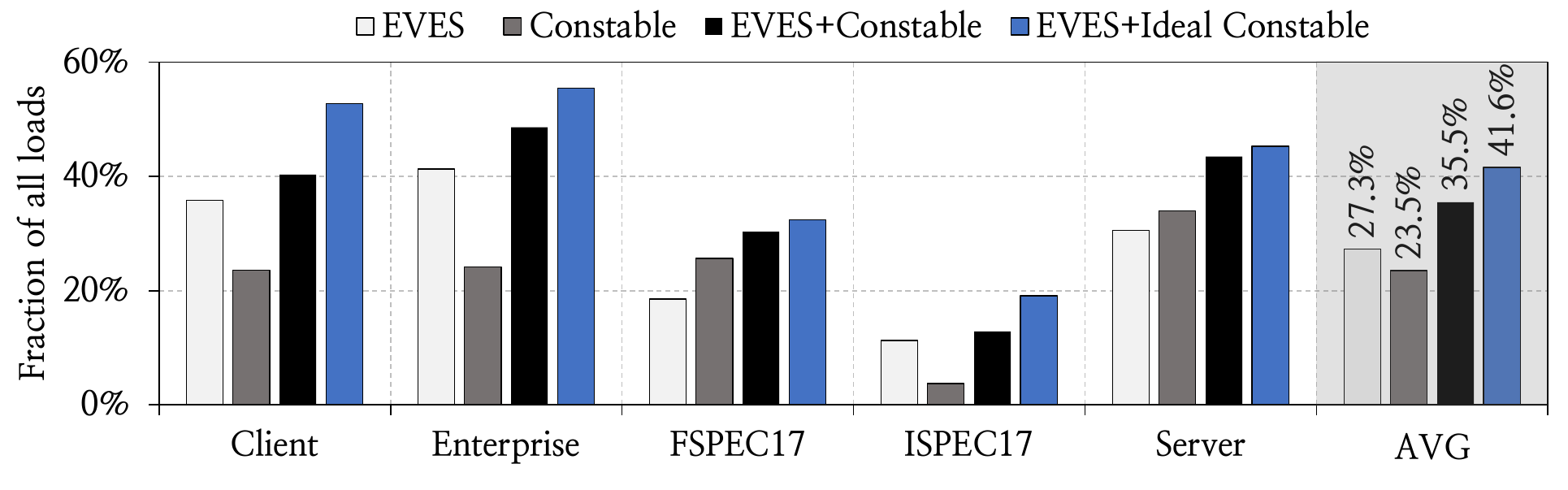}
\caption{Load coverage of Constable \rbd{versus} EVES.}
\label{fig:cst_cov_main}
\end{figure}

\Cref{fig:cst_cov_main} shows the load coverage (i.e., the fraction of load instructions that are either eliminated or value-predicted by Constable or EVES, respectively) of EVES, Constable, and Constable and Ideal Constable combined with EVES in the baseline system. We make three key observations.
First, Constable alone covers $23.5\%$ of the loads, whereas EVES covers $27.3\%$. 
This is because Constable target loads which show \emph{both} value and address locality (i.e., repeatedly fetching same value from same memory address), whereas EVES target loads that show only value locality. 
Despite \rbd{its} lower coverage, Constable matches performance of EVES as Constable mitigates \emph{both} data dependence and resource dependence on covered loads. 
Second, Constable combined with EVES has higher load coverage ($35.5\%$ on average) than EVES alone.
Third, when combined with EVES, Constable provides $85.4\%$ of the coverage of Ideal Constable.
We conclude that Constable covers a significant fraction of the load instructions both by itself and combined with EVES.

\vspace{2pt}
\subsubsection{Coverage of \rbe{Global-Stable} Loads} \label{sec:cst_evaluation_stable_load_coverage}
\begin{figure}[!ht] 
\centering
\includegraphics[width=5.75in]{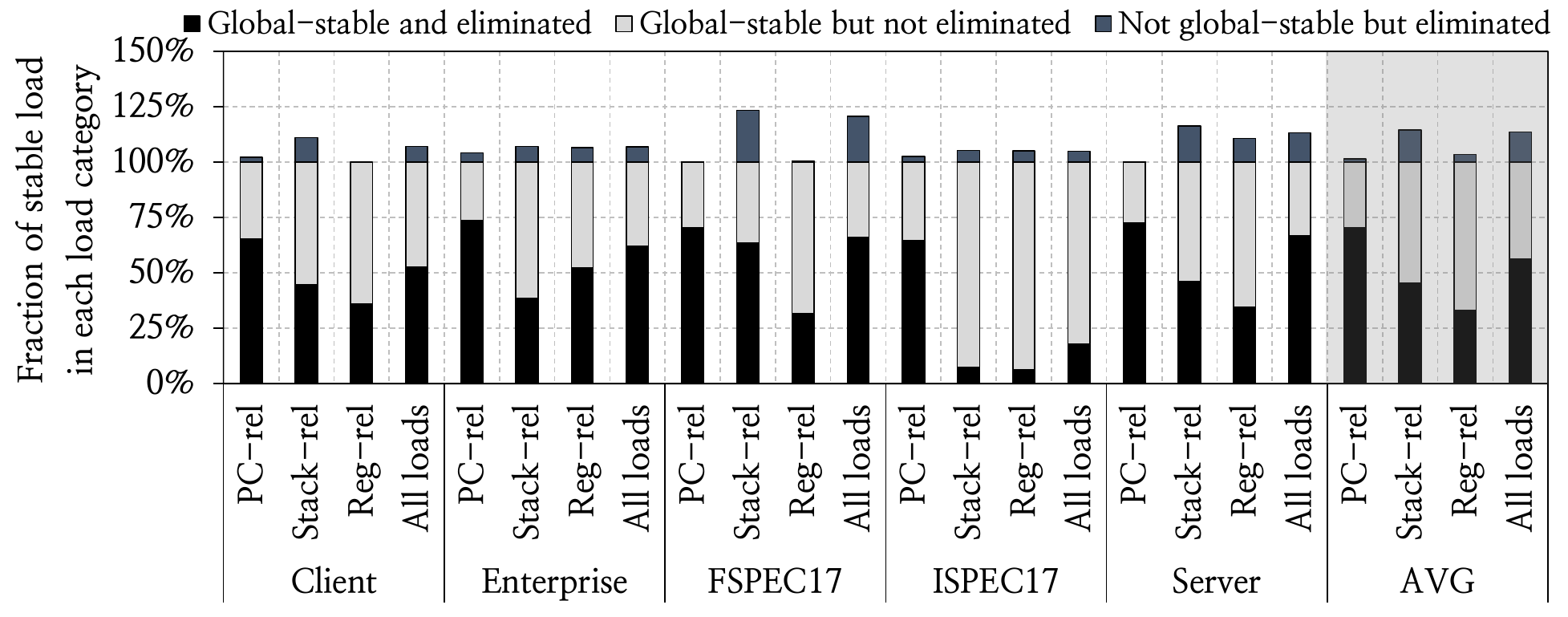}
\caption{Breakdown of eliminated and non-eliminated loads as fractions of \rbe{global-stable} loads.}
\label{fig:cst_cov_breakdown}
\end{figure}

To understand Constable's coverage of \rbe{global-stable} loads (see \mbox{\Cref{sec:cst_headroom}}), \mbox{\Cref{fig:cst_cov_breakdown}} shows the breakdown of loads in each addressing-mode category into three classes: (1) loads that are \rbe{global-stable} and eliminated by Constable, (2) loads that are \rbe{global-stable} but not eliminated, and (3) loads that are not \rbe{global-stable} but eliminated.
We make three key observations.
First, PC-relative and register-relative \rbe{global-stable} loads see the highest and the lowest runtime elimination coverage of $70.2\%$ and $33.2\%$, respectively. 
Second, Constable successfully eliminates $56.4\%$ of all \rbe{global-stable} loads on average at runtime.
For the remaining $43.6\%$ \rbe{global-stable} loads, Constable misses their elimination opportunity due to three key reasons (not shown in the figure):
(a) at least one source architectural register of a \rbe{global-stable} load instruction gets written between its two successive dynamic instances (for $23.3\%$ of all \rbe{global-stable} loads), 
(b) a \emph{silent store}~\mbox{\cite{silent_stores,silent_stores2,silent_stores3,silent_stores4}} occurs between two successive dynamic instances of a \rbe{global-stable} load (for $14.1\%$ of all \rbe{global-stable} loads),
and (c) coverage loss due to other reasons, e.g., stability confidence learning and limited hardware budget for likely-stable load tracking (for $6.2\%$ of all \rbe{global-stable} loads).
Third, on average, $13.5\%$ more loads are eliminated by Constable at runtime which are not identified as \rbe{global-stable}. This is because these loads are not stable across the entire workload trace, but stable in a workload phase to meet the stability confidence threshold and and hence get eligible for elimination.
Based on these results, we conclude that Constable eliminates a significant fraction of the \rbe{global-stable} loads at runtime. 
However many elimination opportunities are still left, which can be unlocked by future works to achieve even higher performance and power efficiency improvements.

\subsection{Impact on Pipeline Resource Utilization} \label{sec:cst_eval_pipe_resource_reduction}

\subsubsection{\textbf{Reduction in RS Allocation}} \label{sec:cst_eval_reduction_rs}
\begin{figure}[!ht] 
\centering
\includegraphics[width=5.75in]{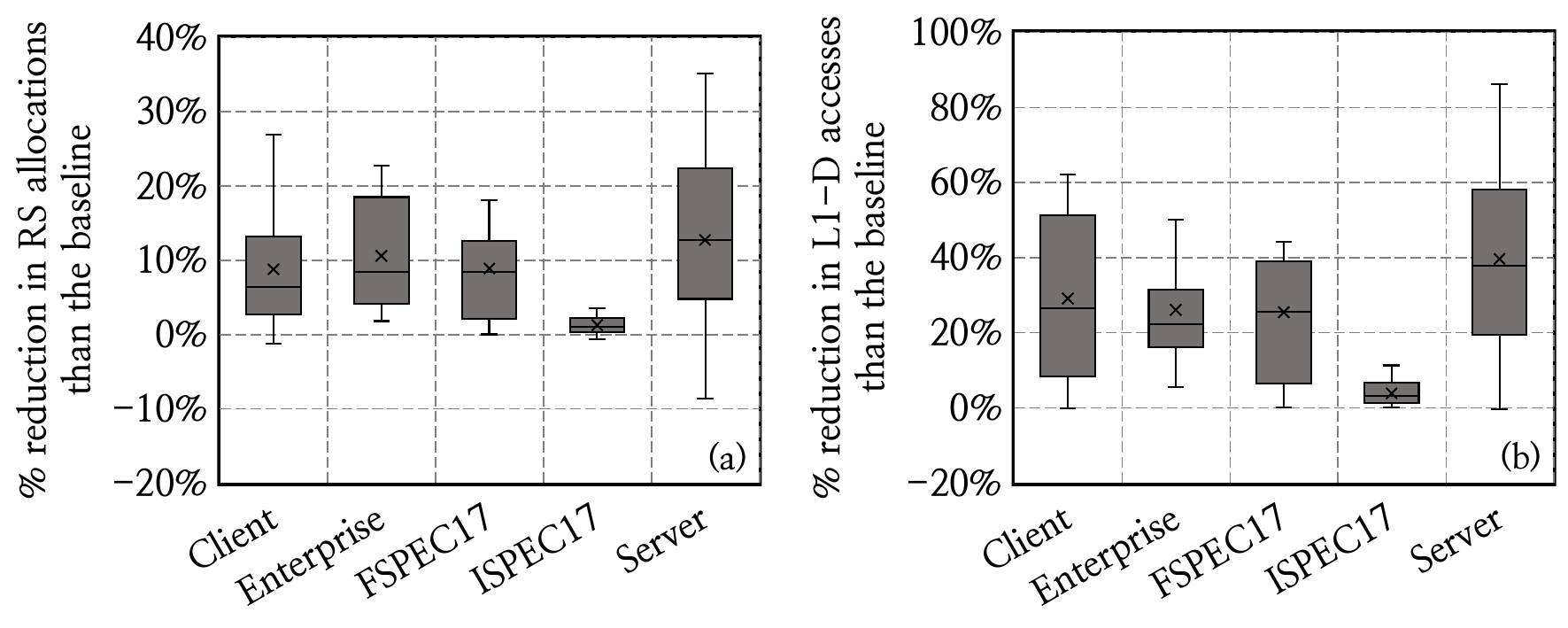}
\caption{Reduction in (a) RS allocations and (b) L1-D accesses.}
\label{fig:cst_pipe_reduction}
\end{figure}

\Cref{fig:cst_pipe_reduction}(a) plots the percentage reduction in RS allocations in a system with Constable over the baseline system as a box-and-whiskers plot. The key observation is that Constable reduces the RS allocation by $8.8\%$ on average (up to $35.1\%$) across all workloads. \texttt{Server} and \texttt{ISPEC17} workloads experience the highest and the lowest average RS allocation \rbd{reductions of} $12.8\%$ and $1.3\%$, respectively. 
\rbd{$37$ of the $90$ workloads experience a reduction in RS allocation by more than $10\%$}.

\subsubsection{\textbf{Reduction in L1-D Access}} \label{sec:cst_eval_reduction_l1d}
\Cref{fig:cst_pipe_reduction}(b) plots the percentage reduction in L1-D accesses in a system with Constable over the baseline as a box-and-whiskers plot. The key observation is that Constable reduces L1-D allocation by $26.0\%$ on average. Similar to the reduction in RS allocation, \texttt{Server} and \texttt{ISPEC17} workloads experience the highest and the lowest average L1-D access reduction of $39.7\%$ and $3.9\%$.

Based on these results, we conclude that, by eliminating load execution, Constable significantly reduces RS allocations and L1-D accesses, both of which aid in \rbd{improving performance (\Cref{sec:cst_const_eval_perf}) and} reducing dynamic power consumption (\Cref{sec:cst_eval_power}).  

\subsection{Power Improvement Analysis} \label{sec:cst_eval_power}

\begin{figure}[!ht] 
\centering
\includegraphics[width=5.75in]{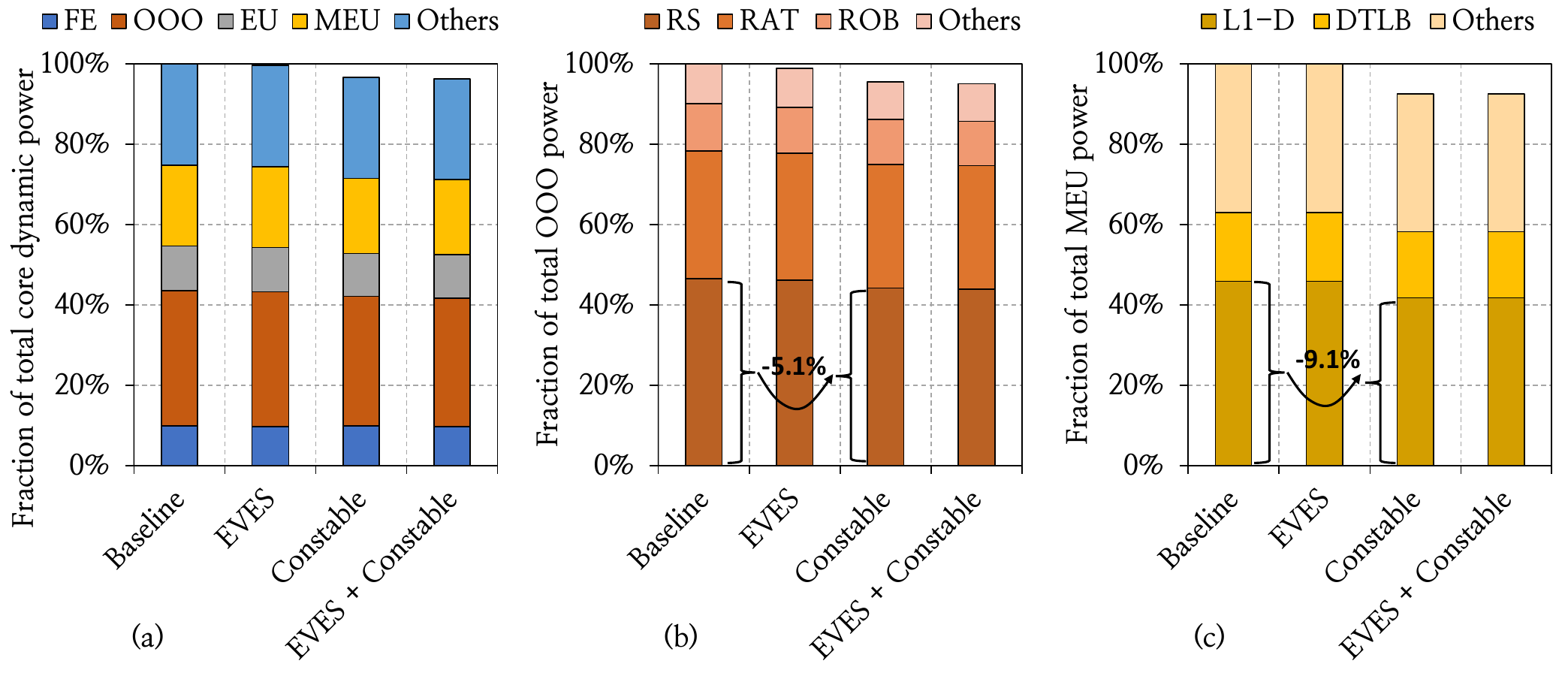}
\caption{(a) Overall core power consumption normalized to baseline. Expanded view of (b) OOO power and (c) MEU power.}
\label{fig:cst_power}
\end{figure}

\Cref{fig:cst_power}(a) shows the core power consumption \rbd{(and its breakdown) in} a system with EVES, Constable, and EVES+Constable normalized to the baseline. 
The key takeaway is that Constable reduces the core power consumption by $3.4\%$ on average over the baseline, whereas EVES reduces power by \rbd{only} $0.2\%$. 
This is because, unlike EVES where the value-predicted load instructions get executed nonetheless, Constable eliminates executing likely-stable loads altogether. 
To understand the distribution of the power benefit across various core structures, we further expand the power consumption \rbd{of} OOO and MEU units in \Cref{fig:cst_power}(b) and (c) respectively. 
As \Cref{fig:cst_power}(b) shows, Constable reduces the power consumed by OOO unit by $4.5\%$ on average over the baseline. 
The RS sub-unit of OOO unit experiences the highest power reduction of $5.1\%$ (marked by braces). 
This is because Constable significantly reduces the number of RS allocations (\Cref{sec:cst_eval_reduction_rs}).
As \Cref{fig:cst_power}(c) shows, Constable also reduces the power consumed by MEU unit by $7.2\%$ on average over the baseline. 
The MEU power reduction is dominated by L1-D cache, which experiences $9.1\%$ reduction in power (marked by braces) on average. This is largely due to the reduction is L1-D accesses (\Cref{sec:cst_eval_reduction_l1d}).
\rbd{We} conclude that Constable, unlike value prediction, reduces the core power by fundamentally eliminating load execution.

\subsection{Performance Sensitivity Analysis} \label{sec:cst_arxiv_ext_eval_sensitivity}

\subsubsection{\textbf{Effect of Load Execution Width Scaling}} \label{sec:cst_arxiv_ext_eval_sen_width}

\begin{figure}[!ht] 
\centering
\includegraphics[width=5.75in]{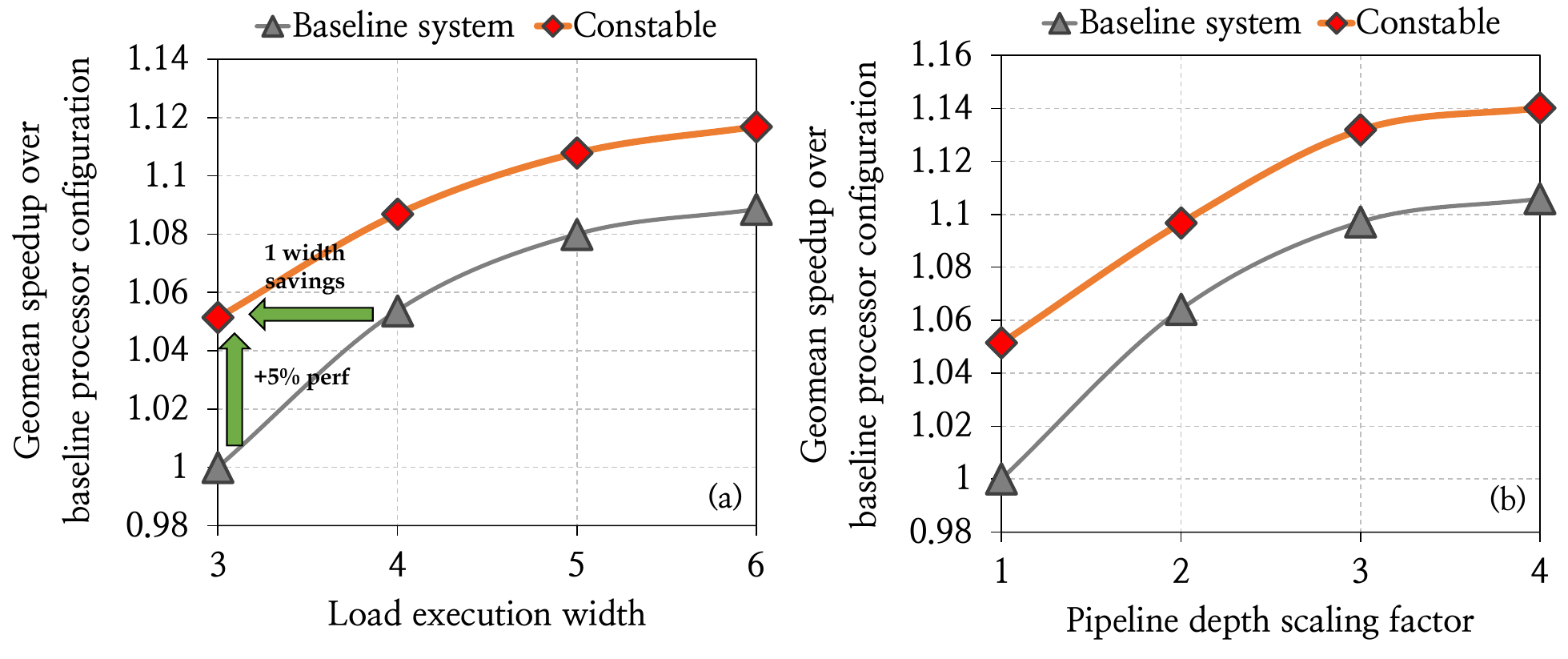}
\caption{Performance sensitivity to (a) load execution width, and (b) pipeline depth.}
\label{fig:cst_sensitivity}
\end{figure}

\Cref{fig:cst_sensitivity}(a) shows the geomean speedup of Constable and the baseline system over the baseline configuration when we increase the load execution width (i.e., increasing both number of AGU and load ports). 
We make two key observations.
First, Constable \emph{consistently} adds performance on top of the baseline system even if we naively scale the load execution width. 
With increasing AGU and load ports (while keeping the pipeline depth resources same), the resource dependence stemming from load reduces. Yet, Constable outperforms the baseline system by $3.5\%$ with $2\times$ load execution width than the baseline configuration.
Second, adding Constable on the baseline system configuration (i.e., with 3 load execution width) essentially provides the similar performance benefit as the baseline system with one extra load execution width, while incurring lower area overhead and \emph{reducing} power consumption.

\subsubsection{\textbf{Effect of Pipeline Depth Scaling}} \label{sec:cst_arxiv_ext_eval_sen_depth}
\Cref{fig:cst_sensitivity}(a) shows the geomean speedup of Constable and the baseline system over the baseline configuration when we scale pipeline depth resources (i.e., size of ROB, RS, LB and SB).
The key takeaway is that Constable \emph{consistently} adds performance on top of the baseline system even if we naively scale the pipeline depth. With $4\times$ depth scaling, Constable improves performance of the baseline system by $3.4\%$ on average.

\subsection{Effect of In-Flight Stores on Elimination Coverage} \label{sec:cst_arxiv_ext_eval_pipe_flush}

\begin{figure}[!ht] 
\centering
\includegraphics[width=5.75in]{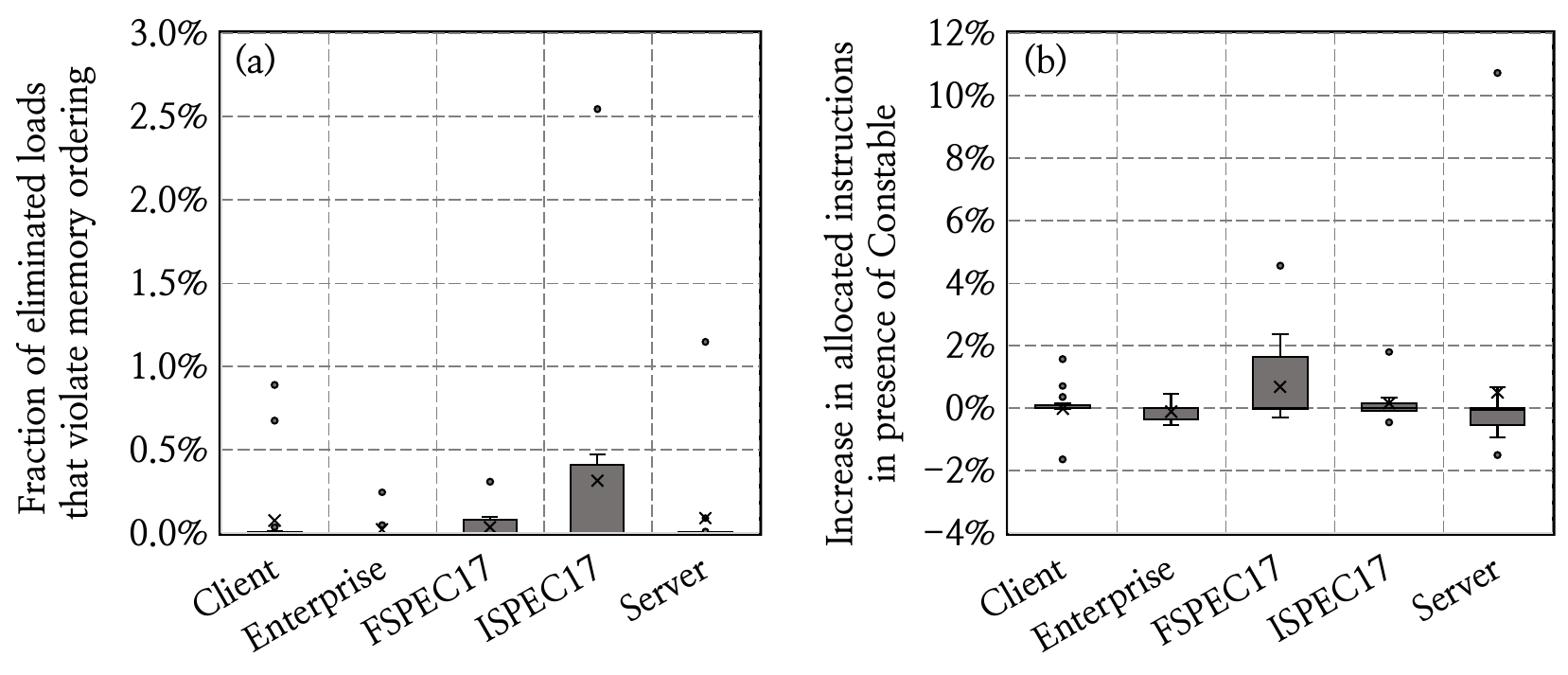}
\caption{(a) Fraction of loads eliminated by Constable that violate memory ordering. (b) Increase in instructions allocated to ROB in presence of Constable.}
\label{fig:cst_elim_nuke}
\end{figure}

When the computed address of an in-flight store instruction matches with that of an eliminated load younger than the store, the existing memory disambiguation logic catches such memory ordering violation and re-executes all instructions younger than (and including) the incorrectly-eliminated load (see~\Cref{sec:cst_design_in_flight_stores}). 
Thus a frequent memory ordering violation by eliminated loads may incur a significant performance and power overhead on Constable.
To understand such overhead, we show the fraction of loads eliminated by Constable that violate memory ordering as a box-and-whiskers plot in \Cref{fig:cst_elim_nuke}(a). 
As we can see, an eliminated load rarely violates memory ordering. On average, only $0.09\%$ of all eliminated loads violate memory ordering. Less than $0.5\%$ of eliminated loads violate memory ordering in $86$ out of $90$ workloads.
This is primarily due to Constable's confidence-based mechanism that considers a load instruction eligible for elimination only if it meets a sufficiently-high stability confidence level threshold (set to 30 in our evaluation).
\Cref{fig:cst_elim_nuke}(b) further shows the increase in instructions allocated to the ROB in presence of Constable as compared to the baseline system to understand the effect of such rare memory ordering violations. As we can see, Constable increases the allocated instructions by only $0.3\%$ on average across all workloads. $79$ out of $90$ workloads observe an increase of less than $1\%$.
Thus, we conclude that Constable observes a very insignificant overhead due to rare memory ordering violations by incorrectly-eliminated loads.

\subsection{Effect of Clean Evictions on Elimination Coverage} \label{sec:cst_arxiv_eval_clean_eviction}

In order to correctly eliminate load instructions in a multi-core system, Constable proposes pinning the CV-bit of a cacheline that is accessed by an eliminated load instruction (see~\Cref{sec:cst_design_multi_core_coherence}). 
However, the change in the coherence protocol may complicate hardware verification.
Another alternative design could be to avoid elimination on every core-private cache eviction. However, this design choice may lose elimination opportunities if the evicted cacheline is clean. In this section, we quantify impact of such elimination opportunity loss on Constable's performance and elimination coverage.

\begin{figure}[!ht] 
\centering
\includegraphics[width=5.75in]{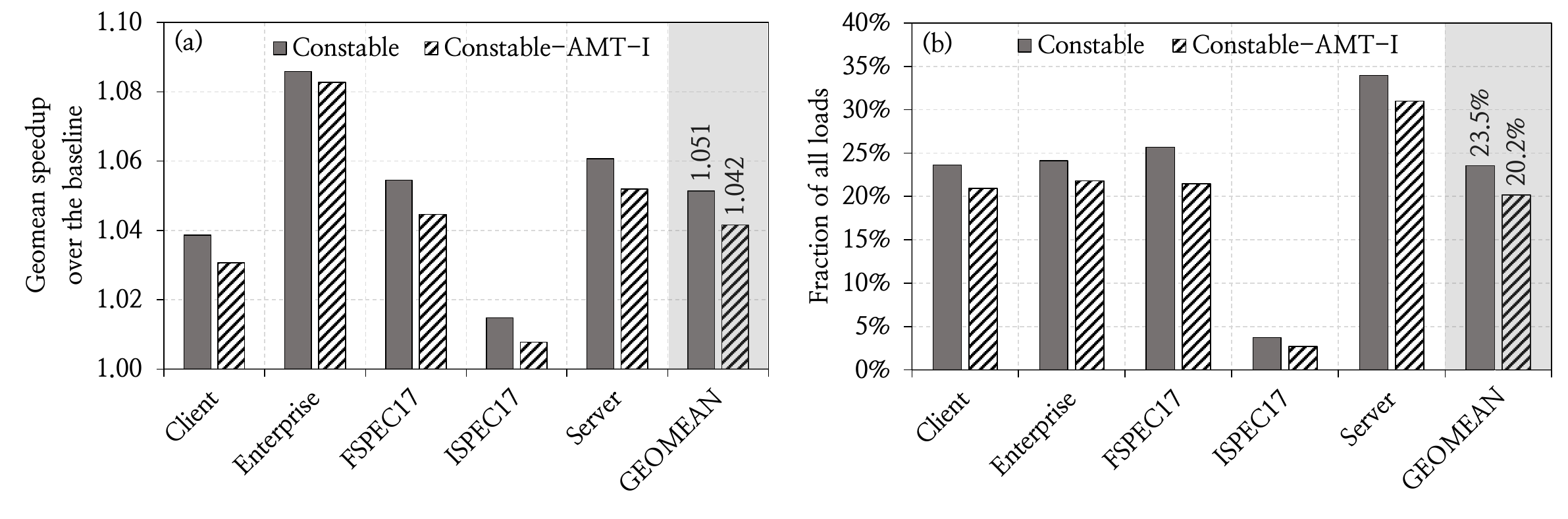}
\caption{(a) Speedup and (b) coverage of Constable with AMT invalidation on L1D eviction compared to a vanilla Constable.}
\label{fig:cst_clean_eviction_effect}
\end{figure}

To understand the effect, we model a Constable variant that looks up AMT for every L1 data (L1-D) cache eviction and invalidates the AMT entry. This prevents Constable from eliminating any further load instructions that access the evicted cacheline. We call this Constable variant Constable-AMT-I.
\Cref{fig:cst_clean_eviction_effect}(a) compares the speedup of Constable-AMT-I with the vanilla Constable. 
Constable-AMT-I loses $0.9\%$ performance improvement than the vanilla Constable on average across all workloads. $11$ out of $90$ workloads observe a performance loss of more than $5\%$ in Constable-AMT-I (with the highest performance loss of $10.4\%$ in \texttt{554.roms\_r}) as compared to vanilla Constable.
The performance loss is primarily attributed to the loss in elimination coverage. As \Cref{fig:cst_clean_eviction_effect}(b) shows, Constable-AMT-I provides $3.4\%$ less load elimination coverage than the vanilla Constable. $17$ out of $90$ workloads observe a coverage loss of more than $5\%$ in Constable-AMT-I (with the highest coverage loss of $27\%$ in \texttt{554.roms\_r}).
Thus, we conclude that CV-bit pinning is a more-performant design choice than avoiding elimination on every core-private cache eviction.

\sectionRB{Constable: Summary}{Summary}{sec:cst_summary}

\noindent We introduce Constable, a purely-microarchitectural technique that safely eliminates the execution of load instructions while breaking the load data dependence.
Our extensive \rbd{evaluation} using a wide range of workloads and system configurations shows that Constable provides significant performance benefit \rbd{and reduced} dynamic power consumption by eliminating load execution.

\vspace{1em}

\subsection{Influence on the Research Community}
Constable has been presented at the the 51st International Symposium on Computer Architecture (ISCA) on June, 2024 and was recognized with the \textit{\textbf{Best Paper Award}} at ISCA 2024~\cite{constable_award}.
We have released \rbfor{an open-source} binary instrumentation tool, called the \textit{Load Inspector}~\cite{load_inspector}, that can identify and provide insights on global-stable loads present in any off-the-shelf x86(-64) binary.

Constable has influenced both academic research and industrial product development. 
A follow-up work~\cite{gsl} has independently verified Constable's key observation of global-stable loads in off-the-shelf programs using a modified version of our Load Inspector tool and has uncovered new insights on why modern compilers fail to eliminate such loads at compile time. 
To the best of our knowledge, the key idea of Constable is in the process of getting transferred to a real-world commercial processor design, and has been the subject matter of a patent application filed by Intel Corporation~\cite{constable_patent}.

As hardware resource scaling is becoming increasingly challenging in today's processor design, we believe and hope that Constable's key observations and insights would inspire future works to explore a multitude of other optimizations that \rbfor{mitigate} ILP loss due to resource dependence \rbd{and load instruction execution}.

\chapter{Conclusions and Future Directions}
\label{chap:conc}

\section{Putting It All Together}

In pursuit of advancing the performance and \rbtwo{energy} efficiency of general-purpose processors, this dissertation argues for a fundamental shift in microarchitectural design: from the conventional data-agnostic approaches to data-driven and data-aware approaches, where microarchitecture would tailor their policies by continuously learning and exploiting the characteristics of application data and system-generated metadata. 
We build a detailed understanding of how the data-agnostic nature of state-of-the-art mechanisms for hiding or tolerating memory latency limits their effectiveness, and use this understanding to propose four new mechanisms, designed from the ground up with data-driven and data-aware principles.
Across all four cases, 
we show that our proposed data-driven and data-aware microarchitectural techniques significantly improve performance and/or energy efficiency over the best conventional techniques.

First, we begin with data prefetching, a fundamental mechanism to hide memory latency by predicting future memory addresses. 
In~\Cref{chap:pythia} we show that state-of-the-art prefetchers are limited by their reliance on a single fixed program feature and their lack of awareness to system-level feedback (e.g., memory bandwidth usage). 
To address these limitations, we formulate prefetching as a reinforcement learning problem, where the prefetcher acts as an autonomous agent that learns to prefetch using multiple program features and system-level feedback, without depending on rigid heuristics. 
We show that our proposed prefetcher, Pythia, consistently outperforms several state-of-the-art prefetchers across a wide range of workloads and system configurations, while incurring only modest area and power overhead.
\rbthr{Our evaluation of Pythia across a large-scale previously-unseen trace corpus (see~\Cref{chap:dpc4}) collected for the 4th Data Prefetching Championship (DPC4) empirically validates that \rbfor{Pythia's  performance benefits generalize well} beyond the workloads used at design time.}

Second, our experience with prefetching reveals that even an advanced prefetcher like Pythia successfully predicts only about half of the load requests that eventually access the main memory. 
For the remaining off-chip loads, we find that a large fraction of their latency is spent accessing the on-chip cache hierarchy to solely determine that they need to go off-chip.
To address this inefficiency, we propose \emph{off-chip prediction}, a fundamentally different form of speculation than prefetching, that predicts which loads will miss the on-chip cache hierarchy and speculatively fetches their data directly from main memory, thereby removing cache access latency from their critical path. 
In~\Cref{chap:hermes}, we design Hermes, which exploits perceptron learning to accurately identify off-chip requests from diverse program features. 
We show that Hermes delivers significant additional performance improvements on top of state-of-the-art prefetchers across a wide range of workloads and system configurations.
\rbthr{Our evaluation of Hermes \rbfor{on} the previously-unseen DPC4 trace corpus empirically validates that \rbfor{Hermes' performance benefits generalize well} beyond the workloads used at design time.}

Third, we further study prefetching and off-chip prediction in unison and find that, while each technique improves performance individually, their simultaneous use often requires careful coordination to avoid negating one another’s benefits, particularly in bandwidth-constrained systems.
To address this, we formulate their coordination as a reinforcement learning problem.
In~\Cref{chap:athena}, we propose Athena, an reinforcement learning-based coordinator that autonomously learns effective policies by observing system-level metrics and the impact of its own decisions on the overall system's performance, without relying on rigid, and often myopic, human-crafted heuristics and thresholds. 
Through extensive evaluation, we show that Athena consistently outperforms multiple state-of-the-art coordination policies across diverse system configurations, prefetcher–OCP combinations, and memory bandwidth settings, while incurring only modest storage overhead.
\rbthr{Moreover, by evaluating Athena \rbfor{on} the previously-unseen DPC4 trace corpus, we empirically validate that \rbfor{Athena's performance benefits generalize well} beyond the workloads used at design time.}

Fourth, we analyze the execution of load instructions in the processor core and observe that a significant fraction of dynamic load instructions repeatedly fetch the same value from the same memory address. 
Executing such redundant loads wastes scarce pipeline resources and limits instruction-level parallelism. 
To address this inefficiency, we propose Constable (see~\Cref{chap:constable}), a purely-microarchitectural technique that dynamically identifies loads with stable address–value behavior and safely eliminates their execution while preserving correctness. 
We show that Constable improves both performance and power efficiency on top of aggressively-optimized out-of-order processors, providing further gains when combined with state-of-the-art load value prediction and simultaneous multithreading.

Together, these four works demonstrate how data-driven and data-aware principles can fundamentally reshape microarchitectural design. 
By leveraging lightweight machine learning methods and exploiting overlooked data characteristics, we show that, techniques that adapt their policies by continuously learning from the vast amount of application data and system-generated metadata, as well as exploit characteristics of application data, indeed deliver performance and energy efficiency improvements that are otherwise untapped by conventional data-agnostic microarchitectural techniques.

\section{Future Research Directions}

While this dissertation primarily focuses on improving performance and \rbtwo{energy} efficiency of traditional general-purpose processors, we believe that the principles and techniques presented here transcend to  various other types of computing systems and paradigms. In this section, we discuss several promising directions for extending the principles more broadly across modern computing systems.

\subsection{Improving and Extending the Proposed Techniques}

Although the four techniques presented in this dissertation significantly advance their respective state-of-the-art, we believe they can be further improved and extended to various aspects of microarchitecture design.
We review several such possibilities in this section.

\subsubsection{Enabling Adaptive Multi-Degree Prefetching}

\rbfor{While Pythia’s reinforcement learning–based formulation establishes a foundation for autonomous, data-driven prefetching, several components of its design remain statically determined at design time through extensive design-space exploration. 
Two key examples are the construction of the set of prefetch actions and the selection of prefetch degree. 
As discussed in~\Cref{sec:py_tuning_action_selection} and shown in~\Cref{tab:pythia_config}, we fix the set of candidate prefetch actions at design time. 
Although this static list yields state-of-the-art performance across a wide range of workloads, it risks missing useful prefetch opportunities when the most beneficial offset is absent from the predefined set. 
Similarly, Pythia currently employs a heuristic-based degree selection policy that adjusts the prefetch degree using static thresholds, which cannot adapt to workload or system dynamics at runtime. 
We believe that future research can explore mechanisms that will allow Pythia to autonomously and adaptively select both its candidate prefetch actions and prefetch degree during execution.}

\subsubsection{Exploiting Off-Chip Prediction for Transparent Offloading to Processing-In-Memory Devices}

Processing-in-memory (PIM) paradigm places computing mechanisms in \rbfiv{memory/storage}~\pum or near~\pnm where the data is stored to reduce/eliminate the data movement bottleneck between the computation units and the memory/storage system. 
Although PIM offers a fundamental solution to address memory bottleneck, its practical adoption remains hindered by the critical challenge of identifying program regions that are suitable for offloading, especially in a manner that is transparent to programs and programmers.
We believe Hermes’s off-chip prediction, in conjunction with cache-reuse prediction~\cite{jimenez2017multiperspective,teran2016perceptron,sdbp,jimenez2010dead,boris_cache4,Mazumdar2021DeadTogether}, presents an opportunity to address this challenge.
More specifically, when a given load instruction is predicted to go off-chip and the data loaded by such instruction will also likely have no reuse, this combination of speculations can serve as a lightweight hint to dynamically offload that load instruction, and any instructions dependent on it, to a PIM-enabled device.
This approach could enable seamless and adaptive exploitation of PIM benefits without requiring explicit code annotations, manual partitioning, or compiler intervention.

\subsubsection{Extending Off-Chip Prediction in Modern Disaggregated Memory Systems}

With the advent of emerging interconnect technologies (e.g., CXL~\cite{cxl}, NVLink~\cite{nvlink}), modern systems employ increasing disaggregated and tiered memory architectures composed of multiple memory levels with various capacity and memory access latency.
In such systems, memory requests typically incur additional latency due to the sequential traversal of each tier in the memory hierarchy~\cite{Ham2024LowOverheadGN,Maruf2022TPPTP,Gu2025PIMIA,Ahn2022EnablingCM,Gouk2023MemoryPW,SongFreqTierLA,Sun2023DemystifyingCM,Maruf2022MULTICLOCKDT,Li2022PondCM,Puri2023DRackSimSC}. 
The fundamental idea behind Hermes, i.e., speculatively bypassing intermediate memory levels by predicting where the requested data will ultimately reside, can be naturally extended to this context. 
Applying such speculation in disaggregated memory systems could yield even greater benefits than in traditional cache hierarchies, by both reducing effective access latency and improving overall \rbtwo{energy} efficiency.

\subsubsection{Extending Stable-Load Elimination Coverage via Compiler-Microarchitecture Co-Optimization}

Our detailed workload analysis and evaluation of Constable discussed in~\Cref{chap:constable} provide two high-level insights. 
First, it shows why global-stable loads remain abundant in real-world workloads even after aggressive compiler optimizations, demonstrating the limitations of current compiler techniques in eliminating such instructions at compile time.
Second, despite its sophisticated hardware mechanisms, Constable is unable to eliminate $43.6\%$ (see~\Cref{sec:cst_evaluation_load_coverage}) of the global-stable loads identified offline, highlighting scenarios where elimination opportunities are left untapped.
These findings open two complementary avenues for future research. 
On the compiler side, more intelligent optimization strategies (such as improved register allocation and profile-guided analysis) can be developed to eliminate global-stable loads during compile time.
On the microarchitectural side, Constable’s mechanisms can be extended to address remaining cases, e.g., by designing hardware support that  effectively handles elimination losses caused by silent stores.

\subsection{New Avenues for ML-Driven and Data-Aware Microarchitectures}

Microarchitectures of modern computing systems (e.g., CPU, GPU, specialized accelerators) employ numerous techniques to enhance their performance and \rbtwo{energy} efficiency.
We believe many of such techniques can be fundamentally reimagined using ML-driven and data-aware approaches presented in this dissertation.
In this section, we highlight several such opportunities, noting that they represent only a subset of the broader possibilities that future research may uncover.

\subsubsection{Coordinating Caching, Prefetching, and Main Memory Scheduling using Multi-Agent Collaborative Reinforcement Learning}

Caching, prefetching, and main memory scheduling policies~\cite{mutlu2008parallelism,kim2010atlas,ebrahimi_fst,ebrahimi2011parallel,subramanian2016bliss,rlmc,mutlu2007stall,Moscibroda2007FairQueuing,ausavarungnirun2012staged,Ghose2013ImprovingMS,usui2016dash,ebrahimi2011parallel,kim2010thread,zhao2014firm,muralidhara2011reducing,subramanian2014blacklisting,usui2015squashsimpleqosawarehighperformance,rixner2000memory,rixner2004memory,Lee2010DRAMAwareLC,subramanian2015application,chang2017understandingphd,zhao2014firm,subramanian2013mise,mutlu2015main,joao2013utility} jointly determine the average memory access latency experienced by a processor.
Although prior works have proposed reinforcement learning (RL)-based agents for each of these components (e.g., caching~\cite{rl_cache,chrome}, prefetching~\cite{pythia,peled_rl,mab}, and memory scheduling~\cite{rlmc,morse}), such agents typically operate in isolation to optimize for their local objective with their local view of the system, without the awareness of the global system behavior. 
Such lack of coordination can lead to suboptimal or even detrimental performance.
We believe that a multi-agent RL-based approach poses a viable solution to co-ordinate caching, prefetching, and main memory scheduling policies to improve overall system's performance, \rbtwo{energy} efficiency, and many other metrics (e.g., fairness), beyond what any single technique can achieve independently.

\subsubsection{Reinforcement Learning for Dynamic Instruction Criticality Prediction}

Instruction criticality prediction has been studied as a means to identify dynamic instructions that disproportionately influence program performance, thereby enabling targeted latency reduction techniques~\cite{fields2001focusing}. 
Traditional approaches rely on handcrafted heuristics (for example, loads feeding to mispredicted branches or other cache-missing loads~\cite{Srinivasan2001LocalityVC}), which are inherently ad-hoc, prone to oversight, and lack adaptability. 
Moreover, criticality prediction lacks a fixed ground truth; whether or not an instruction would be critical evolves based on speculative optimizations applied on that instruction, making critical instruction identification a moving target. 
We believe that re-framing criticality prediction as a reinforcement learning problem could overcome these limitations. 
An RL-based framework could continuously and autonomously learn which instructions are critical by systematically applying optimizations to dynamic instructions and observing their real-time impact on system performance. 
This approach may eliminate dependence on manual heuristics and could surface previously unknown patterns of criticality, providing novel insights to processor architects.

\subsubsection{Exploiting Accurate Speculation Metadata for Lazy Instruction Execution}

Modern processors rely heavily on two fundamental types of speculative techniques that break control dependencies (e.g., branch prediction~\cite{smith1981study,lee1984branch,yeh1991two,tage,perceptron}) and data dependencies (e.g., value prediction~\cite{lipasti1996vp,sazeides_vp,sazeides_vp2,mendelson1997speculative,dfcm,last_n,selective_vp,perais2012revisiting,perais2014eole,perais2014practical,perais2015bebop,fvp,perais2021ssr,eves,rami_ap,rami_composite,kalaitzidis2019value,sakhuja2019combining}, memory renaming~\cite{mrn,mrn2,mrn_classifying,moshovos1997streamlining,moshovos1999speculative}).
Since these mechanisms have been designed to achieve very high accuracy (as their mispredictions cost significant performance penalty), we believe their repetitive correctness can be systematically exploited.
More specifically, when the control- and data-dependence on a given instruction is predicted with high confidence, it can be executed \emph{lazily} (i.e., with a time slack~\cite{fields2001focusing,fields2006using,fields2002slack}) without delaying its dependents.
Such awareness of prediction metadata can potentially free scarce and power-intensive pipeline resources (e.g., ports to execution units and/or L1 data cache) for other non-predictable instructions, thereby improving both performance and energy efficiency.

\subsubsection{Extending Data-Driven and Data-Aware Designs beyond CPUs}

Although we demonstrate Pythia, Hermes, Athena, and Constable in the context of general-purpose processors, the key principles underlying these techniques are broadly applicable to other computing systems, including GPUs, TPUs, and specialized accelerators. 
For example, as GPUs increasingly serve large-scale general-purpose workloads, sophisticated prefetching mechanisms such as Pythia could mitigate the latency of irregular memory accesses. 
Similarly, given the rapid growth of last-level cache capacity in commercial GPUs (e.g., from 4 MB in Nvidia P100 to 40 MB in A100 within four years~\cite{nvidia_ampere_whitepaper}), Hermes can provide significant benefits by eliminating on-chip cache traversal latency for off-chip loads. 
Constable’s principle of exploiting load-value stability also holds promise in GPUs and accelerators, where redundant memory accesses are common, offering an opportunity to reduce pipeline pressure and improve energy efficiency.
We believe that extending data-driven and data-aware designs beyond CPUs presents a promising research direction to address the diverse performance and \rbtwo{energy} efficiency bottlenecks of heterogeneous platforms.

\section{Concluding Remarks}

This dissertation makes two broad contributions. 
First, it develops a detailed understanding of how the data-agnostic nature of conventional microarchitectural techniques limits their ability to mitigate the memory bottleneck. 
Second, it proposes four novel techniques that embody data-driven and data-aware design principles. 
Pythia formulates prefetching as a reinforcement learning problem, enabling adaptive and system-aware prefetch decisions. 
Hermes leverages perceptron learning to accurately predict off-chip memory accesses, thereby removing costly on-chip cache access latency from their critical path.
Athena employs reinforcement learning to synergize prefetching and off-chip prediction, delivering robust performance across diverse workloads and system configurations. 
Constable exploits load-value stability to safely eliminate redundant loads, improving both performance and energy efficiency.
We believe that the insights and techniques presented in this dissertation will not only advance \rbfor{the state-of-the-art} microarchitectural design, but also pave the way for a new generation of data-driven and data-aware systems that can more effectively address the ever-growing memory bottleneck.

\appendix
\cleardoublepage%
\chapter{Comprehensive View of the Author's Contributions}
\label{appendix:otherworks}

During the course of my graduate studies, I led/co-led five successful projects in the broader topic of processor microarchitecture and memory system design. Four of these projects collectively form the core contributions of this dissertation.
First, I started exploring data prefetching and proposed DSPatch~\cite{dspatch}, a bandwidth-aware spatial prefetcher that advanced the state-of-the-art prefetcher performance.
Second, I merged my experience on prefetching with machine learning methods and proposed Pythia~\cite{pythia}, open-sourcing its artifact-evaluated infrastructure~\cite{pythia_github}. 
Pythia has served as both a state-of-the-art baseline for subsequent studies~\cite{mab,duong2024new,pmp,eris2022puppeteer,lin2024pars} and as a reference framework for modeling architectural decision making using machine learning~\cite{singh2022sibyl,rakesh2025harmonia,chrome,chen2025gaze,gogineni2024swiftrl,long2023deep,zhang2022resemble,zhou2023efficient,pandey2023neurocool,yi2025artmem,alkassab2024deepref,huang2023rlop,Xue2025AugurST,Zeng2025ChartenOnlineRL,Xing2025ProactiveDP}.
Third, I continued exploring the intersection of microarchitecture and machine learning and proposed Hermes~\cite{hermes}, along with its artifact-evaluated implementation and infrastructure~\cite{hermes_github,hermes_zenodo}. 
To our knowledge, Hermes is the first work to propose off-chip prediction and to demonstrate the applicability of perceptron-learning for off-chip prediction. 
Hermes influenced multiple follow-up works~\cite{jamet2024tlp,clip,krause2022hbpb,fang2025dhcm,wu2025concurrency,sato2024learning} and was recognized with the \textbf{\textit{Best Paper Award}} at MICRO 2022~\cite{hermes_award}.
Fourth, I shifted the focus from the memory system to the processor core to explore memory-related inefficiencies at the source. This exploration resulted in Constable~\cite{constable}, releasing the open-source Load Inspector~\cite{load_inspector} tool as a part of the project.
Constable was recognized with the \textbf{\textit{Best Paper Award}} at ISCA 2024~\cite{constable_award} for its contribution to the state-of-the-art.
\rbfiv{Most recently, with my mentee Zhenrong Lang, I co-developed Athena~\cite{athena} and released its artifact-evaluated implementation and evaluation infrastructure~\cite{athena_github}.
Athena was recognized with the \textbf{\textit{Distinguished Artifact Award}} at HPCA 2026~\cite{athena_award}.}
Lastly, in our recent IEEE Micro article~\cite{bera2026mldriven}, my co-author Rakesh Nadig and I presented a comprehensive overview of the progress we made over the past half decade in ML-driven microarchitectural design for memory and storage systems.

\vspace{1em}
\noindent Throughout my graduate studies, I also contributed to other research projects. These works can be classified into four groups based on the broader topic of the research. The rest of this section gives a comprehensive summary of such works.

\paraheading{Machine-Learning-Driven Storage System Design.}
In collaboration with Gagandeep Singh and Rakesh Nadig, we explored the application of ML in managing hybrid storage systems (HSS), which employ storage devices with widely varying capacity and access latency. 
Together, we first proposed Sibyl~\cite{singh2022sibyl}, which demonstrates the first RL-based framework to make adaptive data placement decisions by autonomously learning from multiple program/system features (e.g., size and type of the current storage request, remaining capacity of the fast storage device) and system-level feedback (e.g., request latency), without relying on human-designed heuristics or thresholds.
We show that Sibyl not only significantly outperforms the best conventional heuristics-based and learning-based policies, but also provides extensibility to wide range of HSS.
While Sibyl manages the data placement in HSS, it does not address the data migration between different storage devices of an HSS. To this end, we proposed Harmonia~\cite{rakesh2025harmonia}, which demonstrates the first multi-agent RL-based framework for HSS to manage both data placement and data migration in a synergistic way.
Together, Sibyl and Harmonia further reinforce this dissertation's central statement that 
data-driven decision making can provide substantial performance and efficiency gains beyond conventional data-agnostic techniques.

\paraheading{Address Translation Subsystem Design.}
In collaboration with Konstantinos Kanellopoulos, we explored various avenues to accelerate the address translation subsystem of general-purpose processors. 
First, we designed Utopia~\cite{utopia}, an virtual-to-physical address mapping scheme that allows both conventional flexible address mapping and a hash-based restrictive address mapping to co-exist, thereby exploiting the benefits of both worlds in reducing address translation overhead without sacrificing the core virtual memory functionalities (e.g., data sharing).
Second, we \rbfiv{designed} Victima~\cite{victima}, that exploits under-utilized L2 data cache as a victim cache to store evicted second-level translation look-aside buffer (TLB) translation entries. By doing so, Victima converts a significant number of L2 TLB misses from costly page table walks to faster L2 cache hits. 
Third, we \rbfiv{designed} Revelator~\cite{revelator}, which employs a OS-hardware co-designed approach to enable accurate speculative virtual-to-physical address translation. 
Lastly, we open-sourced our flexible development infrastructure that we used for evaluating Utopia, Victima, Revelator, and numerous prior works on OS-architecture co-designed address translation schemes as a standalone simulator named Virtuoso~\cite{virtuoso,virtuoso_github}. 
This \rbfiv{infrastructure} was recognized with the \textbf{\textit{Distinguished Artifact Award}} at MICRO 2023~\cite{victima_award}.

\paraheading{Processing Data Where It Makes Sense.}
I also collaborated in various projects that enable seamless processing of data closer to where they reside.
I significantly contributed to REDUCT~\cite{reduct,proximus}, which enables efficient deep learning inference in multi-core CPUs by enabling near-cache computation.
In collaboration with Alain Denzler, we designed Casper~\cite{casper}, where we extended the key learning from REDUCT to enable efficient stencil kernel computation in CPU via near-cache computation.
In collaboration with Mayank Kabra, we proposed CIPHERMATCH~\cite{ciphermatch}, which accelerates homomorphic encryption (HE)-based string matching via novel data packing scheme and in-storage processing. 
Lastly, in collaboration with Rakesh Nadig, we proposed Conduit~\cite{rakesh2026conduit}, a general-purpose near-data processing framework for SSDs that transparently offloads fine-grained computations to various computation resources available inside an SSD (e.g., SSD controller cores, SSD-internal DRAM chips, SSD's flash chips).

\paraheading{Other Works.}
Together with Ataberk Olgun, we developed Sectored DRAM~\cite{sectored_dram}, which enables fine-grained DRAM data transfer and DRAM row activation in three steps. First, Sectored DRAM predicts which words of a requested cacheblock is likely going to be used during its cache residency. Second, it only activates a smaller set of cells in DRAM that contains the predicted words. Third, it only transfers the predicted words over the DRAM channel. By doing so, Sectored DRAM significantly reduces the overall energy consumption of the system, while simultaneously improving performance.
In collaboration with Jawad Haj-Yahya, we proposed BurstLink~\cite{burstlink}, which improves the efficiency of video streaming in mobile devices by exploiting the full bandwidth of modern display interfaces and remote buffering a decoded frame.

\chapter{Complete List of the Author's Contributions}
\label{appendix:complete_list}
This section lists the author's contributions to the literature in reverse chronological order under three categories: (1)~major contributions that the author led, (2)~co-supervised contributions by the author, and (3)~other contributions.

\section{Major Contributions Led by the Author}
\label{appendix:sec:first_author_contributions}

\begin{enumerate}
    \item Rahul Bera, Rakesh Nadig, Onur Mutlu, \textit{``Machine Learning-Driven Intelligent Memory System Design: From On-Chip Caches to Storage''}, in IEEE Micro, 2026
    
    \item Rahul Bera, Zhenrong Lang, Caroline Hengartner, Konstantinos Kanellopoulos, Rakesh Kumar, Mohammad Sadrosadati, Onur Mutlu, \textit{``Athena: Synergizing Data Prefetching and Off-Chip Prediction via Online Reinforcement Learning''}, in HPCA, 2026. Artifact available: \url{https://github.com/CMU-SAFARI/Athena}. \textbf{\textit{Distinguished Artifact Award}}.
    
    \item Rahul Bera, Adithya Ranganathan, Joydeep Rakshit, Sujit Mahto, Anant V. Nori, Jayesh Gaur, Ataberk Olgun, Konstantinos Kanellopoulos, Mohammad Sadrosadati, Sreenivas Subramoney, Onur Mutlu, \textit{``Constable: Improving Performance and Power Efficiency by Safely Eliminating Load Instruction Execution''}, in ISCA, 2024. Toolkit available: \url{https://github.com/CMU-SAFARI/Load-Inspector}. \textbf{\textit{Best Paper Award}}. Subject matter of US patent US20260003626A1, filed: 2024-06-27.

    \item Rahul Bera, Konstantinos Kanellopoulos, Shankar Balachandran, David Novo, Ataberk Olgun, Mohammad Sadrosadati, Onur Mutlu, \textit{``Hermes: Accelerating Long-Latency Load Requests via Perceptron-Based Off-Chip Load Prediction''}, in MICRO, 2022. Artifact available: \url{https://github.com/CMU-SAFARI/Hermes}. \textbf{\textit{Best Paper Award}}.

    \item Rahul Bera, Konstantinos Kanellopoulos, Anant V. Nori, Taha Shahroodi, Sreenivas Subramoney, Onur Mutlu, \textit{``Pythia: A Customizable Hardware Prefetching Framework using Online Reinforcement Learning''}, in MICRO, 2021. Artifact available: \url{https://github.com/CMU-SAFARI/Pythia}

    \item Rahul Bera, Anant V. Nori, Onur Mutlu, Sreenivas Subramoney, \textit{``DSPatch: Dual Pattern Spatial Prefetcher''}, in MICRO, 2019. Subject matter of US patent US11874773B2, filed: 2019-12-28; published: 2021-03-25.
\end{enumerate}

\section{Co-Supervised Contributions}
\label{appendix:sec:co_supervised_contributions}

\begin{enumerate}
    \item Konstantinos Kanellopoulos, \underline{Rahul Bera}, Kosta Stojiljkovic, Nisa Bostanci, Can Firtina, Rachata Ausavarungnirun, Rakesh Kumar, Nastaran Hajinazar, Mohammad Sadrosadati, Nandita Vijaykumar, Onur Mutlu, \textit{``Utopia: Efficient Address Translation using Hybrid Virtual-to-Physical Address Mapping''}, in MICRO, 2023. Artifact available: \url{https://github.com/CMU-SAFARI/Utopia}.
    
    \item Anant V Nori, \underline{Rahul Bera}, Shankar Balachandran, Joydeep Rakshit, Om J Omer, Avishaii Abuhatzera, Kuttanna Belliappa, Sreenivas Subramoney, \textit{``REDUCT: Keep it Close, Keep it Cool!: Effcient Scaling of DNN Inference on Multi-Core CPUs with Near-Cache Compute''}, in ISCA, 2021. Subject matter of US patent US12405890B2, filed: 2021-12-23; published: 2023-06-29.
\end{enumerate}

\section{Other Contributions}
\label{appendix:sec:other_contributions}

\begin{enumerate}
    \item Rakesh Nadig, Vamanan Arulchelvan, \underline{Rahul Bera}, Taha Shahroodi, Gagandeep Singh, Andreas Kosmas Kakolyris, Ismail Emir Yuksel, Mohammad Sadrosadati, Jisung Park, Onur Mutlu, \textit{``Harmonia: Enhancing Data Placement and Migration in Hybrid Storage Systems via Multi-Agent Reinforcement Learning''}, in ICS, 2026
    
    \item Rakesh Nadig, Vamanan Arulchelvan, Mayank Kabra, Harshita Gupta, \underline{Rahul Bera}, Nika Mansouri Ghiasi, Nanditha Rao, Qingcai Jiang, Andreas Kosmas Kakolyris, Yu Liang, Mohammad Sadrosadati, Onur Mutlu, \textit{``Conduit: Programmer-Transparent Near-Data Processing Using Multiple Compute-Capable Resources in SSDs''}, in HPCA, 2026
    
    \item Konstantinos Kanellopoulos, Konstantinos Sgouras, Harsh Songara, Andreas Kosmas Kakolyris, Vlad-Petru Nitu, \underline{Rahul Bera}, Rakesh Kumar, Onur Mutlu, \textit{``Revelator: Rapid Data Fetching via OS-Driven Hash-based Speculative Address Translation''}, arXiv (under peer-review), 2025
    
    \item Konstantinos Kanellopoulos, Konstantinos Sgouras, F. Nisa Bostanci, Andreas Kosmas Kakolyris, Berkin Kerim Konar, \underline{Rahul Bera}, Mohammad Sadrosadati, Rakesh Kumar, Nandita Vijaykumar, Onur Mutlu, \textit{``Virtuoso: Enabling Fast and Accurate Virtual Memory Research via an Imitation-based Operating System Simulation Methodology''}, in ASPLOS, 2025. Artifact available: \url{https://github.com/CMU-SAFARI/Virtuoso}.
    
    \item Mayank Kabra, Rakesh Nadig, Harshita Gupta, \underline{Rahul Bera}, Manos Frouzakis, Vamanan Arulchelvan, Yu Liang, Haiyu Mao, Mohammad Sadrosadati, Onur Mutlu, \textit{``CIPHERMATCH: Accelerating Homomorphic Encryption-Based String Matching via Memory-Efficient Data Packing and In-Flash Processing''}, in ASPLOS, 2025
    
    \item Ataberk Olgun, F. Nisa Bostanci, Geraldo F. Oliveira, Yahya Can Tugrul, \underline{Rahul Bera}, A. Giray Yaglikci, Hasan Hassan, Oguz Ergin, Onur Mutlu, \textit{``Sectored DRAM: An Energy-Efficient High-Throughput and Practical Fine-Grained DRAM Architecture''}, in TACO, 2024. Artifact available: \url{https://github.com/CMU-SAFARI/Sectored-DRAM}.
    
    \item Konstantinos Kanellopoulos, Hong Chul Nam, Nisa Bostanci, \underline{Rahul Bera}, Mohammad Sadrosadati, Rakesh Kumar, Davide Basilio Bartolini, Onur Mutlu, \textit{``Victima: Drastically Increasing Address Translation Reach by Leveraging Underutilized Cache Resources''}, in MICRO, 2023. Artifact available: \url{https://github.com/CMU-SAFARI/Victima}. \textbf{\textit{Distinguished Artifact Award}}.
    
    \item Alain Denzler, Geraldo F. Oliveira, Nastaran Hajinazar, \underline{Rahul Bera}, Gagandeep Singh, Juan Gómez-Luna, Onur Mutlu, \textit{``Casper: Accelerating Stencil Computation using Near-cache Processing''}, in IEEE Access, 2023
    
    \item Gagandeep Singh, Rakesh Nadig, Jisung Park, \underline{Rahul Bera}, Nastaran Hajinazar, David Novo, Juan Gómez-Luna, Onur Mutlu, \textit{``Sibyl: Adaptive and Extensible Data Placement in Hybrid Storage Systems Using Online Reinforcement Learning''}, in ISCA, 2022
    
    \item Jawad Haj-Yahya, Jisung Park, \underline{Rahul Bera}, Juan Gómez Luna, Efraim Rotem, Taha Shahroodi, Jeremie Kim, Onur Mutlu, \textit{``BurstLink: Techniques for Energy-Efficient Conventional and Virtual Reality Video Display''}, in MICRO, 2021.
    
\end{enumerate}

\chapter{Curriculum Vitae of the Author}
\label{appendix:cv}

\section*{Education}
\begin{description}[
  style=multiline,
  leftmargin=2.6cm,
  align=parleft
]
\item[Sept 2019 -\\ Dec 2025] \textbf{ETH Zürich}, Doctor of Science\\
Advisor: Prof. Dr. Onur Mutlu\\
Thesis: \emph{``Mitigiating the Memory Bottleneck with Machine-Learning-Driven and Data-Aware Microarchitectural Techniques''}
\item[Jul 2014 -\\ Jan 2017] \textbf{Indian Institute of Technology, Kanpur}, Master of Technology\\
Advisor: Prof. Dr. Mainak Chaudhuri\\
Thesis: \emph{``Adaptive Prefetch Filter to Mitigate Prefetcher Induced Pollution''}.
\item[Jul 2010 -\\ Jun 2014] \textbf{Jadavpur University}, Bachelor of Engineering\\
Thesis: \emph{``Design \& Analysis of Logarithmic Multipliers and Dividers using VHDL''}.
\end{description}

\section*{Professional Experience}
\begin{description}[
  style=multiline,
  leftmargin=2.6cm,
  align=parleft
]
\item[Mar 2023 -\\ Aug 2023] \textbf{Processor Architecture Research Lab, Intel Labs}\\
\textit{Graduate Research Intern}\par
Mentors: Anant V. Nori and Sreenivas Subramoney.\\
Worked on dynamic instruction elimination for high-performance and power-efficient processors.
\item[Feb 2017 -\\ Aug 2019] \textbf{Processor Architecture Research Lab, Intel Labs}\\
\textit{Architecture Researcher}\par
Mentors: Anant V. Nori and Sreenivas Subramoney.\\
Worked on high-performance prefetcher design and near-cache computation for DNN inference.
\item[May 2015 -\\ Dec 2015] \textbf{Advanced Micro Devices}\\
\textit{Co-op Engineer}\par
Mentor: Dr. Kanishka Lahiri.\\
Developed simulator to model AMD Data Fabric for performance projections of extremely-threaded SoCs.
\end{description}

\section*{Honors and Recognitions}
\begin{description}[
  style=multiline,
  leftmargin=2.6cm,
  align=parleft
]
\item[2026] \textbf{Distinguished artifact award at HPCA 2026} for ``Athena: Synergizing Data Prefetching and Off-Chip Load Prediction via Online Reinforcement Learning''.
\item[2025] \textbf{MLCommons ML and Systems Rising Stars}, awarded in recognition of research at the intersection of machine learning and systems.
\item[2024] \textbf{Best paper award at ISCA 2024} for ``Constable: Improving Performance and Power Efficiency by Safely Eliminating Load Instruction Execution''.
\item[2023] \textbf{Distinguished artifact award at MICRO 2023} for ``Victima: Drastically Increasing Address Translation Reach by Leveraging Underutilized Cache Resources''.
\item[2022] \textbf{Best paper award at MICRO 2022} for ``Hermes: Accelerating Long-Latency Load Requests via Perceptron-Based Off-Chip Load Prediction''.
\end{description}

\section*{Key Publications}
\textit{Please visit} \url{https://dblp.org/pid/250/2580.html} \textit{for a complete list of publications.}

\begin{description}[
  style=multiline,
  leftmargin=2.6cm,
  align=parleft
]

\item[HPCA\\2026] \textbf{Athena: Synergizing Data Prefetching and Off-Chip Load Prediction via Online Reinforcement Learning}\\
\textit{\underline{Rahul Bera}, Zhenrong Lang, Caroline Hengartner, Konstantinos Kanellopoulos, Rakesh Kumar, Mohammad Sadrosadati, Onur Mutlu}\\
\textcolor{red}{\textbf{Distinguished artifact award at HPCA 2026.}}

\item[HPCA\\2026] \textbf{Conduit: Programmer-Transparent Near-Data Processing Using Multiple Compute-Capable Resources in SSDs}\\
\textit{Rakesh Nadig, Vamanan Arulchelvan, Mayank Kabra, Harshita Gupta, \underline{Rahul Bera}, Nika Mansouri Ghiasi, Nanditha Rao, Qingcai Jiang, Andreas Kosmas Kakolyris, Yu Liang, Mohammad Sadrosadati, Onur Mutlu}

\item[ASPLOS\\2025] \textbf{Virtuoso: Enabling Fast and Accurate Virtual Memory Research via an Imitation-based Operating System Simulation Methodology}\\
\textit{Konstantinos Kanellopoulos, Konstantinos Sgouras, F. Nisa Bostanci, Andreas Kosmas Kakolyris, Berkin Kerim Konar, \underline{Rahul Bera}, Mohammad Sadrosadati, Rakesh Kumar, Nandita Vijaykumar, and Onur Mutlu}

\item[ASPLOS\\2025] \textbf{CIPHERMATCH: Accelerating Homomorphic Encryption-Based String Matching via Memory-Efficient Data Packing and In-Flash Processing}\\
\textit{Mayank Kabra, Rakesh Nadig, Harshita Gupta, \underline{Rahul Bera}, Manos Frouzakis, Vamanan Arulchelvan, Yu Liang, Haiyu Mao, Mohammad Sadrosadati, and Onur Mutlu}

\item[ISCA\\2024] \textbf{Constable: Improving Performance and Power Efficiency by Safely Eliminating Load Instruction Execution}\\
\textit{\underline{Rahul Bera}, Adithya Ranganathan, Joydeep Rakshit, Sujit Mahto, Anant V. Nori, Jayesh Gaur, Ataberk Olgun, Konstantinos Kanellopoulos, Mohammad Sadrosadati, Sreenivas Subramoney, and Onur Mutlu}\\
\textcolor{red}{\textbf{Best paper award at ISCA 2024.}}

\item[MICRO\\2023] \textbf{Victima: Drastically Increasing Address Translation Reach by Leveraging Underutilized Cache Resources}\\
\textit{Konstantinos Kanellopoulos, Hong Chul Nam, F. Nisa Bostanci, \underline{Rahul Bera}, Mohammad Sadrosadati, Rakesh Kumar, Davide Basilio Bartolini, and Onur Mutlu}\\
\textcolor{red}{\textbf{Distinguished artifact award at MICRO 2023.}}

\item[MICRO\\2023] \textbf{Utopia: Efficient Address Translation using Hybrid Virtual-to-Physical Address Mapping}\\
\textit{Konstantinos Kanellopoulos, \underline{Rahul Bera}, Kosta Stojiljkovic, Can Firtina, Rachata Ausavarungnirun, Nastaran Hajinazar, Jisung Park, Nandita Vijaykumar, and Onur Mutlu}

\item[MICRO\\2022] \textbf{Hermes: Accelerating Long-Latency Load Requests via Perceptron-Based Off-Chip Load Prediction}\\
\textit{\underline{Rahul Bera}, Konstantinos Kanellopoulos, Shankar Balachandran, David Novo, Ataberk Olgun, Mohammad Sadrosadati, and Onur Mutlu}\\
\textcolor{red}{\textbf{Best paper award at MICRO 2022.}}

\item[ISCA\\2022] \textbf{Sibyl: Adaptive and Extensible Data Placement in Hybrid Storage Systems Using Online Reinforcement Learning}\\
\textit{Gagandeep Singh, Rakesh Nadig, Jisung Park, \underline{Rahul Bera}, Nastaran Hajinazar, David Novo, Juan Gómez Luna, Sander Stuijk, Henk Corporaal, and Onur Mutlu}

\item[MICRO\\2021] \textbf{Pythia: A Customizable Hardware Prefetching Framework Using Online Reinforcement Learning}\\
\textit{\underline{Rahul Bera}, Konstantinos Kanellopoulos, Anant V. Nori, Taha Shahroodi, Sreenivas Subramoney, and Onur Mutlu}

\item[MICRO\\2021] \textbf{BurstLink: Techniques for Energy-Efficient Conventional and Virtual Reality Video Display}\\
\textit{Jawad Haj-Yahya, Jisung Park, \underline{Rahul Bera}, Juan Gómez Luna, Efraim Rotem, Taha Shahroodi, Jeremie Kim, and Onur Mutlu}

\item[ISCA\\2021] \textbf{REDUCT: Efficient Scaling of DNN Inference on Multi-core CPUs with Near-Cache Compute}\\
\textit{Anant V. Nori, \underline{Rahul Bera}, Shankar Balachandran, Joydeep Rakshit, Om J. Omer, Avishaii Abuhatzera,
Kuttanna Belliappa, and Sreenivas Subramoney}

\item[MICRO\\2019] \textbf{DSPatch: Dual Spatial Access Prefetcher}\\
\textit{\underline{Rahul Bera}, Anant V. Nori, Onur Mutlu, and Sreenivas Subramoney}

\end{description}

\section*{Key Patents}
\begin{description}[
  style=multiline,
  leftmargin=2.6cm,
  align=parleft
]
\item[2024] \textbf{Methods \& Apparatus for Efficiently Processing Load Instruction Sequences}\\
\textit{Adithya Ranganathan, \underline{Rahul Bera}, Joydeep Rakshit, Sujit Mahto, Anant V. Nori, Jayesh Gaur, and Sreenivas Subramoney}\\
US Patent US20260003626A1.
\item[2023] \textbf{Method \& Apparatus for Leveraging Simultaneous Multithreading for Bulk Compute Operations}\\
\textit{Anant V. Nori, \underline{Rahul Bera}, Shankar Balachandran, Joydeep Rakshit, Om Ji Omer, Sreenivas Subramoney, Avishaii Abuhatzera, and Belliappa Kuttanna}\\
US Patent US20230205692A1.
\item[2021] \textbf{Apparatuses, Methods, and Systems for Dual Spatial Pattern Prefetcher}\\
\textit{\underline{Rahul Bera}, Anant V. Nori, and Sreenivas Subramoney}\\
US Patent US20210089456A1.
\item[2020] \textbf{Adaptive Spatial Access Prefetcher Apparatus and Method}\\
\textit{\underline{Rahul Bera}, Anant V. Nori, Sreenivas Subramoney, and Hong Wang}\\
US Patent US10713053B2.
\end{description}

\section*{Invited Talks}
\begin{description}[
  style=multiline,
  leftmargin=2.6cm,
  align=parleft
]
\item[May 2025] \textbf{Mitigating the Memory Bottleneck with Data-Driven and Data-Aware Microarchitectures}\\
\textit{Stanford University, Meta, and Tenstorrent}.
\item[Nov 2024] \textbf{Data-Driven and Data-Aware Microarchitectures for Mitigating Memory Bottleneck in High-Performance Computing System}\\
\textit{Apple Lonestar Design Center, Austin, US}.
\item[Nov 2024] \textbf{Data-Driven and Data-Aware Microarchitectures for Mitigating Memory Bottleneck in High-Performance Computing System}\\
\textit{AMD Research and Advanced Development, Austin, US}.
\item[Sept 2024] \textbf{A Case for Data-Aware Microarchitecture for Alleviating Memory Bottleneck}\\
\textit{Huawei Research Center, Zürich, Switzerland}.
\item[Nov 2022] \textbf{Taming the Memory Wall via Machine Learning Assisted Microarchitecture Design}\\
\textit{Huawei Research Center, Zürich, Switzerland}.
\item[Nov 2022] \textbf{Hermes: Accelerating Long-Latency Load Requests via Perceptron-Based Off-Chip Load Prediction}\\
\textit{Processor Architecture Research Lab, Intel Labs, India}.
\end{description}

\bibliographystyle{IEEEtranS} %
\balance
\begin{singlespace}
\bibliography{
    backmatter/01_main, 
    backmatter/02_more, 
    backmatter/10_jimenez, 
    backmatter/11_boris,
    backmatter/12_aamer, 
    backmatter/13_chris, 
    backmatter/14_pradip
}
\end{singlespace}

\bookmarksetup{startatroot}
\end{document}